\documentclass[12pt,a4paper,fleqn]{book}
\usepackage[english]{babel,varioref}
\usepackage[english]{babel}
\usepackage{epsfig,latexsym,fancyheadings,amssymb}

\catcode`@=11
\def\citer{\@ifnextchar
[{\@tempswatrue\@citexr}{\@tempswafalse\@citexr[]}}
\def\@citexr[#1]#2{\if@filesw\immediate\write\@auxout{\string\citation{#2}}\fi
  \def\@citea{}\@cite{\@for\@citeb:=#2\do
    {\@citea\def\@citea{--\penalty\@m}\@ifundefined
       {b@\@citeb}{{\bf ?}\@warning
       {Citation `\@citeb' on page \thepage \space undefined}}%
\hbox{\csname b@\@citeb\endcsname}}}{#1}}
\catcode`@=12

\newcommand{\slt}{\!\!\!/}
\newcommand{\eins}{{\rm 1\hspace{-0.75ex}1}}
\newcommand{\ket}[1]{\mbox{$|#1\rangle$}}
\newcommand{\bra}[1]{\mbox{$\langle #1|$}}

\renewcommand{\chaptermark}[1]{\markboth{\sc \chaptername~\thechapter. #1}{}}
\renewcommand{\sectionmark}[1]{\markright{\sc \thesection\ #1}}
\begin{document}
\newcommand{\strute}{\rule{0mm}{0mm}}

\def\bdm{\begin{displaymath}}
\def\edm{\end{displaymath}}
\def\beq{\begin{equation}}
\def\eeq{\end{equation}}
\def\beqa{\begin{eqnarray}}
\def\eeqa{\end{eqnarray}}
%
\newcommand{\dv}[1]{\varepsilon_{\rho}(#1,m_{d})}
\newcommand{\uu}[1]{u(\vec p_{#1},s_{#1})}
\newcommand{\uub}[1]{\bar u(\vec p_{#1},s_{#1})}
\newcommand{\uuc}[1]{u^{c}(\vec p_{#1},s_{#1})}
\newcommand{\uua}[1]{u^{\dagger}(\vec p_{#1},s_{#1})}
\newcommand{\vv}[1]{v(\vec p_{#1},s_{#1})}
\newcommand{\vvb}[1]{\bar v(\vec p_{#1},s_{#1})}

\newcommand{\SpinNorm}[1]{\sqrt{\frac{E_{#1}+M}{2M}} }
\newcommand{\Spinor}[1]{ \left(\begin{array}{c}\ME \\ \frac{\vec\sigma 
\cdot \vec p_{#1}}{E_{#1} + M} \end{array}\right)\chi_{#1}}
\newcommand{\SpinorC}[1]{ \left(\begin{array}{c} \frac{\vec\sigma 
\cdot \vec p_{#1}}{E_{#1} + M} \\ \ME \end{array}\right) \chi_{#1}^{c}}
\newcommand{\SpinorA}[1]{\chi_{#1}^{\dagger} \left(\ME ,-\frac{\vec\sigma 
\cdot \vec p_{#1}}{E_{#1} + M} \right)}
%
%
\newcommand{\ME}{1\hspace*{-4pt}1}
\newcommand{\PolVektor}{\varepsilon_{\rho}(d,m_{d})}
\renewcommand{\slash}{/\hspace*{-.5em}}
\newcommand{\vecp}[1]{\mbox{$\vec  #1^{\; '}$}}
\newcommand{\brao}{\mbox{$\langle$}}
\newcommand{\keto}{\mbox{$\rangle$}}
\newcommand{\braket}[2]{\mbox{$\langle #1 | #2 \rangle$}}
%
\newcommand{\vecq}[1]{\vec#1^{\,\,2}}
\newcommand{\betrag}[1]{\,|\vec#1|\,}
\newcommand{\grenze}[1]{\;\rule[-8pt]{.5pt}{24pt}%
\raisebox{-8pt}{$\,\scriptstyle #1$}}
%
\newcommand{\dreivec}[3]{\left( \begin{array}{c} #1\\#2\\#3\end{array}
\right)}
\newcommand{\vsprodukt}[2]{\hspace*{2pt} #1 \hspace*{-3pt} 
                          \cdot\hspace*{-3pt} #2\hspace*{2pt}}
\newcommand{\sprodukt}[2]{\hspace*{2pt}\vec #1 \hspace*{-3pt} 
                          \cdot\hspace*{-3pt}\vec #2\hspace*{2pt}}
\newcommand{\kprodukt}[2]{\hspace*{2pt}(\vec #1 \hspace*{-3pt} 
                          \times \hspace*{-3pt}\vec #2 \,)\hspace*{2pt}}
\newcommand{\spatr}[3]{\hspace*{2pt}(\vec #1 \hspace*{-3pt} 
                      \times \hspace*{-3pt}\vec #2)
                      \hspace*{-3pt}\cdot\hspace*{-3pt}\vec #3\hspace*{2pt}}
\newcommand{\spatl}[3]{\hspace*{2pt}\vec #1\hspace*{-3pt}\cdot
                       \hspace*{-3pt}
                         (\vec #2 \hspace*{-3pt} \times \hspace*{-3pt}#3)
                       \hspace*{2pt}}
\newcommand{\DiracC}[2]{\left(F^{#1}_{1}(q^{2})\gamma_{#2} 
            + \frac{i F_{2}^{#1}(q^{2})}{2M}\sigma_{#2\nu}q^{\nu} \right)}
\newcommand{\PropagatorP}{\frac{\gamma_{\kappa}z^{\kappa}_{p} + M}
                                { t - M^{2}}}
\newcommand{\PropagatorN}{\frac{\gamma_{\kappa}z^{\kappa}_{n} + M}
                                { u - M^{2}}}
\newcommand{\VertexP}[1]{\left(\gamma^{#1} A(t) - \frac{1}{2}(z^{#1}_{p}
                         - p^{#1}_{n})B(t)\right)} 
\newcommand{\VertexN}[1]{\left(\gamma^{#1} A(u) - \frac{1}{2}(z^{#1}_{n}
                         - p^{#1}_{p})B(u)\right)} 
%
\def\gammao{\left( \begin{array}{cccc}
1 & 0 &  0 & 0  \\
0 & 1 &  0 & 0  \\
0 & 0 & -1 & 0  \\
0 & 0 & 0 & -1  \end{array}\right) }
\def\gammai{\left( \begin{array}{cccc}
0 & 0 &  0 & 1  \\
0 & 0 &  1 & 0  \\
0 & -1 & 0 & 0  \\
-1 & 0  & 0 & 0  \end{array}\right) }
\def\gammaii{\left( \begin{array}{cccc}
0 & 0 &  0 & -i  \\
0 & 0 &  i & 0  \\
0 & i &  0 & 0  \\
-i & 0 & 0 & 0  \end{array}\right) }
\def\gammaiii{\left( \begin{array}{cccc}
0 & 0 &  1 & 0  \\
0 & 0 &  0 & -1  \\
-1 & 0 & 0 & 0  \\
0 & 1 & 0 & 0  \end{array}\right) }
\def\gammav{\left( \begin{array}{cccc}
0 & 0 &  0 & 1  \\
0 & 0 &  1 & 0  \\
0 & 1 & 0 & 0  \\
1 & 0 & 0 & 0  \end{array}\right) }
\def\gammac{\left( \begin{array}{cccc}
0 & 0 &  0 & -1  \\
0 & 0 &  1 & 0  \\
0 & -1 & 0 & 0  \\
1 & 0 & 0 & 0  \end{array}\right) }
\def\gammavec{\left( \begin{array}{cc}
0 & \vec\sigma  \\
-\vec\sigma & 0  \end{array}\right) }
\def\sigmaD{\left( \begin{array}{cc}
\vec\sigma & 0 \\
0 & \vec\sigma   \end{array}\right) }
%
%
\newcommand{\hsection}[2]{\setcounter{equation}{0}%
         \section{{#1}}\markboth{ {#1} \hspace*{-\textwidth}
         \underline{\hspace{\textwidth}}}%
{\underline{\hspace*{\textwidth}}\hspace*{-\textwidth} {#1} }%
         \begin{flushright}{\it\Large #2\\[20pt]}\end{flushright}}%
\newcommand{\hsubsection}[1]{\subsection{{#1}}%
           \markright{ {\underline{\hspace*{\textwidth}}%
\hspace*{-\textwidth} {#1} }}}
%
\newlength{\breite}     \breite 25mm
\newlength{\bildbreite} \bildbreite 50mm
\newlength{\textbildbreite} \textbildbreite 80mm

\newcommand{\calX}[2]{
\setlength{\breite}{\textwidth}%
\addtolength{\breite}{-25mm}%
\begin{minipage}[t]{25mm}%
\vspace*{10mm}%
    \psfig{figure=disk_kth:[gbeck.graph.lib]#1.psp,height=5mm,%
bbllx=0bp,bblly=0bp,bburx=100bp,bbury=100bp}  %
\rule{25mm}{0mm}%
\end{minipage}%
\begin{minipage}[t]{\breite} #2 \end{minipage}
}

\newcommand{\bildlinks}[4]{\strute\newline\begin{minipage}[b]{\bildbreite}%
    \psfig{figure={disk_kth:[gbeck.prom.pic]#1},width=\bildbreite} %
    \stepcounter{figure}%
    {\footnotesize%
    \strute\newline%
    \begin{abb}{\bf:} #2 \label{abb:#3}\hfill\end{abb} }%
    \strute\newline%
    \end{minipage}%
    \ifx\Empty\labelset\marginpar{abb:\\#3}\fi%
    \hfill\begin{minipage}[b]{\textbildbreite} #4 \end{minipage}%
\newline}
\newcommand{\bildrechts}[4]{\strute\vspace*{2mm}\strute\newline%
    \begin{minipage}[b]{\textbildbreite}%
    #4 \end{minipage}\hfill%
    \ifx\Empty\labelset\marginpar{abb:\\ #3}\fi%
    \begin{minipage}[b]{\bildbreite}
    \psfig{figure={disk_kth:[gbeck.prom.pic]#1},width=\bildbreite} 
    \stepcounter{figure}%
    {\footnotesize%
    \begin{abb}{\bf:} #2 \label{abb:#3}\hfill\end{abb}}%
    \vspace*{2pt}
    \end{minipage}\newline}%
\newcommand{\einfig}[4]{\strute\newline\stepcounter{figure}%
            \centerline{\begin{minipage}[t]{6cm}\vspace*{10pt}%
            \psfig{figure=#1,width=6cm,angle=#2}%
            \end{minipage}}\\*[10pt]
            \ifx\Empty\labelset\marginpar{abb:\\ #3}\fi%
            \begin{minipage}[t]{\textwidth}%
            {\center{%
            \begin{abb}{\bf:} #4 \label{abb:#3}\end{abb} }}%
            \end{minipage}\strute\vspace*{10pt}\strute\newline%
            }
\newcommand{\zweifig}[4]{\stepcounter{figure}\strute\newline
            \begin{minipage}[t]{\textwidth}\vspace*{10pt}%
            \psfig{figure=#1,width=\textwidth,angle=#2}%
            \end{minipage}\\*[10pt]%
            \ifx\Empty\labelset\marginpar{abb: \\ #3}\fi%
            \begin{minipage}[t]{\textwidth}%
            \begin{abb}{\bf:} #4 \label{abb:#3}\end{abb} %
            \end{minipage}\strute\vspace*{10pt}\strute\newline%
            }

\pagenumbering{roman}
\begin{titlepage}
\begin{center}
 \sc{
  {\Large
    Rescattering Effects in Incoherent \\
    Photoproduction of $\pi$-Mesons off Deuterium\\
    in the $\Delta(1232)$ Resonance Region\\[3pt]

  }}
  \vspace{4.2cm}
    {\large DISSERTATION}\\
    \vspace{0.4cm}
    zur Erlangung des Grades  \\
     \vspace{0.2cm}
     ,,Doktor der Naturwissenschaften'' \\
    \vspace{0.2cm}
    am Fachbereich Physik \\
   \vspace{0.2cm}
    der Johannes Gutenberg--Universit\"at Mainz \\
  \vspace{4cm}
    vorgelegt von\\
    \vspace{0.3cm}
    {\Large Eed M. Darwish}\\
    \vspace{0.3cm}
    geboren in Sohag, \"Agypten\\ \vspace{1.5cm}
    \vfill
    Institut f\"ur Kernphysik \\
    \vspace{0.2cm}
    Johannes Gutenberg-Universit\"at Mainz \\
    \vspace{0.2cm}
    Mai 2002    
  \end{center}
\end{titlepage}
\newpage

\vspace*{10cm}

\begin{tabular}{ll}
Dekan:                & Professor Dr.\ Hartmut Backe      \\[\bigskipamount]
1.\ Berichterstatter: & Professor Dr.\ Hartmuth Arenh\"ovel\\ 
2.\ Berichterstatter:  & Professor Dr.\ Martin Reuter   \\[\bigskipamount]
Datum der Einreichung:  &  27.\ Mai 2002        \\ 
Datum der m\"undlichen Pr\"ufung:  &  10.\ Juli 2002   \\[\bigskipamount]
\end{tabular}

\vspace*{1cm}

\hspace*{-0.5cm}{\underline {\bf Hinweis}}\\
Gegen\"uber der urspr\"unglichen, am 27.\ Mai 2002 im Dekanat des Fachbereichs
Physik der Johannes Gutenberg-Universit\"at Mainz eingereichten Arbeit sind in
der vorliegenden Version einige zwischenzeitlich gefundene Druckfehler
entfernt worden.\\

\hspace*{-0.5cm}Mainz, Juli 2002\\

\hspace*{11cm}{\rm Eed M.\ Darwish}

\pagestyle{empty}
\newpage
\pagestyle{empty}
{\phantom H}
\vspace*{5cm}
\begin{center}
{\large{\sc{In the name of GOD}}}\\
~\\
{\it{most Graciousa and most Merciful}}
\vspace{2cm}

{\large{\sc{Dedicated to the memory of my Parents}}}\\
~\\
{\large{\sc{\&}}}\\
~\\
{\large{\sc{to my Wife and my Children}}}\\
~\\
{\it{for their Patience, Forbearance and Encouragement}}
\end{center}
\cleardoublepage

\begin{center}
{\LARGE{\sc Abstract}}
\vspace{1cm}
\end{center}
%
Incoherent photoproduction of pions on the deuteron in the $\Delta$(1232)
resonance region is investigated in order to study the effect of
nucleon-nucleon ($NN$) and pion-nucleon ($\pi N$) rescattering in the final
state. The elementary $\gamma N\rightarrow \pi N$ production amplitude is
taken in the effective Lagrangian approach and contains besides the standard
pseudovector Born terms the resonance contribution from the $\Delta(1232)$
excitation. It yields for the elementary reaction a good agreement with the
experimental data from MAMI, TRIUMF and TAPS.

\vspace*{-0.2cm}~\\
Pion photoproduction on the deuteron is dominated by the impulse approximation
where the pion is produced on one of the nucleons neglecting all final state
interactions. The comparison of the impulse approximation with the
available experimental data shows a significant overestimation of the data.

\vspace*{-0.2cm}~\\
Therefore, the major point of concern of this thesis was the inclusion of 
rescattering effects in the final $\pi NN$ system which we have limited to the
leading order contributions of two-particle interactions in the $NN$- and $\pi
N$-subsystems. As models for the relevant two-body interactions we have used
separable approximations which fit the phase shift data for $NN$ and $\pi N$
scattering. We found that the influence of $NN$- and $\pi N$-rescattering
effects on total and differential cross sections is significant. Inclusion of
such effects leads to a much improved agreement with the existing experimental
data.

\vspace*{-0.2cm}~\\
Since several experiments to measure the spin asymmetry of the total
photoabsorption cross section, which determines the Gerasimov-Drell-Hearn
(GDH) sum rule, are now being performed or planned at different laboratories
around the world (MAMI, ELSA, LEGS, GRAAL and TJNAF), the contribution of
incoherent pion photoproduction to the spin asymmetry for 
the deuteron is evaluated with inclusion of $NN$ and $\pi N$ rescattering
contributions. The effect of final state rescattering on the spin asymmetry
for the deuteron is found to be quite important and should be included in
forthcoming theoretical studies. 
 
\vspace{24pt}
\noindent
Mainz, May 27, 2002\\
Supervisor: Prof.\ Dr.\ H.\ Arenh\"ovel 

\cleardoublepage

\begin{center}
{\LARGE{\sc Zusammenfassung}}
\vspace{1cm}
\end{center}
%
In der vorliegenden Arbeit wurde die inkoh\"arente Photoproduktion von
$\pi$-Mesonen am Deuteron im Bereich der $\Delta$(1232)-Resonanz unter
Ber\"ucksichtigung der Endzustandswechselwirkung von Nukleon-Nukleon ($NN$)
und Pion-Nukleon ($\pi N$)-R\"uckstreuung untersucht. Die verwendete
elementare Amplitude der Photopionproduktion am freien Nukleon
ber\"ucksichtigt Born-Terme in pseudovektorieller $\pi N$-Kopplung sowie den
$\Delta$(1232)-Resonanzbeitrag. Dur\-ch Anpassung der Modellparameter erh\"alt
man eine gute \"Ubereinstimmung mit den experimentellen Daten von MAMI, TRIUMF
und TAPS.

\vspace*{-0.2cm}~\\
Bez\"uglich der inkoh\"arenten Reaktion am Deuteron ergibt sich im Rahmen
 der Sto{\ss}\-n\"aherung qualitativ eine zufriedenstellende \"Ubereinstimmung
 mit dem Experiment f\"ur geladene Pionen hinsichtlich der totalen und
 differentiellen Wirkungsquerschnitte. Es zeigt sich aber eine systematische 
\"Uber\-sch\"atz\-ung von etwa 30 Prozent f\"ur neutrale Pionen. Daher ist es
 das Ziel dieser Arbeit, den Einflu{\ss} der bisher vernachl\"assigten
$NN$-End\-zu\-stands\-wechsel\-wirkung und der $\pi N$-R\"uckstreuung auf die
Pionphotoproduktion zu untersuchen. 

\vspace*{-0.2cm}~\\
Bei der Behandlung der Drei-Teilchen-Dynamik im Endzustand besch\-r\"an\-ken
wir uns auf Beitr\"age von Zwei-Teilchen-Wechselwirkungen in den $NN$-
bzw. $\pi N$-Subsystem\-en. Als Modell f\"ur die $NN$- und $\pi
N$-Wechselwirkung haben wir separable Potentiale benutzt, wobei die freien
Parameter durch einen Fit der Streuphasendaten bestimmt sind. Dabei zeigt es
sich, da{\ss} der Einflu{\ss} der $NN$- und $\pi N$-R\"uckstreueffekte auf die
totalen und differentiellen Wirkungsquerschnitte signifikant ist. Die
Einbeziehung solcher Effekte f\"uhrt zu einer verbesserten Beschreibung der
experimentellen Daten.

\vspace*{-0.2cm}~\\
Als Beispiel einer Polarisationsobservablen haben wir die in die
Gerasimov-Drell-Hearn-Summenregel (GDH) einflie{\ss}ende Beitr\"age der
inkoh\"arente $\pi$-Produktion zur Spin-Asymmetrie des
totalen Absorptionsquerschnitts am Deuteron unter Ber\"ucksichtigung der
$NN$- und $\pi N$-End\-zu\-stands\-wechsel\-wirkung untersucht. Wir haben
gefunden, dass der Einflu{\ss} der Endzustandswechselwirkungen f\"ur die
Spin-Asymmetrie und das GDH-Integral wichtig ist und in den
zuk\"unftigen theoretischen Arbeiten ber\"ucksichtigt werden mu{\ss}.
 
\vspace{24pt}
\noindent
Mainz, Mai 27, 2002\\
Betreuer: Prof.\ Dr.\ H.\ Arenh\"ovel 

\cleardoublepage
\newpage
\pagestyle{fancy}
\setcounter{page}{1}

\tableofcontents

\addcontentsline{toc}{chapter}{List of Figures}
\listoffigures
\addcontentsline{toc}{chapter}{List of Tables} 
\listoftables
\chapter{Introduction and Motivation\label{chap:1}}
\pagenumbering{arabic}
%
\section{Introduction\label{chap:1:1}}
%
The present work is concerned with photoproduction of $\pi$-meson on
deuterium. The $\pi$--meson or pion is the lightest of all the strongly
interacting particles and plays a central role in the physics of strong
interactions. It was proposed in 1935 by Yukawa \cite{Yukawa35} as the carrier
of the strong interaction between nucleons (proton, neutron) in analogy with
the photon in the electromagnetic interaction. The existence of the
$\pi$-meson was experimentally confirmed in 1947 by Powell and his collaborators
\cite{Powell47} and the pion is now known to exist in three charge states,
$\pi^+$, $\pi^-$ and $\pi^0$, with masses of 139.6 MeV/$c^2$ for charged pions
and 135.0 MeV/$c^2$ for the neutral pion.

\vspace*{-0.2cm}~\\
Photo- and electroproduction of pions on a proton have been studied
thoroughly both theoretically and experimentally. Beyond the pion threshold
energy almost all reactions on the nucleon, either initiated by photons, pions
or other hadrons with energies of a few hundred MeV (in the center-of-mass)
are dominated via the formation of the first excited state of the nucleon, the
$\Delta$(1232) resonance ($J=3/2,~I=3/2$), which almost exclusively decays
into a pion plus a nucleon. 

\vspace*{-0.2cm}~\\
During the last years, pseudoscalar meson
production in electromagnetic reactions on light nuclei has become a very
active field of research in medium-energy nuclear physics with respect to the
study of hadron structure. For the following reasons the deuteron plays an
outstanding role besides the free nucleon. The first one is that the deuteron
is the simplest nucleus on whose structure we have abundant information and a
reliable theoretical understanding, i.e.\ the structure of the deuteron is
very well understood in comparison to heavier nuclei. Furthermore, the small
binding energy of nucleons in the deuteron, which from the kinematical point
of view provides the case of a nearly free neutron target, allows one to
compare the contributions of its constituents to the electromagnetic and
hadronic reactions to those from free nucleons in order to estimate
interaction effects.

\vspace*{-0.2cm}~\\
The basic interaction in photoproduction is as follows: a photon is incident
on a target nucleus and interacts with its constituents. As a result, a
pseudoscalar meson is produced along with other particles. Two kinds of 
processes depending on the nature of the other
particles produced in this interaction are found: coherent and incoherent 
processes. 

\vspace*{-0.2cm}~\\
In the coherent process, the meson is produced with the
target nucleus maintaining its initial character. Thus, the 
interaction starts with a photon and some nucleus, and ends up with a meson
and the same nucleus, i.e.\ $\gamma X_A\rightarrow\pi X_A$,
where $A$ is the mass number of the target nucleus. The process is labeled
``coherent'' because all nucleons in the nucleus participate in the process
coherently, leading to a coherent sum of the individual nucleon contributions.

\vspace*{-0.2cm}~\\
In the incoherent process, the nucleus ruptures and thus
  fails to maintain its initial identity. The meson is produced in association
with a nucleon (or an excited state of the nucleon) and some new recoil
``daughter'' hadronic system. Thus, the interaction 
starts with a photon and some nucleus and ends up with a meson, a free
nucleon (or an excited state of it) and a new hadronic system, i.e.\ $\gamma
X_A\rightarrow\pi NX_{A-1}$. The process is labeled as ``incoherent'' because
it occurs in kinematic and physical circumstances similar to those of the
process that produces a meson from a free nucleon.
%
\section{Review of Previous Work\label{chap:1:2}}
%
The electromagnetic production of pions on the free nucleon, including
photoproduction and electroproduction, has long been studied 
since the pioneering work of Chew, Goldberger, Low and 
Nambu (CGLN) \cite{Chew57}. As a result, an enormous amount of
knowledge has been accumulated. Recently, theoretical interest in these
reactions 
was revived by the new generation of high-intensity and high duty-cycle
electron accelerators. With the developments of these new facilities, it is
now possible to obtain accurate data for meson electromagnetic production,
including spin-dependent observables. Extensive work during these more than
fourty years (see for example \cite{Kroll54}-\cite{Maid})
indicates that, below 500 MeV incident photon energy, the mechanisms of the
$\gamma N\rightarrow \pi N$ reaction are dominated by the Born terms and the
$\Delta(1232)$ excitation. 

\vspace*{-0.2cm}~\\
In this work, we go a step further by studying photoproduction of pions on the
deuteron. First investigations on photoproduction of pions on the deuteron go
back to the early fifties \cite{Chew51,Lax52} with view on the general
structure of spin flip and no spin flip amplitudes. Later, a more systematic
calculation of pion photoproduction on the deuteron was done by Laget
\cite{Laget81,Laget78,Laget77} and Blomqvist and Laget \cite{Blomqvist77}. In 
their work the influence of pion rescattering and $NN$ final state
interaction is included within a diagramatic ansatz. They used the elementary
photoproduction operator of Blomqvist and Laget \cite{Blomqvist77} as input in
their calculations. At the time of these calculations, a
comparison with experimental data was possible only for $\pi^-$ production
since data for $\pi^{+}$ and $\pi^0$ production in the $\Delta$(1232)
resonance region were not available. The agreement of their predictions
including final state interactions with the experimental data of the reaction
$\gamma d\rightarrow \pi^- pp$ \cite{Benz73} is quite good. They 
found that the final state interaction effects are small for the charged
pion photoproduction reactions in comparison to the neutral
channel. In recent years, experimental data for $\pi^0$ photoproduction on the
deuteron have become available \cite{Krusche99}. As mentioned in Ref.\
\cite{Krusche99}, the predictions from
\cite{Blomqvist77,Laget81,Laget78,Laget77} are significantly above the data. A
possible reason for this may be that they used the Blomqvist and 
Laget parametrization \cite{Blomqvist77} of the elementary photoproduction
amplitude which is not able to describe the neutral pion
photoproduction from the proton. 

\vspace*{-0.2cm}~\\
Pion photoproduction on the deuteron in the impulse approximation has 
been studied by Schmidt {\it et al.} \cite{Schmidt9695} neglecting all kinds
of final state interactions and other two-body operators. They constructed an
effective Lagrangian model for the process on the free nucleon and used it in
their calculations on the deuteron. Since data for 
$\pi^+$ and $\pi^0$ photoproduction on the deuteron were absent at the time of
these calculations, the authors could not compare their
predictions with experimental data for these channels. A comparison with
experimental data for $\pi^-$ production, showed a 
slight overestimation of the data. They reported that the reason for that is 
an overestimation of the elementary reaction on the neutron. A 
comparison between their predictions and the recently measured experimental
data for $\pi^0$ photoproduction on the deuteron \cite{Krusche99} is now
possible. One can see in \cite{Krusche99} that this prediction can hardly
provide a reasonable description of the data for this channel. As already
noted in 
\cite{Laget81,Levchuk96,Levchuk00N}, the effect of $NN$
rescattering is important in the incoherent pion photoproduction on the
deuteron, especially for $\pi^0$ production. 

\vspace*{-0.2cm}~\\
Levchuk {\it et al.} \cite{Levchuk96} studied quasifree $\pi^0$ 
photoproduction from the neutron via the $d(\gamma,\pi^0)np$ reaction using the
elementary photoproduction operator of Blomqvist and Laget 
\cite{Blomqvist77}. The contributions from the pole diagrams as well as
one-loop diagrams both with $np$ and $\pi N$ rescattering were taken into
account. They wanted to explore the possibility of measuring the
$E_{1+}/M_{1+}$ ratio via photoproduction from quasifree production on the
neutron.  
The isospin $I=3/2$ component of this ratio characterizes the relative
strength of the recently much discussed (see e.g.\ \cite{KrahnBeck96700,HDT})
quadrupole $E2$-excitation of the $\Delta$ resonance. The idea was that the
$n(\gamma,\pi^0)n$ reaction would be very useful for the isospin separation of
the multipoles. In agreement with the results from the Laget model 
\cite{Laget81}, Levchuk {\it et al.} find that the largest effects disturbing
the extraction of the multipoles for quasifree neutrons arise from the $np$
final state interaction. They predict that these effects lead to a strong
reduction of the cross section at pion forward angles, but are much less
important for backward angles. They found also that the correction due to
$np$ rescattering decreases with 
increasing pion angle and becomes to be less than 8$\%$ at $\theta_{\pi}\ge
90^0$. Furthermore, they pointed out that the contribution of the proton pole
diagram and the one of $\pi N$ rescattering are negligible. The
experimental data from Ref.\ \cite{Krusche99} for the 
$d(\gamma,\pi^0)np$ reaction qualitatively support this prediction since the
disagreement with the spectator approach is most severe at pion forward angles
but less pronounced at backward angles. However, a comparison of the data to
the Laget model including final state interactions shows some unexplained
reduction of the cross section at backward angles. 

\begin{figure}[htb]
  \centerline{\epsfxsize=15cm
    \epsffile{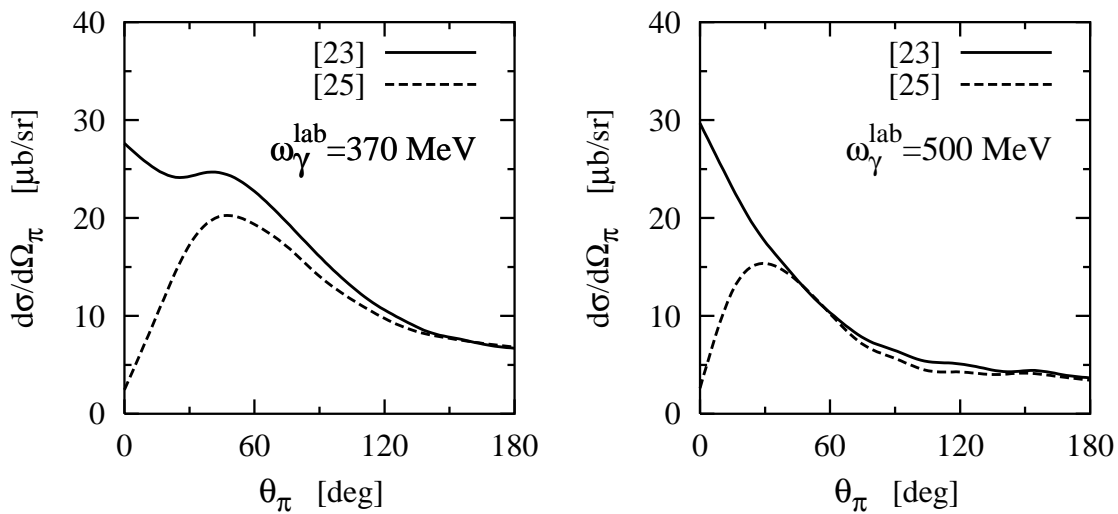}}
  \vspace*{-0.2cm}
  \caption{\small Differential cross section for $\pi^-$ photoproduction on
    the deuteron in the impulse approximation from \cite{Schmidt9695} (solid
  curve) in comparison with the results from \cite{Levchuk00N} (dashed
  curve).}
  \label{fig:1.2}
\end{figure}

\vspace*{-0.2cm}~\\
Recently and during the calculations of this work, Levchuk {\it et al.}
\cite{Levchuk00N} modified their theoretical predictions which have been done
in \cite{Levchuk96} using a more realistic version of the elementary
production operator and including the charged pion production channels which
are not included in their old calculations \cite{Levchuk96}. The elementary
production operator is taken in on-shell form and calculated using the SAID
\cite{Said} and MAID \cite{Maid} multipole analyses. The authors in
\cite{Levchuk00N} studied the semi-inclusive reaction $d(\gamma,\pi)NN$ in the
$\Delta$ resonance region including pole diagrams and one-loop diagrams with
$NN$ and $\pi N$ rescattering in the final state. Their predictions for total
and differential 
cross sections including final state interactions show good agreement with the
experimental data. However, Fig.\ \ref{fig:1.2} shows that a big difference between the
theoretical predictions from \cite{Levchuk00N} (dashed curve) and
\cite{Schmidt9695} (solid curve) is found for charged pion production
reactions at pion forward angles in the impulse approximation. Furthermore, 
the predictions from \cite{Schmidt9695} for the total cross sections for 
charged pion channels are found to be higher than the ones from
\cite{Levchuk00N} in the impulse approximation.
%
\section{Motivation\label{chap:1:3}}
%
The main motivation for studying photoproduction of pions on light nuclei is
to obtain information on the elementary process on the neutron. The main goal of this thesis is to investigate incoherent pion photoproduction on the
deuteron in the $\Delta$($1232$) resonance region in order to study the effect
of $NN$ and $\pi N$ rescattering in the final state. We include besides the pure
impulse approximation, the two-body $t$-matrices from $NN$ and $\pi N$ rescattering
in the final state. 

\vspace*{-0.2cm}~\\
As already mentioned above, a big difference is found between the
theoretical predictions from \cite{Levchuk00N} and \cite{Schmidt9695} in the
impulse approximation. Therefore, it is interesting to check in this work
where this big difference comes from. Since the authors in \cite{Levchuk00N}
pointed out that the main difference between their calculation and the one of
Ref.\ \cite{Laget81} is that a more realistic version of the elementary
production operator is used, we will also examine in this work the use of
different pion photoproduction operators.

\begin{figure}[htb]
  \centerline{\epsfxsize=12cm
    \epsffile{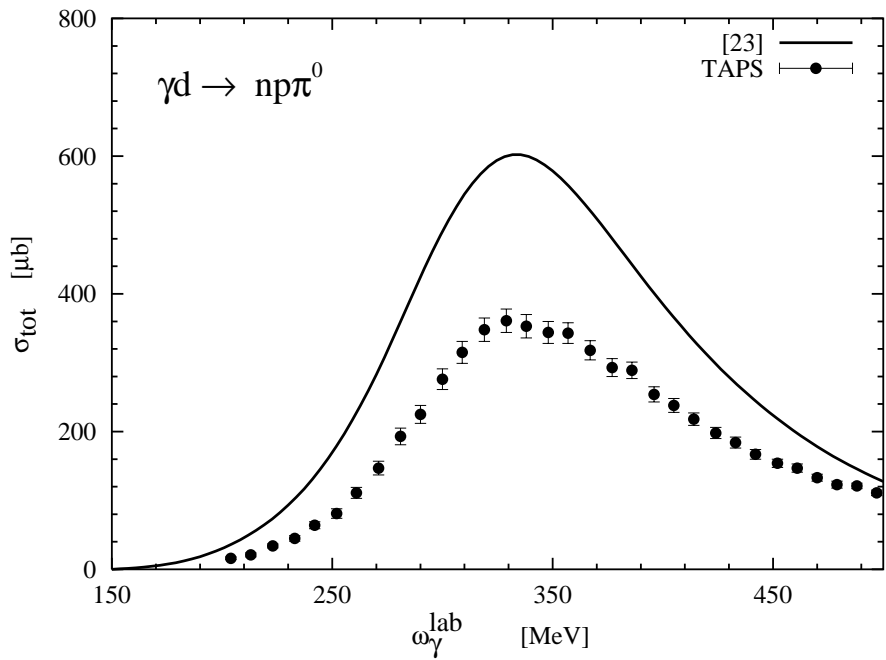}}
  \vspace*{-0.2cm}
  \caption{\small Total cross section for $\pi^0$ photoproduction on the
    deuteron in the impulse approximation from \cite{Schmidt9695} as a
  function of the photon energy in the laboratory frame. Data points are from
  TAPS \cite{Krusche99}.}
  \label{fig:1.1}
\end{figure}

\vspace*{-0.2cm}~\\
Furthermore, the comparison between the theoretical prediction from
\cite{Schmidt9695} for the total cross section of $\pi^0$ photoproduction on
the deuteron in the impulse approximation and the TAPS data \cite{Krusche99}
(see Fig.\ \ref{fig:1.1}) gives a clear indication that the effects of final
state interaction may be important. It was found that the final state
interaction effects are significant in the case of coherent pion
photoproduction on the deuteron (see for example 
\cite{Bosted78,Blaazer95,Wilhelm9592,Kamalov97}). This means that one needs a
reliable description for the rescattering process.

\vspace*{-0.2cm}~\\
Recently, several experiments to measure the spin asymmetry of the total
photoabsorption cross section, which determines the Gerasimov-Drell-Hearn
(GDH) sum rule, on the proton and on the deuteron are now performed or planned
at different laboratories around the world (MAMI, ELSA, LEGS, GRAAL and
TJNAF). The A2 collaboration has prepared an experiment on the proton and
on the deuteron at the Mainzer Microtron MAMI, that shall be completed for the
higher photon energies at ELSA in Bonn. 
This makes the
theoretical investigation of the spin asymmetry and the corresponding GDH
integral particularly interesting. Therefore, we investigate in this work the
influence of $NN$ and $\pi N$ rescattering effects on the  spin asymmetry and
GDH sum rule for the deuteron.

\vspace*{-0.2cm}~\\
Therefore, our aim is to construct a model for the reaction $\gamma d \rightarrow \pi NN$ in the
$\Delta$(1232) resonance region by incorporation the leading contributions
from $NN$ and $\pi N$ final
state interactions in time-ordered perturbation theory. 
%
\section{Outline of this Thesis\label{chap:1:4}}
%
This thesis is organized as follows. In chapter \ref{chap:2}, we 
present the effective Lagrangian model of the elementary pion photoproduction
process on the free nucleon which we use as
input in our calculations on the deuteron. We explain how the
reaction operator for this process is constructed. The nonresonant amplitudes and the
contribution of the $\Delta$(1232) resonance are given in an arbitrary frame of
reference. We end this chapter with a 
discussion of our results for differential and total
cross sections and compare with experimental data. 

\vspace*{-0.2cm}~\\
Chapter \ref{chap:3} is devoted to the central topic of this work. The treatment of the
$\gamma d\rightarrow\pi NN$ amplitude, based on time-ordered perturbation
theory with the inclusion of $NN$ and $\pi N$ final state
interactions is developed in this chapter. The transition matrix element is explicitly described with the inclusion of the $NN$- and $\pi N$-rescattering.  

\vspace*{-0.2cm}~\\
In chapter \ref{chap:4}, we will present our main results together with a comparison with experimental data and other theoretical predictions. We 
discuss in this chapter the effects of rescattering on total and differential
cross sections. The contribution of incoherent single pion photoproduction to the
spin asymmetry and the corresponding GDH sum rule for the
deuteron is also discussed. Finally, we conclude and summarize our results in
chapter \ref{chap:5}. Future considerations are also given in this chapter.

\vspace*{-0.2cm}~\\
For the convenience of the reader, eight appendices are given at the end of
the thesis. Appendix \ref{appendixA}, \ref{appendixB} and \ref{appendixC}
contain the general notations, formalism and 
useful formulas for the process on the nucleon. In appendix \ref{appendixD}, we give the
parametrization of the deuteron wave functions for the Bonn potential which we use in our calculations. In order to compare our results for 
differential cross section of $\pi^0$ photoproduction reaction on the deuteron with the experimental data from
Ref.\ \cite{Krusche99}, transformation formulas from the laboratory frame of
the deuteron to the $\gamma N$ center-of-mass frame are given in appendix
\ref{appendixE}. In appendix \ref{appendixF}, we study in detail the $NN$ and
$\pi N$ scattering matrices which we use as input in our predictions and also give a
solution for the two-body 
scattering matrix using separable two-body interactions. The parameters of the
form factors of these separable models are given in appendix \ref{appendixG}
and \ref{appendixH}. 

\chapter{Pion Photoproduction on the Nucleon\label{chap:2}}
%
Starting point of the construction of an operator for pion photoproduction 
in the two-nucleon space is the study of the elementary process, i.e.\ pion 
photoproduction on the free nucleon. This process is 
usually labeled as "elementary" to distinguish it from the same process from a
nucleus. In the elementary process a photon is
absorbed by a free nucleon (a proton or a neutron) to yield a $\pi$-meson in
addition to a nucleon. 

\vspace*{-0.2cm}~\\
The electromagnetic production of pions on the free nucleon, including
photoproduction and electroproduction, has long been studied since the
pioneering work of Chew, Goldberger, Low and Nambu (CGLN) \cite{Chew57}. An
enormous amount of knowledge has been accumulated as a result. Kroll and
Rudermann  
\cite{Kroll54} were the first to derive model-independent predictions in the 
threshold region, so-called low-energy theorems (LET), by applying gauge and 
Lorentz invariance to the reaction $\gamma N\rightarrow\pi N$. The general 
formalism for this process was developed by Chew {\it et al.} \cite{Chew57} 
(CGLN-amplitudes). Fubini {\it et al.} \cite{Fubini65} extended the earlier 
predictions of LET by including also the hypothesis of a partially conserved 
axial current (PCAC). In this way they succeeded in describing the threshold 
amplitude as a power series in the ratio $m=\frac{m_{\pi}}{M_N}$ up to terms 
of order $m^2$. Berends {\it et al.} \cite{Berends671} analysed the existing 
data in terms of a multipole decomposition and presented tables of the various 
multipole amplitudes constructing in the region up to excitation energies of 
500 MeV. 

\vspace*{-0.2cm}~\\
For more than twenty years, the standard model for pion photoproduction on
the nucleon has been the model of Olsson and Osypowski \cite{Olsson75,Olsson78}, who emphasized the importance of Watson's theorem as a
requirement to be obeyed by the electromagnetic multipoles below
$\pi\pi$-threshold. In
practice, a model that has been more extensively used for comparison with
data is the one of Blomqvist and Laget \cite{Blomqvist77} which is a 
non-relativistic reduction of the model of Olsson and Osypowski. It has been
constructed in a general frame of reference, but Blomqvist and Laget used different $\Delta$
parametrizations for neutral and charged pion photoproduction. These
parametrizations give a satisfactory fit to the amplitude for charged pion
photoproduction, but it is not able to describe the neutral pion
photoproduction from the proton. More than ten years ago, another model for pion photoproduction on
the nucleon in the $\Delta$ region has been proposed by Nozawa, Blankleider
and Lee \cite{Nozawa90}. This model has been extended to pion
electroproduction and has also proved to be successful for photo- and
electroproduction on the free nucleon.

\vspace*{-0.2cm}~\\
Garcilazo and Moya de Guerra \cite{Garcilazo93} have been constructed a model
for photo- and electroproduction on the nucleon. This model is applicable from 
threshold through the first and second resonance regions. It contains the
Born terms, the $\rho$ and $\omega$ vector mesons, the $\Delta$ and the Roper
resonances as well as the resonances $S_{11}$, $D_{13}$, $S_{31}$ and
$D_{33}$. They found a good agreement with experimental data up to 1 GeV. 
Recently, a unitary isobar model for pion photo- and electroproduction off the
nucleon has been developed for nuclear applications at photon
energies up to 1 GeV \cite{Maid}. This model contains Born terms, vector
mesons and 
nucleon resonances and is constructed in the $\pi N$ center-of-mass
frame. Within this model they have obtained good agreement with experimental
data for pion photo- and electroproduction on the free nucleon.

\vspace*{-0.2cm}~\\
In this work we will examine the various observables for pion
photoproduction on the free nucleon using the effective Lagrangian model of
Schmidt {\it et al.} \cite{Schmidt9695} which we will briefly outline in this chapter. The main advantage of this
model is that it has been constructed in a general frame of reference and
therefore can be applied directly to the electromagnetic
photoproduction of $\pi$-mesons on nuclei. This model contains besides the
standard pseudovector Born terms the resonance contribution from the
$\Delta$(1232) excitation. Kinematical
and other useful formulas for pion photoproduction on the free nucleon, which
diagramatically shown in Fig.\ \ref{fig:2.1}, are given in appendix 
\ref{appendixB}.
\begin{figure}[htb]
  \centerline{\epsfxsize=8cm
  \epsffile{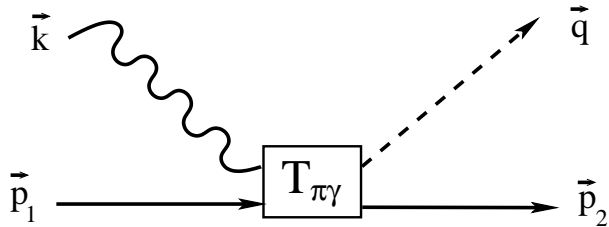}} 
  \caption{\small The elementary process $\gamma N \rightarrow \pi N$. A solid,
    dashed and wavy line represents a nucleon, pion and photon, respectively.}
  \label{fig:2.1}
\end{figure} 
%
\section{The Photoproduction Operator\label{chap:2:2}}
%
The ${\mathcal T}$-operator describing transitions between asymptotically
free states is given in terms of the interaction Hamiltonian ${\mathcal
  H}_{int}$ between all the involved particles as follows
\cite{Joachain79}
\begin{eqnarray}
{\mathcal T} & = & {\mathcal H}_{int} + {\mathcal H}_{int}\frac{1}
        {E-{\mathcal H}+i\epsilon} {\mathcal H}_{int}\, ,
\label{eq2.1}
\end{eqnarray}
which can also be re-written as 
\begin{eqnarray}
{\mathcal T} & = & {\mathcal H}_{int} + {\mathcal H}_{int}\frac{1}
        {E-{\mathcal H}_0+i\epsilon} {\mathcal T}\, ,
\label{eq2.1.1}
\end{eqnarray}
where ${\mathcal H}_0$ is the free Hamiltonian. The on-shell matrix element 
$T_{fi}$ of Eq.\ (\ref{eq2.1.1}) is given in terms of the Hamilton operator 
${\mathcal H}_{int}$ by 
\begin{eqnarray}
T_{fi} &=& \langle f \hspace*{-0.1cm}\mid\hspace*{-0.1cm} {\mathcal T} 
           \hspace*{-0.1cm}\mid\hspace*{-0.1cm} i\rangle ~=~ \langle f \hspace*{-0.1cm}\mid\hspace*{-0.1cm} {\mathcal H}_{int} 
           \hspace*{-0.1cm}\mid\hspace*{-0.1cm} i\rangle + 
           \sum_{\alpha} \langle f \hspace*{-0.1cm}\mid\hspace*{-0.1cm} 
           {\mathcal H}_{int} \hspace*{-0.1cm}\mid\hspace*{-0.1cm} 
           \alpha\rangle \frac{1}{E-E_{\alpha}+i\epsilon} T_{\alpha i}\, , 
\label{eq2.2.2}
\end{eqnarray}
where the energy eigenvalues $E_{\alpha}$ are given by
\begin{eqnarray}
{\mathcal H}_0\hspace*{-0.1cm}\mid\hspace*{-0.1cm}\alpha\rangle &=& 
       E_{\alpha}\hspace*{-0.1cm}\mid\hspace*{-0.1cm}\alpha\rangle\, .
\label{eq2.3}
\end{eqnarray}
Making one-iteration approximation and keeping terms from the second order 
we obtain the following expression for the $T_{fi}$-matrix
\begin{eqnarray}
T^{(2)}_{fi} &=& \langle f \hspace*{-0.1cm}\mid\hspace*{-0.1cm} {\mathcal H}_{int} 
           \hspace*{-0.1cm}\mid\hspace*{-0.1cm} i\rangle + 
           \sum_{\alpha} \langle f \hspace*{-0.1cm}\mid\hspace*{-0.1cm} 
           {\mathcal H}_{int} \hspace*{-0.1cm}\mid\hspace*{-0.1cm} 
           \alpha\rangle \frac{1}{E-E_{\alpha}+i\epsilon} \langle \alpha 
           \hspace*{-0.1cm}\mid\hspace*{-0.1cm} {\mathcal H}_{int} 
           \hspace*{-0.1cm}\mid\hspace*{-0.1cm} i\rangle\, . 
\label{eq2.2}
\end{eqnarray}
The complete Fock space of the system contains $N$-, $\pi N$-, $\pi\pi N$-,
$NN\bar N$-, $\pi NN\bar N$-, $\Delta$-, $\pi\Delta$-, etc states. 

\begin{figure}[htb]
  \centerline{\epsfxsize=14cm 
    \epsffile{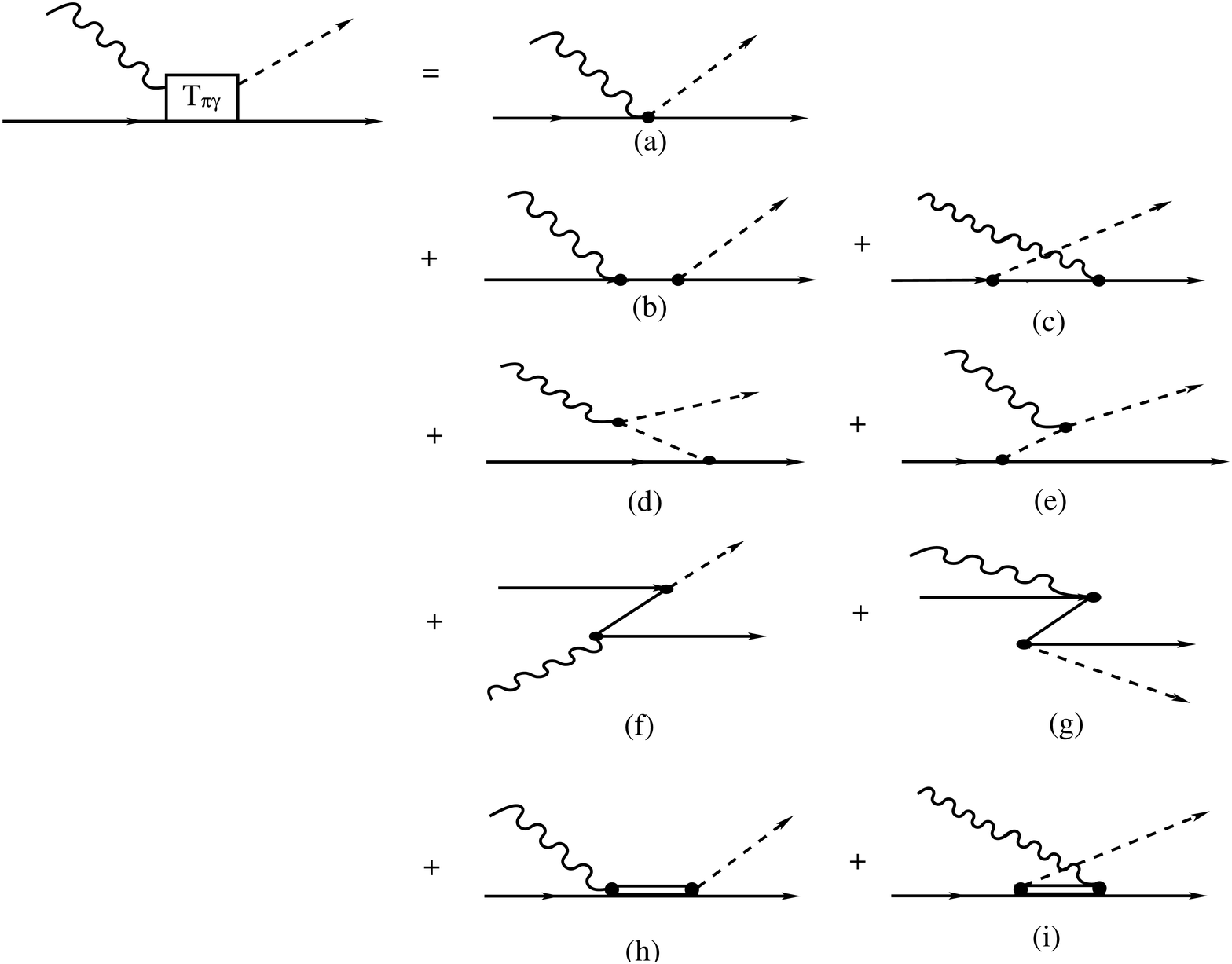}}
  \caption{\small Diagrams for the elementary process $\gamma N 
    \rightarrow \pi N$: (a) the Kroll-Rudermann graph, (b) and 
    (c) the two time-ordered contributions to the direct and crossed 
    nucleon pole graph, (d) and (e) the two time-ordered contributions 
    to the pion pole graph, (f) and (g) the Z-graphs and (h) and (i) the 
    $\Delta(1232)$ resonance graphs. Lines descriptions as in 
    Fig.\ \ref{fig:2.1}.}
  \label{fig:2.2}
\end{figure}

\vspace*{-0.2cm}~\\
For the process in our case, the form of the initial photon-nucleon state
$\hspace*{-0.1cm}\mid\hspace*{-0.1cm} i\rangle$ and the final pion-nucleon
state $\hspace*{-0.1cm}\mid\hspace*{-0.1cm} f\rangle$ is specified by the
asymptotical states as follows
\begin{eqnarray}
\hspace*{-0.1cm}\mid\hspace*{-0.1cm} i \rangle &=& \hspace*{-0.1cm}\mid
        \hspace*{-0.1cm} N,\gamma; {\vec p}_1m_t,{\vec k}{\vec\epsilon}
        \rangle\, ,\\
\hspace*{-0.1cm}\mid\hspace*{-0.1cm} f \rangle &=& \hspace*{-0.1cm}\mid
        \hspace*{-0.1cm} N^{\prime},\pi; {\vec p}_2m_{t^{\prime}},{\vec q}
        \mu\rangle\, ,
\label{eq2.45}
\end{eqnarray}
where $\vec{p}_1$, $\vec{p}_2$, $\vec{k}$ and $\vec{q}$ are the momenta of
initial and final nucleon, photon and meson, respectively. The isospin
projection of the produced pion is given by $\mu$, the polarization vector of
the incoming photon by ${\vec\epsilon}$ and $m_t$ and $m_{t^{\prime}}$ are the
isospin projection of the initial and final nucleon, respectively. The states
of all particles are covariantly normalized (see appendix
\ref{appendixC:1}). The individual terms of the $T_{fi}$-matrix for pion
photoproduction on the free nucleon are shown in Fig.\ \ref{fig:2.2}. In the
following we will evaluate the full interaction Hamiltonian of Eq.\
(\ref{eq2.2}) in more details.
%
\section{Interaction Hamiltonian\label{chap:2:1}}
%
The general form of the interaction Hamiltonian of the involved
particles, i.e.\ nucleon, $\Delta$, pion and photon is described by the
operator ${\mathcal H}_{int}$ which is given by 
\begin{eqnarray}
{\mathcal H}_{int} &=& {\mathcal H}_{em} + {\mathcal H}_{\pi N} + 
        {\mathcal H}_{\pi N\Delta} \, ,
\label{eq2.6}
\end{eqnarray}
where ${\mathcal H}_{em}$, ${\mathcal H}_{\pi N}$ and ${\mathcal H}_{\pi N\Delta}$ are the Hamiltonians
of the electromagnetic interaction, the $\pi N$-coupling and the coupling of
the $\Delta$ resonance to the 
$\pi N$ system, respectively. In the following we begin by evaluating the Hamiltonians which describe the 
interaction between pions, nucleons and photons as well as the contribution of
the $\Delta$(1232) resonance.
%
\subsection{The Electromagnetic Interaction\label{chap:2:1:1}}
\begin{figure}[htb]
  \centerline{\epsfxsize=12cm 
    \epsffile{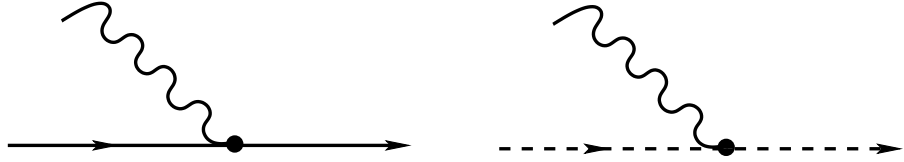}}
  \caption{\small The $\gamma N$- and $\gamma \pi$-vertices.} 
  \label{fig:2.3.1}
\end{figure}

%
The electromagnetic interaction ${\mathcal H}_{em}$ contains in our case the
coupling of the photon field to the free nucleon, pion and $\Delta$ fields
\begin{eqnarray}
{\mathcal H}_{em} &=& {\mathcal H}_{\gamma N} + {\mathcal H}_{\gamma\pi} + 
{\mathcal H}_{\gamma\pi N} + {\mathcal H}_{\gamma N\Delta}\, .
\label{hem}
\end{eqnarray}
The Hamiltonian corresponding to the absorption of a photon at a nucleon or a
pion (see Fig.\ \ref{fig:2.3.1}) are given, respectively, by
\begin{eqnarray}
{\cal H}_{\gamma N} & = & - e \int d^{3} x \bar{\Psi} ( \vec{x} )
        \vec{A}( \vec{x} ) \cdot \vec{\gamma} \Psi ( \vec{x} ) 
\label{eq2.after12_1}
\end{eqnarray}  
and
\begin{eqnarray}
{\cal H}_{\gamma\pi} & = & \frac{1}{2} \int d^{3} x  \sum_{\mu}
        (-)^{\mu}  i e \mu \vec{A}( \vec{x} ) \left[ \left(
        \vec{\nabla} \Phi_{\mu}\left( \vec{x} \right) \right)
        \Phi_{-\mu}\left( \vec{x} \right) - \left( \vec{\nabla}
        \Phi_{-\mu}\left( \vec{x} \right) \right)
        \Phi_{\mu}\left( \vec{x} \right)\right]\, ,\nonumber \\
& & 
\label{eq2.after12_2}
\end{eqnarray} 
where $\vec{A}(\vec{x})$, $\Psi (\vec{x})$ and $\Phi_{\mu} (\vec{x})$ are the
field operators of the photon, nucleon and pion, respectively. More
information about these operators is given in appendix \ref{appendixC}.
$\vec{\gamma}$ represent the Dirac matrices and $e$ denotes the elementary charge.

\vspace*{-0.2cm}~\\
Using field quantization (see appendix \ref{appendixC}) the matrix elements
corresponding to the two diagrams of Fig.\ \ref{fig:2.3.1} are given by 
\begin{eqnarray}
\langle N^{\prime};\vec{p}_2m_t^{\prime} \hspace*{-0.1cm}\mid\hspace*{-0.1cm}
{\cal H}_{\gamma N} \hspace*{-0.1cm}\mid\hspace*{-0.1cm} N, \gamma ; 
\vec{p}_1m_t,\vec{k}{\vec\epsilon} \rangle & = & -(2\pi )^{3} \delta ^{3} 
(\vec{p}_2- \vec{p}_1-\vec{k}) \nonumber \\
         & & \hspace*{-2.5cm}\times \frac{1}{2M_N}\langle m_t^{\prime} \mid
         \left(\hat{e} \left(\vec{p}_2+\vec{p}_1\right) 
        + i \left(\hat{e}+\hat{\kappa}\right) \vec{\sigma}\times \vec{k}\right)
        \cdot \vec{\epsilon} \mid m_t\rangle 
\label{eq2.15}
\end{eqnarray} 
and 
\begin{eqnarray}
\langle \pi ;\vec{q}^{\,\prime}\mu^{\prime} \hspace*{-0.1cm}\mid
\hspace*{-0.1cm} {\cal H}_{\gamma\pi} \hspace*{-0.1cm}\mid\hspace*{-0.1cm}
\pi , \gamma ; \vec{q}\mu ,\vec{k}{\vec\epsilon} \rangle & = & 
- (2\pi)^{3} \delta ^{3} (\vec{q}^{\,\prime}-\vec{q}-\vec{k}) 
\delta_{\mu \mu^{\prime}} e \mu (\vec{q} + \vec{q}^{\,\prime}) 
\cdot \vec{\epsilon}\, ,
\label{eq2.14}
\end{eqnarray} 
where $M_N$ is the nucleon mass and $\vec{\sigma}$ are the Pauli spin
matrices. $\hat e$ and $\hat\kappa$ denote nucleon charge and
anomalous part of the nucleon magnetic moment, respectively. These are 
isospin operators of the nucleon and are given by 
\begin{eqnarray}
\hat{e} & = & \frac{e}{2}\left( \eins + \tau_{0} \right)~ , \nonumber \\
\hat{\kappa} & = & \frac{e}{2}\left[ \kappa_{p} \left( \eins + \tau_{0}
        \right) + \kappa_{n} \left( \eins - \tau_{0} \right)\right]\, ,
\label{eq2.after15}
\end{eqnarray}
where $\kappa_p=\frac{1}{2}(\kappa_s+\kappa_v)=1.79$ and 
$\kappa_n=\frac{1}{2}(\kappa_s-\kappa_v)=-1.91$ are the anomalous magnetic 
moments of the proton and the neutron\footnote{The magnetic moments of the
  proton and neutron are $\mu_p=1+\kappa_p$ and $\mu_n=\kappa_n$,
  respectively.} in units of nuclear magnetons, respectively, 
and $\kappa_s=-0.12$ and $\kappa_v=3.70$. The coupling constant 
$\frac{e^2}{4\pi}=\frac {1}{137}$.

\begin{figure}[htb]
  \centerline{\epsfxsize=5cm 
    \epsffile{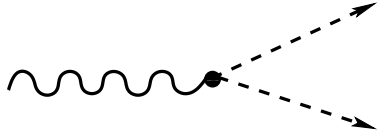}}
  \caption{\small The $\pi\gamma$-vertex.}
  \label{fig:2.3}
\end{figure}

\vspace*{-0.2cm}~\\
In addition to the matrix elements of Eqs.\ (\ref{eq2.15}) and (\ref{eq2.14}),
the following matrix element of ${\cal H}_{\pi \gamma}$ between a state of
two-pion and one photon (see Fig.\ \ref{fig:2.3}) is considered 
\begin{eqnarray}
\langle \pi , \pi^{\prime};\vec{q}\mu , \vec{q}^{\,\prime}\mu^{\prime}
\hspace*{-0.1cm}\mid\hspace*{-0.1cm} {\cal H}_{\pi \gamma} 
\hspace*{-0.1cm}\mid\hspace*{-0.1cm} \gamma;\vec{k}{\vec\epsilon} \rangle & = &
- (2\pi )^{3} \delta ^{3} (\vec{q}^{\,\prime}+\vec{q}-\vec{k}) 
\delta_{-\mu\mu^{\prime}} e \mu (\vec{q} + \vec{q}^{\,\prime}) 
\cdot \vec{\epsilon}\, .
\label{eq2.16}
\end{eqnarray} 
This matrix element will be used later in the construction of the amplitude of
diagrams (d) and (e) in Fig.\ \ref{fig:2.2}. The pionic current terms, which are given in Eqs.\ (\ref{eq2.14}) and 
(\ref{eq2.16}), contribute only to the photoproduction of charged pions.

\begin{figure}[htb]
  \centerline{\epsfxsize=6cm 
    \epsffile{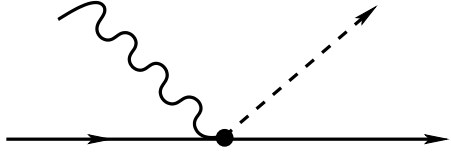}}
  \caption{\small The $\gamma \pi N$-vertex.}
  \label{fig:2.5}
\end{figure}

\vspace*{-0.2cm}~\\
The $\gamma\pi N$ Hamiltonian is given by 
\begin{eqnarray}
{\cal H}_{\gamma\pi N} & = & \frac{if_{\pi N}}{m_{\pi}}\sum_{\mu=\pm 1,0} 
\int d^{3}x \bar{\Psi} (\vec{x})
\vec{\gamma}\cdot\vec{A}(\vec{x})\gamma_{5}\left[\hat{e},\tau^{+}_{\mu}
\right]\Psi (\vec{x}) \Phi_{\mu} (\vec{x})\, ,
\label{eq2.20}
\end{eqnarray}
where $m_{\pi}$ is the pion mass and $\vec{\tau}$ represent the isospin
matrices. We used the $\pi N$ coupling constant 
$\frac{f_{\pi N}^{2}}{4\pi}=0.0735$ which is given in Ref.\ \cite{Arndt90} by
fitting the $\pi N$ scattering data. The $\gamma\pi N$ Hamiltonian is linear in photon and pion fields. This 
leads to a vertex, in which both photon and pion couple to the nucleon. 
The matrix element of ${\cal H}_{\gamma\pi N}$ for the diagram in
Fig.\ \ref{fig:2.5} is given by\footnote{This term contributes only to the photoproduction
  of charged pions.}
\begin{eqnarray}
\langle N^{\prime},\pi;\vec{p}_2m_t^{\prime},\vec{q} \mu \hspace*{-0.1cm}\mid
\hspace*{-0.1cm} {\cal H}_{\gamma\pi N}
\hspace*{-0.1cm}\mid\hspace*{-0.1cm} N,\gamma ;\vec{p}_1m_t,\vec{k}{\vec
\epsilon} \rangle & = & (2\pi)^{3} \delta ^{3} (\vec{p}_2+\vec{q}-\vec{p}_1-
\vec{k}) \nonumber \\
& & \times~\frac{if_{\pi N}}{m_{\pi}} \langle m_t^{\prime} \mid 
\vec{\sigma}\cdot\vec{\epsilon}\left[\hat{e},\tau^{+}_{\mu}\right]\mid m_t 
\rangle\, . 
\label{eq2.21}
\end{eqnarray} 

\begin{figure}[htb]
  \centerline{\epsfxsize=4cm 
    \epsffile{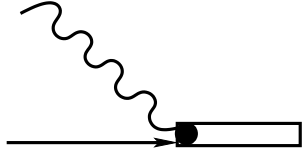}}
  \caption{\small The $\gamma N\Delta$-vertex.}
\label{gndelta2}
\end{figure}

\vspace*{-0.2cm}~\\
Now we evaluate the fourth term in Eq.\ (\ref{hem}). In the description of the $\gamma
N\Delta$-vertex (see Fig~\ref{gndelta2}) one has to take into account the
magnetic dipole $M1$ and a possible electric quadrupole $E2$ excitation of the
$\Delta$ resonance 
\begin{eqnarray}
{\cal H}_{\gamma N\Delta} & = & {\cal H}^{M1}_{\gamma N\Delta}+
        {\cal H}^{E2}_{\gamma N\Delta} \, .
\label{eq2.30}
\end{eqnarray}

\vspace*{-0.2cm}~\\
Since the strength of the electric quadrupole excitation $E2$ is much smaller
than the magnetic dipole one (see for example \cite{Wilhelm9592,Davidson91})
we will neglect it in this work. Following Weber and Arenh\"ovel 
\cite{Weber78}, Wilhelm and Arenh\"ovel \cite{Wilhelm9592} and Schmidt {\it et
  al.} \cite{Schmidt9695}, the $\gamma N\Delta$ vertex reads 
\begin{eqnarray}
\label{gndelta}
{\cal H}_{\gamma N\Delta} & = & \frac{ieG^{M1}_{\Delta
    N}(W_{\pi N})}{2M_{N}}~\vec{\sigma}_{\Delta N}\cdot(\vec{k}\times\vec{\epsilon})~\tau_{\Delta N,0}\, .
\end{eqnarray}
Here $W_{\pi N}$ denotes the invariant mass of the $\pi N$-subsystem and it is
given by 
\begin{eqnarray}
  W_{\pi N} & = & E_{N}(q_{c.m.}) + \omega_{\pi}(q_{c.m.})\, ,
\label{wpinsub}
\end{eqnarray}
where $E_N=\sqrt{M_N^2 +q_{c.m.}^2}$ and
$\omega_{\pi}=\sqrt{m_{\pi}^2+q_{c.m.}^2}$  with the c.m.\ pion momentum $q_{c.m.}$. The transition spin (isospin)
operator $\vec{\sigma}_{N\Delta}=\vec{\sigma}_{\Delta N}^{\dagger}$
($\vec{\tau}_{N\Delta}=\vec{\tau}_{\Delta N}^{\dagger}$) is normalized as
\begin{eqnarray}
\langle \frac{3}{2}\mid\mid\sigma_{\Delta N}(\tau_{\Delta N})\mid\mid\frac{1}{2}\rangle &
= & - \langle \frac{1}{2}\mid\mid\sigma_{N\Delta}(\tau_{N\Delta})\mid\mid\frac{3}{2}
\rangle~~=~~ 2\, .
\end{eqnarray}

\vspace*{-0.2cm}~\\
The energy dependent and complex coupling $G^{M1}_{\Delta N}(W_{\pi N})$ 
is given as in \cite{Wilhelm9592} by  
\begin{equation}
G_{\Delta N}^{M1}(W_{\pi N}) = \left\{ \begin{array}{cc}
\mu^{M1}(W_{\pi N}) e^{i\Phi^{M1}(W_{\pi N})} & {\rm ~~~for} \quad W_{\pi N} >
 m_{\pi}+M_{N} \\
\hspace*{-2.7cm} 0 & \hspace*{-3.0cm} {\rm else} \end{array} \right. \, ,
\label{complexcoup}
\end{equation}
where $\mu^{M1}(W_{\pi N})$ is given by
\begin{equation}
\mu^{M1}(W_{\pi N}) = \mu_0 + \mu_2\left(\frac{q_{\Delta}}{m_{\pi}}\right)^2 +
\mu_4\left(\frac{q_{\Delta}}{m_{\pi}}\right)^4
\end{equation}
and the phase $\Phi^{M1}(W_{\pi N})$ by \cite{Koch84}
\begin{equation}
\Phi^{M1}(W_{\pi N}) = \frac{q_{\Delta}^3}{a_1 + a_2 q_{\Delta}^2}\, .
\end{equation}
$q_{\Delta}$ is the on-shell pion momentum in the $\pi N$ c.\ m.\ frame
on the top of the resonance, i.e., when the invariant mass $W_{\pi N}$ of the
$\pi N$ state equals the mass of the $\Delta$ resonance
\begin{equation}
W_{\pi N} = \omega_{\pi}(q_{\Delta}) + E_{N}(q_{\Delta}) ~=~ M_{\Delta}\, .
\end{equation}
It is given by 
\begin{equation}
q_{\Delta} = \sqrt{\frac{ \left( W^2_{\pi N}-m^2_{\pi}-M^2_N
    \right)^2-4m^2_{\pi}M^2_N}{4W^2_{\pi N}}} \, .
\end{equation}
The free parameters $\mu_0=4.16$, $\mu_2=0.542$, $\mu_4=-0.0757$, $a_1=0.185$
fm$^{-3}$ and $a_2=4.94$ fm$^{-1}$ are fitted to the experimental data 
for the $M_{1+}^{3/2}$-multipole of pion photoproduction
\cite{Said,Wilhelm9592}.
%
\subsection{The $\pi N$ Interaction\label{chap:2:1:2}}
\begin{figure}[htb]
  \centerline{\epsfxsize=10cm 
    \epsffile{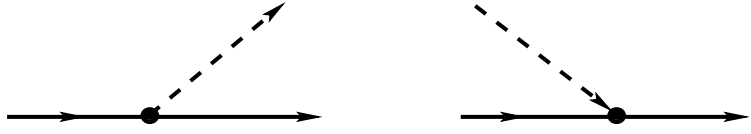}}
  \caption{\small The $\pi N$-vertices.}
  \label{fig:2.4}
\end{figure}

\begin{figure}[htb]
  \centerline{\epsfxsize=4cm 
    \epsffile{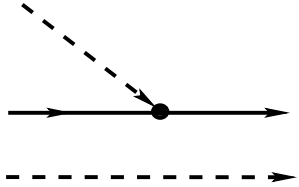}}
  \caption{\small The $\pi N$-vertex in case of the pion
    absorption.}
  \label{fig:2.pab}
\end{figure}

%
The $\pi N$ interaction operator in pseudovector coupling is given by 
\begin{eqnarray}
{\cal H}_{\pi N} & = & -\frac{f_{\pi N}}{m_{\pi}}\int d^{3}x\bar{\Psi}(\vec{x})
        \vec{\gamma}\cdot \gamma_{5}\vec{\tau}\cdot\Psi(\vec{x})\vec{\nabla}
        \vec{\Phi}(\vec{x})\, .
\label{eq2.17}
\end{eqnarray} 
This operator is linear in the pion field operator
$\vec{\Phi}(\vec{x})$. Therefore, only one pion can be produced or absorped at
the $\pi N$-vertex. Thus, only two possible diagrams as
shown in Fig.\ \ref{fig:2.4} contribute. The evaluation of these two graphs
yields the matrix elements 
\begin{eqnarray}
\langle N^{\prime},\pi;\vec{p}_2m_t^{\prime},\vec{q}\mu \hspace*{-0.1cm}\mid
\hspace*{-0.1cm} {\cal H}_{\pi N} \hspace*{-0.1cm}\mid\hspace*{-0.1cm} 
N;\vec{p}_1m_t \rangle & = & -(2\pi)^{3}\delta^{3}(\vec{p}_2+\vec{q}-\vec{p}_1)
\nonumber \\ 
& & \times~\frac{if_{\pi N}}{m_{\pi}}~
\langle m_t^{\prime}\mid\vec{q}\cdot\vec{\sigma}\tau^{+}_{\mu}\mid m_t \rangle
\label{eq2.18}
\end{eqnarray} 
for the emission of a pion and  
\begin{eqnarray}
\langle N^{\prime},\pi;\vec{p}_2m_t^{\prime},\vec{q}\mu \hspace*{-0.1cm}\mid
\hspace*{-0.1cm} {\cal H}_{\pi N} \hspace*{-0.1cm}\mid\hspace*{-0.1cm} 
N,\pi_{1},\pi_{2};\vec{p}_1m_t,\vec{q}_1 \mu_{1},\vec{q}_2\mu_{2} 
\rangle & = & -(2\pi)^{6}~\frac{2if_{\pi N}}{m_{\pi}} \nonumber \\ 
& &  \hspace*{-7.7cm}  \times~\left[
\delta^{3} (\vec{p}_2-\vec{q}_1-\vec{p}_1) 
\delta^{3} (\vec{q}-\vec{q}_2)\delta_{\mu\mu_{2}} \omega_{\vec{q}_2}
(-)^{\mu_{1}}\langle m_t^{\prime}\mid 
\tau_{\mu_{1}}\vec{q}_1\cdot\vec{\sigma}\mid m_t \rangle \right. \nonumber \\
& &  \hspace*{-8.5cm} \quad\qquad + \left.\delta ^{3} (\vec{p}_2-\vec{q}_2-
\vec{p}_1) \delta ^{3} (\vec{q}-\vec{q}_1)\delta_{\mu\mu_{1}} 
\omega_{\vec{q}_1}(-)^{\mu_{2}} \langle m_t^{\prime}\mid
\tau_{\mu_{2}}\vec{q}_2\cdot\vec{\sigma} \mid m_t \rangle \right]
\label{eq2.19}
\end{eqnarray} 
for the absorption of a pion, where
$\omega_{\vec{q}_i}=\sqrt{m_{\pi}^2+q_i^2}$ is 
the energy of the pion with momentum $\vec{q}_i$. Note, that in case of the pion absorption the vertex is evaluated
for the transition of a two to a one pion state (see Fig.\ \ref{fig:2.pab})
which will appear later in the construction of the matrix elements.
%
\subsection{The $\pi N\Delta$-Vertex\label{chap:2:1:4}}
\begin{figure}[htb]
  \centerline{\epsfxsize=4cm 
    \epsffile{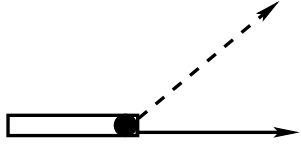}}
  \caption{\small The $\pi N\Delta$-vertex.}
  \label{pindelta}
\end{figure}

%
Here we will evaluate the $\pi N\Delta$-vertex which
contribute to the amplitude of the $\Delta$ resonance. For the $\pi N\Delta$-vertex which
diagramatically is given in Fig.\ \ref{pindelta} we use
\cite{Wilhelm9592,Weber78}
\begin{eqnarray}
{\cal H}_{\pi N\Delta} & = & -\frac{i}{m_{\pi}} F_{\Delta}(q^2)
~(-)^{\mu}~\vec{\tau}_{N\Delta,-\mu}~\vec{\sigma}_{N\Delta}\!\cdot\!\vec{q}\, .
\label{pindelta1}
\end{eqnarray}

\vspace*{-0.2cm}~\\
We have introduced a hadronic monopole form factor 
\begin{equation}
F_{\Delta}(q^2) = f_{\pi N\Delta}\frac{\Lambda^{2}_{\Delta} + q^2_{\Delta}}
        {\Lambda^{2}_{\Delta}+q^{2}}\, .
\label{formfacdel}
\end{equation}
The coupling constant $\frac{f^2_{\pi N\Delta}}{4\pi}=1.393$ and the cutoff
$\Lambda_{\Delta} = 315$ MeV are fixed in Refs.\ \cite{Wilhelm9592,Poepping87} 
to fit the $\pi N$ scattering phase shift in the $P_{33}$ channel and is also
used in the calculations of this work. 

\vspace*{-0.2cm}~\\
Now, using the Hamilton operator (\ref{eq2.6}) and taking into
account all possible intermediate states $\mid\hspace*{-0.2cm}\alpha\rangle$ in (\ref{eq2.2}), we can
calculate the on-shell $T_{fi}$-matrix for pion photoproduction on the free nucleon by
constructing the lowest order diagrams as shown in
Fig.\ \ref{fig:2.2}.
%
\section{Construction of the Amplitude\label{chap:2:3}}
%
Using the previously mentioned electromagnetic and hadronic vertices, it is possible now to calculate the $T_{fi}$-matrix from Eq.\ (\ref{eq2.2})
for photoproduction of pions on the free nucleon. Obviously, only the
Kroll-Rudermann term\footnote{Named also {\it seagull} or {\it contact} term.} 
($T_{fi}^{\rm KR}$) contributes to the first term of the
right-hand side of Eq.\ (\ref{eq2.2}). 
%
\subsection{The Born Terms\label{chap:2:3:1}}
%
First, we consider the nonresonant amplitudes. These
are referred to as the Born terms and they are dominant at low energy and for
charged pion photoproduction still provide $50\%$ of the cross section in the
energy region of the $\Delta$(1232) resonance. 

\vspace*{-0.2cm}~\\
Using graphs (a) to (e) in Fig.\ \ref{fig:2.2} one finds in addition to
the Kroll-Rudermann term (graph (a)) the direct and crossed
nucleon pole terms (graphs (b) and (c)) and the two pion pole terms
(graphs (d) and (e)). These terms are given, respectively,
by 
\begin{eqnarray} 
T_{fi}^{\rm KR} & = & (2\pi)^{3} \delta ^{3}
(\vec{p}_2+\vec{q}-\vec{p}_1-\vec{k})
\frac{if_{\pi N}}{m_{\pi}} \vec{\sigma}\cdot\vec{\epsilon}
~[\hat{e},\tau^{+}_{\mu}]\, ,
\label{eq2.23}
\end{eqnarray}
\begin{eqnarray} 
T_{fi}^{\rm N} & = & -(2\pi)^{3} \delta ^{3}
(\vec{p}_2+\vec{q}-\vec{p}_1-\vec{k}) 
\frac{if_{\pi N}}{2m_{\pi}} \nonumber \\
& & \times~\left(\frac{\tau^{+}_{\mu}\vec{\sigma}\cdot\vec{q}
\left(2(\vec{p}_2+\vec{q})\cdot\vec{\epsilon} ~\hat{e} + 
i\vec{\sigma}\cdot\vec{k}\times\vec{\epsilon}~(\hat{e}+\hat{\kappa})
\right)}{E_{\vec{p}_2+\vec{q}}(\omega_{\vec{q}}+
E_{\vec{p}_2}-E_{\vec{p}_2+\vec{q}})} \right.\nonumber \\
& & \left. + ~ \frac{\left(2\vec{p}_2\cdot\vec{\epsilon}~\hat{e}
+ i\vec{\sigma}\cdot\vec{k}\times\vec{\epsilon}~(\hat{e}+\hat{\kappa})
\right)\tau^{+}_{\mu}\vec{\sigma}\cdot\vec{q}}{E_{\vec{p}_2-\vec{k}}
(E_{\vec{p}_2}-E_{\vec{p}_2-\vec{k}}-\omega_{\gamma})} \right) \, ,
\label{eq2.24}
\end{eqnarray}
\begin{eqnarray} 
T_{fi}^{\pi} & = &  (2\pi)^{3} \delta^{3}(\vec{p}_2+\vec{q}-\vec{p}_1-\vec{k})
\frac{if_{\pi N}}{m_{\pi}}\frac{\vec{q}\cdot\vec{\epsilon} 
\vec{\sigma}\cdot(\vec{q}-\vec{k})}{\omega_{\vec{q}-\vec{k}}}\nonumber \\
& & \times~\left(\frac{1}{\omega_{\vec{q}}-\omega_{\vec{q}-
\vec{k}}-\omega_{\gamma}} + \frac{1}{\omega_{\gamma}-\omega_{\vec{q}-
\vec{k}}-\omega_{\vec{q}}}\right) [\hat{e},\tau^{+}_{\mu}]\, ,
\label{eq2.25}
\end{eqnarray}
where $E_{\vec p} = \sqrt{M_N^2 + \vec{p}\,^2}$ and 
$\omega_{\vec p} = \sqrt{m_{\pi}^2 + \vec{p}\,^2}$ are the energies 
of a nucleon and a pion with momentum $\vec{p}$, respectively. 

\vspace*{-0.2cm}~\\
To take the anti-nucleon terms in the propagators of direct and
crossed nucleon pole graphs into account, we consider also the two $Z$-graphs
given by diagrams (f) and (g) in Fig.\ \ref{fig:2.2}. The matrix element using
these $Z$-graphs ($T_{fi}^{\rm Z}$) is given by
\begin{eqnarray} 
T_{fi}^{\rm Z} & = & (2\pi)^{3}\delta^{3}(\vec{p}_2+\vec{q}-\vec{p}_1-\vec{k})
\frac{i f_{\pi N}}{m_{\pi}}M_{N}\omega_{\vec{q}}~\vec{\sigma}\cdot\vec{\epsilon}
\nonumber \\
& & \times\left(\frac{\tau_{\mu}^{+}\hat{e}}{E_{\vec{p}_2+\vec{q}}
(E_{\vec{p}_2+\vec{q}}+E_{\vec{p}_2}+\omega_{\vec{q}})}
+ \frac{\hat{e}\tau_{\mu}^{+}}{E_{\vec{p}_2-\vec{k}}
(E_{\vec{p}_2-\vec{k}}+E_{\vec{p}_2}-\omega_{\gamma})}
\right)\, .
\label{eq2.26}
\end{eqnarray}

\vspace*{-0.2cm}~\\
We would like to note that for a better description of the real part of the
$M_{1+}^{3/2}$ multipole (see section \ref{chap:2:4}) the suppression of the nonresonant background by the
form factor is essential. Therefore, it is necessary to introduce a form factor
\begin{eqnarray}
F_B(q) &=& \frac{\Lambda_B^2-m_{\pi}^2}{\Lambda_B^2 + q^2}\, ,
\label{formf}
\end{eqnarray}
with the cutoff $\Lambda_B=800$ MeV to obtain a better description. 
%
\subsection{The $\Delta$(1232) Resonance Term\label{chap:2:3:2}}
%
The dominant non-Born contribution for photon energies up to 500 MeV is that of
the $P_{33}$ pion-nucleon resonance, the $\Delta$(1232) resonance. Now, we
will evaluate the contribution of the $\Delta$(1232) resonance corresponding
to the vertices given in Eqs.\ (\ref{gndelta}) and (\ref{pindelta1}).
Using the nonrelativistic form of the $\Delta$ propagator, 
the various diagrams involving an intermediate
$\Delta$(1232) (see graphs (h) and (i) in Fig.\ \ref{fig:2.2}) can be
calculated. We obtain the following expression for the $s$
and $u$ channel contributions in center-of-mass frame 
\begin{eqnarray}
  T_{fi}^{\Delta} & = & (2\pi)^{3}\delta^{3}(\vec{p}_2+\vec{q}-\vec{p}_1
  -\vec{k}) \nonumber \\
  & & \hspace{-0.7cm}\times~\left(\frac{F_{\Delta}\left(q^2 \right)}{m_{\pi}} 
    \frac{e f_{\pi N\Delta} G^{M1}_{\Delta N}(W_{\pi N})}
    {2\sqrt{E_{{\vec p}_1}E_{{\vec p}_2}}}\frac{\{\tau_{\mu}^{\dagger},
      \tau_{0}\}- 
      \frac{1}{2}[\tau_{\mu}^{\dagger},\tau_{0}] }{3}
    \frac{\vec{\sigma}_{N\Delta} \cdot \vec{q} \vec{\sigma}_{\Delta N} \cdot 
      \vec{k}\times\vec{\epsilon}}{W_{\pi N}-M_{\Delta} +\frac{i}{2}
      \Gamma_{\Delta}(W_{\pi N})}\right. \nonumber \\
  & &
  \left. \hspace{-0.2cm} + \frac{F_{\Delta}\left(0\right)}{m_{\pi}}
  \frac{e f_{\pi N\Delta} G^{M1}_{\Delta N}(0)}
  {2\sqrt{E_{{\vec p}_1}E_{{\vec p}_2}}}
  \frac{\{\tau_{\mu}^{\dagger},\tau_{0}\}+
  \frac{1}{2}[\tau_{\mu}^{\dagger},\tau_{0}]}{3}
\frac{ \vec{\sigma}_{\Delta N}\cdot
  \vec{k}\times\vec{\epsilon} \vec{\sigma}_{N\Delta}\cdot
  \vec{q}}{E_{\vec{p}_2} - \omega_{\gamma} -
  E^{\Delta}_{\vec{p}_2-\vec{k}} }\right) \, ,
\label{eq2.38}
\end{eqnarray}
where $M_{\Delta}$ is the mass of the $\Delta$ resonance. 

\vspace*{-0.2cm}~\\
The energy dependent width of the $\Delta$ resonance above pion threshold
$\Gamma_{\Delta}(W_{\pi N})$ is given by  
\begin{equation}
\Gamma_{\Delta}(W_{\pi N}) = \left\{ \begin{array}{cc}
\frac{1}{6\pi}\frac{M_{N}}{\omega_{\vec{q}}+M_{N}}\frac{q^3}{m_{\pi}^{2}}
f^{2}_{\pi N\Delta} F^{2}_{\Delta}(q^2) & {\rm ~~~for} \quad W_{\pi N} > 
m_{\pi}+M_{N} \\
\hspace*{-3.5cm} 0 & \hspace*{-3.0cm} {\rm else} \end{array} \right. \, .
\label{reswidth}
\end{equation}

\vspace*{-0.2cm}~\\
In the $u$ channel contribution which is given by the second part in
Eq.\ (\ref{eq2.38}) we take the values of the form factor 
$F_{\Delta}$, the electromagnetic coupling $G_{\Delta N}^{M1}$ and the width
of the resonance $\Gamma_{\Delta}$ at pion threshold. 
In the expressions of the hadronic form factors (Eq.\ (\ref{formfacdel})), 
the complex coupling $G^{M1}_{\Delta N}$ (Eq.\ (\ref{complexcoup})) and the
width 
of the $\Delta$ resonance (Eq.\ (\ref{reswidth})) we use the
c.m.\ pion momentum as given by the invariant mass of the $\pi N$-subsystem
(see Eq.\ (\ref{wpinsub})).

\vspace*{-0.2cm}~\\
In order to build the elementary pion photoproduction operator into the
two-nucleon system, which will be necessary for the calculations on the
two-nucleon space in the forthcoming chapters, we must construct the 
elementary amplitude in a general frame of reference. The Born terms given in
section \ref{chap:2:3:1} are already constructed in an arbitrary frame. In order to build the 
resonance term of Eq.\ (\ref{eq2.38}) into the two-nucleon system we must
re-write Eq.\ (\ref{eq2.38}) in an arbitrary frame of reference. In
Ref.\ \cite{Wilhelm9592} the following Galilei invariant transformations are
found for the photon and pion momenta which have to be replaced by the
relative photon-nucleon momentum 
\begin{eqnarray}
\vec{k} \longrightarrow \vec{k}_{\gamma N}(\vec{k},\vec{p}_1) & := & 
\frac{M_{N}\vec{k} - (M_{\Delta}-M_N)\vec{p}_1}{M_{\Delta}}
\label{eq2.39}
\end{eqnarray}
and respective pion-nucleon momentum
\begin{eqnarray}
\vec{q} \longrightarrow \vec{q}_{\pi N}(\vec{q},\vec{p}_2) 
        & := & \frac{M_{N}\vec{q}-\omega_{\vec{q}}\vec{p}_2}{M_{N}+
        \omega_{\vec{q}}} \, .
\label{eq2.39a}
\end{eqnarray}

\vspace*{-0.2cm}~\\
Now, all contributions for the elementary photoproduction operator on the free
nucleon, which in Fig.\ \ref{fig:2.2} are graphically represented, are
calculated. The full $T_{fi}$-matrix is
then given in a simple form by\footnote{In this work no further contributions 
are used. This means in particular that terms from the
$\omega$- and $\rho$-meson exchange in the $t$-channel are not considered.} 
\begin{eqnarray}
T_{\pi\gamma} & = & T_{fi}^{\rm KR}+T_{fi}^{\rm N}+T_{fi}^{\pi}+T_{fi}^{\rm Z}
+ T_{fi}^{\Delta}\, .
\label{eq2.40}
\end{eqnarray}
%
\section{Results for Elementary Process\label{chap:2:4}}
%
In this section we will examine the various observables for pion
photoproduction on the free nucleon using the effective Lagrangian model of
Schmidt {\it et al.} \cite{Schmidt9695}. This model contains besides the
standard pseudovector Born terms the resonance contribution from the
$\Delta$(1232) excitation. The parameters of
the $\Delta$ resonance are fixed by fitting the experimental 
data \cite{Said} for the $M_{1+}^{3/2}$ multipole which is dominant in the
$\Delta$(1232) region. Due to the use of a
constant $\Delta$ mass in the 
$\Delta$ propagator and a different cutoff $\Lambda_{\Delta}$ we had to
increase $G^{M1}_{\Delta N}$ from \cite{Wilhelm9592} by a factor of $1.15$ to
fit the experimental $M_{1+}^{3/2}$ multipole. 
\begin{figure}[htb]
  \centerline{\epsfxsize=15cm 
    \epsffile{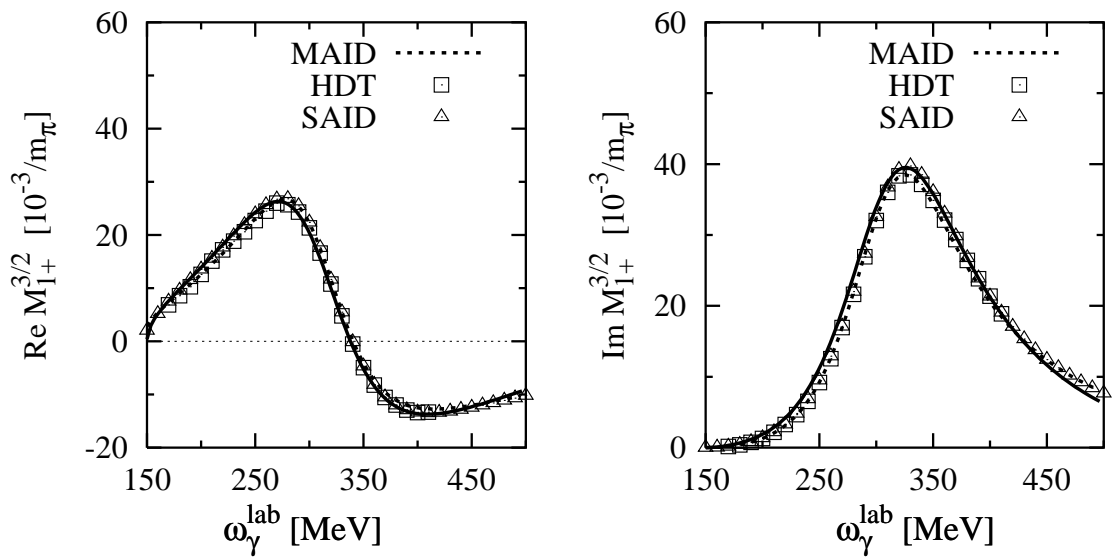}}
\vspace*{-0.5cm}
  \caption{\small Real and imaginary parts of the $M_{1+}^{3/2}$ multipole in
    comparison with the calculation using the MAID program \cite{Maid}. The
    data points show the results of the VPI analysis \cite{Said} (solution:
    September 2000) and the Mainz dispersion analysis (HDT) \cite{HDT}.}
  \label{multipoles}
\end{figure}

\vspace*{-0.2cm}~\\
Fig.\ \ref{multipoles} shows our fit for the real and
imaginary parts of the $M_{1+}^{3/2}$ multipole in comparison with the
MAID analysis \cite{Maid}, the Mainz
dispersion analysis (HDT) \cite{HDT} and the VPI analysis from the SAID
program \cite{Said}. We see that the agreement of our results (solid curves)
in comparison with data and theoretical 
calculation from MAID (dashed curves) is good.
%
\subsection{Differential Cross Section\label{chap:2:4:1}}
%
The c.m.\ differential cross section for the transition from an initial
photon-nucleon state $\mid\hspace*{-0.1cm} i\rangle$ to a final 
pion-nucleon state $\mid\hspace*{-0.1cm} f\rangle$ is given by
\begin{eqnarray}
\frac{d\sigma}{d\Omega_{\pi}^{c.m.}} &=& 
        \frac{1}{64\pi^2}\frac{\tilde q M_N^2}{\omega_{\gamma}W_{\pi N}^2}
        \sum_{m_{\gamma}m_sm_{s^{\prime}}}\hspace*{-0.1cm}\mid\hspace*{-0.1cm} 
        T_{\pi\gamma}\hspace*{-0.1cm}\mid\hspace*{-0.1cm} ^2 \, ,
\label{dcsel}
\end{eqnarray}
where $\omega_{\gamma}$ is the photon energy in the laboratory frame and
$m_{\gamma}=\pm 1$. The magnetic quantum numbers of the target 
and the recoiling nucleons 
are respectively $m_s$ and $m_{s^{\prime}}$. The invariant $\pi N$ mass
$W_{\pi N}$ is given in the c.m.\ frame as 
\begin{eqnarray}
W_{\pi N} &=& 
        E_{{\vec p}_1} + \omega_{\gamma} \, ,
\end{eqnarray}
and the absolute value of the pion momentum $\tilde q$ is given by
\begin{figure}[htb]
\centerline{\epsfxsize=14.5cm 
\epsffile{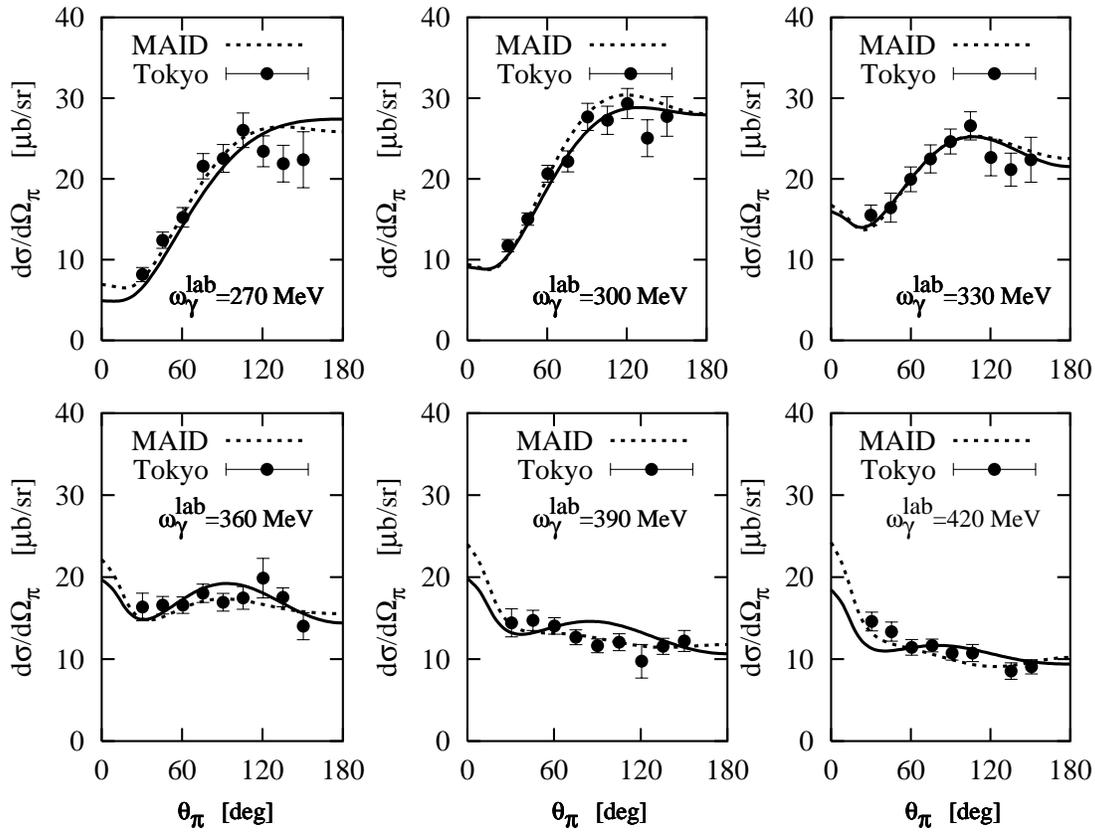}}
\vspace*{-0.2cm}
\caption{\small Differential cross section for $\gamma n\rightarrow p\pi^-$
  at six different values of the photon energy in the laboratory frame. The
        solid curve shows the result of our calculations and the dotted one 
        shows the results using the MAID program \cite{Maid}. The 
        experimental data are from Tokyo \cite{Fujii77}.}
\label{gnppim1}
\end{figure}
\begin{eqnarray}
\tilde q &=& 
        \sqrt{ \left( \frac{m^2_{\pi}+2\omega_{\gamma}W_{\pi N}}{2W_{\pi
        N}}\right)^2  - m^2_{\pi} }\, .
\end{eqnarray}
\begin{figure}[htb]
\centerline{\epsfxsize=14.5cm 
\epsffile{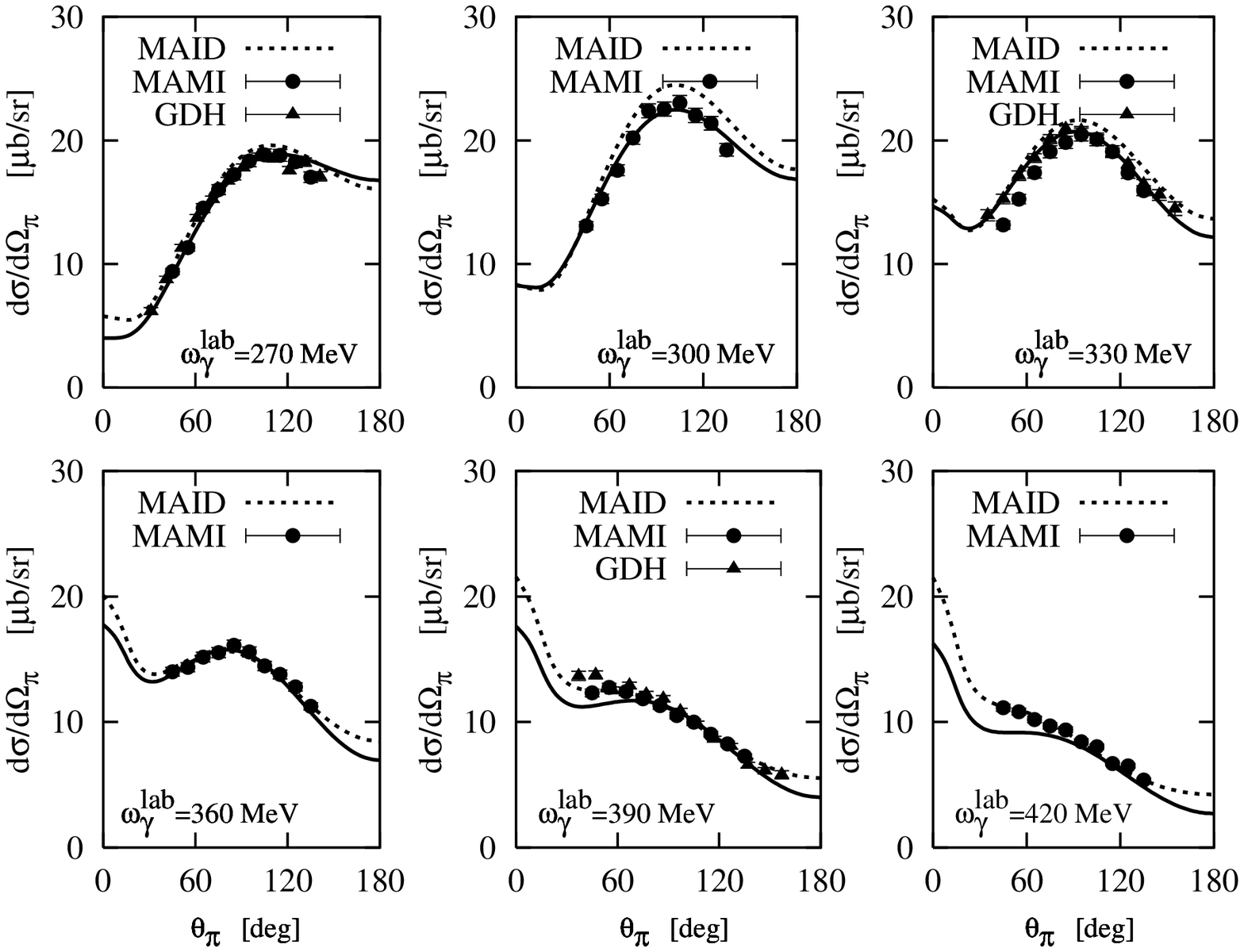}}
\vspace*{-0.2cm}
\caption{\small Differential cross section for $\gamma p\rightarrow n\pi^+$. 
Lines description as in Fig.\ \ref{gnppim1}. The experimental data are from
        GDH \cite{Ilia00} and MAMI \cite{KrahnBeck96700}.}
\label{gnppip1}
\end{figure}
\begin{figure}[htb]
\centerline{\epsfxsize=14.5cm 
\epsffile{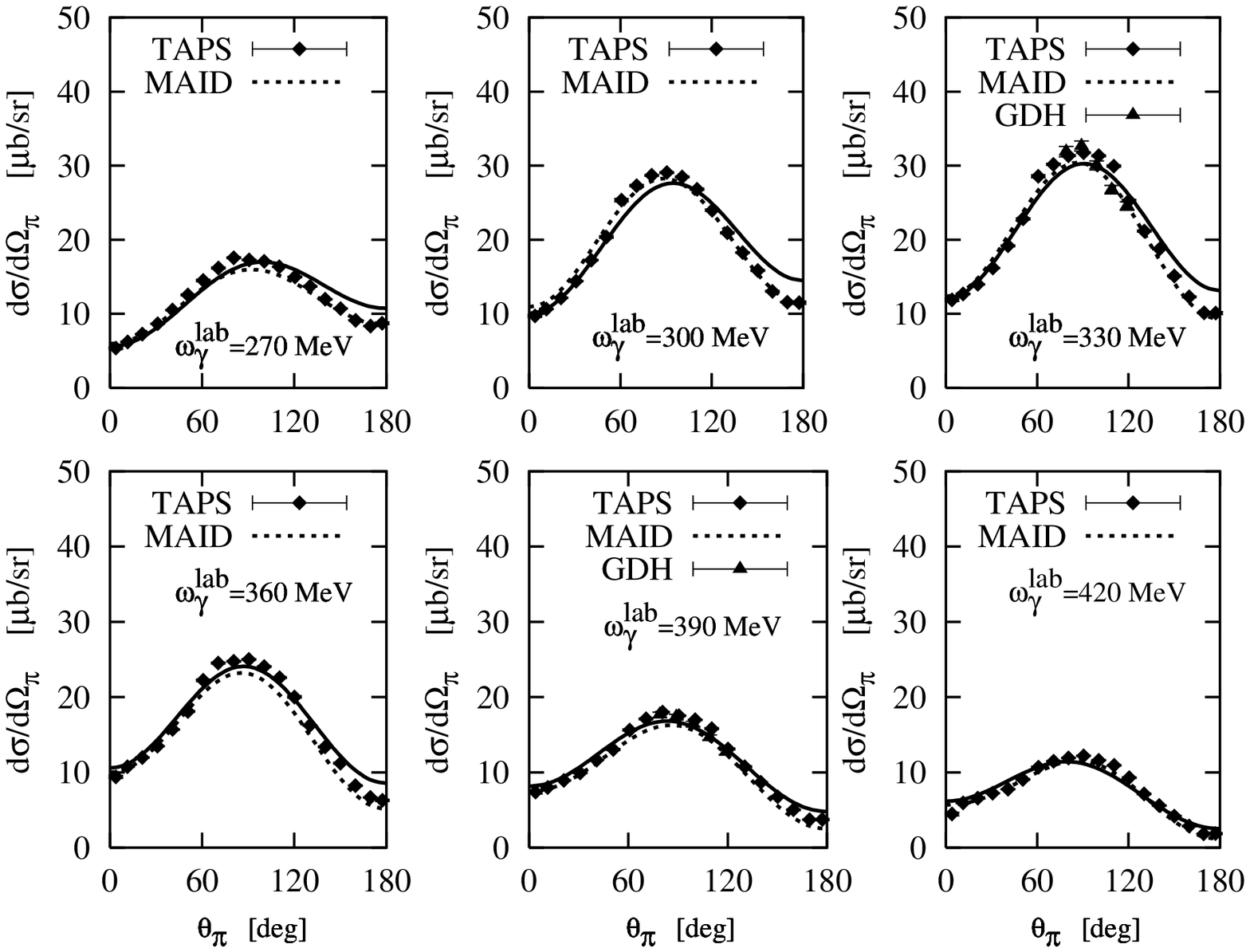}}
\vspace*{-0.2cm}
\caption{\small Differential cross section for $\gamma p\rightarrow p\pi^0$. 
Lines description as in Fig.\ \ref{gnppim1}. The experimental data are from
        TAPS \cite{Leukel00} and GDH \cite{Ilia00}.}
\label{gnppin1}
\end{figure}
\begin{figure}[htb]
  \centerline{\epsfxsize=11cm 
    \epsffile{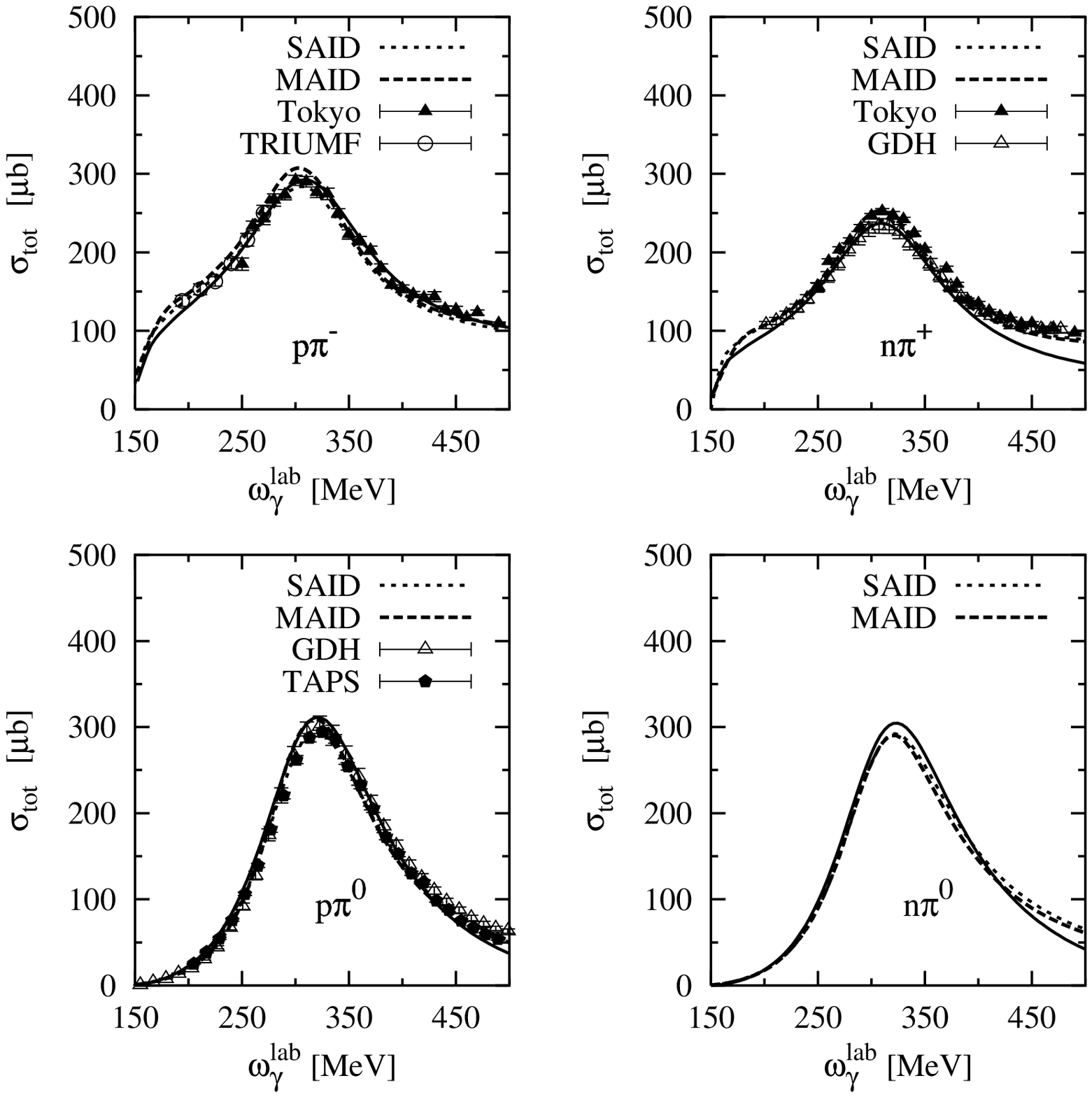}}
  \vspace*{-0.2cm}
  \caption{\small Total cross sections for pion photoproduction on the nucleon
    as a function of photon energy for the four physical reactions. The solid
    curve shows the result of our calculation, the dashed curve shows the 
    results using the MAID program \cite{Maid} and the dotted one shows the
    results of the VPI analysis of the SAID program \cite{Said}. The
    experimental data are from GDH \cite{Ilia00}, Tokyo \cite{Fujii77}, TRIUMF
    \cite{Bagheri88} and TAPS \cite{Haerter96}.}
  \label{tot}
\end{figure}

\vspace*{-0.2cm}~\\
In Figs.\ \ref{gnppim1}--\ref{gnppin1} we compare our results (solid curves)
for the differential cross sections with the MAID analysis 
\cite{Maid} (dashed curves) and with old and new experimental data from Mainz
and Bonn. Most interesting for our analysis are the recent experimental data
from the Mainz Microtron MAMI. Differential cross sections of both $\pi^{+}$
and $\pi^{0}$ photoproduction off the proton have been measured
\cite{KrahnBeck96700,Leukel00,Ilia00}, with high precision at all angles and photon energies for both channels. To get the
full isospin decomposition we also had to include data for $\pi^{-}$
photoproduction, for which we took the differential cross sections from \cite{Fujii77}. In general, we obtain a good agreement with the experimental
data for pion photoproduction on both the proton and neutron, especially in
the energy range of the $\Delta$(1232) region. 

\vspace*{-0.2cm}~\\
For charged pion channels we see from
Figs.\ \ref{gnppim1} and \ref{gnppip1} a similar shape,
but we find a somewhat different situations. Fig.\ \ref{gnppim1} shows the
result of our calculation of $\pi^-$ 
photoproduction on the neutron at six different values for the photon energy in
comparison with the data from \cite{Fujii77} and also with the MAID analysis \cite{Maid}. It can be seen that we fit this experiment
quite well, in particular for a photon energy of 330 MeV which is very near to the 
$\Delta$(1232) region. Furthermore, the comparison with the
MAID analysis at this energy is good. For
higher and lower photon energies we note small discrepancies. In the case of high energies these discrepancies come from the fact
that no other resonances besides the $\Delta$(1232) are included in our calculation. It is also
interesting to point out the importance of the Born terms in the charged
pion photoproduction reactions in comparison to the contribution of the
$\Delta$ resonance. These terms play an important role in the case of low
photon energies. 

\vspace*{-0.2cm}~\\
In Fig.\ \ref{gnppip1} we compare our predictions for $\pi^{+}$ photoproduction on
the proton with the more recent experimental data from the Mainz experiment
\cite{KrahnBeck96700} and the GDH experiment \cite{Ilia00} for different
photon energies. We see that the agreement with both experiment and
MAID analysis \cite{Maid} is good. 
We think that the agreement with experiment is in our case better 
than the MAID analysis for low and intermediate photon energies. For high energies small
discrepancies are found since we assume a pure magnetic dipole transition in
the $\gamma N\Delta$ coupling. Moreover, the $D_{13}$ resonance
contributes with a non-vanishing term in this region.

\vspace*{-0.2cm}~\\
In case of the $\pi^0$ photoproduction on the proton, shown in 
Fig.\ \ref{gnppin1}, the situation is 
much more satisfactory. We see that the agreement of our calculation with the
most recent experimental data from \cite{Leukel00} and \cite{Ilia00}
is good. Small discrepancies between our calculation and the
MAID analysis appear which very likely are due to the fact that no
other resonances are included in our calculation.  
%
\subsection{Total Cross Section\label{chap:2:4:2}}
%
The total cross sections for the different pion channels are
shown in Fig.\ \ref{tot} and compared with experimental data. The total cross 
sections for $\gamma p\rightarrow\pi^+ n$ and  $\gamma n\rightarrow\pi^- p$ 
have a similar structure. In particular, the $\Delta$ peak is seen exactly 
at the same energy, 300 MeV.

\vspace*{-0.2cm}~\\
The comparison of our calculation with experimental data is carried out for 
photon energies up to 500 MeV. For higher photon energies the parameterization
of the $\Delta$ resonance possesses no more valid. In general, we
obtain a good agreement with the recent experimental data using the small value
$\frac{f^2_{\pi N}}{4\pi}=0.069$ for the pion-nucleon coupling constant. The
agreement with the experimental data from \cite{Fujii77} and \cite{Bagheri88} for $\pi^-$ photoproduction on the neutron is
satisfactory. In case of the $\pi^+$ photoproduction, the agreement with the recent
experimental data from \cite{Ilia00} is good up to photon energy of about
400 MeV. For high energies the $D_{13}$ resonance, which is not included in
our calculation, has non-vanishing contribution. The $\pi^{+}$ data from
\cite{Fujii77} are slightly overestimated in the resonance region by our
calculation and also by the MAID analysis. 

\vspace*{-0.2cm}~\\
At low photon energies, the charged pion reactions differ significantly in
magnitude from the neutral pion reactions due to the Kroll-Rudermann term
which does not contribute to the neutral pion reactions. The agreement of our
calculation with the experimental data from \cite{Haerter96} and \cite{Ilia00}
in case of the $\pi^0$ photoproduction on the proton is good and give a clear
indication that the model which we used in our predictions can be applied
directly to calculate the electromagnetic photoproduction of pions on the
deuteron which we will study in the forthcoming chapters.

\chapter{Photoproduction of $\pi$-Mesons on the Deuteron\label{chap:3}}
%
In the previous chapter, we have constructed the elementary photoproduction
operator on the free nucleon in a general frame of reference to use it as
input in our calculations during the present chapter. This
chapter is concerned with incoherent pion photoproduction on the deuteron
which can be produced through the following three reactions
\begin{eqnarray}
  & & \gamma d\rightarrow pp\pi^- \, , ~~~~~~ \gamma d\rightarrow nn\pi^+ \, , 
~~~~~~ \gamma d\rightarrow pn\pi^0 \, . \label{eq:3.1}
\end{eqnarray}

\vspace*{-0.2cm}~\\
Let ($\omega_{\gamma}$,$\vec{k}$), ($E_{\vec{d}}$,$\vec{d}$),
($\omega_{\vec{q}}$,$\vec{q}$), ($E_{\vec p_1}$,$\vec{p}_1$) and
($E_{\vec p_2}$,$\vec{p}_2$) be the four-momenta of
the incoming photon, deuteron, the outgoing pion and two nucleons,
respectively. The general expression for the unpolarized differential cross
section of pion photoproduction reaction on the deuteron is given using the
conventions of Bjorken and Drell \cite{Bjorken64} by 
\begin{eqnarray}
\label{eq:3.2}
d\sigma = (2\pi)^{-5}\delta^{4}\left( k+d-p_{1}-p_{2}-q\right)
\frac{1}{|\vec{v}_{\gamma}-\vec{v}_{d}|} \frac{1}{2}
\frac{d^{3}q}{2\omega_{\vec{q}}} \frac{d^{3}p_{1}}{E_{\vec{p}_{1}}}
\frac{d^{3}p_{2}}{E_{\vec{p}_{2}}}
\frac{M_{N}^{2}}{2\omega_{\gamma}2E_{\vec{d}}}\nonumber \\
 & & \hspace{-9cm}\times~\frac{1}{6}\sum_{sm_st,m_{\gamma}m_d} 
|\mathcal M^{(t\mu)}_{sm_s,m_{\gamma}m_d}|^{2} \, , 
\end{eqnarray}
where $m_{\gamma}$ is the photon polarization, $m_{d}$ the spin 
projection of the deuteron, $s$ the total spin of the two outgoing
nucleons, $m_{s}$ its spin projection, $t$ the total isospin of the two
outgoing nucleons, $m_{t}$ ($=-\mu$) its isospin projection and $\vec{v}_{\gamma}$ and
$\vec{v}_{d}$ are the velocities of the photon and the deuteron,
respectively. $\mu$ denotes the pion charge. The states of all particles are covariantly normalized (see
appendix 
\ref{appendixC:1}). The reaction amplitude is given by
\begin{eqnarray}\label{eq:3.3}
\mathcal M^{(t\mu)}_{sm_s,m_{\gamma}m_d} & = &
\bra{\vec{p_1}\vec{p_2},sm_{s},t,-\mu}\mathcal
M(\vec{q},\mu,\vec{k},m_{\gamma})\ket{\vec{d}, m_{d}}\, .
\end{eqnarray}

\vspace*{-0.2cm}~\\
Choosing the $z$-axis in the direction of the incoming photon and isolating
the azimuthal dependence of the direction of pion momentum, we obtain the
following general form for the reaction matrix  
\begin{eqnarray}\label{eq:3.31}
\mathcal M^{(t\mu)}_{sm_s,m_{\gamma}m_d}(\theta_{\pi},\phi_{\pi}) & = &
\mathcal
O^{(t\mu)}_{sm_s,m_{\gamma}m_d}(\theta_{\pi})
~e^{i(m_{\gamma}+m_d)\phi_{\pi}}\, . 
\end{eqnarray}
Using parity conservation one can show that
the $\mathcal M$-matrix elements obey the following symmetry relation
\begin{eqnarray}\label{sym}
\mathcal O^{(t\mu)}_{s,-m_s,-m_{\gamma},-m_d} & = &
(-)^{s+m_s+m_{\gamma}+m_d} ~\mathcal O^{(t\mu)}_{s,m_s,m_{\gamma},m_d}\, .
\end{eqnarray}
This symmetry relation reduces the number of complex amplitudes from 24 to 12
independent ones. For their determination one needs 23 real observables since
a overall phase remains arbitrary.

\vspace*{-0.2cm}~\\
The unpolarized differential cross section in Eq.\ (\ref{eq:3.2}) is a product
of two
terms. The first term is a kinematical one describing the invariant phase
space. The second
one describes the dynamics of the process and is determined by 
the transition matrix element $\mathcal M$. In the following we disscuss
these two terms in more detail.
%
\section{Kinematics\label{chap:3:1}}
\begin{figure}[htb]
  \centerline{\epsfxsize=14cm \epsffile{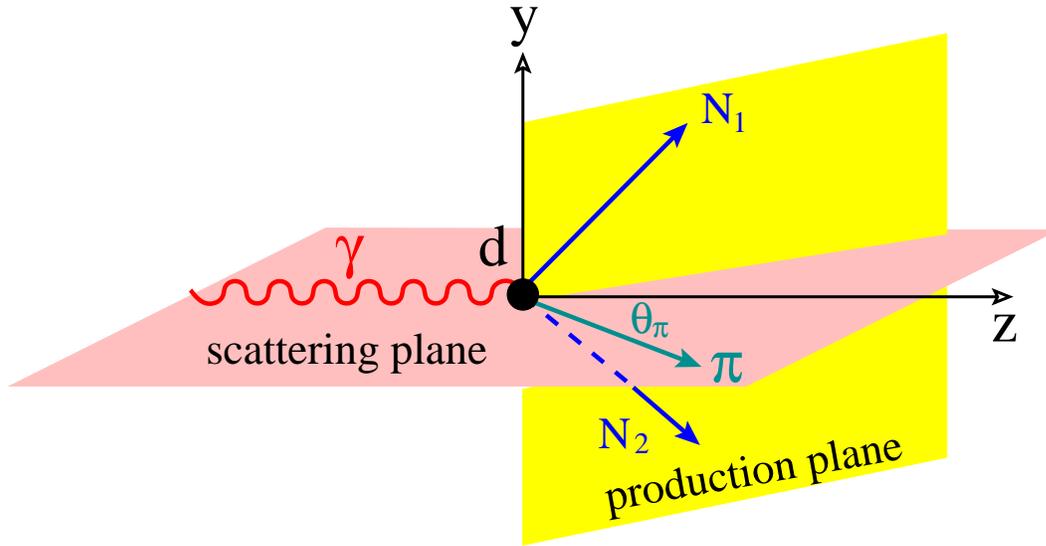}} 
  \caption{\small Kinematics in the laboratory frame for the reaction 
    $\gamma d\rightarrow \pi NN$.}
  \label{labsys}
\end{figure}

%
Calculations of total and differential cross sections are carried out in the
rest frame of the deuteron (laboratory frame) which is shown diagramatically in
Fig.\ \ref{labsys}. In this frame the relative velocity of both particles in the initial state is unity
in units of $c$.
The coordinate system is chosen to have a right-hand orientation with $z$-axis
along the momentum $\vec k$ of the incoming photon and $y$-axis parallel to
$\vec k\times\vec q$. The $x$-$z$-plane is then given by the outgoing pion,
i.e. $\phi_{\pi} = 0$, where $\phi_{\pi}$ is azimuthal angle of the pion. 

\vspace*{-0.2cm}~\\
The four-momenta of the incoming photon and deuteron
are given in the laboratory frame by $k^{\mu}$=($\omega_{\gamma},\vec{k}$) and
$d^{\mu}$=($M_d$,$\vec 0$), respectively, where $M_d$ is the deuteron
mass and $\omega_{\gamma}$ is the photon energy in the laboratory frame. 
Energy-momentum conservation gives  
\begin{eqnarray}
\vec{k} & = &  \vec{q} + \vec{p}_1 + \vec{p}_2 
\label{eq:3.4}
\end{eqnarray}
and
\begin{eqnarray}
\omega_{\gamma} + M_d & = &  \omega_{\vec q} + E_{{\vec p}_1} + E_{{\vec p}_2}
~=~ E_{\gamma d}\, ,
\label{eq:3.44}
\end{eqnarray}
where  $E_{\gamma d}$ is the total energy of the system.

\vspace*{-0.2cm}~\\
The invariant mass of the $\gamma d$-system is given by
\begin{eqnarray}
W_{\gamma d} & = &
\sqrt{\left( k^{\mu}+d^{\mu}\right)\left( k_{\mu}+d_{\mu}\right)}\nonumber\\
& = &\sqrt{(M_{d}+\omega_{\gamma})^{2}-\vec{k}^{2}} \nonumber \\
& = & \sqrt{M_{d}^{2}+2M_{d}\omega_{\gamma}}\, .
\label{eq:3.5}
\end{eqnarray}

\vspace*{-0.2cm}~\\
The final state is determined by the four momenta of the two outgoing
nucleons and the outgoing pion. Energies of these particles are
then fixed by 
\begin{eqnarray}
E_{\vec p_i} &=& \sqrt{M_{N}^{2} + \vec{p}_{i}^2}\, ,\hspace*{1cm} 
(i\in\{ 1,2\})\, , \nonumber \\
\omega_{\vec{q}} &=& \sqrt{m_{\pi}^{2} + \vec{q}^2}\, .
\label{eq:3.6}
\end{eqnarray}
Finally, four-momentum conservation reduces
the number of the  
independent variables to five, out of nine. In this work, the
pion momentum $q$, its angles $\theta_{\pi}$ and $\phi_{\pi}$, the polar angle
$\theta_{p_r}$ and the azimuthal angle 
$\phi_{p_r}$ of the relative momentum $\vec p_r$ of the two outgoing nucleons are selected
as independent variables. We prefer this choice of
variables, because in this case the kinematical factor, i.e. the phase space
factor, does not have any singularities on the boundary of the available phase
space, when $p_r\rightarrow 0$ (see also \cite{Schmidt9695}). This means in 
particular that with another selection of the free kinematical variables 
singularities in the phase space, i.e.\ kinematical factor, at the boundary of the physical region can occur (see for 
example \cite{Byckling73}).

\vspace*{-0.2cm}~\\
The total and relative momenta of the final $NN$-system are given 
respectively by
\begin{eqnarray}
\vec{P} = \vec{p}_{1} + \vec{p}_{2}
\label{eq:3.7}
\end{eqnarray}
and
\begin{eqnarray}
\vec p_r = \frac{1}{2}\left(\vec{p}_{1} - \vec{p}_{2}\right)\, .
\label{eq:3.8}
\end{eqnarray}
The inverted relations read
\begin{eqnarray}
\vec{p}_1 & = & \frac{1}{2}\vec{P} + \vec{p}_{r}\\
\vec{p}_2 & = & \frac{1}{2}\vec{P} - \vec{p}_{r}\, .
\label{p12}
\end{eqnarray}
The absolute value of the relative momentum $\vec p_r$ is given by 
\begin{eqnarray}
p_r = \frac{1}{2}\sqrt{ \frac{E_{NN}^{2}(W_{NN}^{2}-4
M_{N}^{2}) }{E_{NN}^{2}-P^{2}\cos^{2}\theta_{Pp_r}} }\, ,
\label{eq:3.11}
\end{eqnarray}
where
\begin{eqnarray}
E_{NN} & = & \omega_{\gamma}+M_{d}-\omega_{\vec{q}} ~ = ~
E_{{\vec p}_1}+E_{{\vec p}_2}~ , \nonumber \\
W_{NN}^{2} & = & E_{NN}^{2}-P^{2}
\label{eq:3.12}
\end{eqnarray}
and $\theta_{Pp_r}$ is the angle between $\vec{P}$ and $\vec p_r$. The absolute
value of the total momentum $P$=$\mid\hspace*{-0.13cm}\vec
P\hspace*{-0.12cm}\mid$ is given from Eqs.\ (\ref{eq:3.4})  and (\ref{eq:3.7})
by 
\begin{eqnarray}
P & = & \mid\vec k - \vec q\mid\, .
\label{eq:3.112}
\end{eqnarray}

\vspace*{-0.2cm}~\\
The main features of the processes (\ref{eq:3.1}) will be investigated by
considering the partially integrated differential cross sections $d^3\sigma
/(d\Omega_{\pi} dq)$ and $d^2\sigma /d\Omega_{\pi}$, which are obtained from
the fully exclusive cross section 
\begin{eqnarray}
\frac{d^5\sigma}{d\Omega_{p_r} d\Omega_{\pi} dq} = \frac{\rho_{s}}{6}
\sum_{sm_st,m_{\gamma}m_{d}} |\mathcal M^{(t\mu)}_{sm_s,m_{\gamma}m_d}|^{2}
\label{eq:3.13}
\end{eqnarray}
by appropriate integration. The phase space factor $\rho_{s}$ depends on the
selection of the five independent variables in Eq.\ (\ref{eq:3.2}). It is
expressed in terms of relative and total momenta of the two final nucleons as
follows  
\begin{eqnarray}
\rho_{s} & = & \frac{1}{(2\pi)^{5}}\frac{p_r^{2}M_{N}^{2}}
{\left| E_{\vec p_2} (p_r+\frac{1}{2} P 
\cos\theta_{Pp_r}) + E_{\vec p_1}
(p_r-\frac{1}{2} P \cos\theta_{Pp_r}) \right| }
\frac{q^{2}}{16\omega_{\gamma}M_{d}\omega_{\vec{q}}} \, ,
\label{eq:3.14}
\end{eqnarray}
whereas $q$ between $0$ and $q_{max}$. Since the limit value of both the
numerator and the denominator are zero when $p_r \rightarrow 0$, we can apply
the L'Hospital's rule to obtain the exact limit value for the phase space as
follows   
\begin{eqnarray}
\lim_{p_r \rightarrow 0} \rho_{s} & = &
\lim_{p_r \rightarrow 0} \frac{
\frac{\partial}{\partial p_r} p_r^{2} } {
\frac{\partial}{\partial p_r}\left|E_{\vec{p}_{2}}(p_r+\frac{1}{2}
P \cos\theta_{Pp_r}) +
E_{\vec{p}_{1}}(p_r-\frac{1}{2} P
\cos\theta_{Pp_r}) \right| } \nonumber \\
& & \hspace*{1cm}\times~\frac{1}{(2\pi)^{5}}\frac{M_N^2q^{2}}{16\omega_{\gamma}M_{d}\omega_{\vec{q}}}\nonumber \\
& = & 0\, .
\label{eq:3.15}
\end{eqnarray}
The existence of this limit value ensures that the phase space factor has no
singularities on the boundary of the physical region.
%
\section{Two-Nucleon Wave Function\label{chap:3:3}}
%
Using a covariant normalization (see appendix \ref{appendixC:1}) the deuteron
wave function in the momentum space is given in the rest frame of the deuteron
by 
\begin{eqnarray}\label{deuteron1}
\Psi_{dm_{d}}(\vec{p}) & = & \frac{E_{\vec{p}}}{M_{N}}~\sum_{m_s}
\chi_{1m_{s}}~\tilde{\Psi}_{m_{d}m_{s}}(\vec{p}) \, ,
\end{eqnarray}
where
\begin{eqnarray}
\tilde{\Psi}_{m_{d}m_{s}}(\vec{p}) = (2\pi)^{\frac{3}{2}}\sqrt{2M_{d}}
\sum_{\stackrel{L=0,2}{m_{L}}} i^{L}u_{L}(p)Y_{Lm_{L}}(\hat{p}) 
(Lm_{L}1m_{s}|1m_{d})\, .
\end{eqnarray}
 
\vspace*{-0.2cm}~\\
For the radial wave functions $u_L(p)$ we use the parametrization of the Bonn potential (full model) \cite{Machleidt8789} (see
appendix \ref{appendixD})
\begin{eqnarray}
u_{0}(p) & = & \sqrt{\frac{2}{\pi}}\,\sum_{i=1}^{n_{u}}\frac{C_{i}}
{p^{2}+m_{i}^{2}} \, , \\
u_{2}(p) & = & \sqrt{\frac{2}{\pi}}\,\sum_{i=1}^{n_{w}}
\frac{D_{i}} {p^{2}+m_{i}^{2}}\, .
\end{eqnarray}

\vspace*{-0.2cm}~\\
In case of the impulse approximation (see section \ref{chap:3:4:1}), we use
for the $NN$ final state a complete antisymmetric $NN$ plane wave.  
For the total spin and isospin part of the two nucleon wave
functions we use a coupled basis ($|s m_{s}\rangle |t m_{t}\rangle $). The
corresponding antisymmetric final $NN$ wave function can formally be written as
\begin{equation}
|\vec{p}_{1}\vec{p}_{2},s m_{s},t m_{t} \rangle =
  \frac{1}{\sqrt{2}}\left(
    |\vec{p}_{1}\rangle^{(1)}|\vec{p}_{2}\rangle^{(2)} - (-)^{s+t}
    |\vec{p}_{2}\rangle^{(1)}|\vec{p}_{1}\rangle^{(2)}\right)|s
  m_{s}\rangle |t m_{t}\rangle\, ,
\label{nn_anti}
\end{equation}
where $\ket{\vec{p}_{i}}^{(j)}$ 
describes a state in which a nucleon "$j$" has the momentum
$\vec{p}_{i}$. Only the $t = 1$ channel contributes in the case of charged
pions whereas for $\pi^{0}$ production both $t = 0$ and $t = 1$ channels have
to be taken into account.
%
\section{Matrix Elements for $\gamma d\rightarrow \pi NN$ \label{chap:3:4}}
\begin{figure}[ht]
  \centerline{\epsfxsize=9cm
    \epsffile{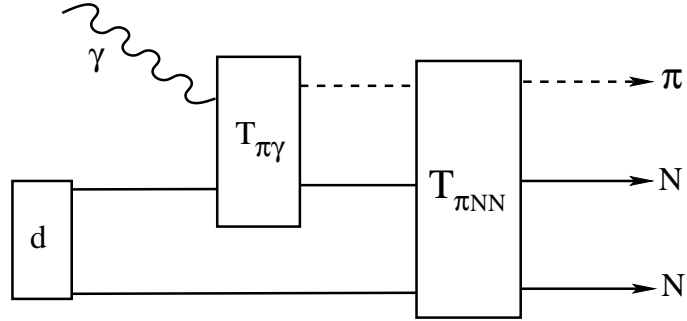}}
   \caption{\small Diagramatic representation of the $\gamma d\rightarrow \pi
    NN$ amplitude including rescattering in the final state and neglecting all 
    contributions of two-body currents.}
  \label{all_prozess}
\end{figure}

\begin{figure}[ht]
  \centerline{\epsfxsize=13cm \epsffile{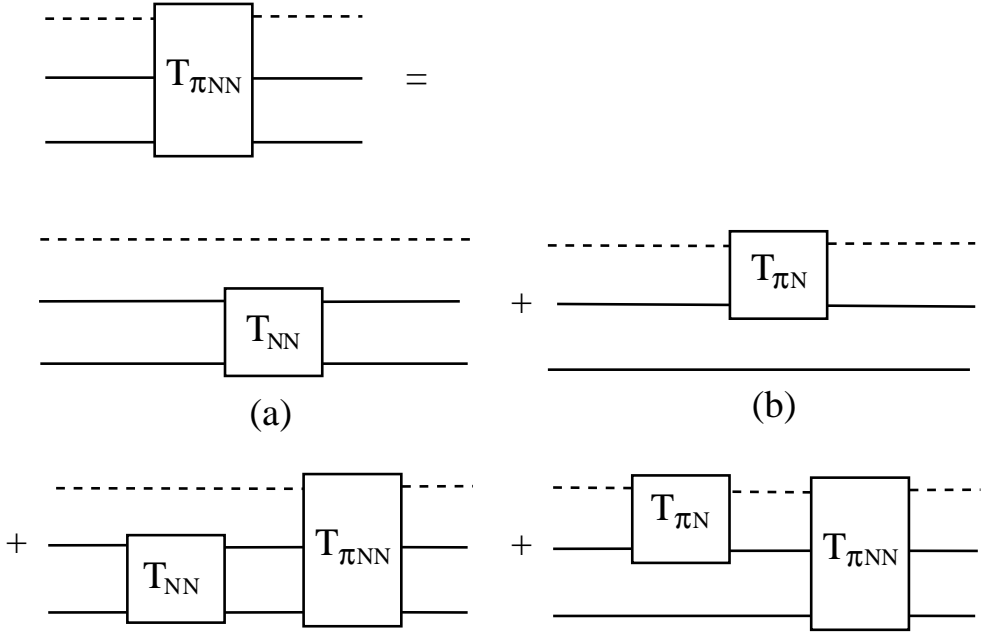}}
  \caption{\small Diagramatic representation of final state interactions of
    the reaction $\gamma d\rightarrow \pi NN$: (a) and (b) are "driving terms"
    from $NN$ and $\pi N$ rescattering in the final state, respectively.}
  \label{fsi_only}
\end{figure}

%
The transition matrix elements $\mathcal M^{(t\mu)}_{sm_s,m_{\gamma}m_d}$ are
calculated in the frame of time-ordered perturbation theory, using the
elementary operator introduced in chapter \ref{chap:2} and including $NN$ and
$\pi N$ rescattering in the final state. Scattering reactions are described in
terms of probability amplitudes relating 
the initial and final asymptotic states of the combined system of projectile
and target. In this section we consider the transition matrix element
$\mathcal M$ of Eq.\ (\ref{eq:3.2}) which describes the dynamics of the
processes (\ref{eq:3.1}). These processes are diagramatically shown in
Fig.\ \ref{all_prozess}. There, all contributions of two-body currents are
neglected. All possible rescattering diagrams are shown separately in
Fig.\ \ref{fsi_only}. The $\mathcal M$-matrix has the general form
\begin{eqnarray}\label{general}
\mathcal M  & = &
\bra{\pi
  NN;\vec{q}\vec{p_1}\vec{p_2},\alpha}\epsilon_{\mu}J^{\mu}(0)\ket{\gamma
  d;\vec{k}\vec{d},\beta}\, , 
\end{eqnarray}
where $J^{\mu}(0)$ is the current operator, $\alpha$ and $\beta$ stand for the
quantum numbers of the states asymptotically.

\vspace*{-0.2cm}~\\
We include in this work besides the pure impulse approximation (IA), the
driving terms from $NN$- and $\pi N$-rescattering, so that the full transition
matrix element reads   
\begin{eqnarray}
\label{three}
\mathcal M^{(t\mu)}_{sm_s,m_{\gamma}m_d} & = &
\mathcal M_{sm_s,m_{\gamma}m_d}^{(t\mu)~IA} + 
\mathcal M_{sm_s,m_{\gamma}m_d}^{(t\mu)~NN} + 
\mathcal M_{sm_s,m_{\gamma}m_d}^{(t\mu)~\pi N}\, , 
\end{eqnarray}
in an obvious notation. 

\vspace*{-0.2cm}~\\
Now, we will consider successively all the three terms of Eq.\ (\ref{three}),
i.e., the impulse approximation, the $NN$ final state interaction and the $\pi
N$ rescattering in more detail.
%
\subsection{The Impulse Approximation (IA)\label{chap:3:4:1}}
\begin{figure}[htb]
  \centerline{\epsfxsize=15cm  \epsffile{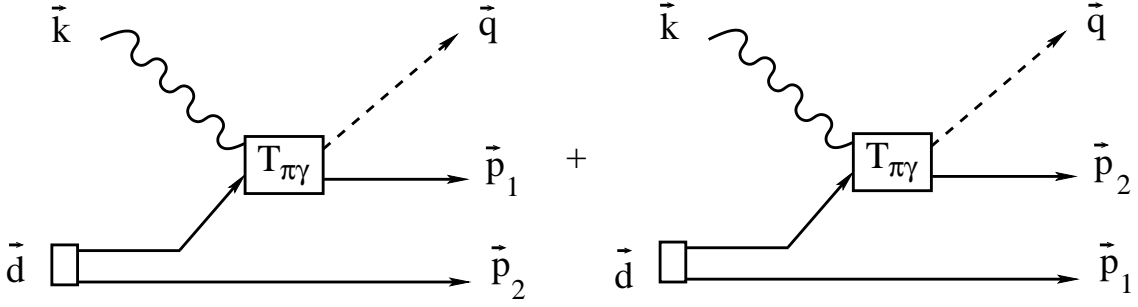}}
  \caption{\small Feynman diagrams for the reaction $\gamma d\rightarrow \pi
    NN$ in the impulse approximation.}
  \label{ia}
\end{figure}

%
In order to qualitatively explain the approximations concerning the
rescattering terms which are of important interest in this work and discussed
below, we would like to demonstrate here some features of the reaction
amplitude keeping in (\ref{three}) only the IA term. In case of the
$\gamma d\rightarrow\pi NN$ reaction, the impulse approximation leads to the
so-called spectator nucleon model. The Feynman diagrams of this model are
shown in Fig.\ \ref{ia}. The corresponding transition matrix ${\mathcal
  M}^{IA}$ is then given by 
\begin{eqnarray}
\mathcal M^{IA} = {T}_{\pi\gamma}^{(1)}\eins^{(2)}
+\eins^{(1)}{T}_{\pi\gamma}^{(2)}\, ,
\label{g17}
\end{eqnarray}
where the upper index indicates on which of the nucleons the elementary
production operator acts. The operator ${\mathcal M}^{IA}$ contains
contributions of pure single nucleon terms.

\vspace*{-0.2cm}~\\
Using the initial and final states from section \ref{chap:3:3} and the
operator of the pion photoproduction in the two-nucleon space from
Eq.\ (\ref{g17}) we can write now the ${\mathcal M}^{IA}$-matrix element 
using a covariant normalization for all particles (see appendix \ref{appendixC:1}) as follows 
\begin{eqnarray}
\label{volle t-matrix}
\bra{\vec{p}_{1}\vec{p}_{2},s\,m_{s},t,-\mu}{\mathcal
  M}^{IA}(\vec{k},m_{\gamma},\vec{q},\mu) \ket{\vec{d},\,m_{d}} & = & 
\frac{1}{2}\int \frac{d^{3}p_1^{\prime}}{(2\pi)^{3}}\int
  \frac{d^{3}p_2^{\prime}} 
{(2\pi)^{3}}\,\frac{M_{N}}{E_{\vec p^{\,\prime}_1}}
\frac{M_{N}}{E_{\vec p^{\,\prime}_2}} \nonumber \\
 & &\hspace*{-4cm}
\times \sum_{m_{s^{\prime}}}\bra{\vec{p_1}\vec{p_2},s\,m_{s},t,-\mu}{\mathcal
  M}^{IA}(\vec{k},m_{\gamma},\vec{q},\mu) \ket{\vec p^{\,\prime}_1
\vec p^{\,\prime}_2,1m_{s^{\prime}},00}  \nonumber \\
& & \hspace*{-4cm}\times ~\langle\vec p^{\,\prime}_1\vec p^{\,\prime}_2,1m_{s^{\prime}},\,00|\vec{d},m_d 
\rangle\, .
\end{eqnarray}

\vspace*{-0.2cm}~\\
In the laboratory frame one finds for the matrix
element (\ref{volle t-matrix}) the following expression
\begin{eqnarray}
\label{volle t-matrix1}
\lefteqn{\bra{\vec{p_{1}}\vec{p_{2}},s\,m_{s},t,-\mu}{\mathcal M}^{IA}
(\vec{k},m_{\gamma},\vec{q},\mu)\ket{\vec{d}=0, m_{d},\,00} =} \nonumber \\
& &
 \frac{1}{\sqrt{2}}\sum_{m_{s^{\prime}}}
\bra{s\,m_{s},t,-\mu} \left[\left(\bra{\vec{p}_{1}}
{T}^{(1)}_{\pi\gamma}\ket{-\vec{p}_{2}}-(-)^{s+t}\bra{
\vec{p}_{1}}{T}^{(2)}_{\pi\gamma}\ket{-\vec{p}_{2}}\right)
\tilde{\Psi}_{m_{s^{\prime}}\,m_d}(\vec{p}_{2})  \right. \nonumber \\
& & \, \; \;
+\left.\left(\bra{\vec{p}_{2}}
{T}^{(2)}_{\pi\gamma}\ket{-\vec{p}_{1}}-(-)^{s+t}\bra{
\vec{p}_{2}}{T}^{(1)}_{\pi\gamma}\ket{-\vec{p}_{1}}\right)
\tilde{\Psi}_{m_{s^{\prime}}\,m_d}(\vec{p}_{1})\right]
\ket{1m_{s^{\prime}},00}\, .
\end{eqnarray}
Note that the upper index is maintained on ${T_{\pi\gamma}}$. The reason for 
that lies in the fact that the ${T_{\pi\gamma}}$-matrix contains spin 
and isospin operators, which still act on the spinors of the $NN$ system. 
Using the symmetry properties 
\begin{eqnarray}
\bra{s\,m_{s}}\eins^{(1)}\,\sigma_{\nu}^{(2)}\ket{s^{\prime}
\,m_{s^{\prime}}} =
(-)^{s-s^{\prime}} \bra{s\,m_{s}}\sigma_{\nu}^{(1)}\,\eins^{(2)}
\ket{s^{\prime}\,m_{s^{\prime}}} \, ,
\end{eqnarray}
which applies also likewise to the $\tau$-matrices, the matrix element of
  Eq.\ (\ref{volle t-matrix1}) is given by the following expression
\begin{eqnarray}
\label{g16}
\mathcal M^{(t\mu)~IA}_{sm_s,m_{\gamma}m_d}  & = &
  \langle \vec{p}_{1}\vec{p}_{2},s m_{s},t,-\mu |\mathcal
  M^{IA}(\vec{k},m_{\gamma},\vec{q},\mu)| 
  \vec{d}=0,m_{d},\,00 \rangle \nonumber \\
 & = & \sqrt{2}\sum_{m_{s^{\prime}}}\langle s
  m_{s},t,-\mu| \Big( \langle
  \vec{p}_{1}| {T}_{\pi\gamma}^{(1)} | - \vec {p}_{2} \rangle
  \tilde{\Psi}_{m_{s^{\prime}},m_{d}}(\vec {p}_{2}) \nonumber \\ 
& &  -~(-)^{s+t} \langle
  \vec{p}_{2}| {T}_{\pi\gamma}^{(1)}|-\vec{p}_{1}\rangle
  \tilde{\Psi}_{m_{s^{\prime}},m_{d}}(\vec{p}_{1})\Big) |1
  m_{s^{\prime}},00\rangle \, . 
\end{eqnarray}

\vspace*{-0.2cm}~\\
The evaluation of the spin and isospin operators becomes very simple, 
because they act now on the nucleon "1" only. Using the
Wigner-Eckart-theorem (see for example \cite{Messiah64}) one finds 
\begin{eqnarray}
\bra{sm_{s}}\sigma^{(1)}_{\nu}\ket{1m_{s^{\prime}}} = (-)^{s-1}~3\sqrt{2}
~(1\nu\,1m_{s^{\prime}}|sm_{s})\left\{ \begin{array}{ccc}
\frac{1}{2} & 1 & \frac{1}{2} \\
1 & \frac{1}{2} & s
\end{array} \right\}\, ,
\end{eqnarray}
\begin{eqnarray}
\bra{tm_{t}}\tau^{\dagger\,(1)}_{\mu}\ket{00} & = & (-)^{\mu}
~\delta_{t,1}\delta_{m_t,-\mu}\, .
\end{eqnarray}
%
\subsection{The $NN$ Final State Interaction\label{chap:3:4:2}}
\begin{figure}[htb]
  \centerline{\epsfxsize=13cm  \epsffile{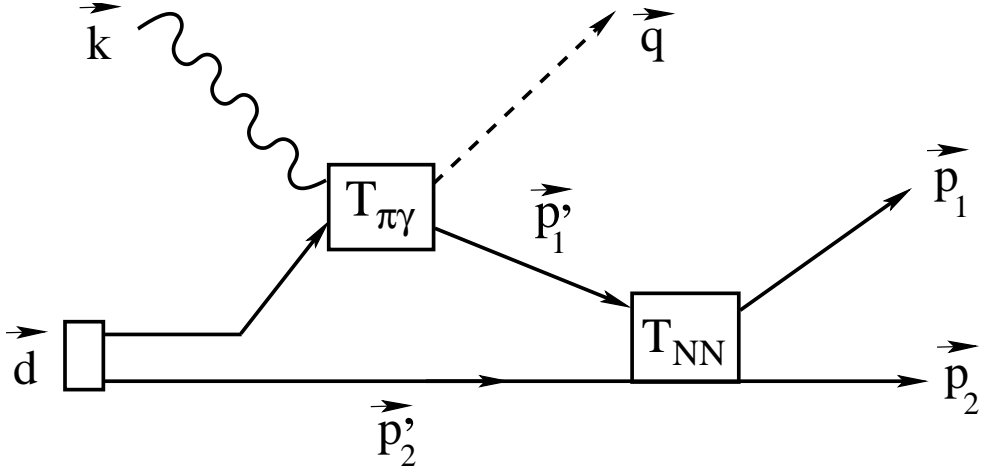}}
  \caption{\small Feynman diagram for the reaction $\gamma d\rightarrow \pi NN$
    including $NN$-rescattering in the final state.}
  \label{nn-fsi}
\end{figure}

%
Now we will consider in addition $NN$-rescattering in the
final state. The corresponding matrix element is given by the second term in
Eq.\ (\ref{three}). The Feynman diagram for this process is shown in
Fig.\ \ref{nn-fsi} for the case when the production operator
acts on nucleon "1". Note that the
second diagram when the operator acts on nucleon "2" is also taken into
account in the calculation. 

\vspace*{-0.2cm}~\\
The transition matrix element with $NN$-rescattering of the diagram in Fig.\
\ref{nn-fsi} is given by
\begin{eqnarray}
\label{tnn-fsi}
\mathcal M^{(t\mu)~NN}_{sm_s,m_{\gamma}m_d} & = & 
\bra{\vec{q},\vec{p_1}\vec{p_2};\mu,sm_s,t,-\mu}\mathcal M^{NN}(\vec{k},
m_{\gamma})\ket{\vec{d}=0,m_d,\,00} \nonumber \\ 
& = & \frac{1}{2(2\pi)^9}\sum_{s^{\prime}m_{s^{\prime}}t^{\prime}\mu^{\prime}}
\int\hspace{-0.2cm}\int\hspace{-0.2cm}\int \frac{d^3q^{\,\prime}}{2\omega_{q^{\,\prime}}} \frac{d^3p^{\,\prime}_1}{E_{p^{\,\prime}_1}/M_N}\frac{d^3p^{\,\prime}_2}{E_{p^{\,\prime}_2}/M_N} \nonumber \\
& & \hspace{-0.3cm}\times~\bra{\vec{q},\vec{p_1}\vec{p_2};\mu,sm_{s},t,-\mu}
\mathcal R_{NN}(W_{NN})\ket{\vec q^{\,\prime},\vec p^{\,\prime}_1 
  \vec p^{\,\prime}_2;\mu^{\prime},s^{\prime}m_{s^{\prime}},
t^{\prime},-\mu^{\prime}} \nonumber \\ 
& & \hspace{-0.3cm}\times ~\mathcal G_{0NN}^{\pi NN}(E_{\gamma d},
\vec q,\vec p^{\,\prime}_1,\vec p^{\,\prime}_2)\nonumber \\
& & \hspace{-0.3cm}\times~\bra{\vec q^{\,\prime},\vec p^{\,\prime}_1
  \vec p^{\,\prime}_2;\mu^{\,\prime},s^{\prime}m_{s^{\prime}},
t^{\prime},-\mu^{\prime}} 
\mathcal M^{IA}(\vec{k},m_{\gamma})\ket{\vec{d}=0, m_{d},\,00}\, .
\end{eqnarray}
Here $\mathcal R_{NN}$ denotes the half-off-shell $NN$-scattering matrix, 
$\mathcal G_{0NN}^{\pi NN}(E_{\gamma d},\vec q,\vec p^{\,\prime}_1,
\vec p^{\,\prime}_2)$ is the free $\pi NN$ propagator 
and $\mathcal M^{IA}$-matrix is given in Eq.\ (\ref{g16}).
%
\subsubsection{3.3.2.1 The Half-Off-Shell $NN$-Scattering
  Matrix\label{chap:3:4:2:1}}
%
The $NN$ dynamics in the final state is determined by the half-off-shell
$NN$-scattering amplitude $\mathcal R_{NN}$. In the 
presence of a spectator meson it is given by 
introducing the relative and total momenta $\vec p_r$ and $\vec P$,
respectively, of the two outgoing nucleons (see Eqs.\ (\ref{eq:3.7}) and
(\ref{eq:3.8})) and using the basis states 
\begin{eqnarray}
\label{big_using_nn}
\langle p_r\theta_r\phi_r\ket{\vec{p}_r,sm_s,tm_t} & = & 
\sum_{JM_J\ell m_{\ell}} (sm_s\,\ell m_{\ell}\mid JM_J)~u_{\ell J}^{st}(p_r)
~Y_{\ell m_{\ell}}(\hat{p}_r)~\ket{\ell sJM_J}~\ket{tm_t}\nonumber \\
   & & 
\end{eqnarray}
by 
\begin{eqnarray}
\label{tnn-hos1}
\bra{\vec{q},\vec{p_1}\vec{p_2};\mu,sm_{s},t,-\mu} \mathcal
R_{NN}(W_{NN})\ket{\vec q^{\,\prime},\vec p^{\,\prime}_1 \vec p^{\,\prime}_2;
\mu^{\prime},s^{\prime}m_{s^{\prime}},t^{\prime},-\mu^{\prime}} & = &
2(2\pi)^9 \delta_{\mu^{\prime}\mu} \delta_{s^{\prime}s} \delta_{t^{\prime}t}\nonumber \\
& &\hspace{-10cm} \times~\omega_{\vec q}~\delta^3(\vec q^{\,\prime}-\vec q)
\delta^3(\vec P^{\,\prime}-\vec P) \sqrt{\frac{E_{\vec p_1} E_{\vec p_2} E_{\vec p^{\,\prime}_1} E_{\vec p^{\,\prime}_2}}{M_N^4}}
\mathcal~ D_{m_sm_{s^{\prime}}}^{st\mu}(W_{NN},\vec p_r,\vec p^{\,\prime}_r)\, ,
\end{eqnarray}
with 
\begin{eqnarray}
\label{tnn-hos}
\mathcal D_{m_sm_{s^{\prime}}}^{st\mu}(W_{NN},\vec p_r,\vec p^{\,\prime}_r)
 & = & \bra{\vec p_r,sm_{s},\,t-\mu} \mathcal R_{NN}(W_{NN})
 \ket{\vec{p}_r^{\,\prime},sm_{s^{\prime}},\,t-\mu} \nonumber \\
 & = & \sum_{J\ell\ell^{\prime}}
 \mathcal
 F_{\ell\ell^{\prime}\,m_sm_{s^{\prime}}}^{Js}(\hat{p}_r,\hat{p}_r^{\,\prime}) 
 ~ \mathcal T_{\ell\ell^{\prime}}^{Jst\mu}(W_{NN},p_r,p_r^{\,\prime})\, ,
\end{eqnarray}
where $\ell$ is the orbital angular momentum and $J$ is the total angular
momentum of the two-nucleon system. The initial and final relative momenta of
the two nucleons in the $NN$-subsystem are given, respectively, by
\begin{eqnarray}
\label{prelative}
\vec p^{\,\prime}_r & = & \frac{1}{2}(\vec p^{\,\prime}_1 - \vec
p^{\,\prime}_2)\, ,
\nonumber \\
\vec p_r & = & \frac{1}{2}(\vec{p_1}-\vec{p_2})\, .
\end{eqnarray}
where $\vec p^{\,\prime}_1 = \vec{p}^{\,\prime}_r + \frac{1}{2}(\vec{k}-\vec{q})$
and $\vec p^{\,\prime}_2 = -\vec{p}^{\,\prime}_r + \frac{1}{2}(\vec{k}-\vec{q})$ with the
integration variable $\vec p^{\,\prime}_r$. The function $\mathcal
F_{\ell\ell^{\prime}\,m_s m_{s^{\prime}}}^{Js}(\hat{p}_r,\hat{p}_r^{\,\prime})$ is
given by  
\begin{eqnarray}
\label{tnn-ff}
\mathcal F_{\ell\ell^{\prime}\ell\,m_sm_{s^{\prime}}}^{Js}(\hat{p}_r,\hat{p}_r^{\,\prime}) & = & 
 \sum_{M_Jm_{\ell}m_{\ell^{\prime}}} (\ell m_{\ell}\,sm_s|JM_J)(\ell^{\prime}
 m_{\ell^{\prime}}\,sm_{s^{\prime}}|JM_J) Y^{\star}_{\ell m_{\ell}}(\hat{p}_r)
 Y_{\ell^{\prime} m_{\ell^{\prime}}}(\hat{p}_r^{\,\prime})\, .\nonumber \\
& & 
\end{eqnarray}
The half-off-shell partial wave amplitudes $\mathcal
T_{\ell\ell^{\prime}}^{Jst\mu}(W_{NN},p_r,p_r^{\,\prime})$ were found for partial waves
with total angular momentum $J\le 3$, which are important in the
energy region under consideration, by numerical solution of the 
Lippmann-Schwinger (LS) equation \cite{Lippmann51}. In appendix
\ref{appendixF} we give a detailed solution of the LS equation for a given
$NN$ potential model. In the calculations presented here, these amplitudes are
obtained from the separable representation of the Paris $NN$ potential
\cite{Haidenbauer84,Haidenbauer85}, which reproduces its on-shell as well as
off-shell properties.
%
\subsubsection{3.3.2.2 The $\pi NN$ Propagator\label{chap:3:4:2:2}}
%
The free $\pi NN$ propagator is given by
\begin{eqnarray}
\label{pinn-propag}
\mathcal G_{0NN}^{\pi NN}(E_{\gamma d},\vec{q},\vec{p}_1^{\,\prime},
\vec{p}_2^{\,\prime}) & = & \frac{1}{E_{\gamma
    d}-\omega_{\pi}(\vec{q})-E_{N_1}(\vec{p}_1^{\,\prime})-
    E_{N_2}(\vec{p}_2^{\,\prime}) + i\epsilon}\, .
\end{eqnarray}
In the nonrelativistic limit it is given by
\begin{eqnarray}
\label{nonrelpropag}
\mathcal G_{0NN}^{\pi NN}(E_{\gamma d},\vec{q},\vec{p}_1^{\,\prime},
\vec{p}_2^{\,\prime}) & = & 
\frac{M_N}{\tilde{p^{\,\prime}}^2 - {p_r^{\,\prime}}^2 + i\epsilon}\, ,
\end{eqnarray}
where $\tilde{p^{\,\prime}}$ is given by
\begin{eqnarray}
\tilde{p^{\,\prime}}^2 & = & M_N\left( E_{\gamma
    d}-\omega_{\pi}(\vec{q})-2M_N-\frac{(\vec{k}-\vec{q})^2}{4M_N} \right)\, .
\end{eqnarray}

\vspace*{-0.2cm}~\\
The magnitude of the relative on-shell momentum of the two nucleons is given by
\begin{eqnarray}
p_0 & = & \sqrt{M_N(W_{NN}-2M_N)}\, .
\end{eqnarray}

\vspace*{-0.2cm}~\\
Collecting the various pieces and substituting Eqs.\ (\ref{tnn-hos1}) and
(\ref{nonrelpropag}) in Eq.\ (\ref{tnn-fsi}), we obtain the following expression for the transition matrix
element with $NN$ rescattering 
\begin{eqnarray}
\label{tnn-fsi-final}
\mathcal M^{(t\mu)~NN}_{sm_s,m_{\gamma}m_d}(\vec k,\vec q,\vec p_1,\vec p_2) & = & 
\sum_{m_{s^{\prime}}}\int d^3\vec p^{\,\prime}_r \sqrt{\frac{E_{\vec p_1}E_{\vec p_2}}{E_{\vec p^{\,\prime}_1}E_{\vec p^{\,\prime}_2}}}
~~\mathcal D_{m_sm_{s^{\prime}}}^{st\mu}(W_{NN},\vec p_r,\vec p^{\,\prime}_r)
 \nonumber
\\
& & \times ~\frac{M_N}{\tilde{p^{\,\prime}}^2 - {p_r^{\,\prime}}^2 + i\epsilon}~\mathcal M^{(t\mu)~IA}_{sm_{s^{\prime}},m_{\gamma}m_d}(\vec k,\vec q,\vec p^{\,\prime}_1,\vec p^{\,\prime}_2)\, ,
\end{eqnarray}
where $\mathcal D_{m_sm_{s^{\prime}}}^{st\mu}(W_{NN},\vec p_r,\vec p^{\,\prime}_r)$
is given by Eq.\ (\ref{tnn-hos}). The three dimensional integral over $\vec
p^{\,\prime}_r$ is carried out numerically.
%
\subsection{The $\pi N$ Final State Interaction\label{chap:3:4:3}}
\begin{figure}[htb]
  \centerline{\epsfxsize=13cm  \epsffile{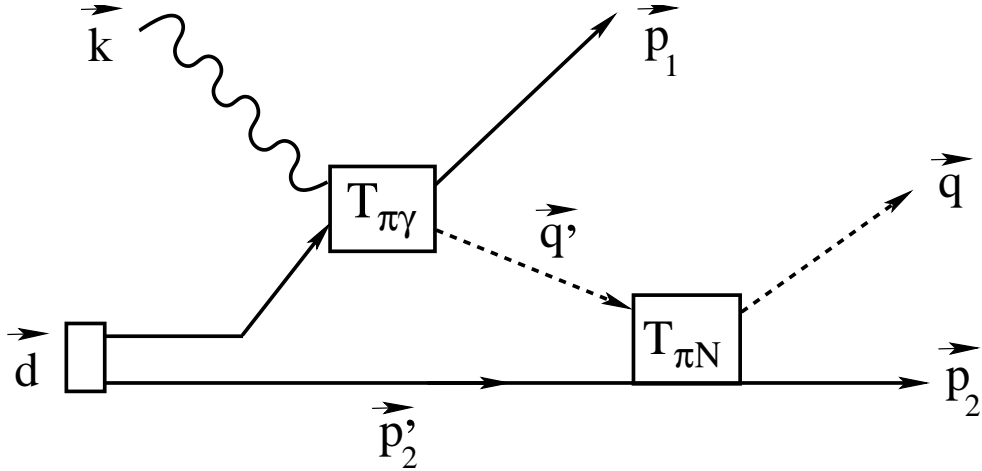}}
  \caption{\small Feynman diagram for the reaction $\gamma d\rightarrow \pi NN$
    including $\pi N$-rescattering in the final state.}
  \label{pin-fsi}
\end{figure}

%
In addition to the $NN$ final state interaction we consider also in
this work the $\pi N$-rescattering in the final state. The corresponding matrix
element in this case is given by the third term in Eq.\ (\ref{three}). The
Feynman diagram for this process is shown in Fig.\ \ref{pin-fsi} for the case
when the production operator acts only on nucleon "1". The second diagram when
the operator acts on nucleon "2" is also taken into account in the calculation.

\vspace*{-0.2cm}~\\
The transition matrix element with $\pi N$-rescattering in the final state
(see Fig.\ \ref{pin-fsi}) is given by
\begin{eqnarray}
\label{tpn-fsi}
M^{(t\mu)~\pi N}_{sm_s,m_{\gamma}m_d} & = & 
\bra{\vec q,\vec{p_1}\vec{p_2};\mu,sm_s,t,-\mu}\mathcal M^{\pi
  N}(\vec{k},m_{\gamma})\ket{\vec{d}=0,m_d,\,00} \nonumber \\ 
& = & \frac{1}{2(2\pi)^9}\sum_{s^{\prime}m_{s^{\prime}}t^{\prime}\mu^{\prime}}
\int\hspace{-0.2cm}\int\hspace{-0.2cm}\int
\frac{d^3q^{\,\prime}}{2\omega_{q^{\,\prime}}}
\frac{d^3p^{\,\prime}_1}{E_{p^{\,\prime}_1}/M_N}\frac{d^3p^{\,\prime}_2}{E_{p^{\,\prime}_2}/M_N}
\nonumber \\
& &
\hspace{-0.2cm}\times~\left\{\bra{\vec{q},\vec{p_1}\vec{p_2};\mu,sm_{s},t,-\mu}\mathcal
  R_{\pi N}(W_{\pi N}(\vec p_2))\ket{\vec q^{\,\prime},\vec p^{\,\prime}_1 \vec
    p^{\,\prime}_2;\mu^{\,\prime},s^{\prime}m_{s^{\prime}},t^{\prime},-\mu^{\prime}} \right.\nonumber \\
& & \hspace{-0.2cm} \left.\ - (-)^{s+t} (\vec p_1 \leftrightarrow \vec p_2)\right\} ~\mathcal G_{0\pi N}^{\pi NN}(E_{\gamma d},\vec
q^{\,\prime},\vec p_1,\vec p^{\,\prime}_2)\nonumber \\
& & \hspace{-0.2cm}\times~\bra{\vec q^{\,\prime},\vec p^{\,\prime}_1 
\vec p^{\,\prime}_2;\mu^{\prime},s^{\prime}m_{s^{\prime}},t^{\prime},-\mu^{\prime}}
\mathcal M^{IA}(\vec{k},m_{\gamma})\ket{\vec{d}=0, m_{d},\,00}\, ,
\end{eqnarray}
where $\mathcal R_{\pi N}$ denotes the half-off-shell $\pi N$-scattering
matrix at the invariant mass 
\begin{eqnarray}
W_{\pi N}(\vec p_2) & = & \sqrt{(E_{\vec p_2} + \omega_{\vec q})^2 - (\vec p_2 + \vec q)^2}
\end{eqnarray}
of the $\pi N$ subsystem, $\mathcal G_{0\pi N}^{\pi NN}(E_{\gamma d},\vec q^{\,\prime},\vec
p_1,\vec p^{\,\prime}_2)$ is the free $\pi NN$ propagator (see Eq.\
(\ref{pinn-propag})) and $\mathcal M^{IA}$-matrix is given in Eq.\
(\ref{g16}).
%
\subsubsection{3.3.3.1 The Half-Off-Shell $\pi N$-Scattering
  Matrix\label{chap:3:4:3:1}}
%
Using the coupled basis states
\begin{eqnarray}
\label{basis_pin}
\ket{\vec{q},\vec{p_1}\vec{p_2};\mu,sm_{s},t,-\mu} & = &
\sum_{\stackrel{m_{s_1}m_{t_1}}{m_{s_2}m_{t_2}}}
(\frac{1}{2}m_{s_1}\frac{1}{2}m_{s_2}\mid sm_s)
(\frac{1}{2}m_{t_1}\frac{1}{2}m_{t_2}\mid t\,-\mu) \nonumber \\
& & \times~\ket{\vec{q},\vec{p_1},\vec{p_2};1\mu,\frac{1}{2}m_{s_1},
\frac{1}{2}m_{t_1},\frac{1}{2}m_{s_2},\frac{1}{2}m_{t_2}}\, ,
\end{eqnarray}
the half-off-shell $\pi N$-scattering amplitude $\mathcal R_{\pi N}$ is given
in the presence of a spectator nucleon by  
\begin{eqnarray}
\label{tpn-hos1}
\bra{\vec{q},\vec{p_1}\vec{p_2};\mu,sm_{s},t,-\mu} 
\mathcal {R}_{\pi N}(W_{\pi N}(\vec p_2))\ket{\vec q^{\,\prime},\vec p^{\,\prime}_1 
\vec
p^{\,\prime}_2;\mu^{\prime},s^{\prime}m_{s^{\prime}},t^{\prime},-\mu^{\prime}}
& = & (2\pi)^3\frac{E_{\vec p_1}}{M_N}\nonumber \\
& & \hspace*{-9cm}\times~\delta^3(\vec p^{\,\prime}_1-\vec p_1)\sum_{m_{s_2}m_{t_2}m_{s^{\prime}_2}m_{t^{\prime}_2}}
\mathcal{A}_{t\mu,t^{\prime}\mu^{\prime}}^{sm_ss^{\prime}m_{s^{\prime}}} 
(m_{s_2},m_{t_2},m_{s^{\prime}_2},m_{t^{\prime}_2}) \nonumber \\
& & \hspace*{-9cm}\times~\bra{\vec{q},\vec{p_2};1\mu,\frac{1}{2}m_{s_2},\frac{1}{2}m_{t_2}}\mathcal {R}_{\pi N}(W_{\pi N}(\vec p_2))
\ket{\vec q^{\,\prime},\vec
  p^{\,\prime}_2;1\mu^{\prime},\frac{1}{2}m_{s^{\,\prime}_2},\frac{1}{2}m_{t^{\,\prime}_2}} \, , \nonumber \\
& &
\end{eqnarray}
where $m_{s_2}$ ($m_{s^{\,\prime}_2}$) and $m_{t_2}$ ($m_{t^{\,\prime}_2}$)
are the spin and isospin projections of the final (initial) nucleon which is
interacting with the pion in the $\pi N$ subsystem, respectively. The function
$\mathcal A$ is given by
\begin{eqnarray}
\label{big_a}
\mathcal{A}_{t\mu,t^{\prime}\mu^{\prime}}^{sm_ss^{\prime}m_{s^{\prime}}}
(m_{s_2},m_{t_2},m_{s^{\prime}_2},m_{t^{\prime}_2}) & = & \sum_{m_{s_1}m_{t_1}}
  \delta_{m_{s^{\prime}_1}m_{s_1}}
\delta_{m_{t^{\prime}_1}m_{t_1}} \nonumber \\
& & \hspace*{-1.0cm}\times~(\frac{1}{2}m_{s_1}\frac{1}{2}m_{s_2}\mid sm_s)
(\frac{1}{2}m_{t_1}\,\frac{1}{2}m_{t_2}\mid t\,-\mu)\nonumber \\
& & \hspace*{-1.0cm}\times~(\frac{1}{2}m_{s^{\prime}_1}\,\frac{1}{2}m_{s^{\prime}_2}\mid
  s^{\prime}m_{s^{\prime}})
(\frac{1}{2}m_{t^{\prime}_1}\,\frac{1}{2}m_{t^{\prime}_2}\mid
 t^{\prime}\,-\mu^{\prime})\, ,
\end{eqnarray}
where $m_{s_1}$ and $m_{t_1}$ are the spin and isospin projections of the
spectator nucleon, respectively. 

\vspace*{-0.2cm}~\\
Coupling the pion and nucleon isospins into the total isospin $\tilde{t}$ of
the pion-nucleon pair according to
\begin{eqnarray}
\label{basis_single}
\ket{\vec{q},\vec{p_2};1\mu,\frac{1}{2}m_{s_2},\frac{1}{2}m_{t_2}} & = & 
\sum_{\tilde{t}\,\tilde{m}_t}(1\mu\,\frac{1}{2}m_{t_2}\mid\tilde{t}\tilde{m}_t)
~\ket{\vec{q},\vec{p_2};\tilde{t}\tilde{m}_t,\frac{1}{2}m_{s_2},\frac{1}{2}m_{t_2}}
\end{eqnarray}
and introducing the relative and total momenta of the final (initial)
pion-nucleon subsystem which are given, respectively, by
\begin{eqnarray}
\label{pinprelative}
\tilde{\vec p_r} & = & \frac{M_N\vec q-m_{\pi}\vec p_2}{M_N+m_{\pi}}
~~\left(\tilde{\vec p^{\,\prime}_r}~ =~ \frac{M_N\vec q^{\,\prime}-m_{\pi}\vec p^{\,\prime}_2}{M_N+m_{\pi}}\right) 
\end{eqnarray} 
and
\begin{eqnarray}
\label{pinptotal}
\tilde{\vec P} & = & \vec q+\vec p_2
~~\left(\tilde{\vec{P}^{\,\prime}}~ =~ \vec{q}^{\,\prime}+\vec p^{\,\prime}_2\right) 
\end{eqnarray}
we can re-write Eq.\ (\ref{tpn-hos1}) in the form
\begin{eqnarray}
\label{tpn-hos2}
\bra{\vec{q},\vec{p_1}\vec{p_2};\mu,sm_{s},t,-\mu} 
\mathcal {R}_{\pi N}(W_{\pi N}(\vec p_2))\ket{\vec q^{\,\prime},\vec p^{\,\prime}_1 
\vec
p^{\,\prime}_2;\mu^{\prime},s^{\prime}m_{s^{\prime}},t^{\prime},-\mu^{\prime}}
& = &2(2\pi)^9\frac{E_{\vec{p}_1}}{M_N}\nonumber\\ 
& & \hspace*{-8cm}\times~\sqrt{\frac{E_{\vec{p}_2}E_{\vec p^{\,\prime}_2}}{M_N^2}}\sqrt{\omega_{\vec q}\omega_{\vec q^{\,\prime}}}\delta^3(\vec p_1-\vec p^{\,\prime}_1)
\delta^3(\tilde{\vec P}-\tilde{\vec P^{\,\prime}}) \nonumber \\
& & \hspace*{-8cm}\times~\sum_{m_{s_2}m_{t_2}m_{s^{\prime}_2}m_{t^{\prime}_2}}
\mathcal{A}_{t\mu,t^{\prime}\mu^{\prime}}^{sm_ss^{\prime}m_{s^{\prime}}}
(m_{s_2},m_{t_2},m_{s^{\prime}_2},m_{t^{\prime}_2}) \nonumber \\ 
& & \hspace*{-8cm}\times~ \sum_{\tilde{t}\tilde{m}_{t}\,\tilde{t^{\prime}}\tilde{m}_{t^{\prime}}}
\mathcal{B}_{\mu\mu^{\prime}\tilde{t}\tilde{t^{\prime}}}^{\tilde{m}_t\tilde{m}_{t^{\prime}}}(m_{t_2},m_{t^{\prime}_2})
\nonumber \\
& & \hspace*{-8cm}\times~ \bra{\tilde{\vec
  p_r},\frac{1}{2}m_{s_2},\tilde{t}\tilde{m}_{t}} \tilde{\mathcal{R}}_{\pi N}(W_{\pi N}(\vec p_2))
\ket{\tilde{\vec{p}_r^{\,\prime}},\frac{1}{2}m_{s^{\prime}_2},\tilde{t}^{\prime}\tilde{m}_{t^{\prime}}}\,.
\end{eqnarray}
The function $\mathcal B$ is given by
\begin{eqnarray}
\label{big_b}
\mathcal{B}_{\mu\mu^{\prime}\tilde{t}
\tilde{t^{\prime}}}^{\tilde{m}_t\tilde{m}_{t^{\prime}}}
(m_{t_2},m_{t^{\prime}_2}) & = &  (1\mu\,\frac{1}{2}m_{t_2}\mid \tilde{t}\tilde{m}_t)\,
(1\mu^{\prime}\,\frac{1}{2}m_{t^{\prime}_2}\mid \tilde{t^{\prime}}
\tilde{m}_{t^{\prime}}) \, .
\end{eqnarray}

\vspace*{-0.2cm}~\\
Using the partial wave expansion 
\begin{eqnarray}
\label{big_using}
\langle \tilde{p}_r \theta\phi \ket{\vec{\tilde{p}}_r,\frac{1}{2}m_{s_2},\tilde{t}\tilde{m}_{t}} & = & 
\sum_{\tilde{J}\tilde{M}_J\tilde{\ell}\tilde{m}_{\ell}} 
(\frac{1}{2}m_{s_2}\,\tilde{\ell}\tilde{m}_{\ell}\mid\tilde{J}\tilde{M}_J)
~\tilde{u}_{\tilde{\ell}\tilde{J}}^{\tilde{t}}(\tilde{p}_r)~\nonumber \\
& & \times~
Y_{\tilde{\ell}\tilde{m}_{\ell}}(\hat{\tilde{p}}_r)~\ket{\tilde{\ell}\frac{1}{2}\,\tilde{J}\tilde{M}_J}~\ket{\tilde{t}\tilde{m}_{t}}\, ,
\end{eqnarray}
where $\tilde{\ell}$, $\tilde{J}$ and $\tilde{M}_J$ denote the relative orbital angular momentum, the
total angular momentum and its $z$-axis projection of the $\pi N$ system,
respectively, we obtain  
\begin{eqnarray}
\label{tpn-vfinal}
\bra{\tilde{\vec
  p_r},\frac{1}{2}m_{s_2},\tilde{t}\tilde{m}_{t}} \tilde{\mathcal{R}}_{\pi N}(W_{\pi N}(\vec p_2))
\ket{\tilde{\vec{p}_r^{\,\prime}},\frac{1}{2}m_{s^{\prime}_2},\tilde{t}^{\prime}\tilde{m}_{t^{\prime}}}
& = &  \mathcal{Q}^{\tilde{t}\tilde{t^{\prime}}}_{\tilde{m}_t\tilde{m}_{t^{\prime}}}(W_{\pi N}(\vec p_2),\tilde{\vec{p}_r},\tilde{\vec{p}^{\,\prime}_r}) \nonumber \\
& & \hspace*{-6cm} ~=~\sum_{\tilde{J}\tilde{M}_J\tilde{\ell}\tilde{m}_{\ell}}
\delta_{\tilde{t}^{\prime}\tilde{t}}
\delta_{\tilde{m}_{t^{\prime}}\tilde{m}_{t}}
\delta_{\tilde{\ell}^{\prime}\tilde{\ell}}
\delta_{\tilde{m}_{\ell^{\prime}}\tilde{m}_{\ell}}
\delta_{\tilde{J}^{\prime}\tilde{J}}\delta_{\tilde{M}_{J^{\prime}}\tilde{M}_J}
(\frac{1}{2}m_{s_2}\,\tilde{\ell}\tilde{m}_{\ell}\mid\tilde{J}\tilde{M}_J)
\nonumber \\
& & \hspace*{-6cm} \times~(\frac{1}{2}m_{s^{\prime}_2}\,\tilde{\ell}^{\prime}\tilde{m}_{\ell^{\prime}}\mid\tilde{J}^{\prime}\tilde{M}_{J^{\prime}})Y_{\tilde{\ell}\tilde{m}_{\ell}}^{\star}(\hat{\tilde{p}}_r)Y_{\tilde{\ell}^{\prime}\tilde{m}_{\ell^{\prime}}}(\hat{\tilde{p}^{\prime}_r})\tilde{\mathcal{R}}_{\pi
  N}^{\tilde{J}\tilde{\ell}\tilde{t}}(W_{\pi N}(\vec
  p_2),\tilde{p}_r,\tilde{p}^{\prime}_r)\,. \nonumber \\
& &  
\end{eqnarray}
Inserting Eq.\ (\ref{tpn-vfinal}) in (\ref{tpn-hos2}), we obtain the
final form for the half-off-shell $\pi N$-scattering amplitude. In the 
calculations presented here, the half-off-shell partial wave amplitudes 
$\tilde{\mathcal{R}}_{\pi N}^{J\ell\tilde{t}}(W_{\pi N}(\vec p_2),\tilde{p}_r,\tilde{p}^{\prime}_r)$ were obtained for all $S$-, $P$- and $D$-waves channels by
solving the Lippmann-Schwinger equation \cite{Lippmann51} for a separable
energy-dependent $\pi N$ potential built in Ref.\ \cite{Nozawa90}. This
potential describes precisely the on-shell as well as the off-shell
properties of the $\pi N$ scattering amplitude. More details about this
potential is given in appendix \ref{appendixF}.

\vspace*{-0.2cm}~\\
Collecting the various pieces, we obtain the following expression for the
amplitude (\ref{tpn-fsi}) with $\pi N$-rescattering in the final state
\begin{eqnarray}
\label{tpn-hos21}
M^{(t\mu)~\pi N}_{sm_s,m_{\gamma}m_d}(\vec k,\vec q,\vec p_1,\vec p_2) & = &
\frac{1}{2}\sum_{s^{\prime}m_{s^{\prime}}t^{\prime}\mu^{\prime}} \int
d^3\vec{p}~\sqrt{\frac{\omega_{\vec
      q}}{\omega_{\vec q^{\,\prime}}}}\sqrt{\frac{E_{{\vec p}_2}}
{E_{{\vec{p}^{\,\prime}}_2}}} \nonumber \\
& & \hspace*{-0.7cm}\times~\sum_{m_{s_2}m_{t_2}m_{s^{\prime}_2}m_{t^{\prime}_2}}\mathcal{A}_{t\mu,t^{\prime}\mu^{\prime}}^{sm_ss^{\prime}m_{s^{\prime}}}
(m_{s_2},m_{t_2},m_{s^{\prime}_2},m_{t^{\prime}_2}) \nonumber \\
& & \hspace*{-0.7cm}\times~\left[\sum_{\tilde{t}\tilde{m}_{t}\,\tilde{t^{\prime}}\tilde{m}_{t^{\prime}}}
\mathcal{B}_{\mu\mu^{\prime}\tilde{t}\tilde{t^{\prime}}}^{\tilde{m}_t\tilde{m}_{t^{\prime}}}(m_{t_2},m_{t^{\prime}_2})
\right. \nonumber \\
& & \hspace*{-0.7cm}\left.\times~\left\{
\mathcal{Q}^{\tilde{t}\tilde{t^{\prime}}}_{\tilde{m}_t\tilde{m}_{t^{\prime}}}(W_{\pi
  N}(\vec p_2),\tilde{\vec{p}_r},\tilde{\vec{p}^{\,\prime}_r})
\mathcal G_{0\pi N}^{\pi
  NN}(E_{\gamma d},\vec q^{\,\prime}(\vec{p}),\vec p_1,\vec
p^{\,\prime}_2(\vec{p}))\right.\right.\nonumber\\
& & \hspace*{-0.7cm}\left.\left.\times~\mathcal M^{(t^{\prime}\mu^{\prime})~IA}_{s^{\prime}m_{s^{\prime}},m_{\gamma}m_d}(\vec k,\vec
q^{\,\prime}(\vec{p}),\vec p_1,\vec p^{\,\prime}_2(\vec{p})) - (-)^{s+t}
(\vec p1\leftrightarrow \vec p_2)\right\}\right] \,, \nonumber \\
& & 
\end{eqnarray}
where $\tilde{\vec{p}^{\,\prime}_r}(\vec{p})$ and $\tilde{\vec{p}_r}$ are given in Eq.\
(\ref{pinprelative}) and 
\begin{eqnarray}
\label{tpn-hos211}
\vec q^{\,\prime}(\vec{p}) & = & \vec p - \tilde{\vec{p}_1}~~~~~~~{\rm with}~~~~
\tilde{\vec{p}_1}~=~ \vec{p}_1 -\vec{k} \nonumber \\
\vec p^{\,\prime}_2(\vec{p}) & = & -\vec{p}\, .
\end{eqnarray}

\vspace*{-0.2cm}~\\
Similarly to the case of $NN$ scattering, the three dimensional integral over
$\vec{p}$ is carried out numerically.

\chapter{Results and Discussion\label{chap:4}}
%
In this chapter we will present and discuss our results for pion
photoproduction on the deuteron in the $\Delta$(1232) resonance region
including $NN$ and $\pi N$ rescattering in the final state. We will discuss
the important observables 
that have been measured experimentally, such as the total and differential
cross sections. Since several experiments to test the Gerasimov-Drell-Hearn
(GDH) sum rule both on
the proton and on the neutron are now in preparation or planned at different
laboratories around the world (MAMI, ELSA, LEGS, GRAAL and TJNAF), the
theoretical investigation of the spin asymmetry and the corresponding GDH
integral are nowadays a very interesting subject of research. This sum rule is
derived under very general assumptions (Lorentz and gauge invariance,
causality, relativity and unitarity). This makes its verification to be 
an important check of our understanding of the hadronic spin 
structure. The inclusion of final state interaction effects may play an
important role on the spin asymmetry and the corresponding GDH sum rule for
the deuteron. Therefore, we will also discuss in this chapter the influence of
rescattering effects on the spin asymmetry and the GDH sum rule for the
deuteron and, for comparison, for the nucleon.

\vspace*{-0.2cm}~\\
The discussion in this chapter is divided into three parts. In the first part, 
we present our results for the total cross section of the processes
(\ref{eq:3.1}) in comparison with experimental data and other theoretical
predictions. In the second part, we discuss the semi-exclusive 
differential cross section $d^2\sigma /d\Omega_{\pi}$. We also compare our results
for the differential cross section with experimental data and other
theories. In 
the third part, we study the influence of $NN$ and $\pi N$ rescattering on the
spin asymmetry and the corresponding GDH sum rule for the deuteron. We would
like to mention that the results presented here are calculated using the
deuteron wave function of the Bonn potential (full model) \cite{Machleidt8789}
(see section \ref{chap:3:3}). 

\vspace*{-0.2cm}~\\
To calculate the semi-exclusive differential cross section $d^2\sigma
/d\Omega_{\pi}$, the fully exclusive differential cross section
$d^{5}\sigma/(d\Omega_{\pi}dq_{\pi}d \Omega_{p_r})$ (see Eq.\ (\ref{eq:3.13}))
was integrated over the pion momentum $q_{\pi}$ and the polar angle
$\theta_{p_r}$ and the azimuthal angle $\phi_{p_r}$ of the relative momentum
$\vec p_r$ of the two outgoing nucleons. The total cross section was
calculated by 
integrating Eq.\ (\ref{eq:3.13}) over all the remaining independent variables,
i.e., the pion momentum $q_{\pi}$, its angles $\theta_{\pi}$ and $\phi_{\pi}$,
the 
polar angle $\theta_{p_r}$ and the azimuthal angle $\phi_{p_r}$ of the
relative momentum of the two outgoing nucleons. These integrations
are carried out numerically. The number of integration points was being
increased until the accuracy of calculated observables becomes good to 1$\%$.
%
\section{Total Cross Section\label{chap:4:1}}
\begin{figure}[htb]
\centerline{\epsfxsize=15.2cm 
\epsffile{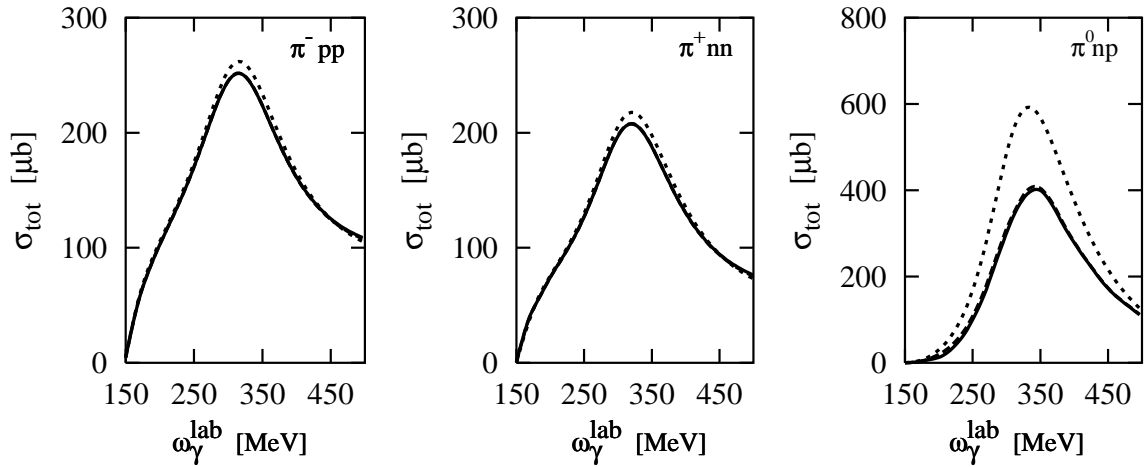}}
\vspace*{-0.2cm}
\caption{\small Total cross sections for $\gamma d\rightarrow\pi NN$
  reaction obtained within the impulse approximation (dotted curves) in
    comparison with the calculation including $NN$ final state interaction
    (dashed curves) and the calculation with additional $\pi N$ rescattering
    (solid curves). The left, middle and right panels represent the total cross
    section for $\gamma d\rightarrow \pi^-pp$, $\pi^+nn$ and $\pi^0np$,
    respectively.}
\label{tot_our}
\end{figure}

%
We start the discussion with presenting our results for the total cross
sections in Fig.\ \ref{tot_our} including $NN$ and $\pi N$ final state
interaction effects. 

\vspace*{-0.2cm}~\\
Fig.\ \ref{tot_our} shows that the differences between the full and the spectator model calculations clearly
demonstrate the importance of rescattering effects, in particular for the
$\pi^0$ channel. One can see that final state
interactions lead to a strong reduction of the total cross section. In the
case of charged pion photoproduction reactions, the final state
interaction effects are small in comparison with the case of neutral pion
photoproduction reaction. The differene between the solid and dashed curves in
Fig.\ \ref{tot_our} 
shows that the $\pi N$ rescattering changes the final results only by a few
percent. A possible explanation for this smallness comes from the fact that
the scattering length of the most important $S$-wave $\pi N$-scattering is
about two orders smaller than the one of $NN$-scattering. In the energy range
of the 
$\Delta$(1232) resonance, one finds the strongest manifestation of
rescattering effects. In the case of $\pi^-$ and $\pi^+$ channels, one sees 
that the $NN$ and $\pi N$ rescattering effects reduce the total cross sections
for both channels in the energy range of the $\Delta$(1232) resonance by about
5$\%$. For lower and higher energies, the influence of  
rescattering effects changes the results for total cross sections of the
charged pion channels with a few percent. With respect to $\pi^0$ channel,
we see from the right panel of Fig.\ \ref{tot_our} that, the influence of
rescattering effects is much more significant than the case of charged pion
channels. It leads to a strong reduction of the total cross section by about
35$\%$ in the energy range of the $\Delta$(1232) resonance.
\begin{figure}[htb]
\centerline{\epsfxsize=15.2cm 
\epsffile{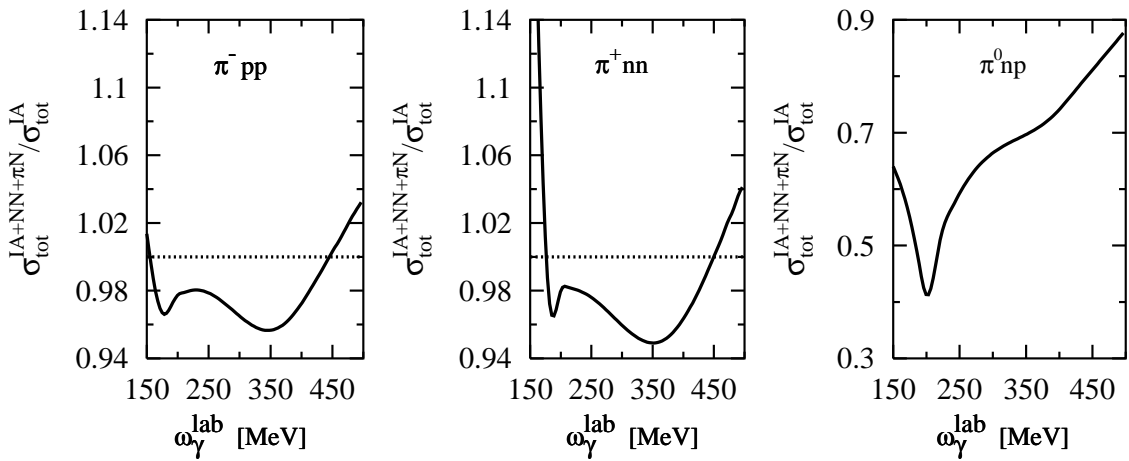}}
\vspace*{-0.2cm}
\caption{\small The ratio $\sigma_{tot}^{IA+NN+\pi N}/\sigma_{tot}^{IA}$  as a
  function of the photon energy in the laboratory frame. The left, middle and
  right panels represent the ratio for $\gamma d\rightarrow
  \pi^-pp$, $\pi^+nn$ and $\pi^0np$, respectively.}
\label{tot_deut}
\end{figure}
\begin{figure}[htb]
\centerline{\epsfxsize=15.2cm 
\epsffile{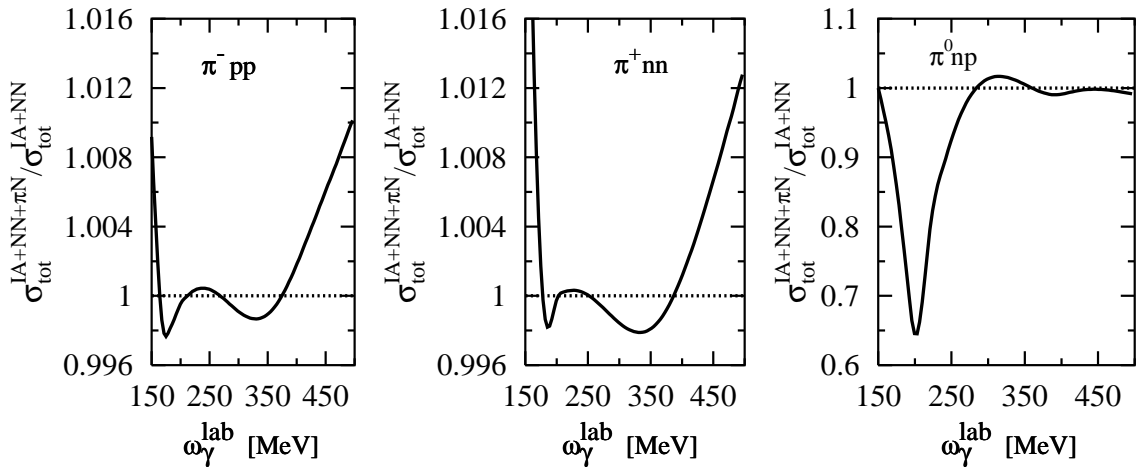}}
\vspace*{-0.2cm}
\caption{\small The ratio $\sigma_{tot}^{IA+NN+\pi N}/\sigma_{tot}^{IA+NN}$ as
  a function of the photon energy. The left, middle and right panels represent
  the ratio for $\gamma d\rightarrow \pi^-pp$, $\pi^+nn$ and $\pi^0np$,
  respectively.}
\label{tot_deut1}
\end{figure}

\vspace*{-0.2cm}~\\
In order to show in greater detail the relative influence of rescattering effects on the total cross
sections, we show in Figs.\ \ref{tot_deut} and \ref{tot_deut1}
the ratios $\sigma_{tot}^{IA+NN+\pi N}/\sigma_{tot}^{IA}$ and
$\sigma_{tot}^{IA+NN+\pi N}/\sigma_{tot}^{IA+NN}$, respectively, as a function
of the photon energy in the laboratory frame. Here $\sigma_{tot}^{IA}$ denotes
the total cross section in the impulse approximation, $\sigma_{tot}^{IA+NN}$
denotes the total cross section including $NN$ rescattering and
$\sigma_{tot}^{IA+NN+\pi N}$ denotes the total cross section with additional
$\pi N$ rescattering (see section \ref{chap:3:4}).

\vspace*{-0.2cm}~\\
Fig.\ \ref{tot_deut} shows the combined effect of $NN$ and $\pi N$ rescattering as a
function of the photon energy. This emphasize the important role of final
state interactions for the total cross sections, especially for $\pi^0$
photoproduction 
on the deuteron. It is obvious that for charged pion channels
the ratio has a similar shape,
but in the case of $\pi^0$ production, a totally different shape is observed. 
At energies near to the pion threshold, the effect of rescattering increases the total cross section for $\pi^+$ production than for $\pi^-$
production. For $\pi^+$ production we found that at 150 MeV the ratio is 1.32,
while it reads 
1.01 for $\pi^-$ production at the same energy. For $\pi^0$ production, the situation is completely
different since final state interactions lead to a strong reduction which
amounts to about 60$\%$ at 200 MeV. This demonstrates that rescattering effects play a
very important role, in particular, in the case of $\pi^0$ photoproduction.  

\vspace*{-0.2cm}~\\
As next we show the relative effect of $\pi N$ rescattering in Fig.\
\ref{tot_deut1}. In general, we obtain a similar shape as
in the case of  $\sigma_{tot}^{IA+NN+\pi N}/\sigma_{tot}^{IA}$ (see Fig.\
\ref{tot_deut}), especially that the peaks have 
approximately the same energy. Fig.\ \ref{tot_deut1} shows
that the influence of $\pi N$ rescattering in the region of the $\Delta$
resonance is much less important than $NN$ rescattering. For energies near
to the pion threshold, we see that the effect of $\pi N$ rescattering
increases the total cross sections for charged pion channels. This increment 
is noticeable in the case of $\pi^+$ production since the ratio 
reads 1.033 at 150
MeV. Furthermore, we see that the effect of $\pi N$ rescattering in the case
of $\pi^0$ production leads to a strong reduction for the total cross section,
in particular at 200 MeV. A very big effect of the $\pi N$ rescattering in the threshold region
was also found in Refs.\ \cite{Bosted78,Benmerrouche98,Koch77,Levchuk00}. As
already mentioned in \cite{Levchuk00}, that the charged pion channels
are also of importance in the threshold region, because a big $\pi^{\pm} N$
rescattering effect is certainly possible.
%
\subsection{Comparison with Experimental Data\label{chap:4:1:2}}
%
Here we compare our results for the total cross sections of the reaction $\gamma
d\rightarrow\pi NN$ including $NN$ and $\pi N$ rescattering in the
final state with experimental data. In our comparison with experiment we
concentrate our discussion on $\pi^{-}$ and $\pi^0$ 
photoproduction on the deuteron, since data for $\pi^{+}$
production in the $\Delta$(1232) resonance region are not available. In the case of $\pi^-$ production we compare our results with the
experimental data from Refs.\ \cite{Benz73,Chiefari75,Asai90}. For $\pi^0$
production we compare our results with the experimental data from \cite{Krusche99}.
\begin{figure}[htb]
\centerline{\epsfxsize=15cm 
\epsffile{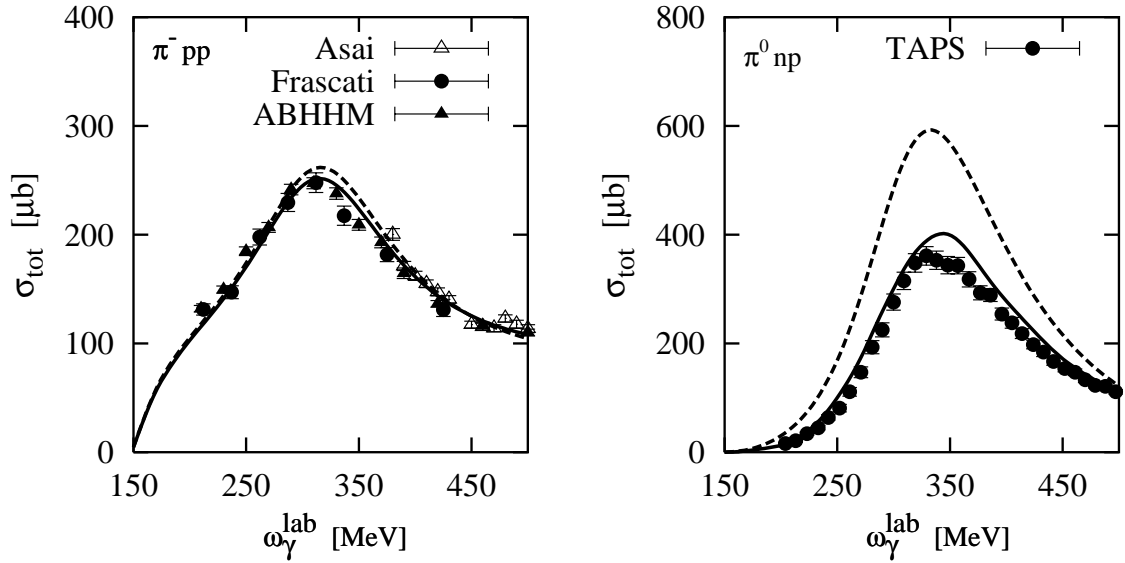}}
\vspace*{-0.2cm}
\caption{\small Total cross sections for $\pi^-$ (left panel) and $\pi^0$
  (right panel) photoproduction on the deuteron. Solid curves show our 
 results with $NN$ and $\pi N$ rescattering in the final state
 and dashed curves show the results in the impulse approximation. The
 experimental data are from \cite{Benz73} (ABHHM), \cite{Chiefari75} (Frascati)
 and \cite{Asai90} (Asai) in case of $\pi^-$ production and from
    \cite{Krusche99} (TAPS) in case of $\pi^0$ production.}
  \label{tot_deut_exp}
\end{figure}

\vspace*{-0.2cm}~\\
Fig.\ \ref{tot_deut_exp} shows our results for the total cross sections for $\pi^-$
and $\pi^0$ photoproduction on the deuteron compared with experimental
data. One readily notes, that the spectator approach can
not describe the experimental data in the $\Delta$(1232) resonace region (see
also \cite{Schmidt9695}), especially in the case of $\pi^0$ production. 
Note, that the contribution of final state interactions in the case of $\pi^0$
production is much larger than that for $\pi^-$ production. Furthermore, final
state interaction effects lead to a strong reduction of total cross 
section. The inclusion of such effects improve the
agreement between experimental data and theoretical predictions. Indeed,
a quite significant contribution from $NN$ rescattering is found. Only in the
center of the peak for $\pi^0$ production our model overestimates the measured
total cross section by about 6$\%$. Moreover, we found that the $\pi N$
rescattering changes the final results only by a few percent. As mentioned
before, this smallness can be explained by the fact that the most important 
$S$-wave $\pi N$ scattering length is about two orders smaller than the one
for $NN$ scattering. 

\vspace*{-0.2cm}~\\
These results clearly show that the rescattering effects are significant and
reduce the 
total cross section in the $\Delta$(1232) resonance region. This means, in
particular with respect 
to a test of theoretical models for pion production amplitudes on the neutron,
that one needs a reliable description of the rescattering process. Compared to
experimental data, one readily finds that the sizable discrepancies without
rescattering are largely reduced and that a reasonable agreement with the data
is achieved. 
%
\subsection{Comparison with other Theoretical Predictions\label{chap:4:1:1}}
%
Pion photoproduction on the deuteron in the impulse approximation has 
been studied by Schmidt {\it et al.} \cite{Schmidt9695} neglecting all kinds
of final state interactions and other two-body operators. Since data for
$\pi^+$ and $\pi^0$ photoproduction on the deuteron were absent at the time of
these calculations, the authors had no possibility to compare their
predictions with experimental data for these channels. A comparison with
experimental data was possible only for $\pi^-$ production, where they found a
slight overestimation of the data. They reported that the reason for that is
an overestimation of the elementary reaction on the neutron. They
mentioned also that the differences between their theoretical predictions and
the experimental data show very clearly that the calculation of pion
photoproduction on nuclei in the nucleon spectator model can only be
considered as a first step towards a more realistic description of this
process. As noted in \cite{Laget81,Levchuk96,Levchuk00N,Levchuk00} 
the effect of $NN$ rescattering is important in the incoherent pion
photoproduction on the deuteron, especially for small pion angles.

\vspace*{-0.2cm}~\\
Laget \cite{Laget81,Laget78,Laget77} and Blomqvist and Laget
\cite{Blomqvist77} have investigated pion photoproduction on the deuteron with 
the inclusion of pion rescattering and $NN$ final state interaction within a
diagramatic ansatz. They used the elementary photoproduction operator of
Blomqvist and Laget \cite{Blomqvist77} as input in their calculations. 
At the time of these calculations, a comparison with experimental data was
possible only for $\pi^-$ production since data for $\pi^{+}$ and $\pi^0$ 
production in the $\Delta$(1232) resonance region were not available. The agreement of their predictions including final state interactions
with the experimental data of the cross sections of the reaction $\gamma
d\rightarrow \pi^- pp$ is quite good. They found that the final state
interaction effects are small for the charged pion photoproduction reactions
in comparison with the case of the neutral channel. As mentioned in
\cite{Krusche99}, these predictions are significantly above the data for
$\pi^0$ photoproduction on the deuteron. A possible reason for this may be that
they used the Blomqvist and Laget parametrization \cite{Blomqvist77} of the
elementary photoproduction amplitude. This amplitude is constructed using
different $\Delta$ parametrizations for neutral and charged pion
photoproduction. It gives a satisfactory fit to the amplitude for charged pion
photoproduction, but it is not able to describe the neutral pion
photoproduction from the proton. An attempt to remedy this defect in \cite{Sabutis83} led to a $\pi^0$ photoproduction amplitude which is not very
suitable for the use in nuclear calculations. 

\vspace*{-0.2cm}~\\
Levchuk {\it et al.} \cite{Levchuk96} studied quasifree $\pi^0$ 
photoproduction from the neutron via the $d(\gamma,\pi^0)np$ reaction using the
elementary photoproduction operator of Blomqvist and Laget
\cite{Blomqvist77}. The contributions from the pole diagrams as well as
one-loop diagrams both with $np$ and $\pi N$ rescattering were taken into
account. They wanted to explore the possibility of measuring the
$E_{1+}/M_{1+}$ ratio via photoproduction from the quasifree neutron. 
The isospin $I=3/2$ component of this ratio characterizes the relative
strength of the recently much discussed (see e.g.\ \cite{KrahnBeck96700,HDT})
quadrupole $E2$-excitation of the $\Delta$ resonance. The idea was that the
$n(\gamma,\pi^0)n$ reaction would be very useful for the isospin separation of
the multipoles. In agreement with the results from the Laget model 
\cite{Laget81}, Levchuk {\it et al.} find that the largest effects disturbing
the extraction of the multipoles for quasifree neutrons arise from the $np$
final state interaction. They predict that these effects lead to a strong
reduction of the cross section at pion forward angles, but are much less
important for backward angles. They found also that the correction due to
$np$ rescattering decreases with 
increasing pion angle and becomes to be less than 8$\%$ at $\theta_{\pi}\ge
90^0$. Furthermore, they pointed out that the contribution of the proton pole
diagram and the one of $\pi N$ rescattering are negligible. The
experimental data from \cite{Krusche99} for the 
$d(\gamma,\pi^0)np$ reaction qualitatively support this prediction since the
disagreement with the spectator approach is most severe at pion forward angles
but less pronounced at backward angles. However, a comparison of the data to
the Laget model including final state interactions shows some unexplained
reduction of the cross section at backward angles. 

\vspace*{-0.2cm}~\\
Recently and during the calculations of this thesis work, Levchuk {\it et al.}
\cite{Levchuk00N} studied the inclusive reaction $d(\gamma,\pi)NN$ in the
$\Delta$ resonance region. This calculation is based on the use of the
diagramatic approach. Pole diagrams and one-loop diagrams with $NN$ and $\pi
N$ rescattering in the final state are considered. The authors in
\cite{Levchuk00N} pointed out that the main difference between their
calculation and the one of Ref.\ \cite{Laget81} is that a more realistic
version of the elementary pion photoproduction operator is used. It is taken
in on-shell form and calculated using the SAID \cite{Said} and MAID
\cite{Maid} multipole analyses. This operator is constructed in the $\gamma N$
c.\ m.\ frame. Therefore, it has to be transformed to an arbitrary frame of
reference to be used as input in calculations on light nuclei. This may be
done by a Lorentz boost of all four momenta on which the elementary
amplitude depends. Their predictions for total and differential cross sections
including final state interactions show good agreement with the experimental
data.

\begin{figure}[htb]
\centerline{\epsfxsize=15cm 
\epsffile{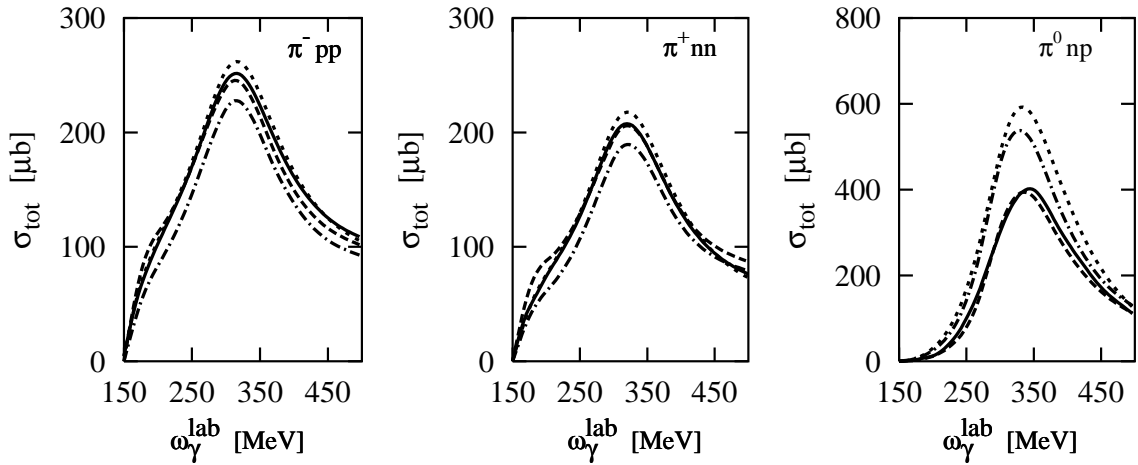}}
\vspace*{-0.4cm}
\caption{\small Total cross sections for pion photoproduction on the deuteron
 obtained within the impulse approximation (dotted curves) in comparison with
    the 
    calculation when $NN$ and $\pi N$ final state interactions are taken into
    account (solid curves). The results of Levchuk {\it et al.}
    \cite{Levchuk00N} as shown by the dash-dotted curves for the impulse
    approximation and by the
    dashed curves with $NN$- and $\pi N$-rescattering.}
  \label{tot_deut_theory}
\end{figure}

\vspace*{-0.2cm}~\\
Now, we compare our results for the total cross section of the
processes (\ref{eq:3.1}) with the theoretical predictions from \cite{Levchuk00N} as shown in Fig.\ \ref{tot_deut_theory}.

\vspace*{-0.2cm}~\\
In the impulse approximation, Fig.\ \ref{tot_deut_theory} shows that our
results for charged pion photoproduction (dotted curves) are higher than the
results including rescattering effects (solid curves). This means that
rescattering effects lead to a reduction of the total cross sections in our
case. In the same figure, we see from the difference between the dash-dotted
and dashed curves that the theoretical predictions from \cite{Levchuk00N} display the opposite situation. In the next section we will
give an explanation for this opposite situation in some detail. In the case of
$\pi^0$ photoproduction, rescattering effects lead in both cases to a strong
reduction of the total cross section. 

\vspace*{-0.2cm}~\\
Including final state interactions we found small discrepancies between our
calculations which are given by the solid curves in Fig.\
\ref{tot_deut_theory} and the predictions
of Ref.\ \cite{Levchuk00N} which are given by the dashed curves in the same
figure. A reason for that may be the use of different pion photoproduction
operators. 
As mentioned in section
\ref{chap:3:4:2} that we used in our calculation the separable representation
of the Paris $NN$ potential \cite{Haidenbauer84,Haidenbauer85} to solve the
Lippmann-Schwinger equation for the $NN$ scattering. In \cite{Levchuk00N}, the
authors used another type of $NN$ potentials which is the Bonn OBE potential
model (OBEPR) \cite{Machleidt8789}. This choice of different $NN$ potential
models may also be a reason for the small discrepancies between both
theoretical predictions. In the case of $\pi N$ scattering, the separable
energy-dependent $\pi N$ potential built in \cite{Nozawa90} is used in
both theoretical predictions.
\newpage
%
\section{Differential Cross Section\label{chap:4:2}}
\begin{figure}[ht]
\centerline{\epsfxsize=11cm 
\epsffile{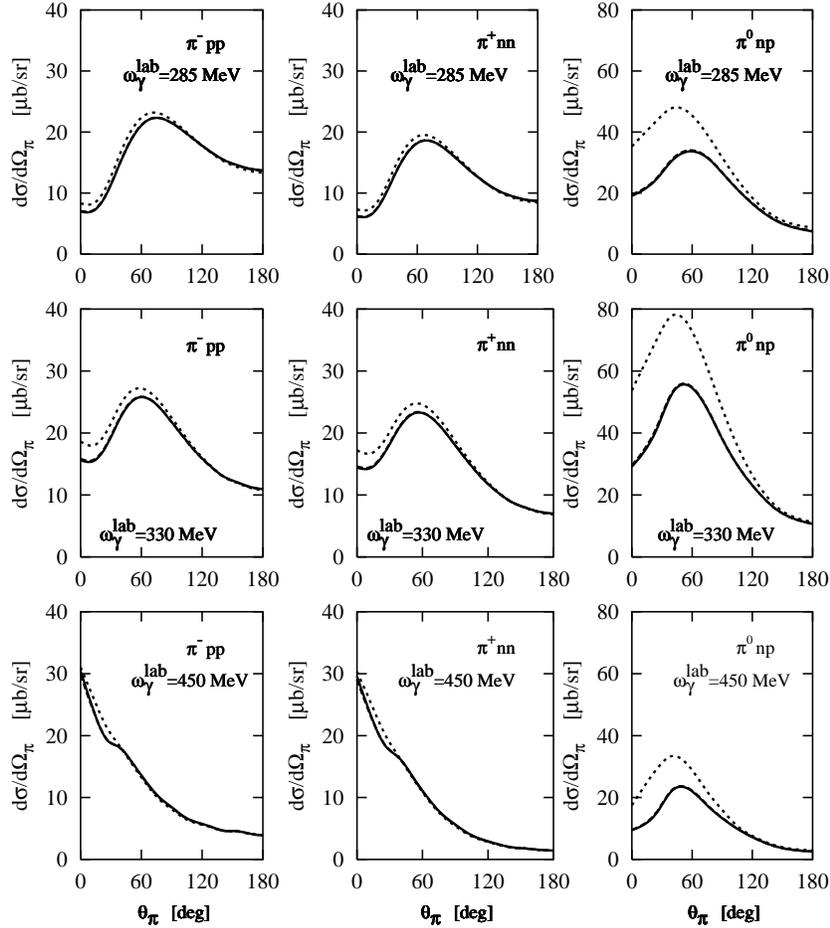}}
\vspace*{-0.1cm}
\caption{\small Differential cross sections for pion photoproduction on the
  deuteron within the impulse approximation (dotted curves) in comparison with
  the calculation including $NN$ final state interaction (dashed curves) and
  the calculation with additional $\pi N$ rescattering (solid curves). The
  left, middle and right panels represent the differential cross section for
  $\gamma d\rightarrow \pi^-pp$, $\pi^+nn$ and $\pi^0np$, respectively.}
\label{diffour}
\end{figure}

%
In this section we discuss our results for the differential cross sections 
and compare with experimental data and other theoretical predictions. We begin
with presenting our results in Fig.\ \ref{diffour}.

\begin{figure}[htb]
\centerline{\epsfxsize=11cm 
\epsffile{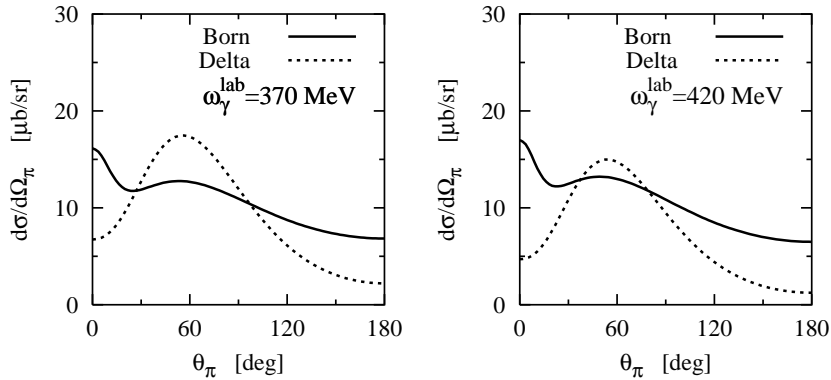}}
\vspace*{-0.3cm}
\caption{\small Differential cross section for $\pi^-$ photoproduction on
  the deuteron in the impulse approximation at different photon energies. The
  solid curve shows the results 
  using only the Born terms and the dotted one shows the contribution of the
  $\Delta$(1232) resonance.}
\label{borndelta}
\end{figure}

\vspace*{-0.2cm}~\\
As in the case of total cross section in the previous section, Fig.\
\ref{diffour} shows that  
the difference between the full and the spectator model calculations clearly
demonstrate the importance of rescattering effects. One sees also that $\pi N$
rescattering changes the final results for the differential cross 
sections only by a few percent. Furthermore, for charged pion channels, one
can see that the effect of final state interactions becomes small for low and
high energies, but it becomes maximal at energies near to the
$\Delta$(1232) resonance. In the case of $\pi^0$ channel, it is obvious that
the effect of rescattering is important for all energies, in particular for
forward pion angles. In the case of both charged and neutral pions, we see
that rescattering effect is quite small for backward pion angles. 
Looking at the right and middle panels in Fig.\ \ref{diffour}, one can see
that at $\theta_{\pi}=0$, the differential cross section increases with
increasing photon energy. This increase comes mainly from the Born terms. We
found that more than 70$\%$ from the values of the differential cross section
at $\theta_{\pi}=0$ comes from the Born terms and less than 30$\%$ comes from
the contribution of the $\Delta$(1232) resonance (see Fig.\ \ref{borndelta}).

\begin{figure}[ht]
\centerline{\epsfxsize=12cm 
\epsffile{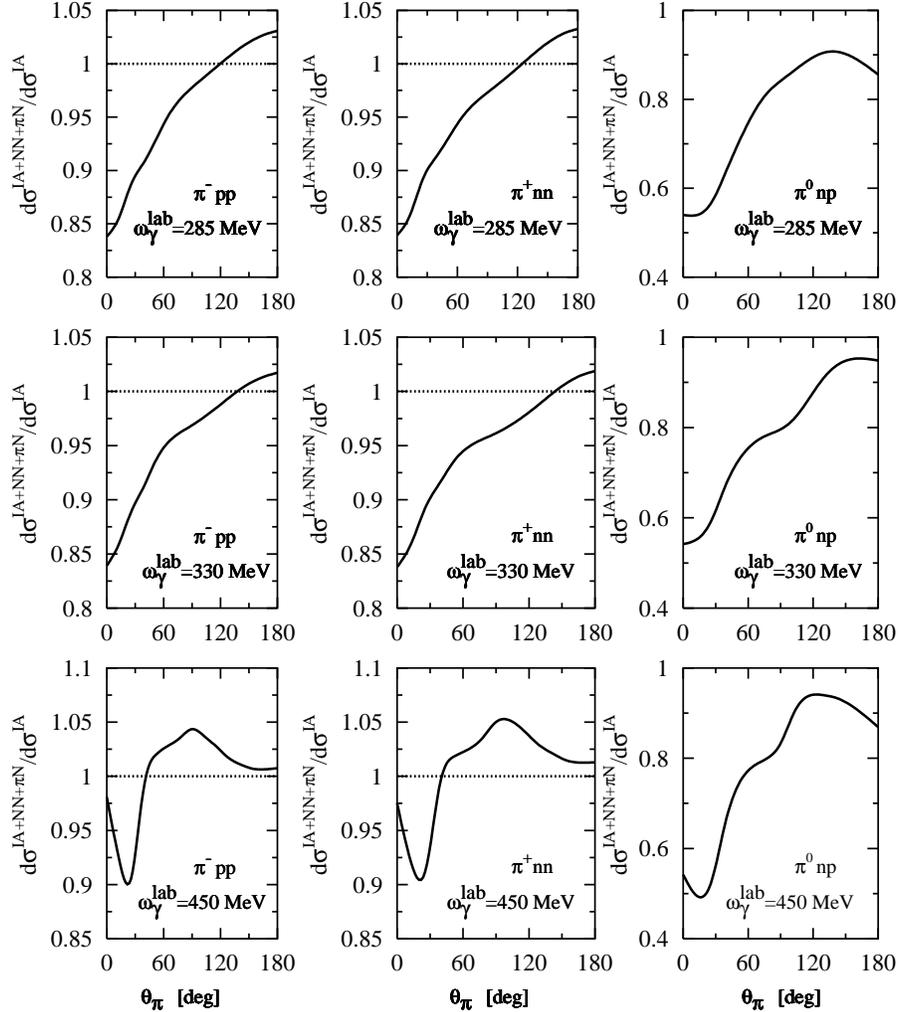}}
\vspace*{-0.2cm}
\caption{\small The ratio $d\sigma^{IA+NN+\pi N}/d\sigma^{IA}$ as
  a function of the pion angle in the laboratory frame at different photon
  energies. The left, middle and right panels represent the ratio for the
  $\gamma d\rightarrow\pi^-pp$, $\pi^+nn$ and $\pi^0np$, respectively.}
\label{diffour_ratio}
\end{figure}

\begin{figure}[htb]
\centerline{\epsfxsize=12cm 
\epsffile{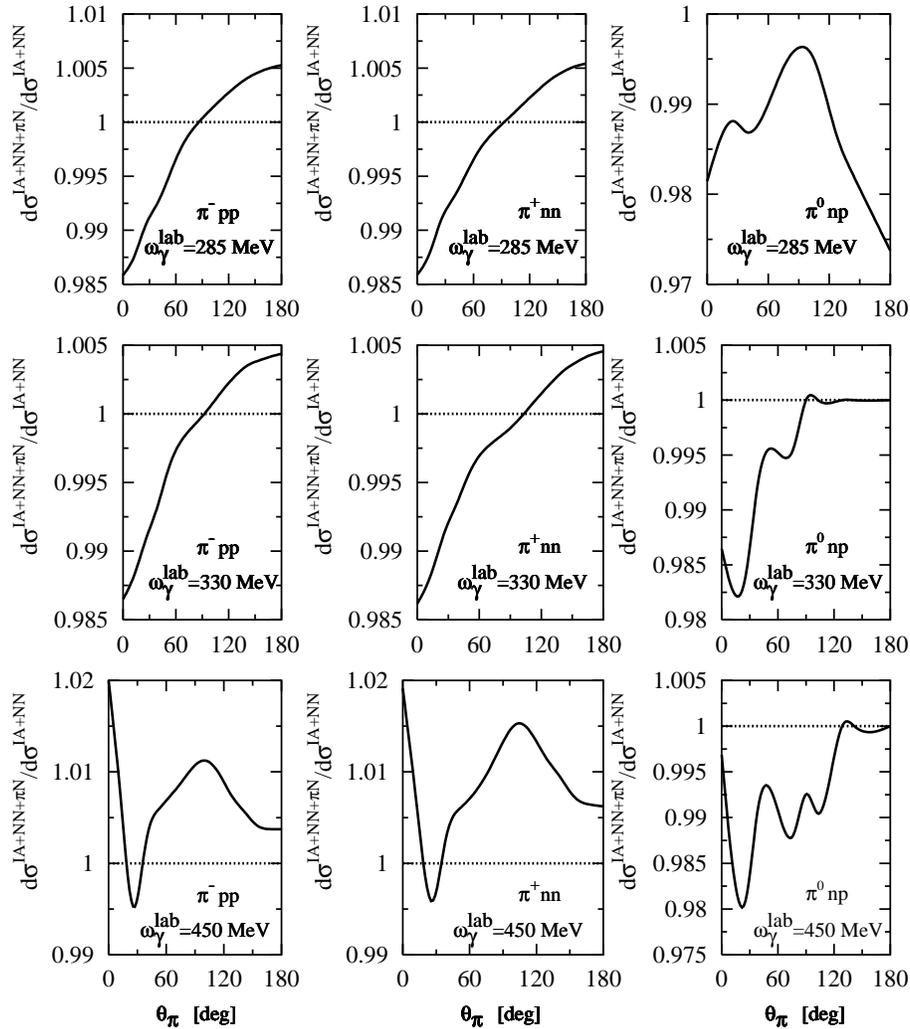}}
\vspace*{-0.2cm}
\caption{\small The ratio $d\sigma^{IA+NN+\pi N}/d\sigma^{IA+NN}$ as
  a function of the pion angle at different photon energies. The left, middle
  and right panels represent the ratio for the $\gamma
  d\rightarrow\pi^-pp$, $\pi^+nn$ and $\pi^0np$, respectively.} 
\label{diffour_ratio1}
\end{figure}

\vspace*{-0.2cm}~\\
In order to show in more detail the relative importance of final state
interactions on the differential cross sections, we show in Figs.\
\ref{diffour_ratio} and \ref{diffour_ratio1} the ratios $d\sigma^{IA+NN+\pi
  N}/d\sigma^{IA}$ and $d\sigma^{IA+NN+\pi N}/d\sigma^{IA+NN}$, respectively,
as a function of the pion angle in the laboratory frame. 

\vspace*{-0.2cm}~\\
Fig.\ \ref{diffour_ratio} shows the relative effect of both $NN$ and $\pi N$
final state interactions by the ratio $d\sigma^{IA+NN+\pi N}/d\sigma^{IA}$ as
a function of the pion angle. As mentioned in the previous section, we see
that the ratio has a similar shape for charged pion channels. It is also
obvious that rescattering effects are important for forward pions and much
less important for backward pions. In the case of 
$\pi^0$ production, we see that the influence of final state interaction
effects on the differential cross section is significant. Rescattering
effects lead to a strong reduction of the differential cross sections, in
particular for small pion angles. The contribution decreases quickly when the
pion angle increases and becomes very small for backward pion angles. 

\vspace*{-0.2cm}~\\
Now, we would like to clarify the relative effect of $\pi N$ rescattering in
Fig.\ \ref{diffour_ratio1} by the ratio $d\sigma^{IA+NN+\pi
  N}/d\sigma^{IA+NN}$. Here one can also see that we obtain a similar shape
for the charged pion channels, but a totally different shape for the $\pi^0$
channel is seen. The effect of $\pi N$ rescattering is seen much less
important than $NN$ rescattering. This effect is less than 1$\%$ for charged
pion channels and less than 5$\%$ for neutral channel. In the energy range of
the $\Delta$(1232) resonance, we find that this effect is maximal at pion
forward angles.
%
\subsection{Comparison with Experimental Data\label{chap:4:2:1}}
\begin{figure}[htb]
\centerline{\epsfxsize=8cm 
\epsffile{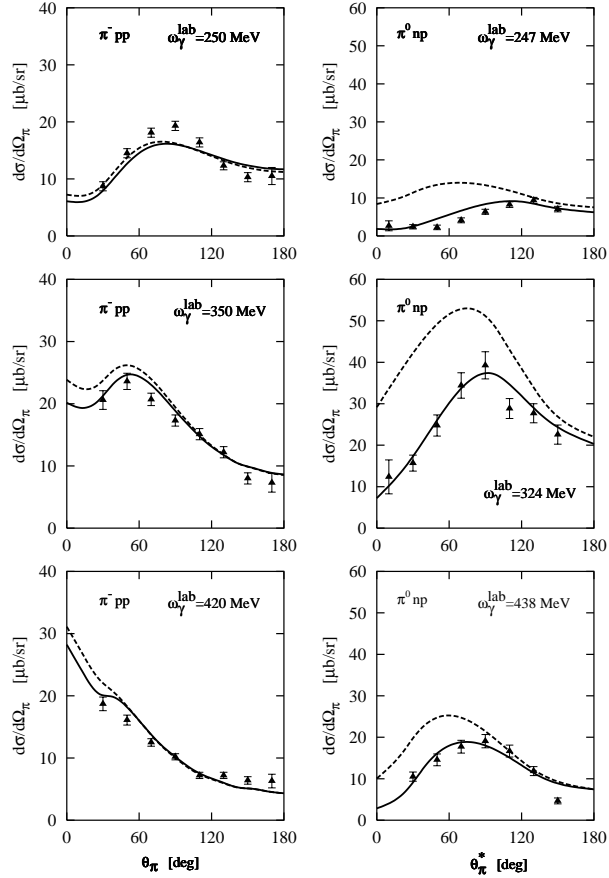}}
\vspace*{-0.1cm}
\caption{\small Differential cross sections for $\pi^-$ (left panels) and
  $\pi^0$ (right panels) photoproduction on the deuteron. Solid
  curves show our results include $NN$ and $\pi N$ rescattering and dashed
  curves show the results in the impulse approximation. The experimental data
  are from \cite{Benz73} for $\pi^-$ production and from \cite{Krusche99} for
  $\pi^0$ production.} 
\label{diffwithexp}
\end{figure}

%
Here we compare our results for the differential cross sections with the
experimental data from \cite{Benz73} for $\pi^-$ production and from
\cite{Krusche99} for $\pi^0$ production. This comparison is shown in Fig.\
\ref{diffwithexp}. As mentioned in the previous section,  
there are no data available for $\pi^+$ production on the deuteron in the
$\Delta$(1232) region so that we can not compare our predictions with
experimental data for this channel. Therefore, we concentrate our discussion
on $\pi^{-}$ and $\pi^0$ production. The effect of final state interactions in
the case of charged pion channels is expected to be quite different in
comparison with the neutral channel.

\vspace*{-0.2cm}~\\
In Ref.\ \cite{Krusche99} the differential cross sections for the reaction
$d(\gamma,\pi^0)np$ are given in the so-called $\gamma N$ c.\ m.\
frame. Therefore, in order to compare our results for the differential cross
sections of the reaction $d(\gamma,\pi^0)np$ with the experimental data from 
\cite{Krusche99} we need to transform the differential cross sections
from the rest frame of the deuteron to the $\gamma N$ c.\ m.\ frame. The
Jacoby determinant which we need for this purpose is given in appendix
\ref{appendixE}. The pion angle in the $\gamma N$ c.\ m.\ frame is denoted by
$\theta_{\pi}^{\,\star}$. 

\vspace*{-0.2cm}~\\
Fig.\ \ref{diffwithexp} shows that the spectator approach can not describe the
experimental data for differential cross sections. One sees also that the
effect of final state interactions leads to a strong reduction of the
differential cross sections, especially for forward pion angles but it is much
less important for backward angles. The inclusion of such effects improves the
agreement between experimental data and our calculations. After including
final state interactions we obtain in general a good agreement with the
experimental data for differential cross sections. Small discrepancies are
found only at backward pion angles. We see that the agreement with the
experimental data of the reaction $\gamma d\rightarrow \pi^-pp$
measured in a bubble chamber experiment \cite{Benz73}, is quite good. The
comparison with the TAPS data \cite{Krusche99} for the reaction $\gamma
d\rightarrow \pi^0np$ yields a quite reasonable description of the
experimental differential cross section, in particular in the energy range of
the $\Delta$(1232) resonance. The effect of final state interactions is found
to be smaller for charged pion channels than in the case of neutral channel.
\begin{figure}[htb]
\centerline{\epsfxsize=11cm 
\epsffile{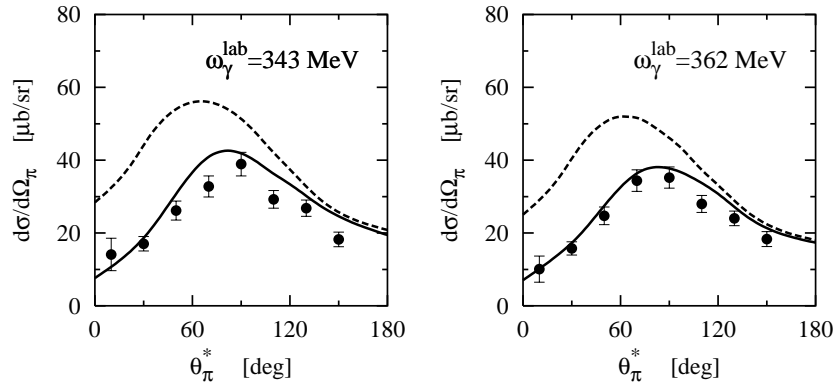}}
\vspace*{-0.1cm}
\caption{\small Differential cross section for $\pi^0$ photoproduction on the
  deuteron as a function of pion angle in the $\gamma N$ c.\ m.\ frame. Solid
  curves show our results include $NN$ and $\pi N$ rescattering and dashed
  curves show the results in the impulse approximation. The experimental data
  are from \cite{Krusche99}.} 
\label{diffwithexppi0}
\end{figure}
\vspace*{-0.2cm}~\\
As shown in Fig.\ \ref{tot_deut_exp} that our predictions for the total cross
section overestimate the experimental data for $\pi^0$ photoproduction in
the energy region between 340--360 MeV by about 6$\%$. In order to see how
good is the agreement with the experimental data for the differential cross
section at this energy region, we predict in Fig.\ \ref{diffwithexppi0} a
comparison between our results and the experimental data from
TAPS \cite{Krusche99}. It is noticeable that our predictions overestimate
the experimental differential cross section at this energy, is
particular at backward pion angles. 
%
\subsection{Comparison with other Theoretical Predictions\label{chap:4:2:2}}
%
In this section we compare our results for the differential cross sections with the theoretical predictions of Levchuk {\it
  et al.} \cite{Levchuk00N}. As in \cite{Levchuk00N}, the pion angle is
given in the $\gamma N$ c.\ m.\ frame in the case of $\pi^0$ channel.

\begin{figure}[ht]
\centerline{\epsfxsize=11cm 
\epsffile{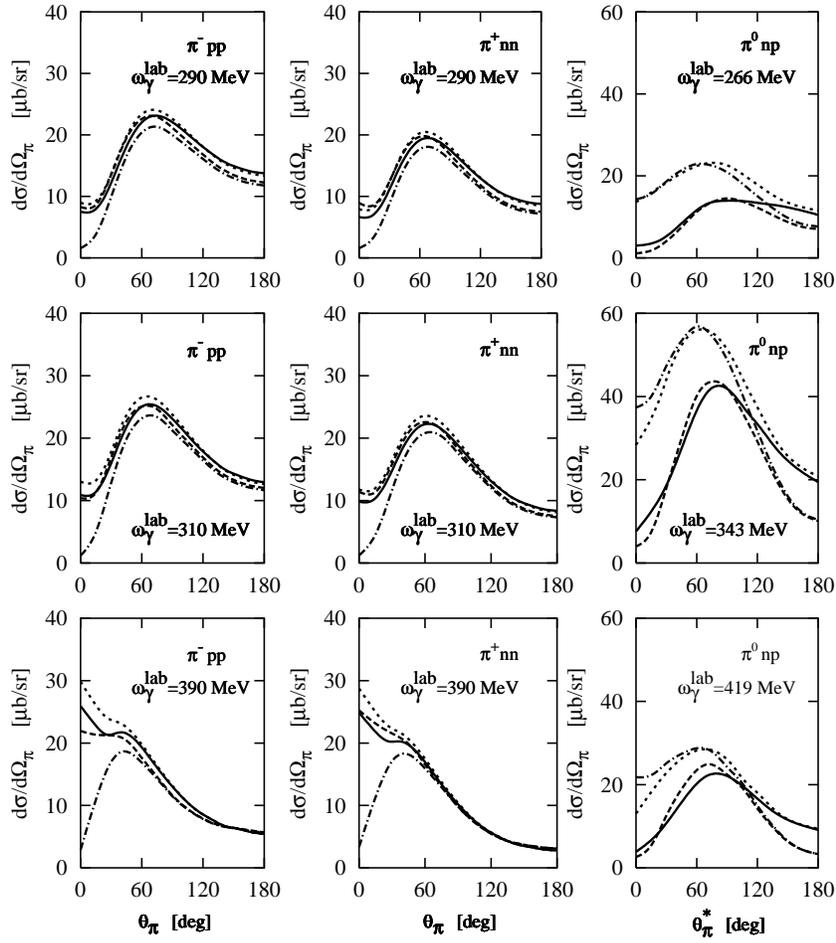}}
\vspace*{-0.1cm}
\caption{\small Differential cross sections for pion photoproduction on the
  deuteron in comparison with the results from \cite{Levchuk00N} at
  different photon energies. Full 
  curves show our results when rescattering effects are included and dotted
  curves show our results in the impulse approximation. Dashed and dash-dotted 
  curves show the results from \cite{Levchuk00N} with and without rescattering
  effects, respectively.}
\label{diffwiththeory}
\end{figure}

\vspace*{-0.2cm}~\\
It is worthwhile first to point out that, in the impulse approximation Fig.\
\ref{diffwiththeory} shows the following interesting features. Without final
state interactions we found a big difference between our calculation and the
prediction from \cite{Levchuk00N} for charged pion channels at pion
forward angles (see the difference between the dotted and the dash-dotted
curves in Fig.\ \ref{diffwiththeory}). Therefore, we checked in this work
where this big difference comes from.

\begin{figure}[htb]
\centerline{\epsfxsize=11cm 
\epsffile{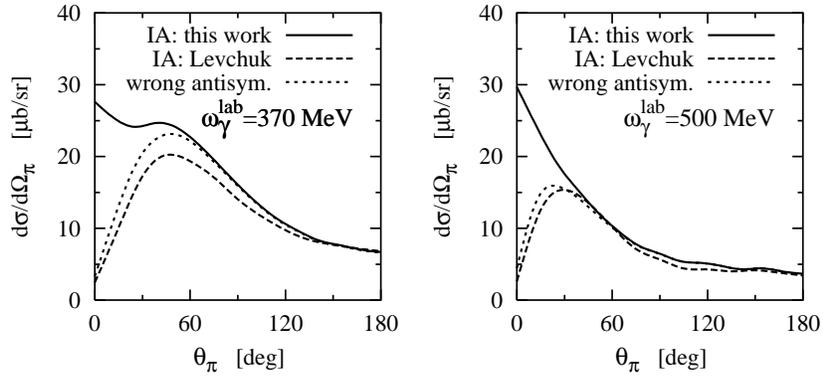}}
\vspace*{-0.3cm}
\caption{\small Differential cross section for $\pi^-$ photoproduction on the
  deuteron in the impulse approximation. The full curve shows the results of
  our calculation and the dashed one shows the results from \cite{Levchuk00N}. The dotted curve shows our results with a wrong $NN$ 
  antisymmetrization for $s=0$ channel, as given in Eq.\
  (\ref{nn_antiwrongd}).}
\label{wronglev}
\end{figure}

\vspace*{-0.2cm}~\\
First, we assume that the authors in Ref.\ \cite{Levchuk00N} used a wrong $NN$
antisymmetrization. That is why we examine using a wrong $NN$
antisymmetrization procedure in our work. We found that if we use a wrong $NN$ antisymmetrization for the $s=0$ channel, i.e.,
\begin{equation}
|\vec{p}_{1}\vec{p}_{2},s m_{s},t m_{t} \rangle =
  \frac{1}{\sqrt{2}}\left(
    |\vec{p}_{1}\rangle^{(1)}|\vec{p}_{2}\rangle^{(2)} + (-)^{s+t}
    |\vec{p}_{2}\rangle^{(1)}|\vec{p}_{1}\rangle^{(2)}\right)|s
  m_{s}\rangle |t m_{t}\rangle\, ,
\label{nn_antiwrongd}
\end{equation}
and a correct one for $s=1$ channel, i.e., 
\begin{equation}
|\vec{p}_{1}\vec{p}_{2},s m_{s},t m_{t} \rangle =
  \frac{1}{\sqrt{2}}\left(
    |\vec{p}_{1}\rangle^{(1)}|\vec{p}_{2}\rangle^{(2)} - (-)^{s+t}
    |\vec{p}_{2}\rangle^{(1)}|\vec{p}_{1}\rangle^{(2)}\right)|s
  m_{s}\rangle |t m_{t}\rangle\, ,
\label{nn_antirightd}
\end{equation}
we obtain results as shown in Fig.\ \ref{wronglev} (dotted curves) with the
  same energy shape as that of Levchuk {\it et al.} \cite{Levchuk00N} (dashed
  curves). This probably means that, they did not use the
  correct $NN$ antisymmetrization in their calculations.

\vspace*{-0.2cm}~\\
Moreover, since we use a coupled basis state for the $NN$ antisymmetrization
in our calculations, we also checked the use of an uncoupled, i.e.,
helicity basis which was used in \cite{Levchuk00N}. We found that both
approaches give the same results. This confirms the $NN$ antisymmetrization 
which we use and one can only suspect that the difference to the results
of Ref.\ \cite{Levchuk00N} may originate from an error in the
antisymmetrization.

\begin{figure}[htb]
\centerline{\epsfxsize=11cm 
\epsffile{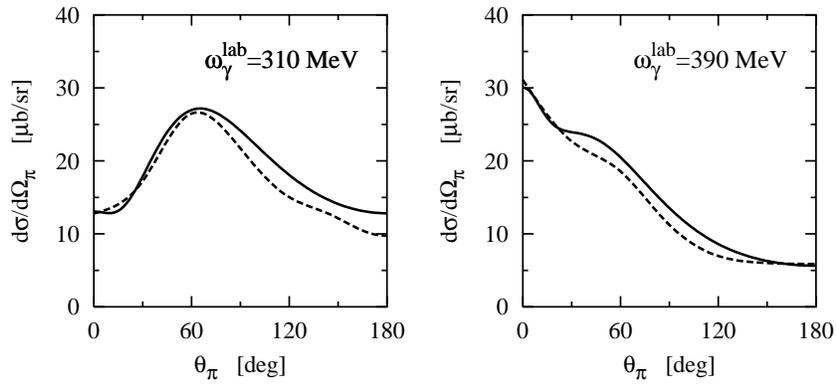}}
\vspace*{-0.3cm}
\caption{\small Differential cross section for $\pi^-$ photoproduction on
  the deuteron in the impulse approximation. The full curve shows our results
  using the elementary production operator of chapter \ref{chap:2} and the
  dashed one shows the results from \cite{Agus} using the elementary
  reaction amplitude from MAID \cite{Maid} analysis.}
\label{agus}
\end{figure}

\vspace*{-0.2cm}~\\
To investigate the influence of $\pi N \rightarrow KY$ rescattering in kaon
photoproduction on the deuteron, the authors in \cite{Agus} used the same
procedures as in Levchuk {\it et al.} \cite{Levchuk00N} to calculate pion
photoproduction on the deuteron in the impulse approximation. 
They obtained
results as shown in Fig.\ \ref{agus} (dashed curves). It is obvious that their
results using the elementary production operator from MAID \cite{Maid}
analysis are in good agreement with our results (solid curves) using the
elementary production operator presented in chapter \ref{chap:2}. It is also
clear that in both cases the differential cross section has a high value
at $\theta_{\pi}=0$ and not as in the case of Ref.\ \cite{Levchuk00N}. This  
suspects that the model in Ref.\ \cite{Levchuk00N} has a kind of error.
\newpage

\vspace*{-0.2cm}~\\
After including final state interaction effects, Fig.\ \ref{diffwiththeory} 
shows that the agreement of our results (solid curves) and the theoretical
predictions from \cite{Levchuk00N} (dashed curves) is good. A small
discrepancy between both calculations is observed. A reason for that may be due to the use of
different pion photoproduction operators. 
\newpage
\newpage
\newpage
%
\section{Spin Asymmetry and Gerasimov-Drell-Hearn Sum Rule\label{chap:4:3}}
%
Recently, several experiments to test the GDH sum rule both on the proton and
on the neutron are now in preparation or planned at different laboratories
around the world (MAMI, ELSA, LEGS, GRAAL and TJNAF). This makes the study of
polarization observables nowadays of great interest in the field of
intermediate 
energy nuclear physics. The spin asymmetry of the total photoabsorption cross
section, entering into the Gerasimov-Drell-Hearn (GDH) sum rule
\cite{Gerasimov66,Drell66}, is of particular interest. Therefore, we 
evaluate in this section the contribution of incoherent single pion
photoproduction including $NN$ and $\pi N$ rescattering to the spin asymmetry
and the corresponding GDH sum rule for the deuteron using the model developed
in chapter \ref{chap:3}.

\begin{figure}[htb]
\centerline{\epsfxsize=10cm 
\epsffile{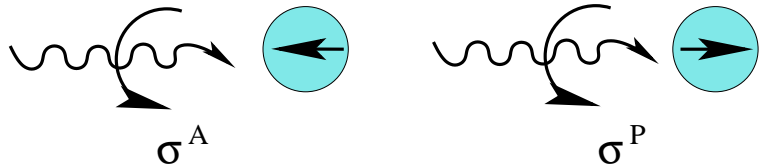}}
\vspace*{-0.3cm}
\caption{\small Illustration of the relative spin orientation of the incoming
  photon and the target nuclei in the GDH sum rule.}
\label{gdhfig}
\end{figure}

\vspace*{-0.2cm}~\\  
The verification of the GDH sum rule is an important issue, which enables a
check of some fundamental physical principles related to the spin of the
nucleon. The GDH sum rule connects the anomalous magnetic moment of a particle
with the energy weighted integral from threshold up to infinity over the spin
asymmetry of the total photoabsorption cross section, i.e., the difference of
the total photoabsorption cross sections for circularly polarized photons on a
target with spin parallel and antiparallel to the spin of the photon (see Fig.\
\ref{gdhfig}). In detail it reads for a particle of mass $M$, charge $eQ$,
anomalous magnetic moment $\kappa$ and spin $S$ 
\begin{eqnarray}
I^{GDH}(\infty) & = & 4\pi^2\kappa^2\frac{e^2}{M^2}\,S
~=~ \int_0^\infty \frac{dk}{k}
\left(\sigma ^P(k)-\sigma ^A(k)\right)\, ,\label{gdh}
\end{eqnarray}
where $\sigma ^{P/A}(k)$ denote the total absorption cross sections for
circularly polarized photons on a target with spin parallel and antiparallel 
to the photon spin, respectively, and the anomalous magnetic moment is 
defined by the total magnetic moment operator of the particle 
\begin{eqnarray}
\vec M = (Q+\kappa)\frac{e}{M}\vec S\,.
\end{eqnarray}

\vspace*{-0.2cm}~\\  
This sum rule gives a very interesting relation between a 
magnetic ground-state property of a particle and an integral property of 
its whole excitation spectrum. In other words, this sum rule shows that the 
existence of a nonvanishing anomalous magnetic moment points directly to an 
internal dynamic structure of the particle. Furthermore, because the
left-hand side of Eq.\ (\ref{gdh}) is positive, it tells us that the
integrated, energy-weighted total absorption cross section of a circularly
polarized photon on a 
particle with its spin parallel to the photon spin is bigger than the one on a
target with its spin antiparallel, if the particle posesses a nonvanishing
anomalous magnetic moment. 

\vspace*{-0.2cm}~\\ 
Since proton and neutron have large anomalous magnetic moments
($\kappa_p=1.79$ and $\kappa_n=-1.91$), one finds correspondingly  
large GDH sum rule predictions for them, i.e., $I^{GDH}_p(\infty)=204.8\,\mu$b 
for the proton and $I^{GDH}_n(\infty)=233.2\,\mu$b for the neutron. 
Applying the GDH sum rule to the deuteron, one finds a very interesting 
feature. The deuteron has isospin zero, ruling out the 
contribution of the large nucleon isovector anomalous magnetic moments to 
its magnetic moment. Therefore, one expects a very small 
anomalous magnetic moment for the deuteron. In fact, the experimental value 
is $\kappa_d=-0.143$ resulting in a GDH prediction of $I^{GDH}_d(\infty) =
0.65\,\mu$b, which is more than two orders of magnitude smaller 
than the nucleon values. 

\vspace*{-0.2cm}~\\
In recent years, Arenh\"ovel {\it et al.} \cite{Kress96} have evaluated
explicitly the GDH sum rule for the deuteron by integrating the difference of
the two total photoabsorption cross sections with photon and deuteron spins
parallel and antiparallel up to a photon energy of 550 MeV. There, three
contributions are included: (i) the photodisintegration channel $\gamma
d\rightarrow np$, (ii) the coherent pion photoproduction $\gamma
d\rightarrow\pi^0 d$ and (iii) the incoherent single pion photoproduction
$\gamma d \rightarrow\pi NN$. In their calculation of the $\gamma d
\rightarrow\pi NN$ contributions to the GDH integral, the authors restricted
themselves to the impulse approximation using the spectator nucleon approach
of Schmidt {\it et al.} \cite{Schmidt9695}. For the total GDH value from
explicit integration up to 550 MeV, they found a  
negative value $I^{GDH}_d(550\,\mbox{MeV})=-183\,\mu$b. However, some
uncertainty lies in the contribution of the incoherent single pion production
channel because final state interactions and other two-body operators are not 
included in the spectator nucleon model of Ref.\ \cite{Schmidt9695}.  
As discussed previously, final state interactions in incoherent single pion
photoproduction on the deuteron play an important role on the differential and
total cross sections. Therefore, the influence of rescattering effects
on the spin asymmetry and the corresponding GDH sum rule has to be
investigated.

\vspace*{-0.2cm}~\\
In this work, the influence of $NN$ and $\pi N$ rescattering effects in
incoherent single pion photoproduction to the spin asymmetry
and the GDH sum rule for the deuteron is investigated. We evaluated explicitly
the GDH sum rule for the deuteron by integrating the difference of the two
total photoabsorption cross sections with photon and deuteron spins parallel
and antiparallel up to a photon energy of 550 MeV. The upper integration limit
of 550 MeV is chosen, because we consider only single pion production. The
contributions which arise from coherent pion photoproduction and
photodisintegration channels are given in Ref.\ \cite{Kress96} to which we
refer for more details. 

\begin{figure}[htb]
\centerline{\epsfxsize=15cm 
\epsffile{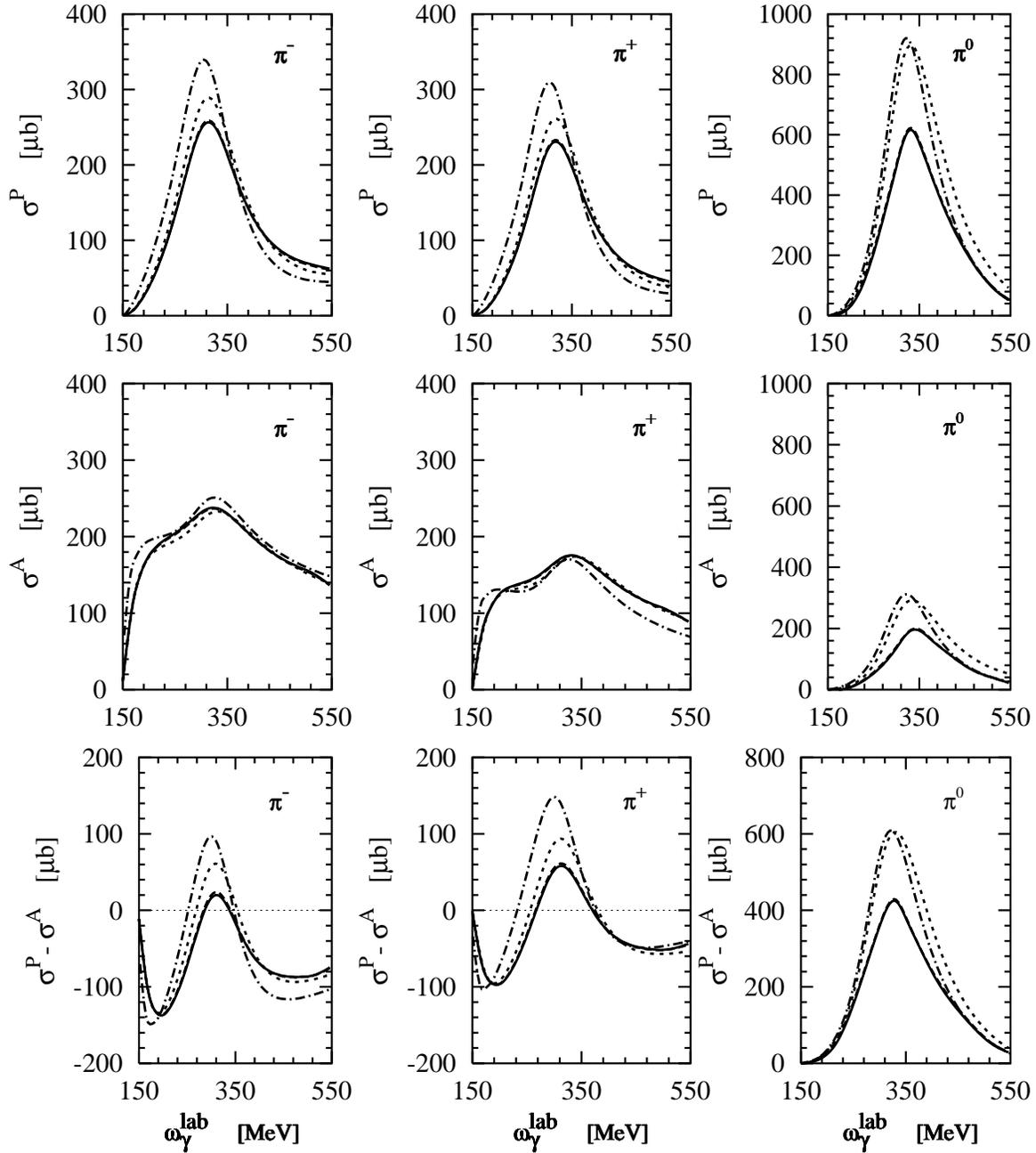}}
\vspace*{-0.1cm}
\caption{\small Total absorption cross sections for circularly polarized
  photons on a target with spin parallel $\sigma ^{P}$ (upper part) and
  antiparallel $\sigma ^{A}$ (middle part) to the photon spin. Lower part
  shows the difference of the cross sections. Dotted curves show the results
  in the impulse approximation, dashed curves show the results including $NN$
  rescattering and solid curves show the results with additional $\pi N$
  rescattering. The dash-dotted curves show the results for $\pi^-$ on the
  neutron (left panels), $\pi^+$ on the proton (middle panels) and $\pi^0$ on
  both the proton and the neutron (right pannels).}
\label{spin_asym_all}
\end{figure}

\vspace*{-0.2cm}~\\ 
Our results for the spin asymmetry are collected in Fig.\ \ref{spin_asym_all}
for the individual contributions from the different charge 
states of the pion. Fig.\ \ref{spin_asym_all} shows our results for the total
photoabsorption cross sections for circularly polarized photons on a target
with spin parallel $\sigma ^{P}$ (upper part) 
and antiparallel $\sigma ^{A}$ (middle part) to the photon spin. In the lower
part of Fig.\ \ref{spin_asym_all} the difference of the cross sections $\sigma
^{P} - \sigma ^{A}$ for incoherent pion photoproduction on the deuteron
including final state interactions is shown. For comparison, we also show our
results on the free nucleon by the dash-dotted curves in Fig.\
\ref{spin_asym_all} using the elementary production amplitude constructed
in chapter \ref{chap:2}. One notes qualitatively a similar
behaviour for the spin asymmetry although the maxima and minima are smaller
and also slightly shifted towards higher energies for the deuteron. The bottom
panels in Fig.\ \ref{spin_asym_all} show also that final state interactions
lead to a strong reduction of the spin 
asymmetry in the energy region of the $\Delta$(1232) resonance. This reduction
becomes more than 35$\%$ for $\pi^0$ photoproduction and becomes more than
15$\%$ for charged pion channels. It is also obvious that $\sigma^P$ is much
greater than $\sigma^A$, in particular in the case of $\pi^0$ production.

\begin{figure}[htb]
\centerline{\epsfxsize=15cm 
\epsffile{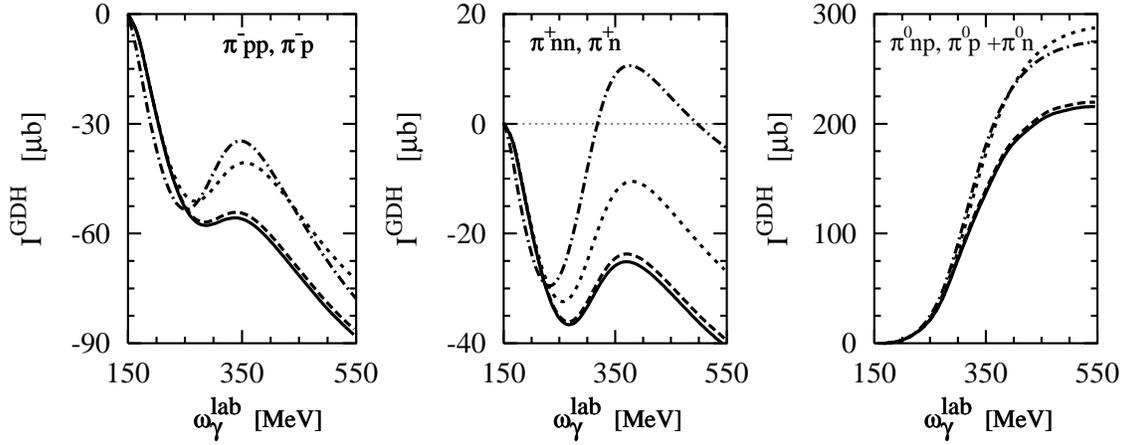}}
\vspace*{-0.3cm}
\caption{\small The Gerasimov-Drell-Hearn integral (see Eq.\ (\ref{gdh})) as a
  function of the upper limit of integration for the different channels of
  incoherent single pion photoproduction on the
  deuteron and the nucleon. Notation of the curves as in Fig.\ 
  \ref{spin_asym_all}.} 
\label{gdh_integral}
\end{figure}

\begin{figure}[htb]
\centerline{\epsfxsize=10cm 
\epsffile{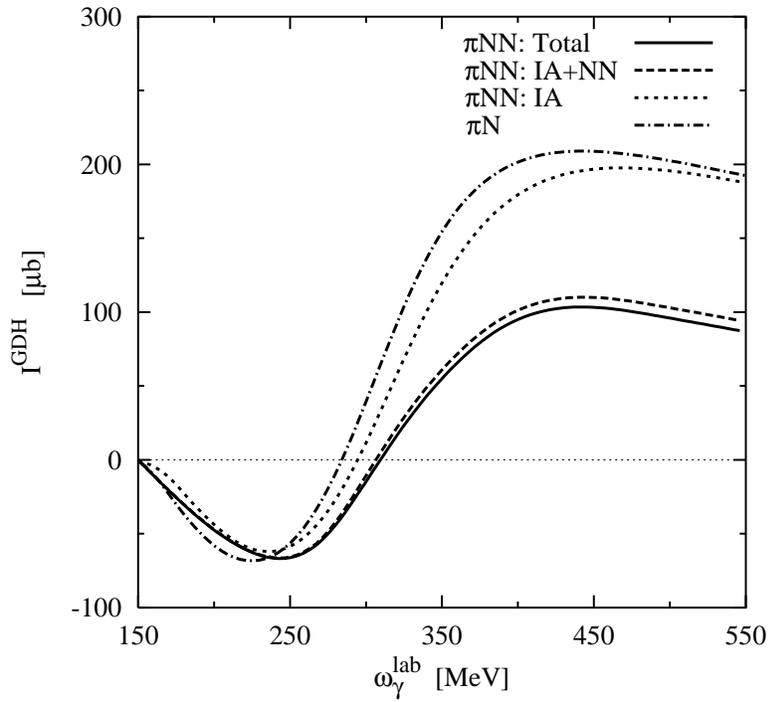}}
\caption{\small Summation of the contributions of the three channels of
  incoherent single pion photoproduction to the GDH sum rule for the deuteron and the nucleon as function of the upper
  integration energy. See caption of Fig.\ \ref{spin_asym_all} for meaning of
  the curves.}
\label{gdh_integral_sum}
\end{figure}

\vspace*{-0.2cm}~\\ 
The results for the GDH sum rule are depicted in Fig.\ \ref{gdh_integral} and
Table \ref{tab1} for the individual contributions from the different charge
states of the pion. Their total sum to the GDH sum rule is shown in Fig.\
\ref{gdh_integral_sum}. Our results on the free nucleon are also shown for
comparison (see the dash-dotted curves in Figs.\ \ref{gdh_integral} and
\ref{gdh_integral_sum}). It is obvious that a large positive contribution to
the GDH sum rule comes from the $\pi^0$ production channel whereas the charged
pions give a negative but - in absolute size - smaller contribution to the GDH
value. Up to an energy of 550 MeV one finds for the total contribution of the
incoherent pion production channels a value $I^{GDH}_{\gamma d \to
  NN\pi}(550\,\mbox{MeV})=87\,\mu$b.

\vspace*{-0.2cm}~\\ 
A very interesting and important result is the large negative contribution 
from the $\pi^{\pm}$ channels and the large positive contribution comes from
the $\pi^0$ channel to the GDH value. Hopefully, this low energy feature of 
the GDH sum rule could be checked experimentally in the near future. At the
same time, precise data on $\sigma^P-\sigma^A$ from a direct measurement is
urgently needed.

\vspace*{-0.2cm}~\\ 
Last but not least, we would like to conclude that the results presented here
for the spin asymmetry and the GDH sum rule can be used as a basis for the
simulation of the behaviour of polarization observables and for an optimal
planning of new polarization experiments of the reaction $\gamma d\rightarrow
\pi NN$.
\begin{table}
\begin{center}
\caption{\small Contributions of incoherent single pion photoproduction to
  the GDH integral for the deuteron integrated up to 550 MeV in $\mu$b.}
\vspace*{0.4cm}
\begin{tabular}{cccc}
\hline\hline\\
reaction&$I^{GDH}_{IA}$&$I^{GDH}_{IA+NN}$&$I^{GDH}_{IA+NN+\pi N}$\\ [2.1ex]
\hline\\
$\gamma d \to pp\pi^-$ &   -73   & -87   & -88 \\    
$\gamma d \to nn\pi^+$ &   -27   & -39   & -41 \\
$\gamma d \to np\pi^0$ &   287   & 220   & 216 \\ [2.1ex]
\hline\\
$\gamma d \to \pi NN$                  &   187   & 94    & 87  \\[2.1ex]
\hline\hline
\end{tabular}
\label{tab1}
\end{center}
\end{table}

\chapter{Summary and Outlook\label{chap:5}}
%
%
%
\section{Summary\label{chap:5:1}}
%
The subject of this work was the investigation of the influence of
final state interaction effects in incoherent single pion photoproduction on
the deuteron in the $\Delta (1232)$ resonance region. The transition matrix
elements are calculated in the frame of time-ordered perturbation theory,
using an elementary production operator, which yields a good agreement with
experimental data, and including final state interaction effects.

\vspace*{-0.2cm}~\\
For the elementary $\gamma N\rightarrow \pi N$ photoproduction amplitude we
used the effective Lagrangian model of Schmidt {\it et al.}
\cite{Schmidt9695}. This amplitude contains besides the standard
pseudovector Born terms the resonance contribution from the $\Delta(1232)$
excitation. The Born terms of the elementary production amplitude are
determined in pseudovector $\pi N$ coupling and supplied with a form
factor. The $\Delta$(1232) resonance is considered in both the $s$- and
$u$-channel. The parameters of the $\Delta$ resonance and the cut-off of the
form factors are fixed on the leading photoproduction multipole amplitudes. 
We found a good agreement between our results for total and differential cross
sections and experimental data.

\vspace*{-0.2cm}~\\
This elementary production operator is then used to study pion photoproduction on the
deuteron. In addition to the impulse approximation, where all kind of final
state interactions and other two-body operators are neglected, we have
included as presumably dominant final state interaction effects two-body
interactions in the $NN$- and $\pi N$-subsystems. As models
for the interaction of the $NN$- and $\pi N$-subsystems we used separable
interactions which are fitted to the phase shift data for $NN$ and $\pi N$
scattering. For $NN$ scattering
we have included all partial waves with total angular momentum $J\le 3$ by
numerical solution of the Lippmann-Schwinger equation \cite{Lippmann51} with
the separable representation of the Paris $NN$ potential
\cite{Haidenbauer84,Haidenbauer85}. In the case of $\pi N$ rescattering we have considered
the $S$-, $P$- and $D$-waves by solving the Lippmann-Schwinger equation for a separable energy-dependent
$\pi N$ potential from \cite{Nozawa90}. 

\vspace*{-0.2cm}~\\
For the total cross section we found that the influence of $NN$ and $\pi N$
rescattering effects on the total cross section is significant.
These effects reduce the total cross sections for the charged pion
photoproduction reactions in the $\Delta$(1232) resonance region by about
5$\%$. In the case of $\pi^0$ photoproduction reaction, rescattering is much
more important, reducing the total cross section by about 35$\%$ on the
maximum. Furthermore, it
was found that $\pi N$ rescattering is much less important compared to $NN$
rescattering. In comparison with 
experimental data, we found that the inclusion of $NN$ and $\pi N$
rescattering effects leads to an improved agreement between experimental data
and theoretical predictions. Only in the maximum of $\pi^0$
production our model overestimates the measured total cross section by about
6$\%$. In comparison with the theoretical predictions
from \cite{Levchuk00N}, we found that for charged pion production the
inclusion of rescattering effects leads to a reduction of the total cross
sections in contrast to the theoretical predictions from \cite{Levchuk00N}. 
This different behaviour may have its origin in the difference already for the
impulse approximation for which \cite{Levchuk00N} predicts a 14$\%$ lower
total cross section. In the case of $\pi^0$ channel, also in \cite{Levchuk00N} a strong reduction
by rescattering has been found. After including final state interactions we
found, however, only small differences between our calculations and the predictions from
\cite{Levchuk00N}.

\vspace*{-0.2cm}~\\
In the case of the differential cross section one may draw the following
conclusions. The inclusion of $NN$ and $\pi
N$ final state interactions reduces the
differential cross sections for the charged pion channels mainly at pion
forward angles by about 15$\%$. The reduction is much stronger for the  
$\pi^0$ channel by about 40$\%$ at pion forward angles. At pion backward
angles the influence is much less important. As already noted for the total
cross section, the $\pi N$ rescattering changes the final results only by a
few  
percent. In comparison with experiment, after including final state
interactions we have obtained a satisfactory agreement with the 
experimental data. Small discrepancies 
were found only at backward pion angles. In comparison with the theoretical
predictions from \cite{Levchuk00N}, without final state interactions we found
a big difference between both calculations for the differential cross sections
for charged pion channels at pion forward angles. After including rescattering effects we have obtained a
good agreement with the theoretical predictions from \cite{Levchuk00N}.
 
\vspace*{-0.2cm}~\\
Finally, we have evaluated in this work the
contribution of incoherent single pion photoproduction to the spin asymmetry
of the total photoabsorption cross section. We found that the inclusion of
final state 
interactions leads to a strong reduction of the spin asymmetry in the energy
region of the $\Delta$(1232) resonance. This reduction amounts to about 35$\%$
for $\pi^0$ photoproduction and about 15$\%$ for 
charged pion channels. The corresponding GDH integral was also evaluated up to
550 MeV. We found that a large positive contribution to the GDH sum rule comes
from the $\pi^0$ production channel whereas the charged pions give a negative
but - in absolute size - smaller contribution to the GDH value. Up to an
energy of 550 MeV we have obtained for the total 
contribution of the incoherent pion production channels a value
$I^{GDH}_{\gamma d \to NN\pi}(550\,\mbox{MeV})=87\,\mu$b.
%
\section{Future Extensions\label{chap:5:2}}
%
%
The studies we have discussed here will serve as the basis for 
further investigations including the dynamics of the $\pi NN$ system in a more
complete way. Undoubtedly, there is still a lot of work to be done both
experimentally and theoretically. Indeed, many challenging and interesting
lessons have yet to be learned before a deep understanding of the incoherent
pion photoproduction process on the deuteron will emerge. In first instance,
this work could be continued by 
the further refinement of the elementary production operator. Modifying the
elementary pion photoproduction operator on the free nucleon above the two
pion threshold with, for example, the inclusion of the $\omega$ and $\rho$
meson exchange in the $t$-channel and contributions at higher energies may 
improve our results for the spin asymmetry and the corresponding GDH sum rule
for the deuteron. Our goal is to have an operator that can describe the
elementary process on the free nucleon reasonably well over a larger energy
region and that is suitable
for applications to nuclear systems.

\vspace*{-0.2cm}~\\
In view of the importance of the first order rescattering it is natural to ask
about 
the role of higher order rescattering terms. Answering this question points to
the necessity of a three-body approach, where the final state interaction is
included to all orders. This may result in a better agreement between
experimental data and theoretical predictions for $\pi^0$ photoproduction. 
Furthermore, the inclusion of such rescattering terms may be
important in studying polarization observables. Studying these observables
will give us much more detailed information and thus will provide much more
stringent tests for theoretical models. In fact we plan to embark on such a
study in the near future.

\vspace*{-0.2cm}~\\
In the long run, one would need also to extend the formalism to the
threshold region for which the elementary production operator has to be
improved. This process is of great interest since experimental
data for the reaction $d(\gamma,\pi^0n)p$ have been measured recently in Mainz
(MAMI/TAPS) and Saskatoon (SAL) \cite{Hornidge01}. Moreover, the formalism
should be extended to investigate coherent and incoherent electroproduction of
pions on the deuteron including final state interaction effects in both the
threshold and the $\Delta$(1232) resonance regions in order to analyze recent
results from MAMI \cite{Merkel00}.

%
\newpage 
\newpage
\addcontentsline{toc}{chapter}{Appendices}
\pagestyle{empty}
\vspace*{8.5cm}
\qquad\qquad\qquad\qquad\qquad\qquad\qquad\qquad\qquad\qquad\qquad{\bf\huge{Appendices}} 
\vspace*{0.2cm}
\hrule
\newpage ~\\
\newpage
\pagestyle{empty}
\begin{appendix}
\def\BIG#1{{\hbox{$\left#1\vbox to 25pt{}\right.$}}}
\renewcommand{\chaptermark}[1]{\markboth{\sc \appendixname\ 
                \thechapter.\  #1}{}}
\renewcommand{\sectionmark}[1]{\markright{\sc \thesection\ #1}{}}
\pagestyle{fancy}
\chapter{General Notations and Conventions\label{appendixA}}
%
\section{Dirac Algebra\label{appendixA:1}}
%
Throughout this work we adopt the natural system of units where 
$\hbar = c = 1$; conversion factor: $\hbar c=197.32696$ MeV.fm\footnote{In the
  calculations of this work we use units of MeV for energies, masses and
  momenta.}. We also follow the convention of Bjorken and Drell
\cite{Bjorken64}, where the contravariant space-time four-vector is defined as
\begin{eqnarray}
x^{\mu} & \equiv & (x^0, x^{1}, x^{2}, x^{3})~ \equiv ~(t, 
        {\vec{x}})~ \equiv ~(t,x,y,z)\, ,
\end{eqnarray}
and the covariant space-time four-vector is given by
\begin{eqnarray}
x_{\mu} & \equiv & (x_0, x_{1}, x_{2}, x_{3})~ \equiv ~(t, -{\vec{x}})~ 
        \equiv ~(t,-x,-y,-z) \nonumber\\
        & = & g_{\mu \nu}~x^{\nu}\, ,
\end{eqnarray}
with the transformation matrix 
\begin{eqnarray}
g_{\mu \nu} = \left( \begin{array}{rrrr} 1 & 0 & 0 & 0\\  
0 & -1 & 0 & 0\\ 0 & 0 & -1 & 0 \\ 0 & 0  & 0 & -1 \end{array} \right)\, .
\end{eqnarray}

\vspace*{0.2cm}~\\
Likewise the contravariant four-momentum is
\begin{eqnarray}
p^{\mu} & \equiv & (p^{0}, p^1, p^2, p^3)~ \equiv ~(E_{p}, {\vec{p}})\, ,
\end{eqnarray}
and the scalar product between two four-momenta is given by
\begin{eqnarray}
p \cdot q & \equiv & p^{\mu} q_{\mu}~ \equiv ~E_{p} E_{q} - {\vec{p}} 
\cdot {\vec{q}}\, .
\end{eqnarray}

\vspace*{0.2cm}~\\
The Dirac matrices are
\begin{eqnarray}
\gamma^{\mu} & \equiv & (\gamma^{0}, {\vec{\gamma}})\, ,
\end{eqnarray}
with the matrix representation 
\begin{eqnarray}
\gamma^{0} ~ = ~ \left( \begin{array}{rr} {\rm 1\hspace{-0.75ex}1} & 0\\  0 & 
-{\rm 1\hspace{-0.75ex}1} \end{array} \right)~ , 
{\vec{\gamma}} ~ = ~ \left( \begin{array}{cc} 0 & {\vec{\sigma}}
\\  -{\vec{\sigma}} & 0 \end{array} \right)\, ,
\end{eqnarray}
where $\gamma_0$ is hermitian and $\gamma_1$, $\gamma_2$ and $\gamma_3$ are
anti-hermitian. The Pauli matrices ${\vec{\sigma}}$ = ($\sigma^{1}$, $\sigma^{2}$, 
$\sigma^{3}$) are denoted by
\begin{eqnarray}
\sigma^{1}~ = ~\left( \begin{array}{rr} 0 & 1\\  1 & 0 \end{array} \right)~ ,
~\sigma^{2}~ = ~\left( \begin{array}{rr} 0 & -i\\  i & 0 \end{array} \right)~ ,
~\sigma^{3}~ = ~\left( \begin{array}{rr} 1 & 0\\  0 & -1 \end{array} \right)\, .
\label{paulimat}
\end{eqnarray}
These matrices satisfy the anticommutation relations
\begin{eqnarray}
\left\{ \sigma^{i}, \sigma^{j} \right\} & \equiv & \sigma^{i} \sigma^{j} + 
\sigma^{j} \sigma^{i}~ = ~2 \delta_{ij}\, ,
\end{eqnarray}
as well as the commutation relations
\begin{eqnarray}
\left[ \sigma^{i}, \sigma^{j} \right] & \equiv & \sigma^{i} \sigma^{j} - 
\sigma^{j} \sigma^{i}~ = ~2i \epsilon_{ijk} \sigma^{k}\ ,
\label{eijk}
\end{eqnarray}
where $\epsilon_{ijk}$ represents the antisymmetric Levi-Civita tensor in
$R^3$.

\vspace*{0.2cm}~\\
The Dirac matrices $\gamma$ satisfy the anticommutation relations
\begin{eqnarray}
\left\{ \gamma^{\mu}, \gamma^{\nu} \right\} & \equiv & \gamma^{\mu} 
\gamma^{\nu} + \gamma^{\nu} \gamma^{\mu}~ = ~2 g^{\mu \nu}\, .
\end{eqnarray}
Important combinations of $\gamma$ matrices are the antisymmetric combination
\begin{eqnarray}
\sigma^{\mu\nu} & = & \frac{i}{2} \left[\gamma^{\mu},\gamma^{\nu}\right]
\end{eqnarray}
with components
\begin{eqnarray}
\sigma^{ij}~ = ~\epsilon_{ijk} \left( \begin{array}{cc} \sigma^k & 0\\  0 & \sigma^k 
\end{array} \right)~ ~~{\rm and}~~ \sigma^{0i}~ = ~i~\left( 
\begin{array}{cc} 0 & \sigma^i\\  \sigma^i & 0 \end{array} \right)\, .
\end{eqnarray}

\vspace*{0.2cm}~\\
Other useful combinations are
\begin{eqnarray}
\gamma^{5} & \equiv & i \gamma^{0} \gamma^{1} \gamma^{2} \gamma^{3}
~=~ \gamma_{5} ~=~ {\textstyle \frac{1}{24}} i\epsilon_{\mu \nu \rho \sigma} 
\gamma^{\mu}\gamma^{\nu} \gamma^{\rho} \gamma^{\sigma} ~=~ 
\left( \begin{array}{rr} 0 & {\rm 1\hspace{-0.75ex}1}\\  {\rm 1\hspace{-0.75ex}1} & 0 \end{array} \right)~ ,\\
i \epsilon_{\mu \nu \rho \sigma} \gamma^{\mu} & = & \gamma_{5} 
(-\gamma_{\nu} \gamma_{\rho}
\gamma_{\sigma} + g_{\nu \rho} \gamma_{\sigma} + g_{\rho \sigma} \gamma_{\nu}
- g_{\nu \sigma} \gamma_{\rho})~ ,\\
\gamma_{5} \sigma^{\mu \nu} & = & {\textstyle \frac{1}{2}} 
i\epsilon^{\mu \nu \rho \sigma} \sigma_{\rho \sigma}~~~ ,\\
\gamma_5 \gamma_{\sigma} & = & -\gamma_{\sigma}\gamma_5 ~=~
{\textstyle \frac{1}{6}} i \epsilon_{\mu \nu \rho \sigma}
\gamma^{\mu}\gamma^{\nu} \gamma^{\rho}\, .
\end{eqnarray}
The antisymmetric Levi-Civita tensor is defined by
\begin{eqnarray}
\epsilon_{\mu \nu \rho \sigma} & = & \left\{ \begin{array}{rl}
+1 & ~~{\rm for ~ an~ even~ permutation~ (e.g.\ 0,1,2,3) }\\
-1 & ~~{\rm for~ an~ odd~ permutation}\\
0 & ~~{\rm if~two~or~more~indices~are~the~same} \end{array} \right. \, .
\label{eijkl}
\end{eqnarray}

\vspace*{0.2cm}~\\
The scalar product between $\gamma$ matrices and a four-momentum is written as
\begin{eqnarray}
 \gamma^{\mu}  p_{\mu} & = & \gamma^{0} p^{0} - 
{\vec{\gamma}} \cdot {\vec{p}}~ \equiv ~p \slt\, .
\end{eqnarray}

\vspace*{0.2cm}~\\
The positive-energy four-component free Dirac spinor has the form
\begin{eqnarray}
u(p,s)~ = ~\biggl(\frac{E_p + m}{2m} \biggr)^{\frac{1}{2}}
\left( \begin{array}{c} \chi_{s} \vspace{2mm}\\ \displaystyle 
\frac{{\vec{\sigma}} \cdot {\vec{p}}}{E_p + m} \chi_{s} \end{array} \right)\, ,
\end{eqnarray}
and the negative-energy four-component Dirac spinor (antiparticle spinor) has 
the form 
\begin{eqnarray}
v(p,s)~ = ~\biggl(\frac{E_p + m}{2m} \biggr)^{\frac{1}{2}}
\left( \begin{array}{c} \displaystyle -\frac{{\vec{\sigma}} \cdot 
{\vec{p}}}{E_p + m} \chi_{s} \\ \chi_{s} \end{array} \right)\, ,
\end{eqnarray}
where $E_p=\sqrt{m^2+p^2}$. They are normalized as
\begin{eqnarray}
\overline{u}(p,s) u(p,s) & = & 1 \, ,\\
\overline{v}(p,s) v(p,s) & = & -1 \, ,
\end{eqnarray}
where $\chi_{s}$ are the two-component Pauli spinors with $\chi_{+\frac{1}{2}}=\left( \begin{array}{c} \displaystyle 1 \\ 0 \end{array} \right)~$ and
$\chi_{-\frac{1}{2}}=\left( \begin{array}{c} \displaystyle 0 \\ 1 \end{array} \right)$, and the Dirac adjoint spinors are defined as
\begin{eqnarray}
\overline{u}(p,s) & = & u^{\dagger} \gamma^{0}\, ,\\
\overline{v}(p,s) & = & v^{\dagger} \gamma^{0}\, .
\end{eqnarray}

\vspace*{0.2cm}~\\
Using Dirac spinors $u$ and $v$, the Dirac equations may be written as
\begin{eqnarray}
(p\slt - m)u(p,s) & = & 0 \, ,\\
(p\slt + m)v(p,s) & = & 0\, ,
\end{eqnarray}
which in terms of the adjoint spinors become
\begin{eqnarray}
\overline{u}(p,s)(p\slt - m) & = & 0\, ,\\
\overline{v}(p,s)(p\slt + m) & = & 0\, .
\end{eqnarray}
%
\section{Normalization of States\label{appendixC:1}}
%
The general form of the state which describes the single particle
$X\in\{N,\Delta,\pi,\gamma\}$ with momentum $\vec{p}$ and quantum numbers\footnote{These quantum numbers indicate about spin, isospin, spin
  projection etc.} $a$ is $\ket{X;\vec{p} a}$. It is also possible to use the form
$\ket{\vec{p};a}$ when we know the type of the used particle. In case of more than
one particle, the state has the form 
$\ket{X_{1},X_{2};\vec{p}_{1} a_{1},\vec{p}_{2} a_{2}}$.

\vspace*{-0.2cm}~\\ 
For all the involved particles a covariantly normalized state is used in this
work. The fermions are normalized as  
\begin{eqnarray}
\braket{N;\vec{p_2}}{N;\vec{p_1}} = (2\pi)^{3}\delta^{3}
(\vec{p_2}-\vec{p_1})\frac{E_{p}}{M_{N}}\, .
\end{eqnarray}
The normalized states for bosons are given by 
\begin{eqnarray}
\braket{\pi ;\vec{q_2}}
{\pi ;\vec{q_1}} = (2\pi)^{3}\delta^{3}
(\vec{q_2}-\vec{q_1})2\omega_{q} & , &
\omega_{q} = \sqrt{m_{\pi}^{2}+\vec{q}^{2}}  \\
\braket{d ;\vec{p_2}}{d ;\vec{p_1}} = (2\pi)^{3}\delta^{3}
(\vec{p_2}-\vec{p_1})2E_{d} & , &
E_{d} = \sqrt{M_{d}^{2}+\vec{p}^{2}} \\
\braket{\gamma ;\vec{k_2}}{\gamma ;\vec{k_1}} = (2\pi)^{3}\delta^{3}
(\vec{k_2}-\vec{k_1})2\omega_{\gamma} & , & 
\omega_{\gamma} = |\vec{k}| = k \, .
\end{eqnarray}  
The completeness relation for fermions then reads 
\begin{eqnarray} 
\eins = \int \frac{d^{3}p}{(2\pi )^{3}}\frac{M_{N}}{E_{p}}
\ket{\vec{p}}\bra{\vec{p}}
\end{eqnarray} 
and for bosons reads
\begin{eqnarray} 
\eins = \int \frac{d^{3}q}{(2\pi )^{3}} \frac{1}{2\omega}
\ket{\vec{q}}\bra{\vec{q}}\, .
\end{eqnarray}
%
\section{Spherical Basis\label{appendixC:3}}
%
For the construction of the physical pions from the pion field $\vec{\Phi}$ 
and for the evaluation of the spin and isospin operators, $\vec{\sigma}$ and 
$\vec{\tau}$, their representation in spherical basis is used. 
The basis vectors are given by 
\begin{eqnarray} 
\hat{e}_{\mu} & = & -\frac{\mu}{\sqrt{2}}\left(\hat{e}_{x}+i\mu \hat{e}_{y}
\right) \, ,\qquad {\rm for} \; \mu = \pm 1 \nonumber \\
\hat{e}_{0} & = & \hat{e}_{z}\, .
\end{eqnarray} 
The spherical components of a given vector are determined by writing the scalar
product 
\begin{eqnarray}
a_{\mu} = \vec{a}\cdot\hat{e}_{\mu} \, .
\end{eqnarray}
If the spherical components of two vectors are given, then the scalar product
of these two vectors in spherical basis is given by
\begin{eqnarray}
\vec{a}\cdot\vec{b} = \sum_{\mu} a_{\mu}(-)^{\mu}b_{-\mu}
\end{eqnarray}
and the vector product is given by
\begin{eqnarray}
(\vec{a}\times\vec{b})_{\mu} = -i\sqrt{2}\sum_{\mu_{a}\mu_{b}}
(1\mu_{a}\, 1\mu_{b}|1\mu)\, a_{\mu_{a}}b_{\mu_{b}}\, .
\end{eqnarray}

\chapter{Useful Formulas for the Elementary Process\label{appendixB}}
%
In this appendix we give the useful formulas for the pion photoproduction 
reaction on the free nucleon. This process is schematically sketched in 
Fig.\ \ref{fig:2.1}. Many of the formulas given here have been published 
elsewhere; our aim has been to collect all of the kinematic equations needed 
in this work. 
%
\section{Kinematics and Relevant Formulas\label{appendixB:1}}
%
The kinematics of pion photoproduction reaction on the free nucleon, 
$\gamma N \rightarrow \pi N$, are characterized by the four-momentum vectors 
$k=(\omega_{\gamma},{\vec k})$ for the incident photon, $p_1=(E_1,{\vec p}_1)$ 
for the initial nucleon, $p_2=(E_2,{\vec p}_2)$ for the final nucleon and 
$q=(\omega_{\vec q},{\vec q})$ for the produced pion. Then the 
kinematical equations (energy-momentum conservations) are given by
\begin{eqnarray}
{\vec p}_1 + {\vec k} &=& {\vec p}_2 + {\vec q} \, , \nonumber \\
p_1^2 &=& p_2^2 ~=~ M_N^2\, , \nonumber \\
q^2 &=& m_{\pi}^2\, ,
\end{eqnarray}
where $M_N$ is the nucleon mass and $m_{\pi}$ is the pion mass.

\vspace*{-0.2cm}~\\
The usual Mandelstam kinematical variables are given by
\begin{eqnarray}
s &=& (p_2 + q)^2 = (p_1 + k)^2 \, , \nonumber \\
t &=& (k - q)^2 = (p_2 - p_1)^2 \, , \nonumber \\
u &=& (p_2 - k)^2 = (p_1 - q)^2 \, ,
\label{mandel1}
\end{eqnarray}
and they satisfy
\begin{eqnarray}
s + t + u &=& 2M_N^2 + m_{\pi}^2 \, .
\end{eqnarray}
\begin{figure}[htb]
\centerline{\epsfxsize=8cm \epsffile{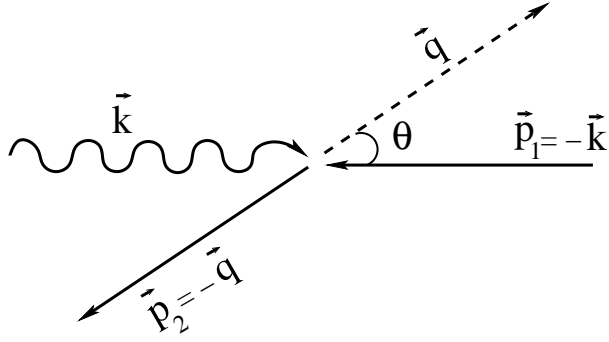}}
\caption{\small Kinematics in the $\pi N$ c.m.\ frame for the reaction 
$\gamma N\rightarrow \pi N$.}
\label{cmsys}
\end{figure}

\vspace*{-0.2cm}~\\
In the $\pi N$ c.m. frame which given graphically in Fig.\ \ref{cmsys} let 
$W$ be the total energy of the system. Then
\begin{eqnarray}
s &=& W^2 \, , \nonumber \\
t &=& 2{\vec k}\cdot{\vec q} - 2\omega_{\gamma}\omega_{\vec q} + m_{\pi}^2 \, , 
                \nonumber \\
u &=& -2{\vec k}\cdot{\vec q} - 2\omega_{\gamma}E_2 + M_N^2 \, ,
\label{mandel}
\end{eqnarray}
where ${\vec p}_1=-{\vec k}$ and ${\vec p}_2=-{\vec q}$ in the c.m. frame. 
The scattering angle $\theta$ is given by
\begin{eqnarray}
\cos~\theta &=& \frac{({\vec k}\cdot{\vec q})}{(\mid{\vec q}\mid\mid{\vec k}
        \mid)}~=~x \, .
\end{eqnarray}
In the following, we will use the following notation to denote the 3-momentum
in the system
\begin{eqnarray}
k &=& \mid{\vec k}\mid,~~\hat{\vec k}~=~\frac{\vec k}{k};~~q~=~\mid{\vec q}
        \mid,~~\hat {\vec q}~=~\frac{\vec q}{q} \, .
\end{eqnarray}

\vspace*{-0.2cm}~\\
The invariant mass of the $\pi N$ subsystem is given using the relativistic 
energy-momentum relations by
\begin{eqnarray}
\label{invgn}
W^2 &=& \left( \sqrt{M_N^2 + \vec p_2^{\, 2}} + \sqrt{m_{\pi}^2 + 
                \vec q^{\, 2}} \right)^2 - ({\vec p}_2 + {\vec q})^2 \nonumber \\
        &=& \left( \sqrt{M_N^2 + q^2} + \sqrt{m_{\pi}^2 + 
                q^2} \right)^2 \nonumber \\
        &=& M_N^2 + 2M_N\omega_{\gamma}^{lab}\, .
\end{eqnarray}

\vspace*{-0.2cm}~\\
The various energies and momenta in the $\pi N$ c.m.\ frame are given in terms 
of the invariant mass of the $\pi N$ subsystem, $W$, by
\begin{eqnarray}
k &=& \frac{W^2 - M_N^2}{2W}~=~\omega_{\gamma} \, ,
\label{kcm}
\end{eqnarray}
\begin{eqnarray}
q &=& \frac{1}{2W}\sqrt{(W^2-M_N^2+m_{\pi}^2)^2 - 4m_{\pi}^2W^2}\, ,
\label{qcm}
\end{eqnarray}
\begin{eqnarray}
\omega_{\vec q} &=& \frac{W^2 - M_N^2 + m_{\pi}^2}{2W} \, ,
\label{eqcm}
\end{eqnarray}
\begin{eqnarray}
E_1 &=& \frac{W^2 + M_N^2 }{2W} \, ,
\label{e1cm}
\end{eqnarray}
\begin{eqnarray}
E_2 &=& \frac{W^2 + M_N^2 - m_{\pi}^2}{2W}\, .
\label{e2cm}
\end{eqnarray}
%
\section{Multipole Decomposition of Amplitudes\label{appendixB:4}}
%
The on-shell $T_{fi}$-matrix element is expanded in the c.m.\ frame to four
independent amplitudes as follows \cite{Chew57}
\begin{eqnarray}
\langle {\vec q}\mu \mid T_{fi}(W)\mid{\vec k}{\vec\epsilon}\rangle &=& 
        \frac{4\pi W}{M_N}\left[ i({\vec\sigma}\cdot{\vec\epsilon})F_1 + 
        ({\vec\sigma}\cdot\hat {\vec q}){\vec\sigma}\cdot(\hat {\vec k}
        \times{\vec\epsilon}) F_2 + 
        i({\vec\sigma}\cdot\hat {\vec k})(\hat {\vec q}
        \cdot{\vec\epsilon})F_3 \right. \nonumber \\  
        && \left. + i({\vec\sigma}\cdot\hat {\vec q})(\hat {\vec q}
        \cdot{\vec\epsilon})F_4\right]\, ,
\label{osmatrix} 
\end{eqnarray}
where $F_i=F_i(W,x)$ with $x={\hat{\vec q}}\cdot{\hat{\vec k}}$ are the well
    known CGLN-amplitudes. ${\hat{\vec k}}$ and ${\hat{\vec
    q}}$ are momentum unit vectors of the photon and meson, respectively.
The isospin decomposition of the CGLN-amplitudes has the form
\begin{eqnarray}
F_i &=& F_i^{(-)}\frac{[\tau^{+}_{\mu},\tau_0]}{2}+
        F_i^{(+)}\frac{\left\{\tau^{+}_{\mu},\tau_0\right\}}{2}+
        F_i^{(0)}\tau_{\mu}^+~~~~({\rm i=1,2,3,4})\, ,
\label{ampsigma}
\end{eqnarray}
with the following commutator and the anti-commutator relations
\begin{eqnarray}
[\tau^{+}_{\mu},\tau_0] & = & 2\mu\tau^{+}_{\mu} ~=~
(-)^{\mu}2\mu\tau_{-\mu}\, , \nonumber \\
\{\tau^{+}_{\mu},\tau_0\} & = & (-)^{\mu}\{\tau_{-\mu},\tau_0\} ~=~
2\delta_{\mu,0}\, ,
\end{eqnarray}
where $\tau_{\mu}$ and $\tau_3$ are the nucleon isospin matrices which act on
the isospinor of the nucleon.

\vspace*{-0.2cm}~\\
With this convention the amplitudes of the pion photoproduction 
process for the four physical channels consist as follows \cite{Ericson88}
\begin{eqnarray}
\langle n\pi^{+} |~F_i~| \gamma p \rangle & = & -\sqrt{2}~[F_i^{(0)} + 
        F_i^{(-)}]\, , \nonumber \\ \label{negsig1}
\langle p\pi^{-} |~F_i~| \gamma n \rangle & = & \sqrt{2}~[F_i^{(0)} - 
        F_i^{(-)}]\, , \nonumber \\
\langle p\pi^{0} |~F_i~| \gamma p \rangle & = & F_i^{(+)} + F_i^{(0)}\, , \nonumber \\
\langle n\pi^{0} |~F_i~| \gamma n \rangle & = & F_i^{(+)} - F_i^{(0)}\, .
\end{eqnarray}

\vspace*{-0.2cm}~\\
The amplitudes $F_i^{(3/2)}$ and $F_i^{(1/2)}$ referring to the final $\pi N$
isospin states with isospin $\frac 32$ and $\frac 12$, respectively, are
defined by
\begin{eqnarray}
F_i^{(3/2)} & = & F_i^{(+)} - F_i^{(-)} ~~~~~~ ({\textstyle {\rm I}=
        \frac{3}{2}})\, ,
\nonumber\\
\label{consisten2}
F_i^{(1/2)} & = & F_i^{(+)} + 2F_i^{(-)} ~~~~~ ({\textstyle {\rm I}=
        \frac{1}{2}})\, .
\end{eqnarray}
The isoscalar amplitudes $F_i^{(0)}$ leads only to a $\pi N$ states with 
isospin $\frac 12$. In terms of these, one can define the neutron and
the proton multipole amplitudes with total isospin ${\rm I}=\frac 12$ as follow
\begin{eqnarray}
F_{i,n}^{(1/2)} &=& F_i^{(0)}-\frac{1}{3}F_i^{(1/2)}\, , \nonumber \\
F_{i,p}^{(1/2)} &=& F_i^{(0)}+\frac{1}{3}F_i^{(1/2)}\, ,
\end{eqnarray}
where the subscript $p$ ($n$) denotes a proton (neutron) target. The four
charge channels in turn are given by
\begin{eqnarray}
F_i(\gamma p \rightarrow n\pi^+) & = & \sqrt{2}~[F_{i,p}^{(1/2)} -
        \frac 13 F_i^{(3/2)}]\, , \nonumber \\
F_i(\gamma n \rightarrow p\pi^-) & = & \sqrt{2}~[F_{i,n}^{(1/2)} +
        \frac 13 F_i^{(3/2)}]\, , \nonumber \\
F_i(\gamma p \rightarrow p\pi^0) & = & F_{i,p}^{(1/2)} +
        \frac 23 F_i^{(3/2)}\, , \nonumber \\
F_i(\gamma n \rightarrow n\pi^0) & = & -~ F_{i,n}^{(1/2)} + \frac 23
        F_i^{(3/2)}\, ,
\end{eqnarray}
where the isospin ${\rm I}=\frac 32$ multipole amplitudes, $F_i^{3/2}$, are the
same for both protons and neutrons.     

\vspace*{-0.2cm}~\\
The partial wave decomposition of $T_{fi}$ defines the multipole 
amplitudes $A_{\ell_{\pm}}(W)$, where $\ell_{\pm}$ is the $\pi N$-orbital 
angular momentum $\ell$ and total angular momentum ($\ell\pm\frac 12$) as well 
as $A=E$ ($A=M$) may characterize an electrical (magnetic) transition. The 
'+', '-' determine whether the nucleon spin $\frac 12$ must be added to or 
subtracted from the orbital momentum to give the total final state momentum. 
One finds the following relations between the multipole and CGLN-amplitudes 
\cite{Olsson75,Drechsel92}
\begin{eqnarray}
E_{\ell +} &=& \frac{1}{2\ell+2}\int_{-1}^{+1}dx\left[P_{\ell}(x)F_1-
P_{\ell+1}(x)F_2+\frac{\ell}{2\ell+1}(P_{\ell-1}(x)-
P_{\ell+1}(x))F_3 \right.\nonumber \\
 & & \left. + \frac{\ell+1}{2\ell+3}(P_{\ell}(x)-P_{\ell+2}(x))F_4 \right] \, , 
\nonumber \\
E_{\ell -} &=& \frac{1}{2\ell}\int_{-1}^{+1}dx\left[P_{\ell}(x)F_1-
P_{\ell-1}(x)F_2+\frac{\ell+1}{2\ell+1}(P_{\ell+1}(x)-P_{\ell-1}(x))F_3 
\right. \nonumber \\ 
 & & \left. + \frac{\ell}{2\ell-1}(P_{\ell}(x)-P_{\ell-2}(x))F_4 \right] \, , \nonumber \\
M_{\ell +} &=& \frac{1}{2\ell+2}\int_{-1}^{+1}dx\left[P_{\ell}(x)F_1-
P_{\ell+1}(x)F_2+\frac{1}{2\ell+1}(P_{\ell+1}(x)-P_{\ell-1}(x))F_3\right] \, , \nonumber \\
M_{\ell -} &=& \frac{1}{2\ell}\int_{-1}^{+1}dx\left[-P_{\ell}(x)F_1+
P_{\ell-1}(x)F_2+\frac{1}{2\ell+1}(P_{\ell-1}(x)-P_{\ell+1}(x))F_3\right] \, .\nonumber \\
  & & 
\end{eqnarray}
Here the $P_{\ell}(x)$ are Legendre polynomials of the first kind.

\vspace*{-0.2cm}~\\
The partial wave decomposition of the CGLN-amplitudes into multipole 
amplitudes $E_{\ell\pm}$, $M_{\ell\pm}$ corresponding to good parity and 
angular momentum states is very convenient for the description of the 
resonance excitation process:
\begin{eqnarray}
F_1 &=& \sum_{\ell=0}^{\infty} \{ (\ell M_{\ell +} + E_{\ell +})
        P^{\prime}_{\ell +1} + \left [ (\ell+1)M_{\ell -} + E_{\ell -}
        \right ] P^{\prime}_{\ell-1}\} \, , \nonumber \\
F_2 &=& \sum_{\ell=0}^{\infty} \left [ (\ell+1)M_{\ell +} + \ell M_{\ell -}
        \right ] P^{\prime}_{\ell} \, , \nonumber \\
F_3 &=& \sum_{\ell=0}^{\infty} \left [ (E_{\ell +} - M_{\ell +})
        P^{\prime\prime}_{\ell +1} + (E_{\ell -} + M_{\ell -})
        P^{\prime\prime}_{\ell-1}\right ] \, , \nonumber \\
F_4 &=& \sum_{\ell=0}^{\infty} (M_{\ell +}-E_{\ell +}-M_{\ell -}-
        E_{\ell -}) P^{\prime\prime}_{\ell}\, ,
\end{eqnarray}
where the $P^{\prime}_{\ell}$ and $P^{\prime\prime}_{\ell}$ are the 
derivatives of Legendre polynomials. 

\chapter{Field Operators\label{appendixC}}
%
This appendix is concerned with the field operators for nucleon, pion and
photon. For more information about field operators and the derivation of
commutator and anti-commutator relations we refer to \cite{WeinbergI}. Quantized nucleon,
pion and photon fields are used in the determination of the
vertices in chapter \ref{chap:2}. The field operator of the nucleon has the form
\begin{eqnarray} 
\Psi (\vec{x}) = \sum_{m_{s},m_{t}} 
\int \frac{d^{3}p}{(2\pi )^{3}}\,\frac{M_{N}}{E_{\vec{p}}}\,\left(
b(\vec{p},m_{s},m_{t})\, u(\vec{p},m_{s})\, e^{i\vec{p}\cdot\vec{x}} 
\right. \nonumber \\
\left. + ~~d^{\dagger}(\vec{p},m_{s},m_{t})\, v(\vec{p},m_{s})\,
  e^{-i\vec{p}\cdot\vec{x}} \right) 
\end{eqnarray} 
or
\begin{eqnarray} 
\bar{\Psi} (\vec{x}) = \sum_{m_{s},m_{t}}
\int \frac{d^{3}p}{(2\pi )^{3}}\,\frac{M_{N}}{E_{\vec{p}}}\,\left(
b^{\dagger}(\vec{p},m_{s},m_{t})\, \bar{u}(\vec{p},m_{s})\, e^{-i\vec{p}
\cdot\vec{x}} \right. \nonumber \\
\left. + ~~d(\vec{p},m_{s},m_{t})\, \bar{v}(\vec{p},m_{s})\, 
e^{i\vec{p}\cdot\vec{x}} \right)\, , 
\end{eqnarray} 
with the following anti-commutator relations for the creation and
annihilation operators, respectively
\begin{eqnarray} 
\{ b(\vec{p_1},m_{s},m_{t}),b^{\dagger}(\vec{p_2},m_{s^{\, \prime}},m_{t
^{\prime}}) \} = (2\pi )^{3} \delta^{3}(\vec{p_2}-\vec{p_1})
\delta_{m_{s},m_{s^{\prime}}}\delta_{m_{t},m_{t^{\prime}}}\,\frac
{E_{\vec{p_1}}}{M_{N}}  \\
\{ d(\vec{p_1},m_{s},m_{t}),d^{\dagger}(\vec{p_2},m_{s^{\prime}},m_{t
^{\prime}}) \} = (2\pi )^{3} \delta^{3}(\vec{p_2}-\vec{p_1})
\delta_{m_{s},m_{s^{\prime}}}\delta_{m_{t},m_{t^{\prime}}}\,\frac
{E_{\vec{p_1}}}{M_{N}}\, .
\end{eqnarray}

\vspace*{-0.2cm}~\\ 
The pion field is given by
\begin{eqnarray} 
\Phi_{\mu} = \int \frac{d^{3}q}{(2\pi)^{3}}\,\frac{1}{2\omega_{\vec{q}}}
\left( a_{-\mu}(\vec{q})e^{i\vec{q}\cdot\vec{x}} + a_{\mu}^{\dagger}(\vec{q})
e^{-i\vec{q}\cdot\vec{x}} \right)
\end{eqnarray}
with the commutator relation
\begin{eqnarray} 
[ a_{\mu}(\vec{q_1}),a_{\mu^{\prime}}^{\dagger}(\vec{q_2}) ] =
(2\pi )^{3} \delta^{3}(\vec{q_2}-\vec{q_1})
\delta_{\mu,\mu^{\prime}}\, 2\omega_{\vec{q}}
\end{eqnarray} 
The photon field in Coulomb gauge $\vec{k}\cdot\vec{\epsilon} = 0$ is
given by 
\begin{eqnarray} 
\vec{A}(\vec{x}) = \sum_{m_{\gamma}} \int\frac{d^{3}k}{(2\pi)^{3}}\,
\frac{\vec{\epsilon}_{m_{\gamma}}}{2\omega_{\gamma}}
\left( a_{m_{\gamma}}(\vec{k})e^{i\vec{k}\cdot\vec{x}} +
a_{m_{\gamma}}^{\dagger}(\vec{k})e^{-i\vec{k}\cdot\vec{x}} \right)\, ,
\end{eqnarray} 
with the commutator relation
\begin{eqnarray} 
[ a_{m_{\gamma}}(\vec{k_1}),a_{m_{\gamma}^{\prime}}^{\dagger}(\vec{k_2}) ]
= (2\pi )^{3} \delta^{3}(\vec{k_2}-\vec{k_1})
\delta_{m_{\gamma},m_{\gamma}^{\prime}}\, 2\omega_{\gamma}\, .
\end{eqnarray} 

\vspace*{-0.2cm}~\\ 
The effect of creation and annihilation operators on the vacuum state is
given by 
\begin{eqnarray} 
b^{\dagger}(\vec{p},m_{s},m_{t} )\ket{0} = \ket{N;\vec{p}\, m_{s}\, m_{t}}\;\;
& , &
b(\vec{p},m_{s},m_{t} )\ket{0} = 0 \, , \\
a_{\mu}^{\dagger}(\vec{q})\ket{0} = \ket{\pi;\vec{q}\,\mu}\quad\quad\quad & , &
\qquad \;\;\; a_{\mu}(\vec{q})\ket{0} = 0 \, , \\
a_{m_{\gamma}}^{\dagger}(\vec{k})\ket{0} = \ket{\gamma;\vec{k}\,m_{\gamma}}\quad\,
\quad & , &
\quad \;\;\;\; a_{m_{\gamma}}(\vec{k})\ket{0} = 0 \, .
\end{eqnarray} 

\chapter{Parametrization of the Deuteron Wave Functions\label{appendixD}}
%
For practical purposes it is more convenient to work with an analytical 
parametrization of the $S$- and $D$-waves in momentum space, $u_0(p)$ and 
$u_2(p)$, respectively, rather than with the discrete sets $u_0(p_i)$ and 
$u_2(p_i)$. The ansatz for the analytic version of the momentum space wave 
functions of the non-relativistic version OBEPR of the Bonn OBE-potential 
\cite{Machleidt8789} is
\begin{eqnarray}
u_0(p) &=& \sqrt{\frac{2}{\pi}}\sum_{i=1}^{n_u}~\frac{C_i}{p^2+m_i^2} \, ,\\
u_2(p) &=& \sqrt{\frac{2}{\pi}}\sum_{i=1}^{n_w}~\frac{D_i}{p^2+m_i^2}\, ,
\end{eqnarray}
with the normalization
\begin{eqnarray}
\int_{0}^{\infty} dp~p^2 \left\{ u_0^2(p)+u_2^2(p)\right\} &=& 1 \, .
\end{eqnarray}

\vspace*{-0.2cm}~\\
In Table \ref{pdwf} we list the coefficients $C_i$ and $D_i$ of the 
parametrized deuteron wave functions $u_0(p)$ and $u_2(p)$ of the Bonn 
OBE-potential. In order to make the $u_0(p)$ and $u_2(p)$ fulfill the 
necessary boundary conditions, the last value of $C_i$ and the last three 
values of $D_i$ must be calculated through the following relations 
\cite{Machleidt8789}
\begin{eqnarray}
C_{n_u} &=& -\sum_{i=1}^{n_u-1}~C_i ~,\nonumber \\
D_{n_w-2} &=& \frac{m^2_{n_w-2}}{(m^2_{n_w}-m^2_{n_w-2})(m^2_{n_w-1}-
m^2_{n_w-2})} ~\nonumber \\
 & & \times ~\left( -m^2_{n_w-1}m^2_{n_w} \sum_{i=1}^{n_w-3} \frac{D_i}{m_i^2} + 
(m^2_{n_w-1}+m^2_{n_w}) \sum_{i=1}^{n_w-3} D_i - \sum_{i=1}^{n_w-3} 
D_im_i^2 \right)\, , \nonumber \\
  & &
\label{cidi}
\end{eqnarray}
and two other relations obtained by circular permutation of ($n_w-2$), 
($n_w-1$) and $n_w$. The masses $m_i$ are chosen to be 
\begin{eqnarray}
m_i &=& \alpha + (i-1) m_0 \, , \nonumber 
\end{eqnarray}
with $\alpha = \sqrt{E_bM_N}=0.231609$ fm$^{-1}$, where $E_b$ is the deuteron 
binding energy, and $m_0 = 0.9$ fm$^{-1}$. This choice ensures the correct 
asymptotic behaviour. In analogy with Ref.\ \cite{Machleidt8789} we use 
$n_u = n_w = 11$. 
\begin{table}
\begin{center}
\caption{\small Coefficients of the parametrized deuteron wave functions for the
  Bonn OBE-potential (full model) \cite{Machleidt8789}. The last $C_i$ and the
  last three $D_i$ are to be computed from Eq.\ (\ref{cidi}) ($n_u=n_w=11$).}
\vspace*{0.3cm}
\begin{tabular}{rr}
\hline\hline\\
$C_i~~~({\rm fm}^{-1/2})$ & $D_i~~~({\rm fm}^{-1/2})$\\
[2.1ex]
\hline\\
 0.90457337 +00                 &   ~0.24133026 ~-01            \\ 
-0.35058661 +00                 &  -0.64430531 +00              \\ 
-0.17635927 +00                 &   0.51093352 +00              \\ 
-0.10418261 +02                 &  -0.54419065 +01              \\ 
 0.45089439 +02                 &   0.15872034 +02              \\ 
-0.14861947 +03                 &  -0.14742981 +02              \\ 
 0.31779642 +03                 &   0.44956539 +01              \\ 
-0.37496518 +03                 &  ~-0.71152863 ~-01            \\ 
 0.22560032 +03                 &  {\rm see Eq.\ (\ref{cidi})}~~ \\ 
-0.54858290 +02                 &  {\rm see Eq.\ (\ref{cidi})}~~ \\ 
{\rm see Eq.\ (\ref{cidi})}~~    &  {\rm see Eq.\ (\ref{cidi})}~~ \\ 
[2.1ex]
\hline\hline
\end{tabular}
\label{pdwf}
\end{center}
\end{table}

\chapter{Transformation of Differential Cross Section\label{appendixE}}
%
In Ref.\ \cite{Krusche99} the differential cross sections for the reaction
$d(\gamma,\pi^0)np$ are given in the so-called photon-nucleon center-of-mass
frame\footnote{Throughout this appendix parameters in this frame are labelled
  by ($^{\star}$).}. This frame corresponds to an
assumption that both nucleons in the deuteron have the same momenta
$-\frac{\vec{k}}{2}$ in the $\gamma d$ c.\ m.\ frame. This means that both
nucleons in the deuteron are at rest in the laboratory frame of the deuteron.

\vspace*{-0.2cm}~\\
In order to compare our theoretical predictions for differential cross sections
of the reaction $d(\gamma,\pi^0)np$ with the experimental data from 
Ref.\ \cite{Krusche99} we need therefore to transform the differential cross
sections from the $\gamma d$ rest frame to the c.\ m.\ frame of the
incident photon and a nucleon at rest. Therefore, we give in this appendix the
Jacoby determinant which we need for this purpose.

\vspace*{-0.2cm}~\\
Since the azimuthal angle remains unchanged by the transformation between the
laboratory and the c.\ m.\ systems, we have \cite{Pilkun79} 
\begin{eqnarray}
\label{difflabcm}
\frac{d\sigma}{d\Omega^{cm}} &=& \frac{d\sigma}{d\Omega^{lab}}\frac{\partial
  Z}{\partial Z^{\star}}\, ,
\end{eqnarray}
where $Z^{\star}(=\cos\theta_{\pi}^{\star}$) and $Z(=\cos\theta_{\pi}^{lab}$)
are the pion angle in the c.m.\ and the $\gamma d$ laboratory frame,
respectively. The derivative of $Z$ with respect to $Z^{\star}$ gives direct
the requested Jacoby determinant.

\vspace*{-0.2cm}~\\
Using the Mandelstam variable $t$ from Eq.\ (\ref{mandel}) one can express
$Z$ through $Z^{\star}$ as follows 
\begin{eqnarray}
\label{zstar}
Z &=&
\frac{Z^{\star}\omega_{\gamma}^{\star}q_{\pi}^{\star}-E_{\pi}^{\star}\omega_{\gamma}^{\star}+E_{\pi}^{lab}\omega_{\gamma}^{lab}}{\omega_{\gamma}^{lab}q_{\pi}^{lab}}\, .
\end{eqnarray}
which gives the following expression for the Jacoby determinant
\begin{eqnarray}
\label{jacoby}
\frac{\partial Z}{\partial Z^{\star}} & = &
\frac{\omega_{\gamma}^{\star}q_{\pi}^{\star}}{\omega_{\gamma}^{lab}} \left [
  \left ( \frac{E_{\pi}^{lab}}{q_{\pi}^{lab}}Z -1 \right ) \frac{\partial
      E_{\pi}^{lab}}{\partial Z} + q_{\pi}^{lab} \right ]^{-1}\, ,
\end{eqnarray}
where $\omega_{\gamma}^{\star}$, $q_{\pi}^{\star}$ and $E_{\pi}^{\star}$ are
the photon energy (Eq.\ \ref{kcm}), pion momentum (Eq.\ \ref{qcm}) and pion
energy (Eq.\ \ref{eqcm}) in the c.m.\ frame, respectively; and 
$\omega_{\gamma}^{lab}$, $q_{\pi}^{lab}$ and $E_{\pi}^{lab}$ are the photon
energy, pion momentum and pion energy in the laboratory frame, respectively.

\vspace*{-0.2cm}~\\
Using the Mandelstam variable $u$ from Eq.\ (\ref{mandel}), one can express the
pion energy in the laboratory frame as follows 
\begin{eqnarray}
\label{allinlab1}
E_{\pi}^{lab} &=& \frac{E_N^{\star}E_{\pi}^{\star}+\omega_{\gamma}^{\star}q_{\pi}^{\star}Z^{\star}}{M_N} \, ,
\end{eqnarray}
where $E_N^{\star}=\sqrt{M_N^2+{\omega_{\gamma}^{\star}}^2}$ is the nucleon energy is the c.m.\ frame. 
Using the well known formulae for momentum and energies from appendix
\ref{appendixB:1} (see also \cite{Byckling73}) the pion energy in the laborartory frame is then given by
\begin{eqnarray}
\label{allinlab}
E_{\pi}^{lab} &=& \frac{\mathcal A\mathcal B + \omega_{\gamma}^{lab} Z
  \mathcal C}{\mathcal A^2 - {\omega_{\gamma}^{lab}}^2 Z^2}\, ,
\end{eqnarray}
where 
\begin{eqnarray}
\label{matha}
\mathcal A = (M_N + \omega_{\gamma}^{lab})\, , 
\end{eqnarray}
\begin{eqnarray}
\label{mathb}
\mathcal B = (M_N\omega_{\gamma}^{lab} + \frac{1}{2}m_{\pi}^2)\, ,
\end{eqnarray}
and
\begin{eqnarray}
\label{mathc}
\mathcal C = \sqrt{\mathcal B^2 + {\omega_{\gamma}^{lab}}^2 m_{\pi}^2 Z^2 -
  m_{\pi}^2 \mathcal A^2}\, .
\end{eqnarray}

\chapter{Two-Body Subsystems\label{appendixF}}
%
In this appendix we give some details of the $NN$ and $\pi N$ potential models
which we use to study the $NN$ and $\pi N$ interaction in the $NN$ and $\pi
N$ subsystems. In the calculations of this work we solve the three dimensional
integral equation of Lippmann-Schwinger \cite{Lippmann51} to
calculate the $\mathcal T$-matrix elements of the two-body scattering and then
show how phase shifts are obtained.

\vspace*{-0.2cm}~\\
The use of separable potentials in the two-body interaction has greatly 
stimulated the theoretical investigation of the three-body problem in the 
last three decades. Most of the important properties of the three-body 
system are already well described by the simplest type of separable models. 
These separable models are most widely used in case of the $\pi NN$ system 
(see for example \cite{Garcilazo90} and references therein). For the $NN$ and $\pi N$ interactions in the
$NN$- and $\pi N$-subsystems we use a specific class of separable potentials
\cite{Nozawa90,Haidenbauer84,Haidenbauer85} which historically have played and still play a major role in the development
of few-body physics and also fit the phase shift data for both $NN$- and $\pi
N$-scattering. The formalism to obtain the 
full two-body $\mathcal T$-matrix and the scattering phase shifts is
given in this appendix in more detail. Momentum space separable potentials
for $NN$- and $\pi N$-interactions are also given.
%
\section{The $NN$ Subsystem\label{appendixF:1}}
%
\subsection{$NN$ Scattering Equation\label{appendixF:1:1}}
\begin{figure}[htb]
  \centerline{\epsfxsize=12cm \epsffile{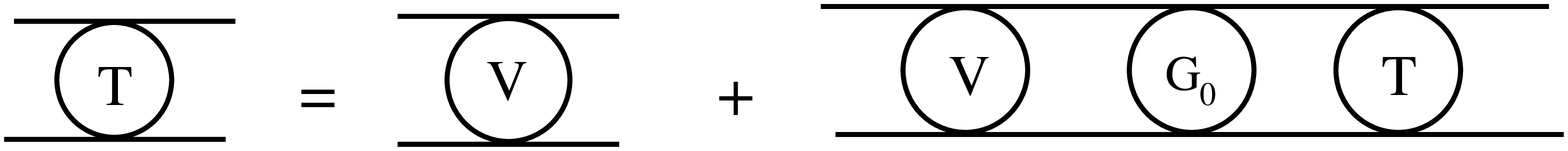}}
  \caption{\small The graphical diagram of the Lippmann-Schwinger equation for 
    $NN$ interaction.}
  \label{ls_eq}
\end{figure}

%
In this section we discuss how the $\mathcal T$-matrix for $NN$-scattering
which we use as input in our predictions in chapter \ref{chap:3} is calculated. Two-nucleon
scattering is described by the Lippmann-Schwinger equation which given
graphically in Fig.\ \ref{ls_eq}.  

\vspace*{-0.2cm}~\\
In partial wave decomposition, represented by the orbital angular 
momentum $\ell$, total angular momentum $J$, spin $s$ and isospin $t$, the
$\mathcal T$-matrix of the two-nucleon scattering may be written as
\begin{eqnarray}
\bra{\ell^{\prime}J^{\prime}s^{\prime}t^{\prime}} \mathcal T(p^{\,
  \prime},p;E) \ket{\ell Jst} & = & \delta_{J^{\prime}J}\delta_{s^{\prime}s}\delta_{t^{\prime}t} \mathcal T_{\ell\ell^{\prime}}^{Jst}(p^{\,
  \prime},p;E)\, ,
\end{eqnarray}
where ${\vec p}$ and ${\vec p^{\,\prime}}$ are the relative 
three-momenta of the two interacting nucleons in the initial and final state, respectively; and 
$p \hspace*{-0.1cm}\equiv \mid\hspace*{-0.1cm}{\vec p}\hspace*{-0.1cm}\mid$ 
and $p^{\,\prime} \hspace*{-0.1cm}\equiv \mid\hspace*{-0.1cm}{\vec 
p^{\,\prime}}\hspace*{-0.1cm}\mid$. $E$ denotes the energy of the two
interacting nucleons in the c.\ m.\ frame and is given by
$E=\frac{p_0^2}{M_N}$, where $p_0$ is the magnitude of the  
initial relative momentum (c.\ m.\ on-shell momentum) which is related to 
the laboratory energy by 
\begin{eqnarray}
T_{lab} & = & \frac{2p_0^2}{M_N}\, .
\end{eqnarray}

\vspace*{-0.2cm}~\\
For a given $NN$ potential $V$, the $\mathcal T$-matrix is obtained from the
Lippmann-Schwinger equation \cite{Lippmann51} which in partial wave
decomposition reads\footnote{We drop the indices $Jst$ for simplicity.}
\begin{eqnarray}
\mathcal T_{\ell}(p^{\, \prime},p;E) & = & 
        V_{\ell}(p^{\,\prime},p) + \int_0^{\infty}dkk^2
        ~V_{\ell}(p^{\,\prime},k) 
        \frac{M_N}{M_N E-k^2+i\epsilon} \mathcal T_{\ell}(k,p;E)  
\label{pw_ls_eq1}
\end{eqnarray}
for single channels. For the coupled channels it reads
\begin{eqnarray}
\mathcal T_{\ell\ell^{\prime}}(p^{\,\prime},p;E) & = & 
        V_{\ell\ell^{\prime}}(p^{\,\prime},p) \nonumber \\
  & & + \sum_{\ell^{\prime\prime}} \int_0^{\infty}dkk^2 ~V_{\ell
        \ell^{\prime\prime}}(p^{\,\prime},k) \frac{M_N}{M_N E-k^2+i\epsilon} 
        \mathcal T_{\ell^{\prime\prime}\ell^{\prime}}(k,p;E)\, ,
\label{pw_ls_eq2}
\end{eqnarray}
where ${\vec k}$ is the relative three-momenta of the two interacting nucleons
in the intermediate state and $k \hspace*{-0.1cm}\equiv
\mid\hspace*{-0.1cm}{\vec k}\hspace*{-0.1cm}\mid$.

\vspace*{-0.2cm}~\\
Using the identity 
\begin{eqnarray}
\frac{1}{x - x_0 + i\epsilon} & = & {\mathcal P} \frac{1}{x - x_0} 
        - i\pi\delta(x - x_0)\, , 
\end{eqnarray}
where ${\mathcal P}$ denotes the principle value integral, one gets the 
$\mathcal T$-matrix elements in partial wave decomposition
\begin{eqnarray}
\mathcal T_{\ell}(p^{\,\prime},p;E) & = & 
        V_{\ell}(p^{\,\prime},p) + {\mathcal P} 
        \int_0^{\infty}dkk^2 ~V_{\ell}(p^{\,\prime},k) \frac{M_N}{p_0^2-k^2} 
        \mathcal T_{\ell}(k,p;E) \nonumber \\
      & & - \frac{1}{2} i\pi M_N p_0 
        V_{\ell}(p^{\,\prime},p_0) \mathcal T_{\ell}(p_0,p;E) 
\label{pw_ls_eq_rs}
\end{eqnarray}
for single channels and 
\begin{eqnarray}
\mathcal T_{\ell\ell^{\prime}}(p^{\,\prime},p;E) & = & 
        V_{\ell\ell^{\prime}}(p^{\,\prime},p) + \sum_{\ell^{\prime\prime}=J\pm 1}{\mathcal P} 
        \int_0^{\infty}dkk^2 ~V_{\ell\ell^{\prime\prime}}
        (p^{\,\prime},k) \frac{M_N}{p_0^2-k^2} 
        \mathcal T_{\ell^{\prime\prime}\ell^{\prime}}(k,p;E) \nonumber \\
     & & - \frac{1}{2} i\pi M_N p_0 
        V_{\ell\ell^{\prime}}(p^{\,\prime},p_0) \mathcal
        T_{\ell\ell^{\prime}}(p_0,p;E)  
\label{pw_ls_eq_rc}
\end{eqnarray}
for coupled channels. These one-dimensional integral equations, 
Eqs.\ (\ref{pw_ls_eq_rs}) and (\ref{pw_ls_eq_rc}), can be solved 
numerically by means of a matrix inversion algorithm. 

\vspace*{-0.2cm}~\\
Let us first consider the partial wave one-dimensional integral
equation for uncoupled channels given by Eq.\ (\ref{pw_ls_eq_rs}). The integral
should now be approximated by a $n$-point integration routine
\begin{eqnarray}
\int_0^{\infty}dk ~F(k) & = & \sum_{i=1}^{n}F(k_i)~w_i\, ,
\end{eqnarray}
where $k_i$ and $w_i$ are the Gaussian integration points and weights, 
respectively. Since the original Gauss points, $\mathcal K_i$ (with weights
$\mathcal W_i$) are in the interval ($-1$,$+1$), we have to map them to the
interval ($0$,$\infty$), the range of our integration. For this purpose we use
the mapping 
\begin{eqnarray}
k_i & = & B ~\tan \left\{ \frac{\pi}{4} (\mathcal K_i + 1)\right\}\, ,
\end{eqnarray}
with the new weights
\begin{eqnarray}
w_i & = & B ~\frac{\pi}{4} \left\{ \frac{\mathcal W_i}{\cos^2[\frac{\pi}{4}
    (\mathcal X_i + 1)]}\right\}\, ,
\end{eqnarray}
where $B=400$ MeV is used. For the Gauss integral we
use $n=16$ mesh points. 

\vspace*{-0.2cm}~\\
To solve Eqs.\ (\ref{pw_ls_eq_rs}) and (\ref{pw_ls_eq_rc}), the principle value
integration has to be replaced by a smooth integrand, which can be achieved by
adding a term of measure zero 
\begin{eqnarray}
\mathcal T_{\ell}(p^{\,\prime},p;E) & = & V_{\ell}(p^{\,\prime},p) +
        \int_0^{\infty}dk \frac{M_N}{p_0^2-k^2} \left\{ k^2
        V_{\ell}(p^{\,\prime},k) \mathcal T_{\ell}(k,p;E) \right. \nonumber\\
    & & \left.\hspace*{-0.3cm} - p_0^2 V_{\ell}(p^{\,\prime},p_0) \mathcal T_{\ell}(p_0,p;E)
        \right\} - \frac{1}{2} i\pi M_N p_0 V_{\ell}(p^{\,\prime},p_0) \mathcal
        T_{\ell}(p_0,p;E). 
\label{zero_term}
\end{eqnarray}
All of the $n$ integration points, $k_1$, $k_2$, $\cdot\cdot\cdot\cdot\cdot$, 
$k_n$, are required to be unequal to $p_0$. If we call $p_0$ the $(n+1)$ 
point ($p_0=p_{n+1}$), then for $p=p_0$ Eq.\ (\ref{zero_term}) can be rewritten as
\begin{eqnarray}
V_{\ell}(p_i,p_{n+1}) & = & \mathcal T_{\ell}(p_i,p_{n+1};E) - \sum_{k=1}^{n}
        \frac{M_Nw_k}{p_{n+1}^2-p_k^2} \left\{ p_k^2 V_{\ell}(p_i,p_k)
        \mathcal T_{\ell}(p_k,p_{n+1};E) \right. \nonumber \\ 
        & & \left. - p^2_{n+1} V_{\ell}(p_i,p_{n+1}) \mathcal
        T_{\ell}(p_{n+1},p_{n+1};E) \right\} \nonumber \\
      & & + \frac{1}{2} i\pi M_N p_{n+1} V_{\ell}(p_i,p_{n+1}) 
        \mathcal T_{\ell}(p_{n+1},p_{n+1};E)\, . 
\label{vzero_term}
\end{eqnarray}
Thus, the scattering equation can be written in matrix form as
\begin{eqnarray}
V_{\ell}(p_i,p_{n+1}) & = & 
        \sum_{j=1}^{n+1} A_{\ell}(p_i,p_j) \mathcal T_{\ell}(p_j,p_{n+1};E)\, , 
\label{vmatform}
\end{eqnarray}
where the dimensionless $A_{\ell}(p_i,p_j)$ matrix is given by 
\begin{eqnarray}
A_{\ell}(p_i,p_j) & = & 
        \delta_{ij} - \frac{M_Nw_jp_j^2}{p^2_{n+1}-p^2_j} V_{\ell}(p_i,p_j) 
        \nonumber \\ 
       & & + \left\{ \left( \sum_{k=1}^n \frac{w_k}{p^2_{n+1}-p^2_k} \right) 
        p_{n+1} + \frac{i\pi}{2} \right\} M_Np_{n+1}
        V_{\ell}(p_i,p_{n+1}) \delta_{n+1,j}\, .\nonumber \\
     & & 
\label{amatrix}
\end{eqnarray}
Choosing the momentum grid such that $p_j\ne p_{n+1}~\forall~j\le n$, 
the matrix $A(p_i,p_j)$ is nonsingular and can be inverted to obtain 
the on-shell or half-off-shell $\mathcal T$-matrix\footnote{For the matrix
  $T(a,b;E)$ we mean by "on-shell" that we are considering elastic
  scattering. This means that the matrix is "on-shell" when $E_a=E_b=E$, where
  $E$ is the non-relativistic kinetic energy. If $E_a\ne E_b=E$ then the
  matrix is "half-off-shell" and it is "full-off-shell" if $E_a\ne E_b \ne
  E$.}
\begin{eqnarray}
\mathcal T_{\ell}(p_i,p_{n+1};E) & = & 
        \sum_{j=1}^{n+1} \left\{
        A_{\ell}(p_i,p_j)\right\}^{-1}V_{\ell}(p_j,p_{n+1})\, . 
\label{solution}
\end{eqnarray}

\vspace*{-0.2cm}~\\
The extension to coupled channels is straightforward. One simply combines 
the points to form a larger ($2n+2$)$\times$($2n+2$) dimensional matrix
\begin{eqnarray}
V(p_i,p_j) & = & 
        \sum_{m=1}^{2n+2} A(p_i,p_m) \mathcal T_{\ell}(p_m,p_j;E)\, . 
\label{solutioncou}
\end{eqnarray}
Here the $i$, $j$ and $m$ labels include both the points $k_1$, $k_2$, 
$\cdot\cdot\cdot\cdot\cdot$, $k_{n+1}$ and the $\ell$-value. For example, 
for $\ell=J-1$ we take the $1\le i\le (n+1)$ points as $k_1$, $k_2$, 
$\cdot\cdot\cdot\cdot\cdot$, $k_{n+1}$. For $\ell=J+1$, the label $i$ ranges 
as $(n+2)\le i\le (2n+2)$ with identical $k$-values, $k_1$, $k_2$, 
$\cdot\cdot\cdot\cdot\cdot$, $k_{n+1}$ $\equiv$ $k_{n+2}$, $k_{n+3}$, 
$\cdot\cdot\cdot\cdot\cdot$, $k_{2n+2}$.

\vspace*{-0.2cm}~\\
So far, we have never mentioned the total isospin of the two-nucleon system,
$t$ (which is either $0$ or $1$). The reason for this is simply that $t$ is not
an independent quantum number. That is, owing to the antisymmetry of the
two-fermion state, the quantum numbers $\ell$, $s$, and $t$ have to fulfill
the condition
\begin{eqnarray}
(-1)^{\ell+s+t} & = & -1\, . 
\label{condition}
\end{eqnarray}
Thus, for a given $\ell$ and $s$, the total isospin $t$ of the two-nucleon
state is fixed.
%
\subsection{Separable $NN$ Potential Model\label{appendixF:1:2}}
%
Now we outline the model which we use to study the $NN$ 
interaction in the $NN$-subsystem. The EST method \cite{Ernst734} for 
constructing separable representations of modern $NN$ potentials has been 
applied by the Graz group \cite{Haidenbauer84,Haidenbauer85} to cast the 
Paris potential 
\cite{Lacombe80} in separable form. In the meantime these so-called PEST potentials have become 
of great use in introducing the features of advanced meson-exchange theory 
\cite{VinhMau79} into calculations of few-body systems like, e.g., of 
nucleon-deuteron scattering \cite{Zankel834}. They have the general 
form 
\begin{eqnarray}
V_{\ell\ell^{\prime}}(p^{\prime},p) &=& \sum_{i,j=1}^{N}~
        g_{\ell i}(p^{\prime})
        ~\lambda_{ij}~ g_{\ell^{\prime}j}(p)\, , 
\label{separable}
\end{eqnarray}
where $N$ and $\lambda$ specify the rank and the strength of the separable 
interaction, respectively, in each partial wave. These separable interactions 
represent a 
good approximation of the on-shell as well as off-shell properties of the 
original Paris potential and show a good fit to the modern $NN$ data base.

\vspace*{-0.2cm}~\\
The form factors of these separable potentials for each partial wave are 
given by
\begin{eqnarray}
g_{\ell i}(p) & = & \sum_{n=1}^{4} 
        \frac{C_{\ell in}p^{\ell +2(n-1)}}{(p^2+\beta^2_{\ell in})^{\ell +n}}
        ~,  \hspace*{3cm}{\rm (\ell\le 1)} \nonumber
\end{eqnarray}
\begin{eqnarray}
g_{\ell i}(p) ~ = ~ \sum_{n=1}^{4} 
        \frac{C_{\ell in}p^{\ell +2(n-1)}}{(p^2+\beta^2_{\ell in})^{\ell
        +n-1}} ~,  \hspace*{3cm}{\rm (2\le \ell \le 3)}
\label{formfactor}
\end{eqnarray}
\begin{eqnarray}
g_{\ell i}(p) ~ = ~ \sum_{n=1}^{4} 
        \frac{C_{\ell in}p^{\ell +2(n-1)}}{(p^2+\beta^2_{\ell in})^{\ell
        +n-2}} ~,  \hspace*{3cm}{\rm (\ell\ge 4)}  \nonumber 
\end{eqnarray}
where $C$ and $\beta$ are fit parameters of the separable interaction. 
The values of these parameters in each partial wave are given appendix
\ref{appendixG}. Let us now discuss the partial wave states in detail.
\begin{figure}[htb]
\centerline{\epsfxsize=14cm \epsffile{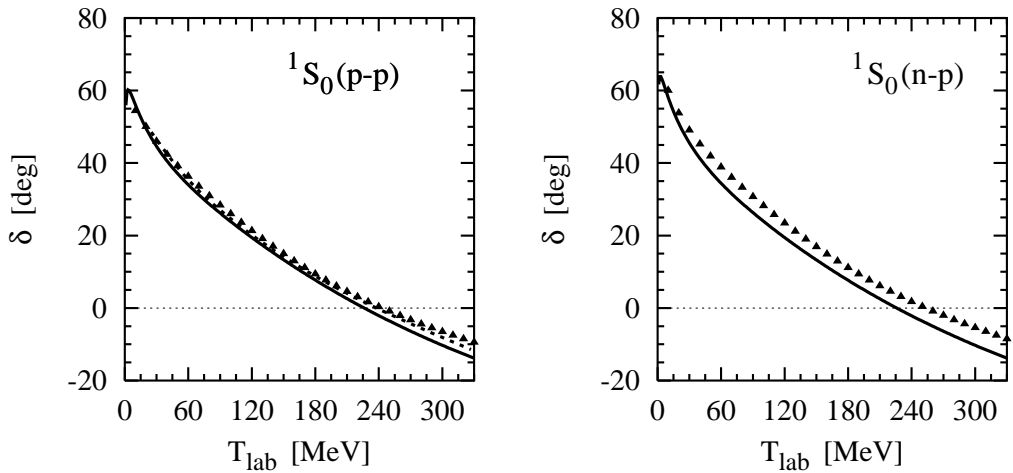}}
\vspace*{-0.2cm}
\caption{\small $^1S_0$ phase shift using a rank-3 separable 
        potential from \cite{Haidenbauer85} (solid curve)
        in comparison with the original Paris potential (dotted curve) which
        is given only for the (p-p) system. 
        The data points are from SAID \cite{Said} (Solution: SP00).}
\label{1s0}
\end{figure}
\begin{figure}[htb]
\centerline{\epsfxsize=15cm \epsffile{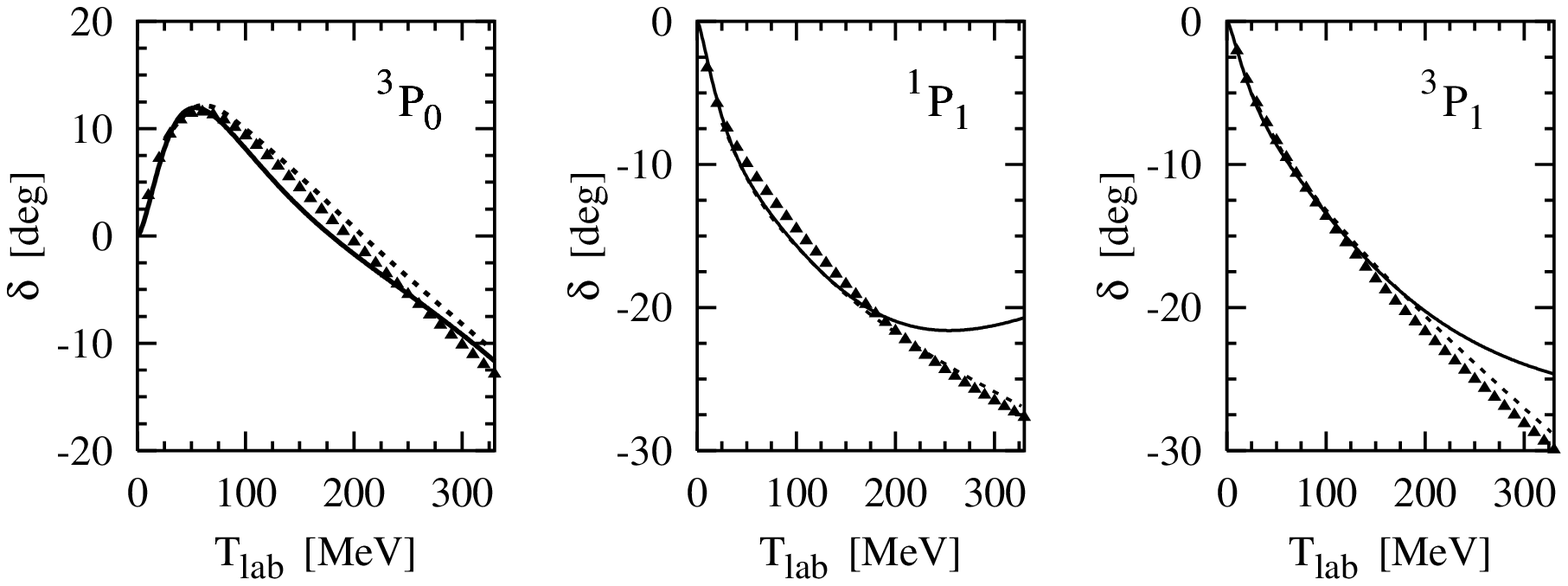}}
\vspace*{-0.4cm}
\caption{\small Same notation as in Fig.\ \ref{1s0} but for the $^3P_0$, 
$^1P_1$ and $^3P_1$ and single channels using a rank-2 separable potential.}
\label{3p03p11p1}
\end{figure}
\begin{figure}[htb]
\centerline{\epsfxsize=14cm \epsffile{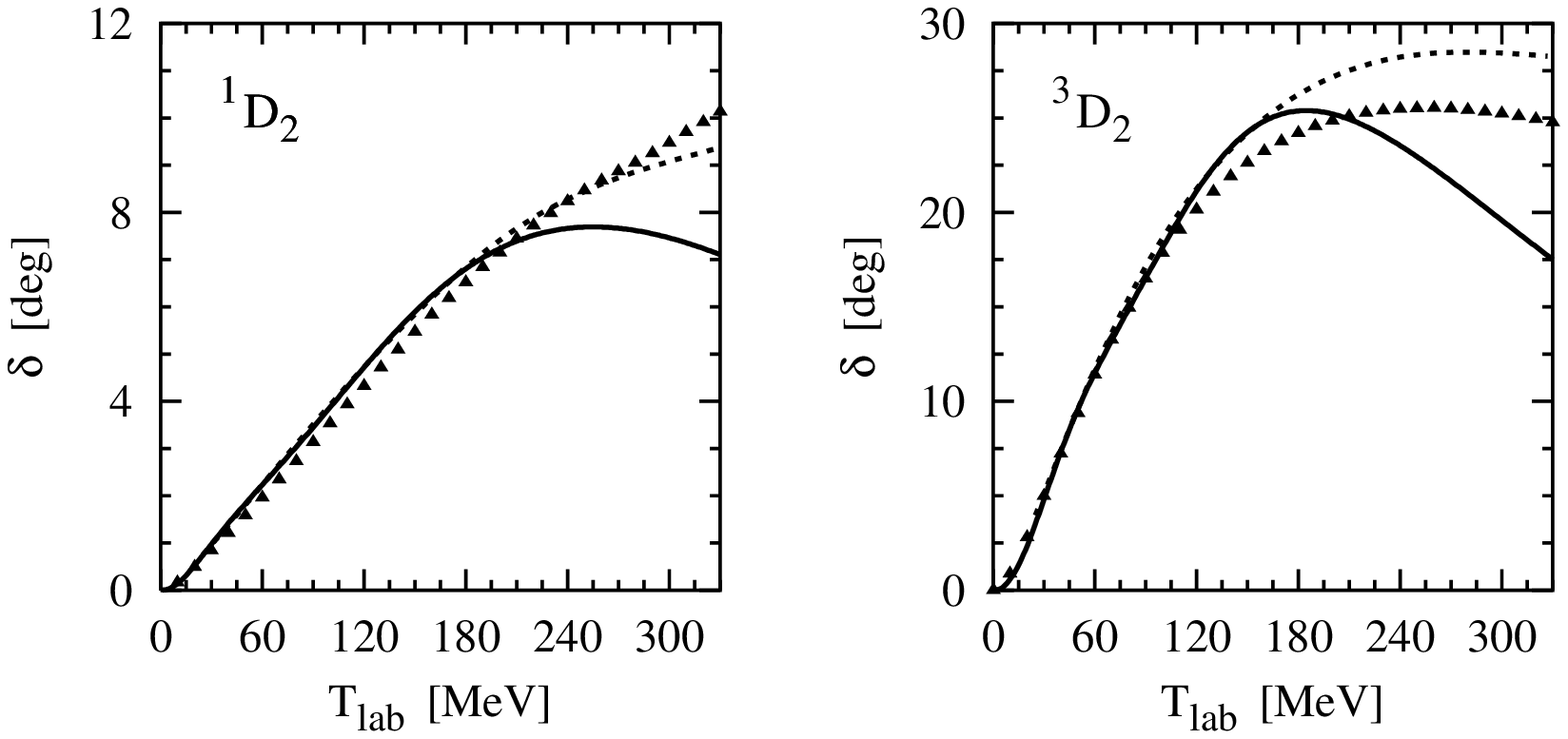}}
\vspace*{-0.1cm}
\caption{\small Same notation as in Fig.\ \ref{1s0} but for the $^1D_2$ 
and $^3D_2$ single channels using a rank-2 separable potential.}
\label{1d23d2}
\end{figure}
\begin{figure}[htb]
\centerline{\epsfxsize=15cm \epsffile{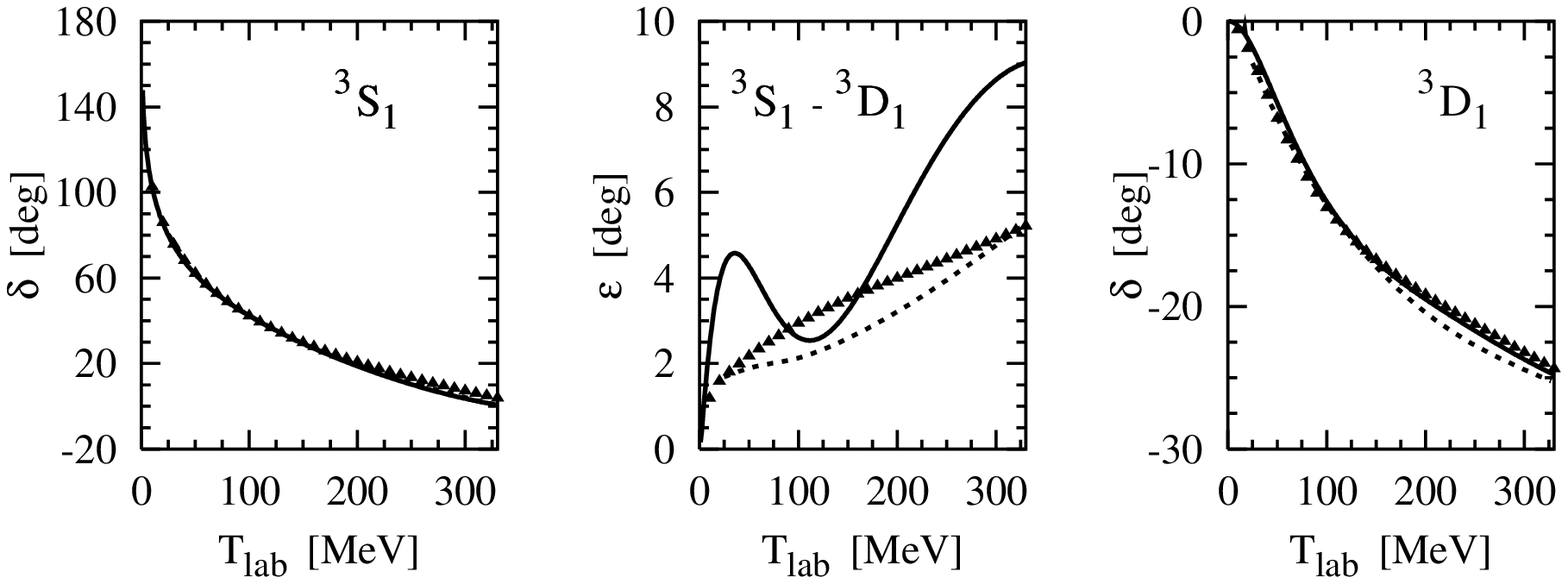}}
\vspace*{-0.3cm}
\caption{\small Same notation as in Fig.\ \ref{1s0} but for the coupled 
        channel $^3S_1$-$^3\hspace*{-0.05cm}D_1$ using a rank-4 separable 
        potential.}
\label{3s1eps3d1}
\end{figure}
\begin{figure}[htb]
\centerline{\epsfxsize=15cm \epsffile{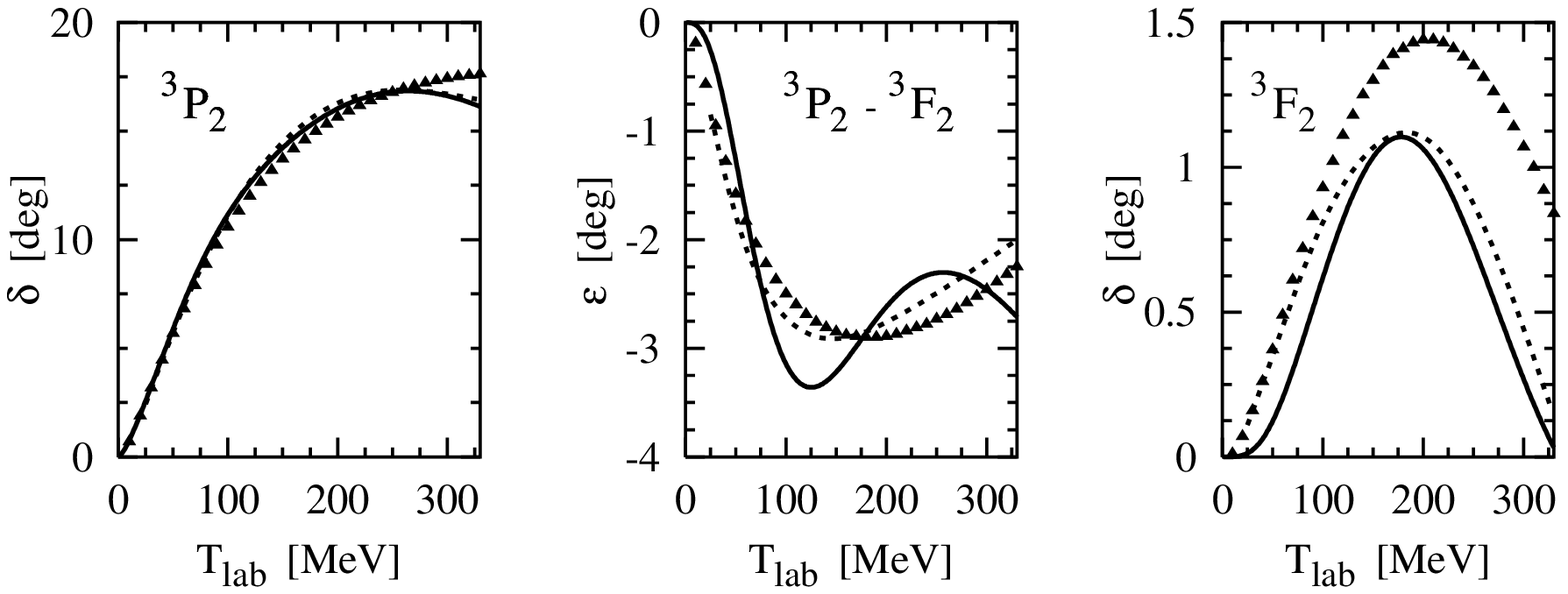}}
\caption{\small Same notation as in Fig.\ \ref{1s0} but for the coupled channel 
        $^3P_2$-$^3\hspace*{-0.05cm}F_2$ using a rank-3 separable potential.}
\label{3p2eps3f2}
\end{figure}
\begin{figure}[htb]
\centerline{\epsfxsize=15cm \epsffile{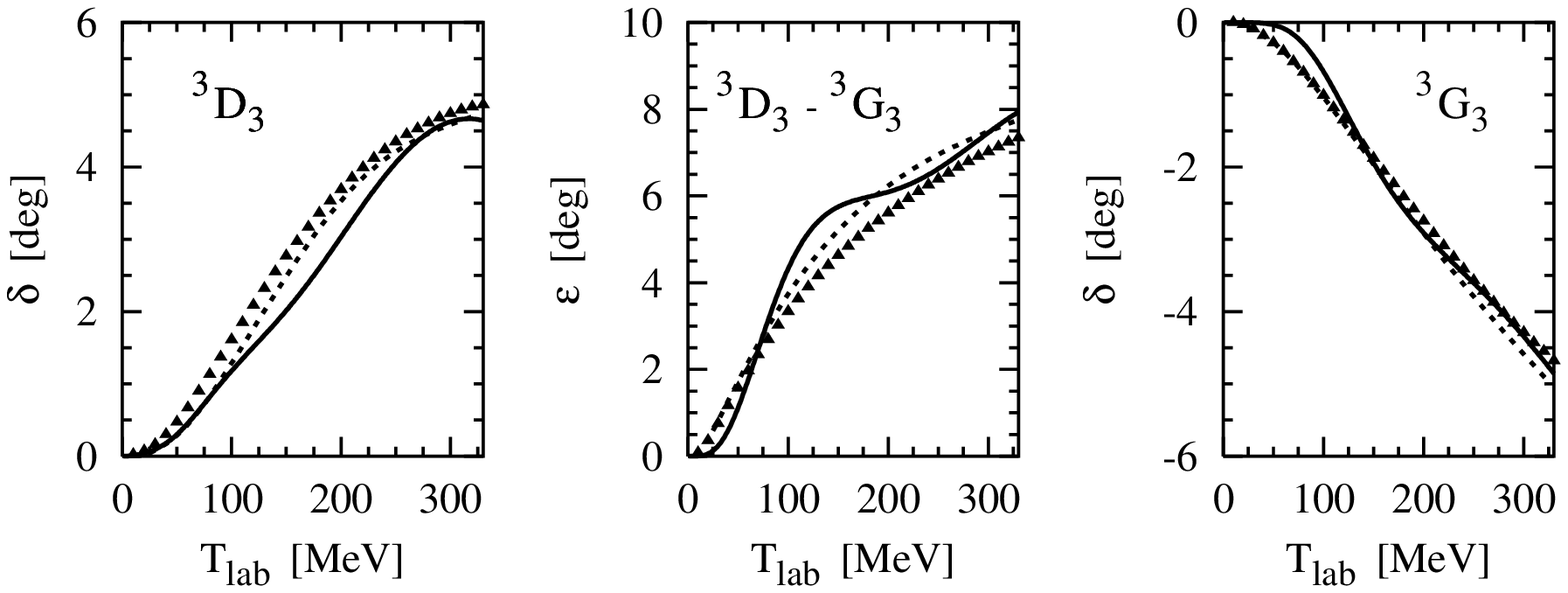}}
\caption{\small Same notation as in Fig.\ \ref{1s0} but for the coupled channel 
        $^3D_3$-$^3G_3$ using a rank-4 separable potential.}
\label{3d3eps3g3}
\end{figure}

\vspace*{-0.2cm}~\\
In case of the $^1S_0$ partial wave the Graz group \cite{Haidenbauer85} have
constructed a modified PEST potential of rank-3. The parameters of this
potential are given in Table \ref{tableG:1} for both the (p-p) and (n-p)
systems. Fig.\ \ref{1s0} shows that the phase shifts of the (p-p)
system\footnote{The Paris potential is given only for the (p-p) system in the
  $^1S_0$ partial wave.} are the same for both PEST3 and the original Paris
potential.

\vspace*{-0.2cm}~\\
For the $^3P_0$, $^3P_1$ and $^1P_1$ partial waves, they constructed a rank-2 separable 
approximations. The potential parameters for these three partial waves are 
given in Table \ref{tableG:2}. In this Table we give the modified 
parameters of the $^3P_0$ partial wave from Ref.\ \cite{Haidenbauer85} which 
we use in this work. The on-shell properties of the PEST 
parametrization for $^3P_0$ are evident from the phase shift for $^3P_0$ 
in the left panel of Fig.\ \ref{3p03p11p1}. Up to $T_{lab}\approx 70$ MeV the 
approximation to the Paris potential is rather good. If we go to higher 
energies, there is some discrepancy in the phase shifts around 
$T_{lab}\approx 100-180$ MeV, but the situation improves again at 
higher energies. For the $^3P_1$ and $^1P_1$ partial waves the separable
potential is good up to $T_{lab}\le 200$ MeV. For $T_{lab}\ge 200$ MeV the
deviation in the $^3P_1$ and $^1P_1$ partial waves from the original Paris
potential increases strongly. 

\vspace*{-0.2cm}~\\
For the $^1D_2$ and $^3D_2$ partial waves the Graz group constructed a rank-2 PEST 
approximation. The properties of these potentials are practically the 
same as discussed above for the uncoupled $P$ waves. The on-shell behaviour 
of the potentials, whose parameters are given in Table \ref{tableG:3}, 
are evident from the phase shifts shown in Fig.\ \ref{1d23d2}

\vspace*{-0.2cm}~\\
The coupled $^3S_1$-$^3\hspace*{-0.05cm}D_1$ partial wave state which acting
in the deuteron is crucial for a proper treatment of the $NN$ 
interaction. But untill now, no satisfactory description of all its 
aspects has been achieved by means of separable forces. The Graz group tried
to remedy this situation by constructing a rank-4 PEST approximation to the 
Paris potential. This separable interaction, whose parameters are given 
in Table \ref{tableG:4}, yields a correct reproduction of the deuteron 
properties as well as a fair description of the scattering domain as 
shown in Fig~\ref{3s1eps3d1}.

\vspace*{-0.2cm}~\\
For the coupled $^3P_2$-$^3\hspace*{-0.05cm}F_2$ partial wave, they constructed
a rank-3 PEST parametrization, whose parameters are given in Table
\ref{tableG:5}. The phase shifts and the mixing parameter are shown in
Fig.\ \ref{3p2eps3f2}. On the other hand, they designed a rank-4 PEST
parametrization for the coupled $^3D_3$-$^3G_3$ partial wave. The parameters
of the PEST potential are given in Table \ref{tableG:6}. The on-shell
properties are shown in Fig.\ \ref{3d3eps3g3}; they exhibit a good agreement
with the Paris potential.

\vspace*{-0.2cm}~\\
From the foregoing presentation of the PEST interactions it is clear that 
these separable representation of the Paris $NN$ potential are likely to be 
relevant in few-body problems. 
%
\subsection{$NN$ Phase Shifts\label{appendixF:1:3}}
%
The full on-shell $S$-matrix in partial wave decomposition is related to the 
on-shell partial wave $T$-matrix by 
\begin{eqnarray}
S(E) & = & 1\hspace*{-0.13cm}1 + 2i~ T(E)\, ,
\label{s_and_t_matrix}
\end{eqnarray}
with 
\begin{eqnarray}
T(E) & := & -\frac{1}{2}\pi p_0 M_N \mathcal T(E)\, ,
\end{eqnarray}
in the nonrelativistic kinematics which we use in this work.

\vspace*{-0.2cm}~\\
The scattering phase shifts for a given partial wave can be calculated 
from the on-energy-shell $K$-matrix (a 2$\times$2 matrix)
\cite{Arndt83,Schwamb99} which is expressed in terms of the $S$- and $T$-matrix by
\begin{eqnarray}
K & := & i~\frac{1 - S}{1 + S} 
   ~ = ~ \frac{T}{1 + i~ T}\, . 
\label{k_and_t_matrix}
\end{eqnarray}
For the single channels, the scattering phase shifts are then determined by
\begin{eqnarray}
\tan\delta & = & \Re e{K}\, .
\label{delta_s}
\end{eqnarray}
For the coupled channels, a unitary transformation is 
needed to diagonalize the two-by-two coupled $K$-matrix. This requires an 
additional parameter, known as the {\it mixing parameter} $\epsilon$. This 
mixing parameter and the phase shifts of the coupled channels are given 
in terms of the $K$-matrix by
\begin{eqnarray}
\sin(2\epsilon) & = & \sqrt{\frac{(2\Re e K_0)^2}{[1 + (\Re e K_0)^2 - 
        \Re e K_+ \Re e K_-]^2 + (\Re e K_+ + \Re e K_-)^2}} 
\end{eqnarray}
and 
\begin{eqnarray}
2\delta_{\pm} & = & \arcsin\left({\frac{\Re e K_+ + \Re e K_-}{2\Re e K_0}
        \sin(2\epsilon)}\right) 
        \pm \arcsin\left({\frac{\Re e K_+ - \Re e K_-}{2\Re e K_0}
        \tan(2\epsilon)}\right)\, , \nonumber\\
      & & 
\end{eqnarray}
where
\begin{eqnarray}
K_0 & := & K_{J+1,J-1} ~=~ K_{J-1,J+1}~, \nonumber\\
K_+ & := & K_{J+1,J+1}~,  \nonumber\\
K_- & := & K_{J-1,J-1}\, .
\end{eqnarray}

\vspace*{-0.2cm}~\\
In this work, all phase shifts are in the Stapp-Ypsilantis-Metropolis (SYM) 
or {\it bar} convention \cite{Stapp57}.
%
\section{The $\pi N$ Subsystem\label{appendixF:2}}
%
Analogous to the case of $NN$-scattering, the formalism to obtain the full 
$\mathcal T$-matrix and the scattering phase shifts for $\pi N$-scattering is
outlined in this section, but not in more details since we give enough details
in case of the $NN$-scattering. A partial wave decomposition of the
separable $\pi N$ potential is performed in this section. The $\pi N$
scattering equation is also given.
%
\subsection{$\pi N$-Scattering Equation\label{appendixF:2:1}}
%
For a given $\pi N$ potential $V$, the $\mathcal T$-matrix for $\pi N$
scattering is obtained from the Lippmann-Schwinger equation which 
in partial wave decomposition, specified by the orbital angular 
momentum $\ell$, total angular momentum $J$ and isospin $t$, reads\footnote{The
  indicies $\ell Jt$ are also dropped here for simplicity.}
\begin{eqnarray}
\mathcal T(p^{\,\prime},p;E) & = & 
        V(p^{\,\prime},p) + \int_0^{\infty}dkk^2 
        ~ V(p^{\,\prime},k) ~\mathcal G_{\pi N}(k,E)~ \mathcal
        T(k,p;E)\, ,   
\label{pw_ls_eq_pin}
\end{eqnarray}
with the $\pi N$ propagator
\begin{eqnarray}
\mathcal G_{\pi N}(k,E) & = & \frac{1}{E-E_{\pi}(k)-E_N(k)+i\epsilon}\, ,   
\label{pw_ls_eq_pinp}
\end{eqnarray}
where ${\vec p}$, ${\vec k}$ and $\vec p^{\,\prime}$ are the $\pi N$ 
relative momentum in the initial, intermediate and final state, 
respectively; and 
$p \hspace*{-0.1cm}\equiv \mid\hspace*{-0.1cm}{\vec p}\hspace*{-0.1cm}\mid$, 
$k \hspace*{-0.1cm}\equiv \mid\hspace*{-0.1cm}{\vec k}\hspace*{-0.1cm}\mid$ 
and $p^{\,\prime} \hspace*{-0.1cm}\equiv \mid\hspace*{-0.1cm}{\vec 
p^{\,\prime}}\hspace*{-0.1cm}\mid$. 
$E=\sqrt{M_N^2+p_0^2}+\sqrt{m_{\pi}^2+p_0^2}$ denotes the total collision
energy with on-shell momentum $p_0$ in the c.\ m.\ frame. 
For our calculations we use relativistic kinematics for both pion and nucleon,
thus 
\begin{eqnarray}
E_{\pi}(k) & = & \sqrt{m_{\pi}^2 + k^2}~,\nonumber \\
E_{N}(k) & = & \sqrt{M_{N}^2 + k^2}\, .
\end{eqnarray}
The partial wave $\mathcal T$-matrix for $\pi N$ scattering is then given by 
\begin{eqnarray}
\mathcal T(p^{\,\prime},p;E) & = & 
        V(p^{\,\prime},p) + \int_0^{\infty}dkk^2 
        ~ V(p^{\,\prime},k) \frac{G(k)}{p_0^2-k^2+i\epsilon} 
        \mathcal T(k,p;E)\, , 
\label{pw_ls_eq_pin1}
\end{eqnarray}
where
\begin{eqnarray}
G(k) & = & 
        \frac{\left[ E_N(p_0) + E_N(k)\right] 
        \left[ E_{\pi}(p_0) + E_{\pi}(k) \right]}
        {\left[ E_N(p_0) + E_N(k) + 
        E_{\pi}(p_0) + E_{\pi}(k) \right] }\, .
\label{func_gk}
\end{eqnarray}

\vspace*{-0.2cm}~\\
To transform this equation into a principle value integral equation, one uses 
the identity 
\begin{eqnarray}
\frac{1}{x^2 - x_0^2 + i\epsilon} & = & {\mathcal P} \frac{1}{x^2 - x_0^2} 
        - i\pi\delta(x^2 - x_0^2)\, , 
\end{eqnarray}
to get the matrix elements in partial wave decomposition as
\begin{eqnarray}
\mathcal T(p^{\,\prime},p;E) & = & 
        V(p^{\,\prime},p) + {\mathcal P} 
        \int_0^{\infty}dkk^2 ~ V(p^{\,\prime},k) \frac{G(k)}{p_0^2-k^2} 
        \mathcal T(k,p;E) \nonumber \\
      & & - \frac{1}{2} i\pi p_0 G(p_0) V(p^{\,\prime},p_0) 
        \mathcal T(p_0,p;E)\, . 
\label{pw_ls_eq_pin2}
\end{eqnarray}
This one-dimensional integral equation can be solved 
numerically for a given $\pi N$ potential model $V(p^{\,\prime},p)$ by using the
matrix inversion method which is explaind in details in case of
$NN$-scattering in the previous section. In this work we use the separable $\pi N$
potential model of Nozawa {\it et al.} \cite{Nozawa90}. This model is given
in more details in the forthcoming section.
%
\subsection{Separable $\pi N$ Potential Model\label{appendixF:2:2}}
%
In the case of the $\pi N$ scattering a large number of dynamical models have been developed over the past few years (see for example
\cite{Nozawa90,Elmessiri98,Fuda95}). Most begin with a separable potential
which is iterated in a Lippmann-Schwinger equation to give the scattering
amplitude, from which phase shifts and observables are obtained. In this work
we use the model of Nozawa {\it et al.} \cite{Nozawa90} in order to study
the $\pi N$ interaction in the $\pi N$-subsystem. This model is consistent
with the existing unitary description of the $\pi NN$ system and treats the
$\pi N$ interaction dynamically, with all $S$-, $P$- and $D$-wave $\pi N$
phase shifts being well reproduced below 500 MeV.

\vspace*{-0.2cm}~\\ 
For partial wave specified by quantum numbers $\ell Jt$, the $\pi N$
scattering equation takes the form given in Eq.\ (\ref{pw_ls_eq_pin}). The $\pi
N$ potential $V(p^{\,\prime},p)$ is given in Ref.\ \cite{Nozawa90} due to
both a vertex interaction $f_{\pi N,B}^0$ and a "background" potential
$\mathcal V$ (which is assumed to be of a separable form) by 
\begin{eqnarray}
V(p^{\prime},p;E)  & = & \sum_{B=N,\Delta} f_{\pi N,B}^0 
~g^0_{B}(E) ~f_{B,\pi N}^0 + \mathcal V\, ,
\label{pinpot}
\end{eqnarray}
where $g^0_{B}(E)$ is the free propagator in subspace\footnote{Subspace
  $S=\sum_{B=N,\Delta} \pi N\oplus B$ describes $\pi N$ scattering without
  coupling to photons.} $B$ and takes the form
\begin{eqnarray}
g^0_{B} & = & \frac{1}{E-m_{0B}}\, , 
\label{g0b}
\end{eqnarray}
with $m_{0B}$ is the bare mass of the baryon $B$. 

\vspace*{-0.2cm}~\\
For $P_{11}$ and $P_{33}$ channels, both terms in the right-hand-side of
Eq.\ (\ref{pinpot}) are taken into account. Thus the total potential matrix
element can be written in terms of {\it baryon} pole and non-pole parts as
follows 
\begin{eqnarray}
V(p^{\,\prime},p;E) & = & f_0(p^{\,\prime})~ g^0_{B}(E) ~
f_0(p)  +  h_0(p^{\,\prime})~ \lambda_0 ~h_0(p)\, , 
\label{vp11p33}
\end{eqnarray}
where $\lambda_0$ is a phase parameter and it is given in Table
\ref{tableG:7}. The form factors $h_0(k)$
and $f_0(k)$ are parametrized as follows
\begin{eqnarray}
h_0(k) & = & \frac{a_1 ~ k^{m_1}}{(k^2+b_1^2)^{n_1}} ~ k^{\ell}\, , 
\label{h0}
\end{eqnarray}
\begin{eqnarray}
f_0(k) & = & \frac{a_2 ~ k^{m_2}}{(k^2+b_2^2)^{n_2}} ~ k^{\ell}\, . 
\label{f0}
\end{eqnarray}
The values of the parameters
$n_1$, $n_2$, $m_1$, $m_2$, $a_1$, $a_2$, $b_1$ and $b_2$ are given in Table
\ref{tableG:7} for each partial wave.

\vspace*{-0.2cm}~\\
By inserting Eq.\ (\ref{vp11p33}) into Eq.\ (\ref{pw_ls_eq_pin}), the $\pi N$
amplitude can then be written as 
\begin{eqnarray}
\mathcal T(p^{\,\prime},p;E) & = & T^{NP}(p^{\,\prime},p;E) +  f(p^{\,\prime})~
g_{B}(E) ~f(p)\, .
\label{tpintwoterms}
\end{eqnarray}
The non-pole $T^{NP}$-matrix is given by
\begin{eqnarray}
T^{NP}(p^{\,\prime},p;E) & = &  h_0(p^{\,\prime})~ \tau_0(E) ~h_0(p)\, , 
\label{tnonploe}
\end{eqnarray}
with
\begin{eqnarray}
\tau_0(E) & = &  \frac{\lambda_0}{1 - \lambda_0 \int dk
    k^2~|h_0(k)|^2~\mathcal G_{\pi
    N}(k,E)}\, . 
\label{tauzero}
\end{eqnarray}
The pole term in Eq.\ (\ref{tpintwoterms}) consists of a dressed form factor
$f(k_{\alpha})$ ($\alpha=i,f$) defined by 
\begin{eqnarray}
f(p_{\alpha}) & = &  f_0(p_{\alpha}) + \tau_0(E)~h_0(p_{\alpha}) ~\int dk k^2~
    h_0(k)~f_0(k)~\mathcal G_{\pi N}(k,E)\, ,
\label{dressedff}
\end{eqnarray}
and a dressed propagator $g_B(E)$ defined by 
\begin{eqnarray}
g_B(E) & = &  \frac{1}{E - m_{0B} - \int dk k^2~ f(k)~f_0(k)~\mathcal G_{\pi
    N}(k,E)}\, .
\label{dressedprog}
\end{eqnarray}

\vspace*{-0.2cm}~\\
For channels other than $P_{11}$ and $P_{33}$, the vertex interaction of
Eq.\ (\ref{pinpot}) does not contribute. The $\pi N$ potential is then assumed
to be of rank 2 separable form 
\begin{eqnarray}
\mathcal V(p^{\,\prime},p) & = & h_1(p^{\,\prime})~ \lambda_1~ h_1(p) +
h_2(p^{\,\prime})~ \lambda_2~ h_2(p)\, .
\label{seprank2}
\end{eqnarray}
The form factors are parametrized as 
\begin{eqnarray}
h_1(k) & = & \frac{a_1 ~ k^{m_1}}{(k^2+b_1^2)^{n_1}} ~ k^{\ell}\, , 
\label{otherchannelsff1}
\end{eqnarray}
\begin{eqnarray}
h_2(k) & = & \frac{a_2 ~ k^{m_2}}{(k^2+b_2^2)^{n_2}} ~ k^{\ell}\, . 
\label{otherchannelsff2}
\end{eqnarray}
As before, the parameters $n_1$, $n_2$, $m_1$, $m_2$, $a_1$, $a_2$, $b_1$ and
$b_2$ are given in Table \ref{tableG:7} for each partial wave.

\vspace*{-0.2cm}~\\
By inserting Eq.\ (\ref{seprank2}) into Eq.\ (\ref{pinpot}), one obtains the
following analytic solution for the $\mathcal T$-matrix
\begin{eqnarray}
\mathcal T(p^{\,\prime},p;E) & = &  h_1(p^{\,\prime})~ \tau_{11}(E) ~h_1(p) + 
h_1(p^{\,\prime})~ \tau_{12}(E) ~h_2(p) \nonumber \\
  & & + h_2(p^{\,\prime})~ \tau_{21}(E) ~h_1(p) +
h_2(p^{\,\prime})~ \tau_{22}(E) ~h_2(p)\, , 
\label{otherchannelstpin}
\end{eqnarray}
where
\begin{eqnarray}
\tau_{11}(E) & = &  \frac{\lambda_1(1-\lambda_2 H_2)}{(1-\lambda_1
  H_1)(1-\lambda_2 H_2) - \lambda_1 \lambda_2 H_{12}^2}\, , 
\label{tau11}
\end{eqnarray}
\begin{eqnarray}
\tau_{12}(E) & = &  \tau_{21}(E) \nonumber \\
  & = & \frac{\lambda_1\lambda_2 H_{12}}{(1-\lambda_1
  H_1)(1-\lambda_2 H_2) - \lambda_1 \lambda_2 H_{12}^2}\, , 
\label{tau12}
\end{eqnarray}
\begin{eqnarray}
\tau_{22}(E) & = &  \frac{\lambda_2(1-\lambda_1 H_1)}{(1-\lambda_1
  H_1)(1-\lambda_2 H_2) - \lambda_1 \lambda_2 H_{12}^2}\, , 
\label{tau22}
\end{eqnarray}
with
\begin{eqnarray}
H_1 & = &  \int dk k^2 ~ |h_1(k)|^2~\mathcal G_{\pi N}(k,E)\, , 
\label{h11}
\end{eqnarray}
\vspace{-0.3cm}
\begin{eqnarray}
H_2 & = &  \int dk k^2 ~ |h_2(k)|^2~\mathcal G_{\pi N}(k,E)\, , 
\label{h22}
\end{eqnarray}
\vspace{-0.3cm}
\begin{eqnarray}
H_{12} & = &  \int dk k^2 ~ h_1(k)~h_2(k)~\mathcal G_{\pi N}(k,E)\, . 
\label{h1221}
\end{eqnarray}
\begin{figure}[htb]
\centerline{\epsfxsize=14cm \epsffile{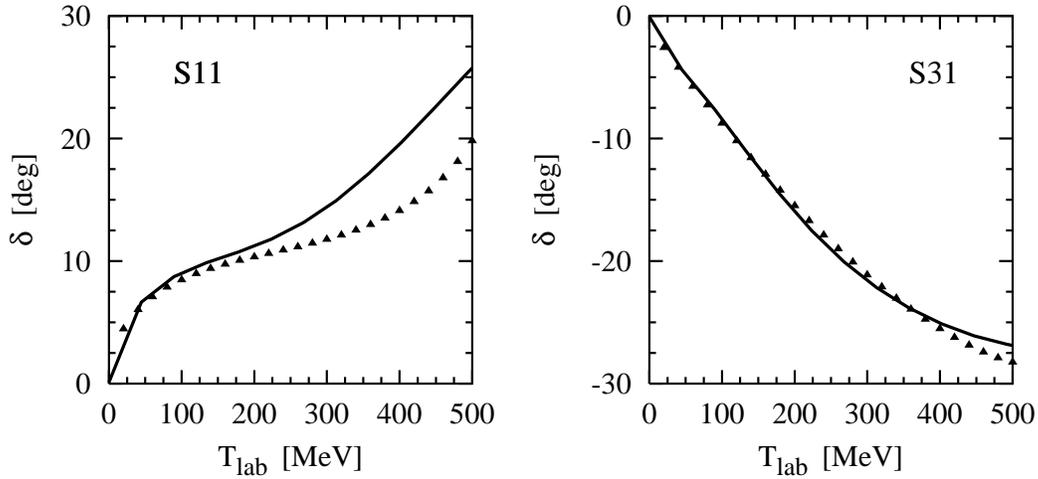}}
\caption{\small The $\pi N$ phase shifts of the $S$ partial waves 
        obtained from the LS equation using the separable potential model of
        Nozawa {\it et al.} \cite{Nozawa90} shown versus the pion laboratory
        energy $T_{lab}$ in MeV. The data points are from the VPI partial wave
        analysis \cite{Said} (Solution: SM99).}
\label{pinphases}
\end{figure}
\begin{figure}[htb]
\centerline{\epsfxsize=13cm \epsffile{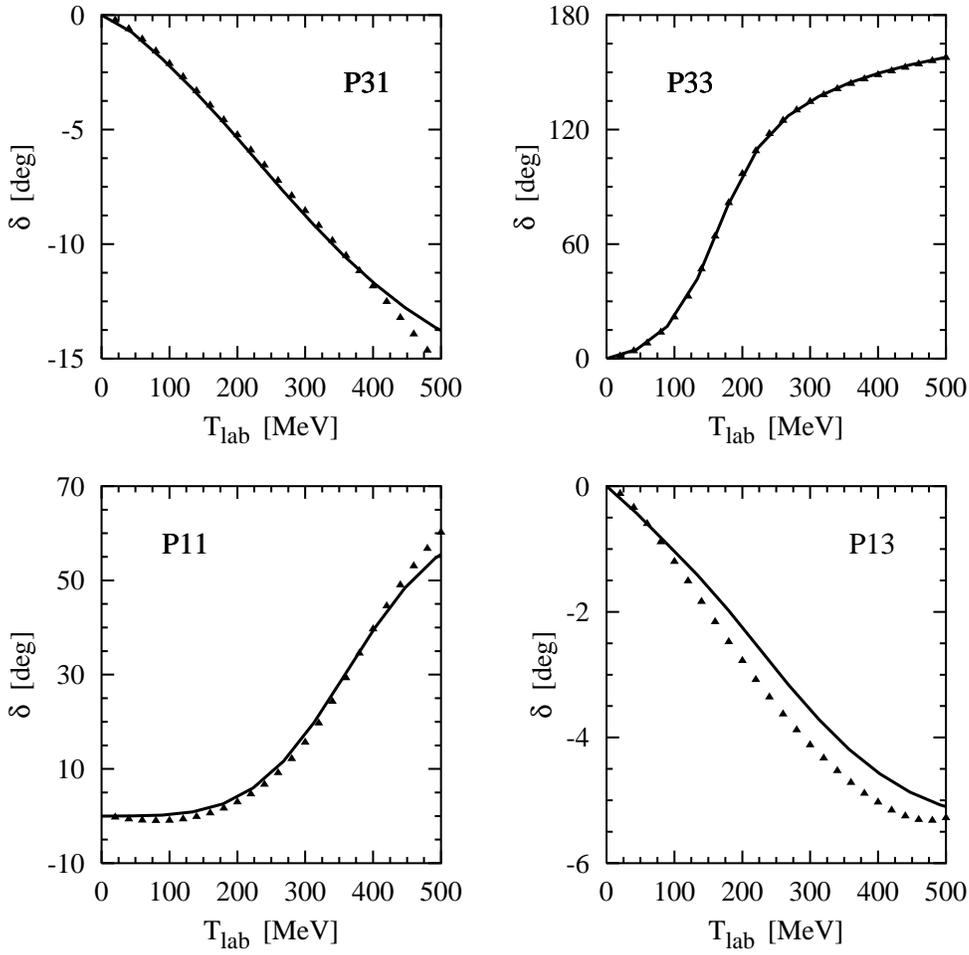}}
\vspace*{-0.2cm}
\caption{\small Same as Fig.\ \ref{pinphases} but for the $P$ partial waves.}
\label{pinphasep}
\end{figure}
\begin{figure}[htb]
\centerline{\epsfxsize=13cm \epsffile{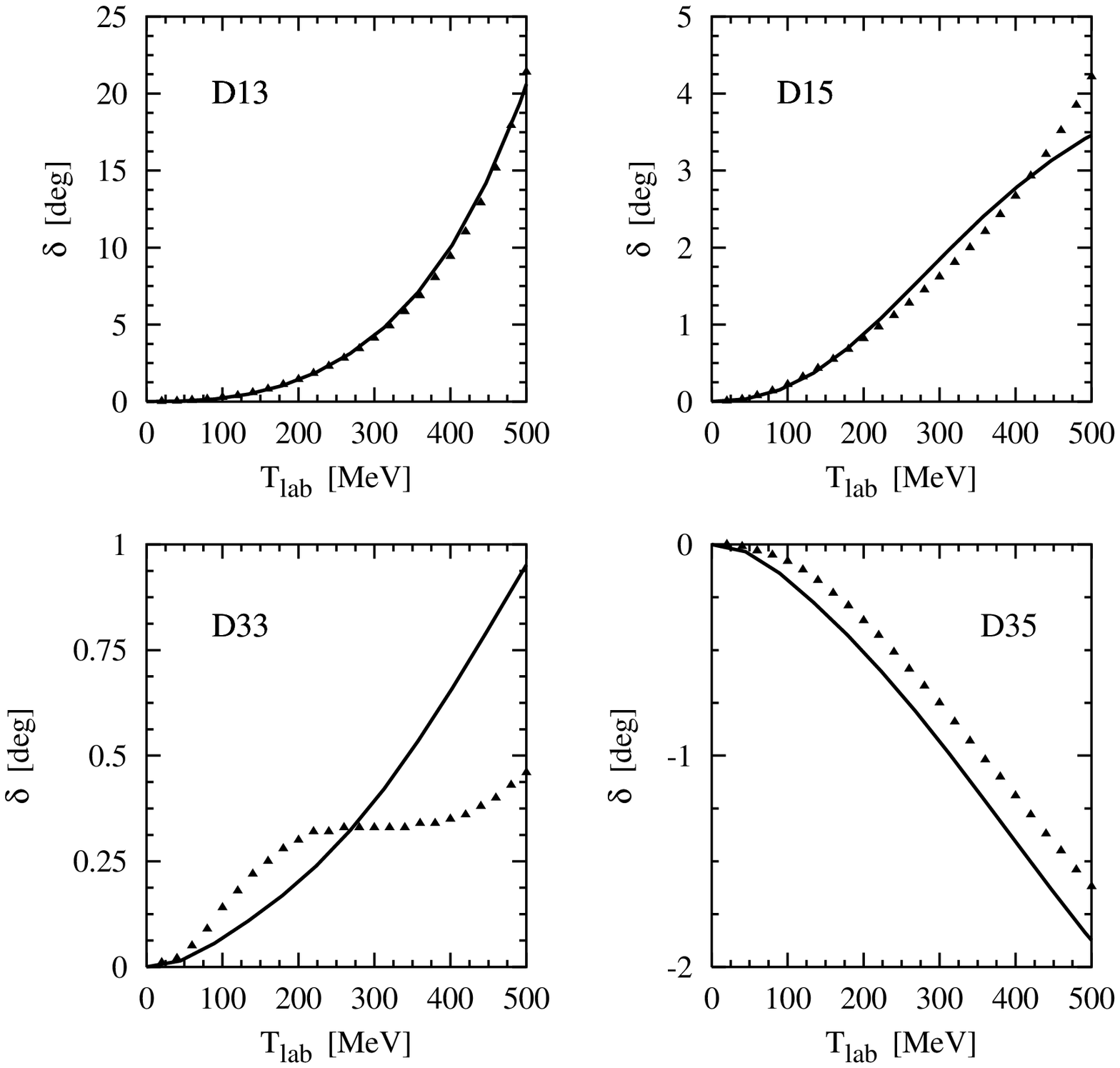}}
\caption{\small Same as Fig.\ \ref{pinphases} but for the $D$ partial waves.}
\label{pinphased}
\end{figure}
%
\subsection{$\pi N$ Phase Shifts\label{appendixF:2:3}}
%
Once the scattering equtaion given in Eq.\ (\ref{pw_ls_eq_pin2}) is solved 
for the on-shell $\mathcal T$-matrix, one can obtain the scattering phase
shifts. The information about the scattering process is commonly represented 
by phase shifts $\delta(E)$ with the on-shell $S$-matrix 
in each channel defined by
\begin{eqnarray}
S(E) & = & e^{2i\delta(E)}\, , 
\label{s_mat_and_del}
\end{eqnarray}
or 
\begin{eqnarray}
S(E) & = & 1 + 2i~ e^{i\delta(E)}~ \sin (\delta(E))\, ,
\label{s_mat_and_del1}
\end{eqnarray}
where $E(p)=\sqrt{M_N^2+p^2}+\sqrt{m_{\pi}^2+p^2}$ is the invariant 
total energy of the two interacting praticles. The partial wave on-shell 
$S$-matrix is related to the partial wave on-shell $\mathcal T$-matrix by
\begin{eqnarray}
S(E) & = & 1 - 2i\pi~\rho (p) ~\mathcal T(E)\, , 
\label{s_and_t_mat}
\end{eqnarray}
where the density of states $\rho (p)$ is given by
\begin{eqnarray}
\rho(p)  & = & \frac{p^2}{\frac{dE(p)}{dp}} ~=~ 
        \frac{p E_N(p) E_{\pi}(p)}{E_N(p) + E_{\pi}(p)}\, . 
\label{rho}
\end{eqnarray}
Combining Eqs.\ (\ref{s_mat_and_del}) and (\ref{s_and_t_mat}) gives
\begin{eqnarray}
\tan (2\delta(E)) & = & \frac{-2\pi~\rho (p)~\Re e (\mathcal
        T(E))} {1 + 2\pi~\rho(p) ~\Im m (\mathcal T(E))}\, . 
\label{del_and_t_mat}
\end{eqnarray}

\vspace*{-0.2cm}~\\
Alternatively, we can express the on-shell $\mathcal T$-matrix in terms of the 
phase shifts as
\begin{eqnarray}
\mathcal T(E) & = & \frac{-1}{\pi\rho(p)} ~e^{i\delta(E)} 
        ~\sin (\delta(E))\, , 
\label{t_and_del}
\end{eqnarray}
whence
\begin{eqnarray}
\tan (\delta(E)) & = & \frac{\Im m (\mathcal T(E))}
        {\Re e (\mathcal T(E))}\, . 
\label{del_and_t_matf}
\end{eqnarray}

\vspace*{-0.2cm}~\\
The phase shifts calculated from the dynamical model of Nozawa {\it et al.}
\cite{Nozawa90} for the more important $\pi N$ partial waves are shown 
in Figs.\ \ref{pinphases}-\ref{pinphased} with the corresponding parameters
being given in Table \ref{tableG:7}. We see that the Lippmann-Schwinger equation gives a
good description of $\pi N$ phase shifts below 500 MeV. We note in particular
the perfect resonance shape of the $P_{33}$ phase shift (see Fig.\ \ref{pinphasep}) corresponding to the 
$\Delta$(1232) resonance of the nucleon ($\delta$=90$^0$ at 
$T_{lab}\simeq$ 180 MeV). The steep rise of the $P_{11}$ phase 
shift for $T_{lab}\rightarrow$ 500 MeV is an indication of the Roper resonance
N(1440). In addition there is some background scattering due to interactions
in the  $S_{11}$ and $S_{31}$ partial waves.

\chapter{Parameters of the $NN$ Separable Potential\label{appendixG}} 
%
We give in this appendix the parameters which we use for the separable $NN$ 
interaction of the Graz group \cite{Haidenbauer84,Haidenbauer85} in each
partial wave.
\begin{table}[ht]
\begin{center}
\caption{\small Parameters of the PEST3 potential in the $^1S_0$ partial wave.}
\vspace*{0.3cm}
\begin{tabular}{llll}
\hline\hline\\
        & $\beta$~~{\rm (fm$^{-1}$)} & $C$~~{\rm (fm$^{0}$)}  & $\lambda$~~{\rm (MeV~fm$^{-1}$)} \\
[2.1ex]
\hline\\
(p-p)   & $\beta_{11}$=1.1115753    &  $C_{11}$=-572.29230  &  $\lambda_{11}$=-0.001021062  \\ 
        & $\beta_{12}$=2.0212320    &  $C_{12}$=-313.62360  &  $\lambda_{12}$= 0.022965747  \\ 
        & $\beta_{13}$=2.6434278    &  $C_{13}$= 12814.143  &  $\lambda_{13}$=-0.013907568  \\ 
        & $\beta_{14}$=4.0877911    &  $C_{14}$=-28477.672  &                                 \\ 
[1.1ex]
        & $\beta_{21}$=1.0297944    &  $C_{21}$=-18.494719  &  $\lambda_{21}$= 0.022965747 \\ 
        & $\beta_{22}$=1.5435994    &  $C_{22}$=-53.628348  &  $\lambda_{22}$=-0.805816820  \\ 
        & $\beta_{23}$=2.6129194    &  $C_{23}$= 925.42408  &  $\lambda_{23}$= 0.825835300  \\ 
        & $\beta_{24}$=4.0837929    &  $C_{24}$=-2040.0020  &                                 \\ 
[1.1ex]
        & $\beta_{31}$=0.8995186    &  $C_{31}$=-2.8837445  &  $\lambda_{31}$=-0.013907568 \\ 
        & $\beta_{32}$=2.6843334    &  $C_{32}$=-84.344154  &  $\lambda_{32}$= 0.825835300  \\ 
        & $\beta_{33}$=3.1505869    &  $C_{33}$= 1163.1547  &  $\lambda_{33}$=-1.295713900  \\ 
        & $\beta_{34}$=3.9826288    &  $C_{34}$=-2018.0523  &                              \\ 
[4.1ex]
(n-p)   & $\beta_{11}$=1.1115753    &  $C_{11}$=-773.80000  &  $\lambda_{11}$=-0.000528400 \\ 
        & $\beta_{12}$=2.0212320    &  $C_{12}$=-424.05244  &  $\lambda_{12}$= 0.015471054  \\ 
        & $\beta_{13}$=2.6434278    &  $C_{13}$= 17326.082  &  $\lambda_{13}$=-0.008800256 \\ 
        & $\beta_{14}$=4.0877911    &  $C_{14}$=-38504.839  &   \\ 
[1.1ex]
        & $\beta_{21}$=1.0297944    &  $C_{21}$=-18.494719  &  $\lambda_{21}$= 0.015471054  \\ 
        & $\beta_{22}$=1.5435994    &  $C_{22}$=-53.628348  &  $\lambda_{22}$=-0.742248760  \\ 
        & $\beta_{23}$=2.6129194    &  $C_{23}$= 925.42408  &  $\lambda_{23}$= 0.770689240\\ 
        & $\beta_{24}$=4.0837929    &  $C_{24}$=-2040.0020  &    \\ 
[1.1ex]
        & $\beta_{31}$=0.8995186    &  $C_{31}$=-2.8837445  &  $\lambda_{31}$=-0.008800256 \\ 
        & $\beta_{32}$=2.6843334    &  $C_{32}$=-84.344154  &  $\lambda_{32}$= 0.770689240  \\ 
        & $\beta_{33}$=3.1505869    &  $C_{33}$= 1163.1547  &  $\lambda_{33}$=-1.252847400 \\ 
        & $\beta_{34}$=3.9826288    &  $C_{34}$=-2018.0523  &   \\ 
[2.1ex]
\hline\hline
\end{tabular}
\label{tableG:1}
\end{center}
\end{table}
\begin{table}[ht]
\begin{center}
\caption{\small Parameters of the PEST2 potential in the uncoupled $P$ waves.}
\vspace*{0.3cm}
\begin{tabular}{llll}
\hline\hline\\
        & $\beta$~~{\rm (fm$^{-1}$)} & $C$~~{\rm (fm$^{0}$)}  & $\lambda$~~{\rm (MeV~fm$^{-3}$)} \\
[2.1ex]
\hline\\
$^3P_0$ & $\beta_{11}$= 1.8691379    &  $C_{11}$=-189.74407   &  $\lambda_{11}$=-0.14665888  \\ 
        & $\beta_{12}$= 3.0499866    &  $C_{12}$= 2096.7281   &  $\lambda_{12}$= 0.40513783   \\ 
        & $\beta_{13}$= 3.5585752    &  $C_{13}$=-3066.2694   &   \\ 
        & $\beta_{14}$= 1.1596292    &  $C_{14}$= 102.75384   &    \\ 
[1.1ex]
        & $\beta_{21}$= 2.5512467    &  $C_{21}$= 59.861550   &  $\lambda_{21}$= 0.40513783  \\ 
        & $\beta_{22}$= 2.6651826    &  $C_{22}$=-681.32014   &  $\lambda_{22}$=-5.47893200   \\ 
        & $\beta_{23}$= 2.9537139    &  $C_{23}$= 5265.5015   &    \\ 
        & $\beta_{24}$= 3.3827130    &  $C_{24}$=-6825.7570   &    \\ 
[4.1ex]
$^3P_1$ & $\beta_{11}$= 1.6225099    &  $C_{11}$= 76.181266  &  $\lambda_{11}$= 0.66973064  \\ 
        & $\beta_{12}$= 2.4712881    &  $C_{12}$=-381.78599  &  $\lambda_{12}$=-0.52203105  \\ 
        & $\beta_{13}$= 2.7120059    &  $C_{13}$= 1471.7803  &    \\ 
        & $\beta_{14}$= 3.1642803    &  $C_{14}$=-1646.4168  &    \\ 
[1.1ex]
        & $\beta_{21}$= 1.7697160    &  $C_{21}$= 61.335643  &  $\lambda_{21}$=-0.52203105  \\ 
        & $\beta_{22}$= 2.4049729    &  $C_{22}$= 1348.5020  &  $\lambda_{22}$= 0.64075503   \\ 
        & $\beta_{23}$= 1.8631357    &  $C_{23}$=-842.10204  &    \\ 
        & $\beta_{24}$= 4.1084782    &  $C_{24}$=-1604.2632  &    \\ 
[4.1ex]
$^1P_1$ & $\beta_{11}$= 1.5652079    &  $C_{11}$= 89.398777  &  $\lambda_{11}$= 0.44514549 \\ 
        & $\beta_{12}$= 2.0058818    &  $C_{12}$=-208.13553  &  $\lambda_{12}$=-0.34067880   \\ 
        & $\beta_{13}$= 2.4100051    &  $C_{13}$= 425.34413  &  \\ 
        & $\beta_{14}$= 4.2981726    &  $C_{14}$=-630.35239  &    \\ 
[1.1ex]
        & $\beta_{21}$= 1.5463024    &  $C_{21}$= 43.477263  &  $\lambda_{21}$=-0.34067880  \\ 
        & $\beta_{22}$= 1.6196926    &  $C_{22}$= 166.46466  &  $\lambda_{22}$= 0.44501514   \\ 
        & $\beta_{23}$= 2.0004143    &  $C_{23}$=-404.03968  &    \\ 
        & $\beta_{24}$= 2.1796197    &  $C_{24}$= 528.13575  &   \\ 
[2.1ex]
\hline\hline
\end{tabular}
\label{tableG:2}
\end{center}
\end{table}
\begin{table}[ht]
\begin{center}
\caption{\small Parameters of the PEST2 potential in the uncoupled $D$ waves.}
\vspace*{0.3cm}
\begin{tabular}{llll}
\hline\hline\\
        & $\beta$~~{\rm (fm$^{-1}$)} & $C$~~{\rm (fm$^{0}$)}  & $\lambda$~~{\rm (MeV~fm$^{-1}$)} \\
[2.1ex]
\hline\\
$^1D_2$ & $\beta_{11}$= 1.1690501    &  $C_{11}$=-7.6541157  &  $\lambda_{11}$=-1.90862530  \\ 
        & $\beta_{12}$= 1.3354235    &  $C_{12}$= 3.8397789  &  $\lambda_{12}$= 1.00851240   \\ 
        & $\beta_{13}$= 3.6311131    &  $C_{13}$=-348.08899  &  \\ 
        & $\beta_{14}$= 2.9623562    &  $C_{14}$= 330.93549  &   \\ 
[1.1ex]
        & $\beta_{21}$= 0.8715792    &  $C_{21}$=-1.0867868  &  $\lambda_{21}$= 1.00851240  \\ 
        & $\beta_{22}$= 1.2853211    &  $C_{22}$=-22.731322  &  $\lambda_{22}$=-0.97543827   \\ 
        & $\beta_{23}$= 3.2457085    &  $C_{23}$= 265.30215  &    \\ 
        & $\beta_{24}$= 3.8839614    &  $C_{24}$=-303.60967  &   \\ 
[4.1ex]
$^3D_2$ & $\beta_{11}$= 1.2055706    &  $C_{11}$=-40.322168  &  $\lambda_{11}$=-0.26674102  \\ 
        & $\beta_{12}$= 1.2960008    &  $C_{12}$=-21.021097  &  $\lambda_{12}$= 0.14305117  \\ 
        & $\beta_{13}$= 1.6239653    &  $C_{13}$= 533.48026  &  \\ 
        & $\beta_{14}$= 1.4452618    &  $C_{14}$=-436.55162  &   \\ 
[1.1ex]
        & $\beta_{21}$= 1.5953203    &  $C_{21}$=-245.06021  &  $\lambda_{21}$= 0.14305117  \\ 
        & $\beta_{22}$= 0.3626991    &  $C_{22}$= 3.0950339  &  $\lambda_{22}$=-0.15817248  \\ 
        & $\beta_{23}$= 1.7370869    &  $C_{23}$= 299.77237  &  \\ 
        & $\beta_{24}$= 0.5873583    &  $C_{24}$= 42.826073  & \\ 
[2.1ex]
\hline\hline
\end{tabular}
\label{tableG:3}
\end{center}
\end{table}
\begin{table}[ht]
\begin{center}
\caption{\small Parameters of the PEST4 potential in the coupled
        $^3S_1$-$^3\hspace*{-0.05cm}D_1$  partial wave.}
\vspace*{0.3cm}
\begin{tabular}{llll}
\hline\hline\\
        & $\beta$~~{\rm (fm$^{-1}$)} & $C$~~{\rm (fm$^{0}$)}  & $\lambda$~~{\rm (MeV~fm$^{-1}$)} \\
[2.1ex]
\hline\\
$L=0$   & $\beta_{11}$= 1.6855291    &  $C_{11}$=-51.009539   &  $\lambda_{11}$=-0.254222420  \\ 
        & $\beta_{12}$= 3.9205339    &  $C_{12}$= 765.24390   &  $\lambda_{12}$=-0.171787720   \\ 
        & $\beta_{13}$= 5.7636840    &  $C_{13}$=-3363.4996   &  $\lambda_{13}$= 0.071723814 \\ 
        & $\beta_{14}$= 6.0419695    &  $C_{14}$= 2783.6926   &  $\lambda_{14}$= 0.005105011  \\ 
[1.1ex]
        & $\beta_{21}$= 1.7877900    &  $C_{21}$=-68.041320   &  $\lambda_{21}$=-0.171787720  \\ 
        & $\beta_{22}$= 2.1759210    &  $C_{22}$= 488.86119   &  $\lambda_{22}$= 0.087724390   \\ 
        & $\beta_{23}$= 2.4705717    &  $C_{23}$=-1199.1640   &  $\lambda_{23}$= 0.094518781 \\ 
        & $\beta_{24}$= 2.7303931    &  $C_{24}$= 871.79604   &  $\lambda_{24}$=-0.079244845 \\ 
[1.1ex]
        & $\beta_{31}$= 1.5971102    &  $C_{31}$=-89.517302   &  $\lambda_{31}$= 0.071723814  \\ 
        & $\beta_{32}$= 9.9678931    &  $C_{32}$= 2582.2164   &  $\lambda_{32}$= 0.094518781   \\ 
        & $\beta_{33}$= 4.5948011    &  $C_{33}$=-3042.3325   &  $\lambda_{33}$=-0.030645992  \\ 
        & $\beta_{34}$= 2.1206347    &  $C_{34}$= 1235.3473   &  $\lambda_{34}$=-0.034997609  \\ 
[1.1ex] 
        & $\beta_{41}$= 4.0268195    &  $C_{41}$=-382.75856   &  $\lambda_{41}$= 0.005105011 \\ 
        & $\beta_{42}$= 5.0466356    &  $C_{42}$= 1005.8091   &  $\lambda_{42}$=-0.079244845  \\ 
        & $\beta_{43}$= 2.5795951    &  $C_{43}$= 2298.7534   &  $\lambda_{43}$=-0.034997609  \\ 
        & $\beta_{44}$= 2.3953665    &  $C_{44}$=-2718.2533   &  $\lambda_{44}$= 0.169553430  \\ 
[4.1ex]
$L=2$   & $\beta_{11}$= 2.6228398    &  $C_{11}$=-410.15751   &  $\lambda_{11}$=-0.254222420  \\ 
        & $\beta_{12}$= 1.8815276    &  $C_{12}$= 156.64127   &  $\lambda_{12}$=-0.171787720   \\ 
        & $\beta_{13}$= 3.8346780    &  $C_{13}$= 888.03985   &  $\lambda_{13}$= 0.071723814 \\ 
        & $\beta_{14}$= 4.9959386    &  $C_{14}$=-792.11994   &  $\lambda_{14}$= 0.005105011 \\ 
[1.1ex]
        & $\beta_{21}$= 1.9098545    &  $C_{21}$= 87.065827   &  $\lambda_{21}$=-0.171787720  \\ 
        & $\beta_{22}$= 1.1170776    &  $C_{22}$= 20.554605   &  $\lambda_{22}$= 0.087724390   \\ 
        & $\beta_{23}$= 1.4259853    &  $C_{23}$=-65.702538   &  $\lambda_{23}$= 0.094518781  \\ 
        & $\beta_{24}$= 3.0734784    &  $C_{24}$=-91.839262   &  $\lambda_{24}$=-0.079244845  \\ 
[1.1ex]
        & $\beta_{31}$= 2.6431889    &  $C_{31}$= 143.62150   &  $\lambda_{31}$= 0.071723814  \\ 
        & $\beta_{32}$= 2.7483825    &  $C_{32}$= 6.4744841   &  $\lambda_{32}$= 0.094518781  \\ 
        & $\beta_{33}$= 1.9283993    &  $C_{33}$=-1384.0981   &  $\lambda_{33}$=-0.030645992  \\ 
        & $\beta_{34}$= 2.2891433    &  $C_{34}$= 1550.1289   &  $\lambda_{34}$=-0.034997609  \\ 
[1.1ex] 
        & $\beta_{41}$= 4.2598369    &  $C_{41}$= 366.50118   &  $\lambda_{41}$= 0.005105011  \\ 
        & $\beta_{42}$= 2.1463834    &  $C_{42}$= 263.87202   &  $\lambda_{42}$=-0.079244845   \\ 
        & $\beta_{43}$= 2.4905282    &  $C_{43}$=-611.66811   &  $\lambda_{43}$=-0.034997609  \\ 
        & $\beta_{44}$= 2.2930841    &  $C_{44}$=-12.606207   &  $\lambda_{44}$= 0.169553430  \\ 
[2.1ex]
\hline\hline
\end{tabular}
\label{tableG:4}
\end{center}
\end{table}
\begin{table}[ht]
\begin{center}
\caption{\small Parameters of the PEST3 potential in the coupled
        $^3P_2$-$^3\hspace*{-0.05cm}F_2$ partial wave.}
\vspace*{0.3cm}
\begin{tabular}{llll}
\hline\hline\\
        & $\beta$~~{\rm (fm$^{-1}$)} & $C$~~{\rm (fm$^{0}$)}  & $\lambda$~~{\rm (MeV~fm$^{-3}$)} \\
[2.1ex]
\hline\\
$L=1$   & $\beta_{11}$= 1.4452936    &  $C_{11}$=-28.609391   &  $\lambda_{11}$= 0.04672503  \\ 
        & $\beta_{12}$= 2.0173835    &  $C_{12}$=-186.15068   &  $\lambda_{12}$= 0.41710484   \\ 
        & $\beta_{13}$= 5.4463467    &  $C_{13}$=-1094.0574   &  $\lambda_{13}$=-0.06092248 \\ 
        & $\beta_{14}$= 2.8490316    &  $C_{14}$= 1022.0712   &    \\ 
[1.1ex]
        & $\beta_{21}$= 1.8993563    &  $C_{21}$= 65.794074   &  $\lambda_{21}$= 0.41710484  \\ 
        & $\beta_{22}$= 2.4547294    &  $C_{22}$= 173.01447   &  $\lambda_{22}$= 0.25492496   \\ 
        & $\beta_{23}$= 1.5499793    &  $C_{23}$=-190.40569   &  $\lambda_{23}$=-0.41481703  \\ 
        & $\beta_{24}$= 5.9773387    &  $C_{24}$=-217.74704   &    \\ 
[1.1ex]
        & $\beta_{31}$= 2.9659955    &  $C_{31}$=-327.40790   &  $\lambda_{31}$=-0.06092248   \\ 
        & $\beta_{32}$= 4.9324835    &  $C_{32}$=-2950.6615   &  $\lambda_{32}$=-0.41481703   \\ 
        & $\beta_{33}$= 2.9051889    &  $C_{33}$= 4127.4473   &  $\lambda_{33}$=-0.13593596 \\ 
        & $\beta_{34}$= 2.0463803    &  $C_{34}$=-1298.0249   &   \\ 
[4.1ex]
$L=3$   & $\beta_{11}$= 1.1667491    &  $C_{11}$=0.96744407   &  $\lambda_{11}$= 0.04672503  \\ 
        & $\beta_{12}$= 1.5406619    &  $C_{12}$= 166.30912   &  $\lambda_{12}$= 0.41710484   \\ 
        & $\beta_{13}$= 2.3802869    &  $C_{13}$=-119.42176   &  $\lambda_{13}$=-0.06092248 \\ 
        & $\beta_{14}$= 4.8335309    &  $C_{14}$= 366.96981   &   \\ 
[1.1ex]
        & $\beta_{21}$= 1.0955323    &  $C_{21}$=-0.6386842   &  $\lambda_{21}$= 0.41710484 \\ 
        & $\beta_{22}$= 1.6020596    &  $C_{22}$=-37.538146   &  $\lambda_{22}$= 0.25492496   \\ 
        & $\beta_{23}$= 4.4585234    &  $C_{23}$=-316.97599   &  $\lambda_{23}$=-0.41481703  \\ 
        & $\beta_{24}$= 1.8254989    &  $C_{24}$= 179.59323   &    \\ 
[1.1ex]
        & $\beta_{31}$= 2.1084981    &  $C_{31}$=-47.364933   &  $\lambda_{31}$=-0.06092248  \\ 
        & $\beta_{32}$= 3.1977404    &  $C_{32}$= 2035.5781   &  $\lambda_{32}$=-0.41481703  \\ 
        & $\beta_{33}$= 3.9573349    &  $C_{33}$=-5925.4169   &  $\lambda_{33}$=-0.13593596  \\ 
        & $\beta_{34}$= 4.5120514    &  $C_{34}$= 5407.1036   &    \\ 
[2.1ex]
\hline\hline
\end{tabular}
\label{tableG:5}
\end{center}
\end{table}
\begin{table}[ht]
\begin{center}
\caption{\small Parameters of the PEST4 potential in the coupled $^3D_3$-$^3G_3$
        partial wave.}
\vspace*{0.3cm}
\begin{tabular}{llll}
\hline\hline\\
        & $\beta$~~{\rm (fm$^{-1}$)} & $C$~~{\rm (fm$^{0}$)}  & $\lambda$~~{\rm (MeV~fm$^{-1}$)} \\
[2.1ex]
\hline\\
$L=2$   & $\beta_{11}$= 1.7275071    &  $C_{11}$=-8.6379311   &  $\lambda_{11}$=-0.09899224  \\ 
        & $\beta_{12}$= 2.5493316    &  $C_{12}$= 178.84657   &  $\lambda_{12}$=-0.39484562  \\ 
        & $\beta_{13}$= 1.9111954    &  $C_{13}$=-114.21507   &  $\lambda_{13}$= 0.04013528 \\ 
        & $\beta_{14}$= 4.0669172    &  $C_{14}$=-86.040765   &  $\lambda_{14}$= 0.11343039  \\ 
[1.1ex]
        & $\beta_{21}$= 1.9189644    &  $C_{21}$=-116.07858   &  $\lambda_{21}$=-0.39484562  \\ 
        & $\beta_{22}$= 2.7381690    &  $C_{22}$= 354.48169   &  $\lambda_{22}$= 0.28242764  \\ 
        & $\beta_{23}$= 1.5396020    &  $C_{23}$= 196.12620   &  $\lambda_{23}$= 0.40048398 \\ 
        & $\beta_{24}$= 1.9159334    &  $C_{24}$=-419.69957   &  $\lambda_{24}$=-0.34677241 \\ 
[1.1ex]
        & $\beta_{31}$= 3.2277180    &  $C_{31}$=-110.31148   &  $\lambda_{31}$= 0.04013528  \\ 
        & $\beta_{32}$= 4.8060158    &  $C_{32}$= 483.29454   &  $\lambda_{32}$= 0.40048398  \\ 
        & $\beta_{33}$= 3.1005090    &  $C_{33}$= 442.80069   &  $\lambda_{33}$=-0.14224121\\ 
        & $\beta_{34}$= 3.7214195    &  $C_{34}$=-903.63957   &  $\lambda_{34}$=-0.38000970 \\ 
[1.1ex] 
        & $\beta_{41}$= 2.8426813    &  $C_{41}$=-364.88710   &  $\lambda_{41}$= 0.11343039 \\ 
        & $\beta_{42}$= 3.3549103    &  $C_{42}$= 1848.1193   &  $\lambda_{42}$=-0.34677241   \\ 
        & $\beta_{43}$= 3.2121201    &  $C_{43}$= 2550.6547   &  $\lambda_{43}$=-0.38000970 \\ 
        & $\beta_{44}$= 2.7766588    &  $C_{44}$=-3985.0257   &  $\lambda_{44}$= 0.50560372  \\ 
[4.1ex]
$L=4$   & $\beta_{11}$= 2.4000943    &  $C_{11}$= 10.076418   &  $\lambda_{11}$=-0.09899224  \\ 
        & $\beta_{12}$= 1.2830437    &  $C_{12}$=-90.394797   &  $\lambda_{12}$=-0.39484562   \\ 
        & $\beta_{13}$= 1.6627122    &  $C_{13}$= 283.81707   &  $\lambda_{13}$= 0.04013528  \\ 
        & $\beta_{14}$= 1.5581999    &  $C_{14}$=-204.52799   &  $\lambda_{14}$= 0.11343039  \\ 
[1.1ex]
        & $\beta_{21}$= 3.9194456    &  $C_{21}$= 10.924992   &  $\lambda_{21}$=-0.39484562  \\ 
        & $\beta_{22}$= 1.5719585    &  $C_{22}$= 82.607174   &  $\lambda_{22}$= 0.28242764  \\ 
        & $\beta_{23}$= 1.6553410    &  $C_{23}$=-132.30266   &  $\lambda_{23}$= 0.40048398 \\ 
        & $\beta_{24}$= 2.0963683    &  $C_{24}$= 47.770331   &  $\lambda_{24}$=-0.34677241 \\ 
[1.1ex]
        & $\beta_{31}$= 4.8032746    &  $C_{31}$=-204.05105   &  $\lambda_{31}$= 0.04013528  \\ 
        & $\beta_{32}$= 3.6976328    &  $C_{32}$=-751.61370   &  $\lambda_{32}$= 0.40048398  \\ 
        & $\beta_{33}$= 2.7214754    &  $C_{33}$= 1886.8354   &  $\lambda_{33}$=-0.14224121 \\ 
        & $\beta_{34}$= 1.9739909    &  $C_{34}$=-954.91540   &  $\lambda_{34}$=-0.38000970 \\ 
[1.1ex] 
        & $\beta_{41}$= 1.8399157    &  $C_{41}$=-0.2429913   &  $\lambda_{41}$= 0.11343039 \\ 
        & $\beta_{42}$= 2.0748484    &  $C_{42}$= 197.58830   &  $\lambda_{42}$=-0.34677241  \\ 
        & $\beta_{43}$= 2.2831760    &  $C_{43}$=-411.75699   &  $\lambda_{43}$=-0.38000970 \\ 
        & $\beta_{44}$= 2.6095142    &  $C_{44}$= 229.35025   &  $\lambda_{44}$= 0.50560372 \\ 
[2.1ex]
\hline\hline
\end{tabular}
\label{tableG:6}
\end{center}
\end{table}

\chapter{Parameters of the $\pi N$ Separable Potential\label{appendixH}} 
%
\vspace*{-1cm}
\begin{table}[ht]
\begin{center}
\caption{\small Parameters of the $\pi N$ separabel potential of Nozawa {\it et al.}
  \cite{Nozawa90} for the $\pi N$ partial waves. These parameters are
  determined by fitting the phase shift data \cite{Hoehler79} up to 500 MeV
  pion laboratory kinetic energy.}
\vspace*{0.3cm}
\begin{tabular}{lllllllllll}
\hline\hline\vspace*{-0.2cm} \\
 $L_{2t,2J}$ & $\ell$ & $m_1$ & $n_1$ & $a_1^{\dag}$ & $b_1^{\S}$ & $m_2$  & $n_2$ & $a_2^{\ddag}$ & $b_2^{\S}$ & $\lambda_0$ \\
[1.1ex]
\hline\\
$S_{11}$ & 0 & 0 & 3 & 100.00 & 2.598 & 2 & 2 & 4.9520 & 2.877 & \hspace{0.12cm}--1  \\ 
$S_{31}$ & 0 & 0 & 2 & 3.0850 & 1.806 & 2 & 2 & 1.9250 & 1.275 & +1  \\ 
$P_{11}$ & 1 & 2 & 3 & 31.623 & 2.665 & 0 & 2 & 0.5793 & 1.185 & \hspace{0.12cm}--1  \\ 
$P_{13}$ & 1 & 0 & 2 & 0.4269 & 1.181 & 2 & 3 & 3.9700 & 1.721 & +1  \\ 
$P_{31}$ & 1 & 0 & 2 & 1.4730 & 1.542 & 2 & 3 & 8.0530 & 1.861 & +1  \\ 
$P_{33}$ & 1 & 0 & 2 & 2.7700 & 1.415 & 0 & 2 & 1.7780 & 1.218 & +1  \\ 
$D_{13}$ & 2 & 0 & 2 & 1.6390 & 2.165$^{\P}$ & 2 & 3 & 9.3120 & 3.263 & \hspace{0.12cm}--1\\ 
$D_{15}$ & 2 & 0 & 2 & 0.2172 & 1.175 & 2 & 3 & 1.0110 & 1.461 & \hspace{0.12cm}--1  \\ 
$D_{33}$ & 2 & 0 & 2 & 0.1306 & 1.128 & 2 & 3 & 1.0810 & 1.972 & \hspace{0.12cm}--1  \\ 
$D_{35}$ & 2 & 0 & 2 & 0.2270 & 1.168 & 2 & 3 & 1.1510 & 1.780 & +1  \\ 
[1.1ex]
\hline\hline
\end{tabular}
\label{tableG:7}
\end{center}
\begin{itemize}
\item[{}]$^{\dag}$  $a_1$ is given in units of (fm)$^{-2n_1+m_1+\ell
  +1}$.
\vspace*{-0.2cm}
\item[{}]$^{\ddag}$ $a_2$ is given in units of (fm)$^{-2n_2+m_2+\ell
  +\frac{1}{2}}$ for $P_{11}$ and $P_{33}$ partial waves, otherwise it is
given in units of (fm)$^{-2n_2+m_2+\ell +1}$.
\vspace*{-0.2cm}
\item[{}]$^{\S}$  $b_1$ and $b_2$ are given in units of (fm)$^{-1}$.
\vspace*{-0.2cm}
\item[{}]$^{\P}$ This value is a missprint in Table 2 of Ref.\
           \cite{Nozawa90}. We would like to thank Professor S.\ Nozawa for
           giving us the correct one.
\end{itemize}
\end{table}

\end{appendix}
\addcontentsline{toc}{chapter}{Bibliography}
\pagestyle{fancy}

\end{document}


\chapter{Polarization Observables\label{chap:3:5}}
%
Polarization observables have the promise of opening a new filed in the
electromagnetic production of pions from protons and nuclear targets. From
this point of view we give in this chapter the formal expressions of all
possible polarization observables in the $\gamma d \rightarrow \pi NN$
reaction with polarized photons and oriented deuterons in terms of the
$\mathcal M$-matrix elements. The importance of this process derives from the fact
that the deuteron, being the simplest nuclear system, plays a similar
fundamental role in nuclear physics as the hydrogen atom plays in atomic
physics.

\vspace*{-0.2cm}~\\
The initial state for photoproduction of $\pi$-mesons on the deuteron is
characterized by the incoming photon polarization $m_{\gamma}$ and the
deuteron spin projection $m_{d}$ with respect to a chosen quantization axis,
in our case for convenience the incoming photon momentum $\vec {k}$. Thus we
associate a frame of reference with $z$-axis in the direction of $\vec {k}$.

\vspace*{-0.2cm}~\\
The final state is described by the two-nucleon relative momentum $\vec{p}_r$
having the spherical coordinates $\theta_r$ and $\phi_r$ with respect to the
chosen frame of reference, by the total spin $s$ and its spin projection $m_s$
on $\vec{p}_r$ and by the pion momentum $\vec{q}$ having the spherical
coordinates $\theta_{\pi}$ and $\phi_{\pi}$. The $x$-$z$-plane is determined by
the outgoing pion, i.e. $\phi_{\pi}=0$. This means that the $\mathcal
M$-matrix is spherical harmonics depending on the pion momentum and its
direction.

\vspace*{-0.2cm}~\\
Choosing the $z$-axis in the direction of the incoming photon and isolating
the azimuthal dependence of the direction of pion momentum, we obtain the
following general form for the reaction matrix  
\begin{eqnarray}\label{eq:3.31}
\mathcal M^{(t\mu)}_{sm_s,m_{\gamma}m_d}(\theta_{\pi},\phi_{\pi}) & = &
\mathcal
O^{(t\mu)}_{sm_s,m_{\gamma}m_d}(\theta_{\pi})
~e^{i(m_{\gamma}+m_d)\phi_{\pi}}\, . 
\end{eqnarray}
Using parity conservation one can show that
the $\mathcal M$-matrix elements obey the following symmetry relation
\begin{eqnarray}\label{sym}
\mathcal O^{(t\mu)}_{s,-m_s,-m_{\gamma},-m_d} & = &
(-)^{s+m_s+m_{\gamma}+m_d} ~\mathcal O^{(t\mu)}_{s,m_s,m_{\gamma},m_d}\, .
\end{eqnarray}
This symmetry relation reduces the number of complex amplitudes from 24 to 12
independent ones. For their determination one needs 23 real observables since
a overall phase remains arbitrary.

\vspace*{-0.2cm}~\\
For a given $\mathcal M$-matrix one can compute the cross section for arbitrary polarized
photons and initial deuterons by applying the density matrix formalism similar
to that given by Arenh\"ovel \cite{Arenhoevel88} for deuteron
photodisintegration. The most general expression for all observables is given
by 
\begin{eqnarray}
\label{observables}
\mathcal O = \rho_s \sum_{\stackrel{sm_st,m_{\gamma}m_d}{s^{\prime}
 m_s^{\prime}t^{\prime},m_{\gamma}^{\prime}m_d^{\prime}}}
(\mathcal M^{(t^{\prime}\mu^{\prime})}_{s^{\prime}m_s^{\prime},m_{\gamma}^{\prime}m_d^{\prime}})^{\star}~~\vec{\Omega}_{s^{\prime}m_s^{\prime}sm_s}
~~\mathcal M^{(t\mu)}_{sm_s,m_{\gamma}m_d}
~\rho^{\gamma}_{m_{\gamma}m_{\gamma}^{\prime}}~\rho^{d}_{m_dm_{d}^{\prime}}~ d\Omega_{\vec{p}_r}~, 
\end{eqnarray}
where $\rho^{\gamma}_{m_{\gamma}m_{\gamma}^{\prime}}$ and
$\rho^{d}_{m_dm_{d}^{\prime}}$ denote the density matrices of initial photon
polarization and deuteron orientation, respectively, and $\vec{\Omega}$ is an
operator associated with the observable which acts in the two-nucleon spin
space. The kinematical phase space factor $\rho_s$ is given in
\cite{Eed02}. For more details about the density matrices we refer to
\cite{Arenhoevel88}.

\vspace*{-0.2cm}~\\
For a given $\mathcal M$-matrix all polarization observables can be expressed
in terms of the following quantities
\begin{eqnarray}
V_{IM} & = &
\frac{1}{\sqrt{3}}~\sum_{m_d^{\prime}m_d}~~\sum_{sm_st,m_{\gamma}}(-)^{1-m_d^{\prime}}  
~\hat{I} ~ \left( \begin{array}{ccc}  1 & 1 & I \\
m_d & -m_d^{\prime} & -M \end{array} \right) \nonumber \\ 
 & & \times \int \rho_s ~ (\mathcal M^{(t\mu)}_{sm_s,m_{\gamma}m_d})^{\star}~~\mathcal M^{(t\mu)}_{sm_s,m_{\gamma}m_d^{\prime}}~d\Omega_{\vec{p}_r}~,
\end{eqnarray}
and
\begin{eqnarray}
W_{IM} & = &
\frac{1}{\sqrt{3}}~\sum_{m_d^{\prime}m_d}~~\sum_{sm_st,m_{\gamma}}(-)^{1-m_d^{\prime}} 
~\hat{I}~ \left( \begin{array}{ccc}  1 & 1 & I \\
m_d & -m_d^{\prime} & -M \end{array} \right) \nonumber \\ 
 & & \times \int \rho_s ~ (\mathcal M^{(t\mu)}_{sm_s,m_{\gamma}m_d})^{\star}~~\mathcal M^{(t\mu)}_{s-m_s,m_{\gamma}-m_d^{\prime}}~d\Omega_{\vec{p}_r}~,
\end{eqnarray}
where $\hat{I} = \sqrt{2I+1}$.

\vspace*{-0.2cm}~\\
The unpolarized differential cross section is then given by
\begin{eqnarray}
\frac{d\sigma}{d\Omega_{\pi}dq} & = & V_{00}~.
\end{eqnarray}
The photon asymmetry for linearly polarized photons is given by
\begin{eqnarray}
{\sum}^{\ell}(\theta_{\pi}) & = & - \left(
  \frac{d\sigma}{d\Omega_{\pi}dq}\right)^{-1}
W_{00}~.
\end{eqnarray}
The target asymmetries are given by
\begin{itemize}
\item The vector target asymmetry:
\begin{eqnarray}
T_{11}(\theta_{\pi}) & = & \left(
  \frac{d\sigma}{d\Omega_{\pi}dq}\right)^{-1} 2~\Im m V_{11}~,
\end{eqnarray}
\item The tensor target asymmetries:
\begin{eqnarray}
T_{2M}(\theta_{\pi}) & = & \left(
  \frac{d\sigma}{d\Omega_{\pi}dq}\right)^{-1} (2-\delta_{M0})~\Re e V_{2M}~,~~
  (M=0,1,2)~.
\end{eqnarray}
\end{itemize}
The photon and target asymmetries are given by
\begin{itemize}
\item Circular:
\begin{eqnarray}
T_{1M}^c(\theta_{\pi}) & = & \left(
  \frac{d\sigma}{d\Omega_{\pi}dq}\right)^{-1} (2-\delta_{M0})~\Re e
  V_{1M}~,~~(M=0,1)~ \nonumber \\
T_{2M}^c(\theta_{\pi}) & = & \left(
  \frac{d\sigma}{d\Omega_{\pi}dq}\right)^{-1} 2~\Im m V_{2M}~,~~(M=0,1,2)~,
\end{eqnarray}
\item Longitudinal:
\begin{eqnarray}
T_{1M}^{\ell}(\theta_{\pi}) & = & \left(
  \frac{d\sigma}{d\Omega_{\pi}dq}\right)^{-1} i ~W_{1M}~,~~(M=0,\pm 1)~,\nonumber \\
T_{2M}^{\ell}(\theta_{\pi}) & = & - \left(
  \frac{d\sigma}{d\Omega_{\pi}dq}\right)^{-1} W_{2M}~,~~(M=0,\pm 1,\pm 2)~.
\end{eqnarray}
\end{itemize}

\addcontentsline{toc}{chapter}{Bibliography}
\pagestyle{fancy}